\documentclass[printer]{tADP2e}

\begin{document}
\doi{10.1080/0001873YYxxxxxxxx}
 \issn{1460-6976}
\issnp{0001-8732}  \jvol{00} \jnum{00} \jyear{2013} \jmonth{March}

\markboth{Taylor \& Francis and I.T. Consultant}{Advances in Physics}

\articletype{REVIEW ARTICLE}

\title{Persistence and First-Passage Properties in Non-equilibrium Systems}

\author{Alan J. Bray$^{\rm a}$$^{\ast}$\thanks{$^\ast$Corresponding author. Email:
\vspace{6pt}}, Satya N. Majumdar$^{\rm b}$ and Gr\'egory Schehr$^{\rm b}$\\\vspace{6pt}  $^{\rm a}${\em{School of Physics and Astronomy, The University of Manchester, Manchester M13 9PL, UK}}; $^{\rm b}${\em{Laboratoire de Physique Th\'eorique et
Mod\`eles Statistiques, UMR 8626, Universit\'e Paris Sud 11 and
CNRS, Bat. 100, Orsay F-91405, France}}\\\vspace{6pt}\received{March 2013}
}

\maketitle

\begin{abstract}
In this review we discuss the persistence and the related first-passage properties in extended 
many-body nonequilibrium systems. Starting with simple systems with one or few degrees of freedom,
such as random walk and random acceleration problems, we progressively discuss
the persistence properties in systems with many degrees of freedom. These systems include
spins models undergoing phase ordering dynamics, diffusion equation, fluctuating interfaces etc.
Persistence properties are nontrivial in these systems as the effective underlying
stochastic process is {\em non-Markovian}. Several exact and approximate methods 
have been developed to compute the persistence of such non-Markov processes over
the last two decades, as reviewed in this article. We also discuss various
generalisations of the local site persistence probability. Persistence in systems
with quenched disorder is discussed briefly. Although the main emphasis
of this review is on the theoretical developments on persistence, we briefly touch upon
various experimental systems as well.
\end{abstract}

\vspace*{1cm}

{\bf pacs:} 05.40.-a, 02.50.-r, 05.70.Ln 

\vspace*{1cm}

{\bf keywords:} persistence, first-passage, random-walks, phase-ordering, non-equilibrium dynamics, fluctuating interfaces 

\newpage

{\bf Table of contents}

\begin{itemize}

\item[1.] {\bf Introduction} 5 \\

\item[2.] {\bf Survival and first-passage for random walks} 9 \\
\hspace*{7pt} {2.1} Sparre Andersen theorem and persistence of a symmetric random walk 10 \\
\hspace*{7pt} {2.2} Generalized Sparre Andersen theorem and persistence of a random walk with a drift 12 \\
\hspace*{7pt} {2.3} Continuous-time random walk (CTRW) 13 \\
\item[3.] {\bf Persistence in single particle systems: Markov and non-Markov processes in continuous time} 14 \\
\hspace*{7pt} {3.1} The Brownian walker: the simplest Markov process 14 \\
\hspace*{14pt} {3.1.1} {\it Brownian walker in one dimension: the backward Fokker-Planck method} 15 \\
\hspace*{14pt} {3.1.2} {\it Brownian walker in higher dimensions} 16 \\
\hspace*{14pt} {3.1.3} {\it Brownian walker in a wedge} 16 \\
\hspace*{14pt} {3.1.4} {\it Brownian walker in a cone} 17 \\
\hspace*{14pt} {3.1.5} {\it Brownian walker in an expanding cage} 18 \\
\hspace*{14pt} {3.1.6} {\it Maximum excursion of a Brownian walk} 20 \\
\hspace*{14pt} {3.1.7} {\it Mean first-passage times of a Brownian walk} 21 \\
\hspace*{7pt} {3.2} The Random acceleration process: the simplest non-Markovian process 22 \\
\hspace*{14pt} {3.2.1} {\it Random acceleration process with partial survival} 23 \\
\hspace*{14pt} {3.2.3} {\it The `windy cliff'} 23 \\
\hspace*{7pt} {3.3} Higher-order processes 26 \\

\item[4.] {\bf Persistence in multi-particle systems} 27 \\
\hspace*{7pt}Ê{4.1} Three-walkers problems 27 \\
\hspace*{7pt}Ê{4.2} Persistence exponents for vicious walkers 28 \\
\hspace*{3pt} {4.3} The trapping reaction 32 \\
\hspace*{14pt} {4.3.1} {\it The target problem} 32 \\
\hspace*{14pt} {4.3.2} {\it The moving target} 34 \\
\hspace*{17pt}{4.3.3} {\it The ``Pascal Principle'' and an upper bound for $Q(t)$} 35 \\
\hspace*{17pt}{4.3.4} {\it A lower bound for $Q(t)$} 36 \\
\hspace*{17pt}{4.3.5} {\it The target problem with a deterministically moving target} 38 \\
\hspace*{17pt}{4.3.6} {\it The lamb and the $N$ lions} 40 \\

\item[5.] {\bf Persistence in coarsening phenomena} 41 \\
\hspace*{7pt}Ê{5.1} Ising and Potts models 42 \\
\hspace*{7pt}Ê{5.2} Spin models in higher dimensions 43 \\
\hspace*{7pt}Ê{5.3} The 1-d Ginzburg-Landau model 44 \\
\hspace*{7pt}Ê{5.4} Coarsening with a conserved order parameter 48 \\

\item[6.] {\bf Persistence of Gaussian sequences and Gaussian processes} 49 \\
\hspace*{7pt}Ê{6.1} Gaussian sequence 50 \\
\hspace*{7pt}Ê{6.2} Gaussian process 52 \\
\hspace*{7pt}Ê{6.3} Gaussian stationary process 53 \\

\item[7.] {\bf Perturbation theory for Non-Markovian Gaussian stationary processes} 55 \\

\item[8.] {\bf The independent interval approximation} 58 \\
\hspace*{7pt}{8.1} {Scaling phenomenon and Lamperti transformation}  61 \\
\hspace*{7pt}{8.2} {Application to the Brownian walker and higher order processes} 61 \\

\item[9.]{\bf Diffusive persistence} 64 \\
\hspace*{7pt}{9.1} {Application to coarsening dynamics} 66 \\
\hspace*{7pt}{9.2} {Connections with random polynomials} 67 \\

\item[10.]{\bf Persistence with partial survival} 69 \\

\item[11.] {\bf Global persistence} 73 \\
\hspace*{7pt}{11.1} Mean-field theory 74 \\
\hspace*{7pt}{11.2} The large-$n$ limit 74 \\
\hspace*{7pt}{11.3} The one-dimensional Ising model 75 \\
\hspace*{7pt}{11.4} $\theta_G$: A new critical exponent 75 \\
\hspace*{7pt}{11.5} The case of a finite initial magnetization for Model A dynamics 78 \\
\hspace*{7pt}{11.6} Global persistence for $T<T_c$ 79 \\
\hspace*{7pt}{11.7} Block persistence for $T < T_c$ 79 \\

\item[12.] {\bf The persistence of manifolds in nonequilibrium critical dynamics} 80 \\
\hspace*{7pt}{12.1} Mean-field theory 82 \\
\hspace*{14pt}{12.1.1} The case $D<2$ 83 \\
\hspace*{14pt}{12.1.2} The case $D>2$ 83 \\
\hspace*{7pt}{12.2} The large-$n$ limit 84\\
\hspace*{14pt}{12.2.1} The case $D<2$ 84 \\
\hspace*{14pt}{12.2.2} The case $D>2$ 85 \\
\hspace*{7pt}{10.3} General scaling theory 85 \\

\item[13.] {\bf Persistence of fractional Brownian motion and related processes} 86 \\

\item[14.] {\bf Persistence of fluctuating interfaces} 89 \\
\hspace*{7pt}{14.1} Linear interfaces: two-time correlation function 94 \\
\hspace*{7pt}{14.2} Linear interfaces: temporal persistence 97 \\
\hspace*{14pt}{14.2.1} {\it Transient regime: $t_0=0$} 98 \\
\hspace*{14pt}{14.2.2} {\it Steady-state regime: $t_0\to \infty$} 99 \\
\hspace*{7pt}{14.3} Linear interfaces: temporal survival probability 101 \\
\hspace*{7pt}{14.4} Nonlinear interfaces: temporal persistence 102 \\
\hspace*{7pt}{14.5} Spatial persistence and spatial survival probability: linear and nonlinear interfaces 104 \\
\hspace*{7pt}{14.6} Persistence properties in flat versus radial geometry 106 \\

\item[15.] {\bf Discrete persistence} 107 \\
\hspace*{7pt}{15.1} The correlator expansion 110 \\

\item[16.] {\bf Persistence in disordered systems} 111 \\
\hspace*{7pt}{16.1} Persistence in the one dimensional Sinai model 112 \\
\hspace*{7pt}{16.2} Persistence of a particle in the Matheron-de Marsily velocity field 116 \\
\hspace*{7pt}{16.3} Rouse chain in a Matheron-de Marsily layered medium: persistence of a tagged 
monomer 119 \\ 

\item[17.] {\bf Various generalisations of persistence} 122 \\
\hspace*{7pt}{17.1} Occupation time and persistent large deviations 122 \\
\hspace*{7pt}{17.2} Persistence of domains and other patterns 125 \\
\hspace*{7pt}{17.3} Spatial structures of persistent sites 126 \\
\hspace*{7pt}{17.4} Persistence in sequential versus parallel dynamics 127 \\

\item[18.] {\bf Persistence in reaction-diffusion models, Voter model, directed percolation} 128 \\
\hspace*{7pt}{18.1} Reaction-diffusion models 128 \\
\hspace*{7pt}{18.2} Voter model 129 \\
\hspace*{7pt}{18.3} Directed percolation 130 \\
\hspace*{7pt}{18.4} Turbulent fluid in $2$ dimensions 131 \\

\item[19.] {\bf Persistence of a stationary non-Markovian non-Gaussian sequence: An exactly solvable case} 131 \\

\item[20.] {\bf Summary and Conclusion} 135 \\

\item[] {\bf Acknowledgements} 138 \\

\item[] {\bf References} 139 \\

\end{itemize}


\newpage

\section{Introduction}

The study  of first-passage  problems by physicists  and mathematicians  
has a long history (see \cite{Redner} for a recent survey of the field). 
To introduce the subject, let us consider tossing a fair coin. The two, 
equally probable, outcomes of each toss are heads (H) and tails (T). 
Suppose we agree that the process will terminate when the first tail appears. 
Then the probability that the process has not yet terminated after $n$ 
tosses is $2^{-n}$ -- the probability to toss $n$ consecutive $H$'s.  
We might call $Q(n)=2^{-n}$ the "persistence probability" for this 
process, i.e.\ it is the probability that the sequence of heads ``persists'' 
for {\em at least} $n$ tosses. The probability that the process terminates 
after exactly $n+1$ tosses is $P_1(n) = Q(n)-Q(n+1) = 2^{-(n+1)}$. We might 
call this the "first-passage probability", i.e.\ it is the probability that the 
{\em first} tail appears at the $(n+1)$th toss. Note that $P_1(n)$ is 
normalised: $\sum_{n=0}^\infty 2^{-(n+1)} = 1$.

\begin{figure}[ht]
\includegraphics[width=\linewidth]{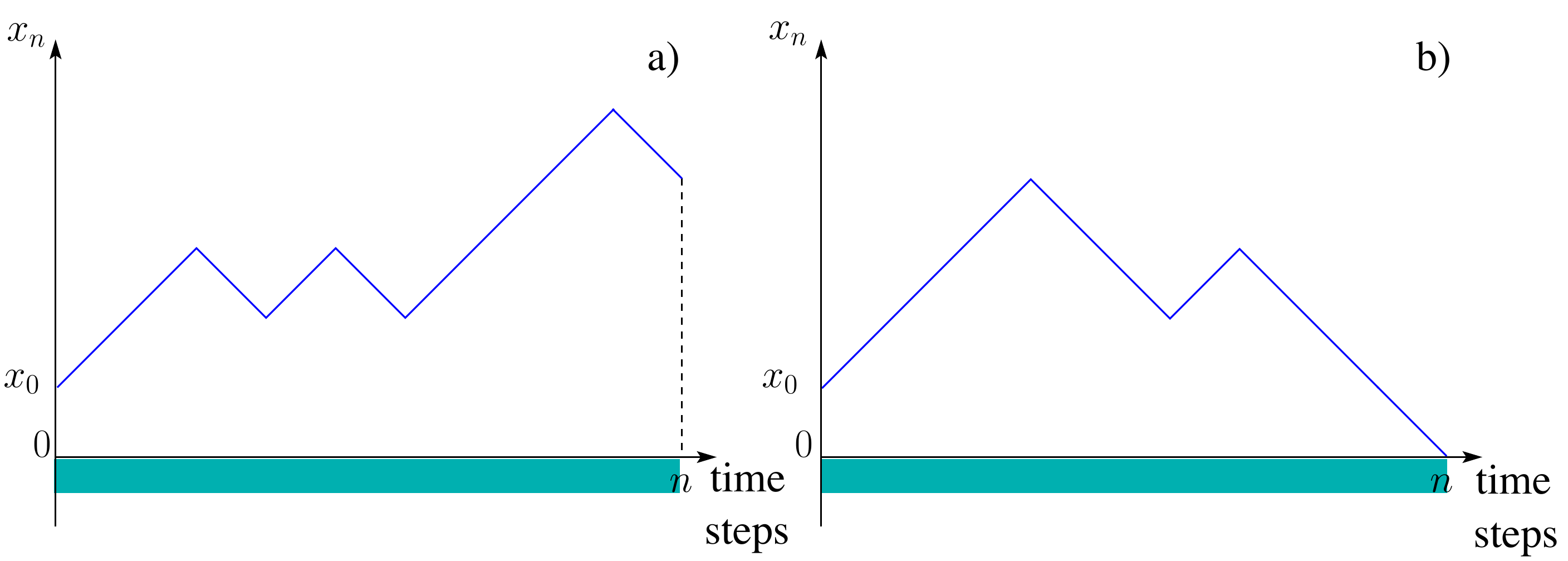}
\caption{{\bf a):} Trajectory of a lattice random walk with an absorbing boundary at the origin, starting from $x_0$, and contributing to the persistence probability $Q(x_0,n)$. {\bf b):} Trajectory of a lattice random walk with an absorbing boundary at the origin, starting from $x_0$, and contributing to $P_1(x_0,n-1)$.}\label{Fig_rw}
\end{figure}

Perhaps the next simplest first-passage problem  is an unbiased random walk  
on a semi-infinite lattice, with an absorbing boundary at  the origin, which means that the walker is removed (or ``dies'') when it reaches the origin (see Fig. \ref{Fig_rw}). 
At each time step the walker moves  one step  to  the left  or  right, with  
equal  probability. A  typical first-passage problem is to compute the 
probability, $P_1(x_0,n)$, that, starting at lattice site $x_0$, the walker 
first reaches the origin at step $n$ [see Fig. \ref{Fig_rw} b)]. The corresponding persistence 
probability is the probability $Q(x_0,n)$ that the walker survives until at  
least step (or time) $n$ having started at $x_0$, and of course $P_1(x,n) = Q(x_0,n)-Q(x_0,n+1)$. A straightforward calculation  \cite{Redner} gives a result  with large-$n$ 
form $Q(x_0,n) \sim c\,x_0/n^{1/2}$ where $c$  is a constant. Introducing the 
terminology that we  will use throughout  this article, we  will call the  
exponent $1/2$, characterising  the asymptotic time  dependence, the  
``persistence exponent'' for this process, represented by the symbol $\theta$. 
Thus $\theta=1/2$ for the unbiased random walk in one dimension. In this article, we will also consider stochastic processes $X(t \geq 0)$, of zero mean, where both space and time
 are continuous. For such a process, the persistence $Q(t)$ is then defined as the probability that $X$
 has not changed sign up to time $t$, the probability density of the first time at which the process crosses $X=0$ being $P_1(t) = -dQ(t)/dt$. 
 
First passage problems have been widely studied by mathematicians since the fifties \cite{SparreAndersen}, often inspired by engineering 
applications \cite{Mfa,Newell,Slepian,BlakeLindsay73}. While remaining an important problem of probability theory (see Ref. \cite{AS12} for a recent review 
from a mathematical point of view), persistence properties have received, in physics, a considerable attention in the context of 
non-equilibrium statistical mechanics of spatially extended systems, both theoretically and experimentally. In various relevant physical 
situations, ranging from coarsening dynamics to fluctuating interfaces or polymer chains, the persistence probability turns out to decay 
algebraically at large times, $Q(t) \sim t^{-\theta}$. The persistence exponent $\theta$ carries interesting and useful information about the 
full history of the stochastic dynamics of the system. For this reason, $\theta$ is usually a nontrivial exponent, the prediction of which 
becomes particularly challenging for non-Markovian processes (for a brief review see \cite{SatyaReview}).

\begin{figure}[ht]
\centering
\includegraphics[width=0.7\linewidth]{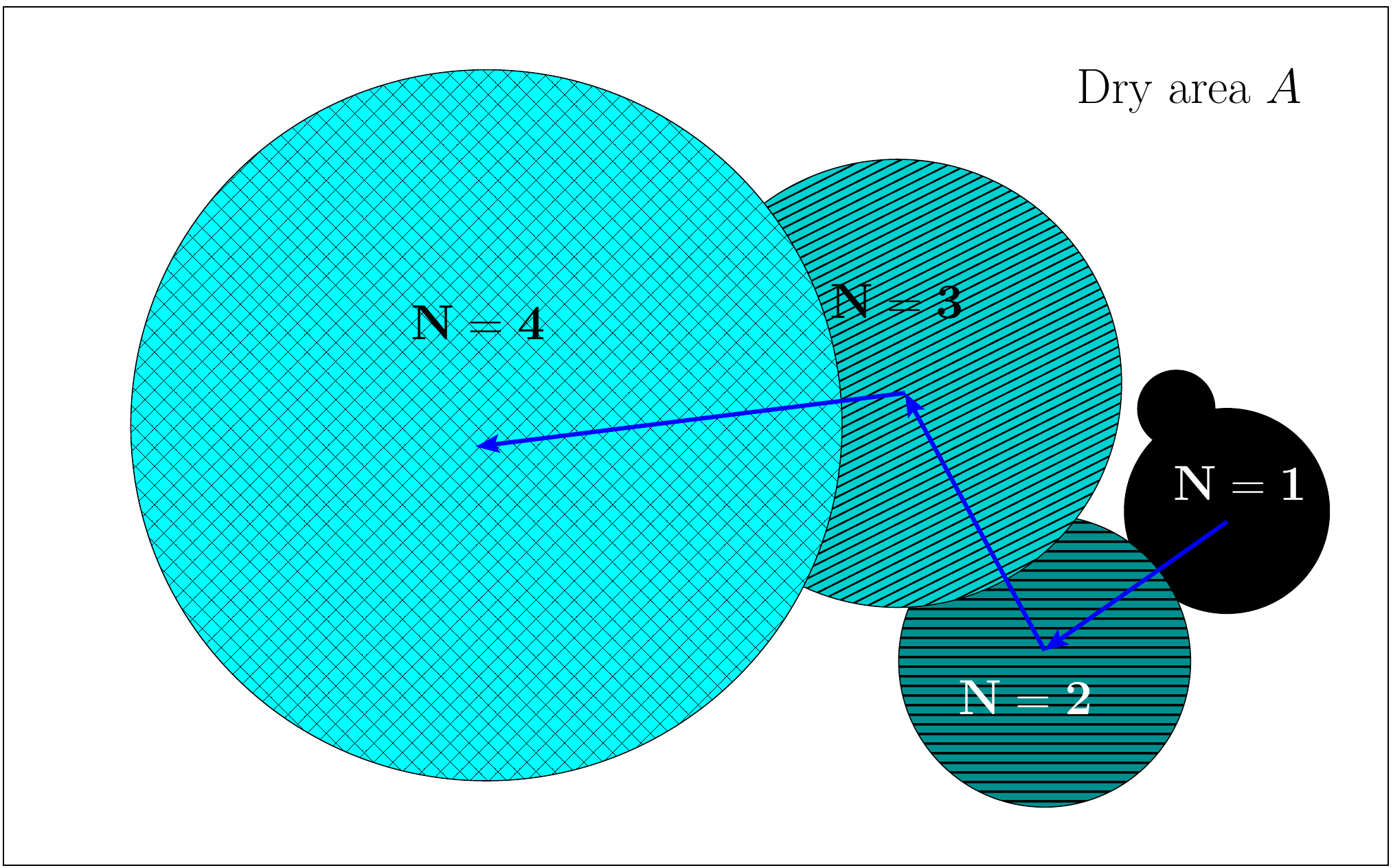}
\caption{A dye spreads over a substrate by the growth and coalescence of droplets. The integer inside the droplet indicates the number of coalescences. The dry area $A$ in white is the area of the substrate which has not been touched by a droplet. Figure inspired by Ref. \cite{droplet_exp95}.}\label{Fig_breath}
\end{figure}

The interest of physicists for persistence properties came from experiments 
performed on the formation of dew, when water vapor condenses on a cold substrate as small droplets that imperfectly wet the substrate: these droplets are called "breath figures" (Fig. \ref{Fig_breath}). 
It was shown in Ref. \cite{droplet_exp95} that the fraction $f_{\rm dry}(t)$ of the surface which was never covered by any droplet decays as a power law $f_{\rm dry}(t) \sim t^{-\theta}$, with $\theta = 
1.0(1)$ \cite{droplet_exp95}. It was realized that $f_{\rm dry}(t)$ is an analogue of a persistence probability, which 
motivated theoretical studies of persistence properties for the coarsening dynamics of ferromagnetic spin models evolving at zero temperature $T=0$ from random initial conditions \cite{BDG1994a,Sta94}. Such situations are paradigmatic instances of phase ordering kinetics, that is the growth of order through domain coarsening when a system is quenched from a homogeneous phase into a broken-symmetry phase. Phase ordering dynamics has been a very active field of investigations since the early sixties \cite{LS61,Wag61,BrayReview,KRBNBook}. 
\begin{figure}[t]
\includegraphics[width=\linewidth]{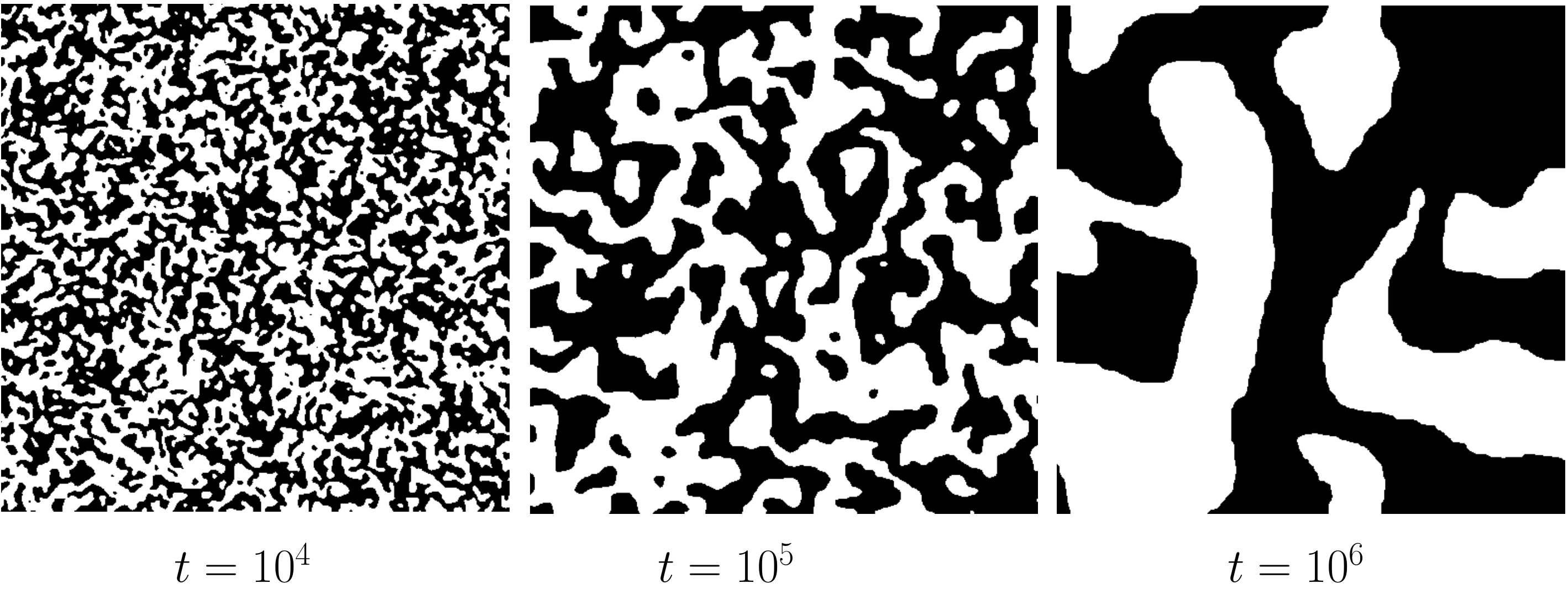}
\caption{Snapshots, at different times $t=10^4, 10^5$ and $10^6$, of a 2-d Ising model on a square lattice of linear size $L=512$ evolving with Glauber dynamics at $T=0$, starting from a completely disordered configuration at $t=0$. In white (respectively in black) are represented the up spins (respectively the down spins).}\label{Fig_coarsening}
\end{figure}
In these coarsening systems, the dynamics is usually characterized by a single length scale $L(t)$ which measures the typical size of the domains (see Fig. \ref{Fig_coarsening}). The growth law of $L(t)$
is governed by the dynamics of domain walls, which involves collective rearrangements over the whole system: it is thus a very slow process. 
In many systems, the growth law is algebraic $L(t) \sim t^{1/z}$, defining the dynamical exponent $z$ \cite{LS61,Wag61,BrayReview,AC79,Hus86}. For example $z=2$ for all systems with short range interactions described by a scalar non conserved order parameter \cite{BrayReview}. Note that the dynamics can be much slower in the presence of quenched disorder where $L(t) \sim (\log t)^{1/\psi}$ \cite{BrayReview,Vil84,HH85}. In many situations $L(t)$ is the only macroscopic length and hence it controls the time dependence of any physical observable. For instance, the two-time correlation function of the order parameter $C(t,t')$ takes, for large times $t,t'$ the scaling form $C(t,t') \sim F_C[L(t)/L(t')]$ \cite{BrayReview}, known under the name of 'simple aging'~\cite{LeticiaLesHouches}. A large body of the work on coarsening dynamics has been the study of the associated scaling functions like $F_C(x)$, and in particular their asymptotic large $x$ behavior, $F_C(x) \sim x^{-\lambda}$ where $\lambda$ is the autocorrelation exponent \cite{FHlambda88}, for which there exists very few exact results \cite{BrayReview}. Characterizing these correlations in coarsening systems is certainly an important and interesting question. However, these correlation functions do not give much information about the history of the evolution process.

The simplest and the most natural way to probe the history of a ferromagnetic system undergoing phase ordering is to focus on the persistence probability $Q(t)$ of the local magnetization, which is the probability that the local spin at site ${\mathbf r}$ has not flipped between time $0$ and time $t$ (note that, for a finite system, $Q(t)$ is independent of $\mathbf r$ far enough from the boundary of the system). Alternatively $Q(t)$ can be viewed as the fraction of spins which have never flipped until time~$t$, and hence is similar to the quantity $f_{\rm dry}(t)$ measured in breath figures experiments (Fig. \ref{Fig_breath}). Numerical simulations were first performed for $q$-states Potts model in low dimension, starting at $t=0$ from a completely random initial condition, and evolving subsequently with Glauber dynamics at $T=0$. They revealed an algebraic decay of $Q(t) \sim t^{-\theta}$ with an exponent which depends both on $q$ and on the dimension of the system. For instance, for $q=2$ corresponding to Ising systems, $\theta = 0.37$ in $d=1$ and $\theta = 0.22$ in $d=2$ \cite{BDG1994a,Sta94}. These results clearly provide evidence that this exponent $\theta$ is indeed nontrivial. Soon after these first numerical results and analytical approaches on related simplified models \cite{BDGEPL, KBR94}, the one-dimensional Glauber dynamics of the $q$-states Potts model at $T=0$ was exactly solved in Ref. \cite{DHP1995a,DHP1995b}, using a relation to a reaction-diffusion model \cite{Der95}, showing that the persistence exponent is indeed a nontrivial function of the parameter $q$, with in particular $\theta = 3/8$ for $q=2$. Although exact results were restricted to $d=1$, approximate analytical methods were developed in Ref. \cite{MajumdarSirePT,MSR00} to compute $\theta$ in any dimensions by exploiting a mapping to a quantum mechanics problem, yielding for instance, for Ising spins, $\theta \approx 0.19$ in $d=2$. It was soon realized that the persistence exponent $\theta$ is actually nontrivial if the underlying stochastic process is non-Markovian. It was indeed shown that such a simple system as the diffusion equation with random initial conditions yields highly nontrivial persistence exponents, for which no exact result is known~
\cite{MajumdarSireBrayCornell,DerridaHakimZeitak}. These surprising results for the persistence exponent of spatially extended systems have subsequently motivated a large body of theoretical works which are reviewed in the rest of the paper.  

Quite remarkably, these theoretical works on the persistence exponent, motivated to a large extent by experiments on breath figures (Fig. \ref{Fig_breath}), 
were shortly followed by several other experiments on various coarsening
systems, which measured a nontrivial persistence exponent. 
%
%
For instance, a quantity similar to $f_{\rm dry}(t)$ in the breath figures -- an analogue of a persistence probability -- was 
measured in experiments on two-dimensional soap froth \cite{soap_exp97a,soap_exp97b}, yielding a persistence exponent slightly larger 
than $\theta=1$, which is the expected exact result for soap froth in $d=2$ dimensions \cite{MS_largeq95,LD97} (or $\theta = d/2$ for arbitrary 
$d>1$). The first experiments on Ising like systems -- like the one discussed in the aforementioned theoretical works -- were performed on 
twisted nematic liquid crystal in two dimensions \cite{YPMS97}. These studies used a liquid crystal sample placed between two glass plates whose surfaces had 
been treated to force the direction of molecular alignment (director field) at the surface of a glass plate to lie parallel to the glass plate 
along a well defined direction. The two glass plates are oriented such that the orientation of the director at one glass surface is orthogonal 
to the orientation of the director at the other glass surface. The director field must thus twist by $\pi/2$, either clockwise or 
counter-clockwise, so that, after a thermal quench from the isotropic to the nematic phase, the liquid crystal organized itself into domains in 
which the director was forced to twist clockwise or counter-clockwise by $\pi/2$ in going from one plate to the other. The boundary between two 
domains of opposite twist consists of a twist disclination line. Regions of opposite twist correspond to Ising model domains in which the spins 
all point up or down. In these experiments, the persistence probability $Q(t)$ that the local order parameter has not switched its state by the 
time $t$ was found to decay algebraically $Q(t) \sim t^{-\theta}$ with a measured persistence exponent $\theta = 0.190(31)$, in good agreement 
with analytical approximation~\cite{MajumdarSirePT} and numerical simulations~\cite{BDG1994a,Sta94}. These first results have been followed by 
a large numbers of other experimental measurements of the persistence probability in a variety of physical systems 
including NMR measurement of persistence in 1-d diffusion in Xenon gases~\cite{WMWC02},
fluctuating step edges on crystals~\cite{Dougherty_exp02,Dougherty_exp03}, advancing combustion fronts~\cite{MMMTA03},
two-dimensional Ostwald ripening~\cite{Stavans09},
reactive-wetting interfaces~\cite{Efraim11} and
liquid crystal turbulence~\cite{TS12}--some of these results will be discussed later in appropriate sections. 

At variance with the original problem of first passages as studied in the mathematics literature, physical situations
usually involve systems containing infinitely many degrees of freedom. It is for these 
systems that the term "persistence" was originally introduced in the physics literature. However, for 
consistency of terminology we will use it also for systems with finitely 
many  degrees of  freedom, and  introduce the "persistence  exponent",    
$\theta$,  to  describe  the   decay  of  survival probability,  for  
cases where  the  decay  has  a power-law form. Before turning to systems with infinitely many degrees of freedom, 
we will first discuss in detail a number of first-passage problems with 
finitely many degrees of freedom, starting in section \ref{section:rw} with the simplest case of random walks, 
where we will review in particular the famous Sparre Andersen theorem.
A very convenient formal approach to such problems is provided by the 
``Backward Fokker-Planck'' (BFP) method, and in section \ref{section:bfp}
we introduce the method, and show how it provides a rather elegant approach 
to calculating persistence exponents in such systems. In particular, we apply the method 
to single particle systems in continuous time, including the Brownian walk and the random acceleration process, $\ddot{x}=\eta(t)$ (with $\eta(t)$ a Gaussian white noise). In section \ref{section:bfp}, we also show that many related processes can be solved exactly using this BFP method. 
Continuing up the hierarchy of increasing complexity, we next discuss the persistence of multi-particle systems in section 4. We will study, in particular, 
the trapping reaction $A + B \to B$, in which the survival probability of a 
single $A$-particle diffusing  in the presence of a sea of diffusing 
$B$-particles is addressed, before reaching the core of this article which is the
study of persistence of fields, i.e systems with infinitely many degrees of freedom. These systems and a variety of theoretical approaches 
are discussed in sections 5 to 19. A short review 
by one of us on some of the themes discussed in this article can be found 
in Ref.~\cite{SatyaReview}.

\section{Survival and first-passage for random walks}\label{section:rw}

Let us consider a simple discrete-time random walker moving on a continuous line. The position $x_n$ of the walker after $n$ steps evolves, for $n \geq 1$ via
\begin{eqnarray}\label{eq:markov}
x_n = x_{n-1} + \eta_n \;,
\end{eqnarray}
starting at $x_0=0$, where the step lengths $\eta_n$'s are independent and identically distributed (i.i.d.) random variables with zero mean and each drawn from a distribution $\phi(\eta)$ which is symmetric, $\phi(\eta) = \phi(-\eta)$. Note that the evolution equation (\ref{eq:markov}) is obviously Markovian since the position $x_n$ at step $n$ depends only on the position at just the previous time $x_{n-1}$ and on the current noise at step $n$, $\eta_n$. Few examples of symmetric jump length distribution $\phi(\eta)$ are
\begin{eqnarray}\label{eq:jump_dist}
({\rm i})&& \; {\phi(\eta) = \tfrac{1}{2} e^{-|\eta|} \; \; ({\rm exponential})} \,, \nonumber \\
({\rm ii})&& \; {\phi(\eta) = \tfrac{1}{\sigma_0 \sqrt{2 \pi}} \exp(-\eta^2/2 \sigma_0^2) \;\; ({\rm Gaussian})} \,, \nonumber \\
({\rm iii})&& \; \phi(\eta) = {\frac{1}{2} \left[ \theta(\eta+1) - \theta(\eta-1) \right] \; ({\rm uniform})} \;, \nonumber \\
({\rm iv})&& \; \phi(\eta) \propto |\eta|^{-1 -\mu} \;, \; \eta \to \infty \; \; ({\rm L\acute{e}vy \; \, random \; walk}) \;, \nonumber \\
({\rm v})&& \; \phi(\eta) = \tfrac{1}{2} \delta(\eta+1) + \tfrac{1}{2} \delta(\eta-1) \; \; ({\rm Lattice \; random \; walk}) \;. 
\end{eqnarray}
Note that in the 4 first examples of (\ref{eq:jump_dist}) the cumulative jump distribution $\Psi(x) = \int_{-\infty}^x \phi(\eta) d\eta$ is a continuous function. In the last example, where the walker is restricted to move on a one-dimensional lattice with unit lattice spacing, $\Psi(x)$ is a non-continuous function. We will see below that this continuity property will play an important role. 

Let us first focus on the first $4$ cases above in (\ref{eq:jump_dist}) where the pdf $\phi(\eta)$ is continuous and symmetric with zero mean. Let ${\hat \phi}(k)= \int_{-\infty}^{\infty} \phi(\eta)\, e^{ik\eta}\, d\eta$ denote the Fourier transform of the jump distribution. It has the following small $k$ behavior
\begin{equation}
{\hat \phi}(k)= 1- (l_\mu\,|k|)^{\mu}+\ldots
\label{smallk.1}
\end{equation}
where $0< \mu\le 2$ and $l_\mu$ represents a typical length scale associated
with the jump. 
The exponent $0<\mu\le 2$ dictates
the large $|\eta|$ tail of $\phi(\eta)$. For jump densities with a finite
second moment $\sigma^2= \int_{-\infty}^{\infty} \eta^2\, \phi(\eta)\,d\eta$,
such as Gaussian, exponential, uniform etc,
one evidently has $\mu=2$ and $l_2=\sigma/\sqrt{2}$. In contrast, $0<\mu<2$ 
corresponds to jump densities with fat tails 
$\phi(\eta)\sim |\eta|^{-1-\mu}$ as $|\eta|\to \infty$.
A typical example is ${\hat \phi}(k)=\exp[-|l_\mu k|^\mu]$ where $\mu=2$ corresponds to the Gaussian jump distribution, while $0<\mu<2$ corresponds
to L\'evy flights (for reviews on these jump processes see \cite{BG90,MK00}).

A quantity that plays a crucial role in the study of persistence properties is $P_n(x)$  which denotes the 
probability density of the position of the symmetric random walk at step $n$.
Using the Markov rule in Eq. (\ref{eq:markov}), it is easy to see that
$P_n(x)$ satisfies the recursion relation
\begin{equation}
P_n(x)= \int_{-\infty}^{\infty} P_{n-1}(x')\, \phi(x-x')\, dx' \;,
\label{ffp.1}
\end{equation}
starting from $P_0(x)=\delta(x)$. This recurrence relation can be trivially solved 
by using Fourier transform to get
\begin{equation}
P_n(x) = \int_{-\infty}^{\infty} \frac{dk}{2\pi}\, \left[{\hat \phi}(k)\right]^n \,
e^{-i\,k\, x}\,.
\label{pdf.x}
\end{equation}
In the limit of large $n$, the small $k$ behavior of ${\hat \phi}(k)$ dominates
the integral on the right hand side (rhs) of Eq. (\ref{pdf.x}). For $\mu=2$, the central limit theorem holds, $x\sim {\sigma}\,n^{1/2}$\;,
and $P_n(x)$ approaches a Gaussian scaling form
\begin{equation}
P_n(x) \to \frac{1}{\sigma\,n^{1/2}}{\cal 
L}_2\left(\frac{x}{\sigma\, n^{1/2}}\right) \;,\quad
{\rm where}\quad {\cal L}_2(y)=\frac{1}{\sqrt{2\pi}}\, \exp(-y^2/2)\, .
\label{clt.1}
\end{equation}

On the other hand, for $0<\mu <2$, substituting 
the small $k$ behavior from Eq.~(\ref{smallk.1}), one easily finds that, typically $x\sim l_\mu n^{1/\mu}$ and $P_n(x)$ approaches the 
scaling form~\cite{BG90}
\begin{equation}
P_n(x) \to \frac{1}{l_\mu\, n^{1/\mu}}\, {\cal 
L}_\mu\left(\frac{x}{l_\mu\, n^{1/\mu}}\right) \;,\quad 
{\rm where}\quad {\cal L}_\mu(y) = \int_{-\infty}^{\infty} \frac{dk}{2\pi}\, 
e^{-|k|^\mu}\, e^{-i\,k\,y}\, .
\label{scaling.1}
\end{equation}  
For $0<\mu<2$, the scaling function ${\cal L}_{\mu}(y)$ decays as a power law for large 
$|y|$~\cite{BG90}
\begin{equation}
{\cal L}_\mu(y) \xrightarrow[y\to \infty]{} \frac{A_\mu}{|y|^{\mu+1}} \;, \quad {\rm 
where}\; A_\mu= \frac{1}{\pi}\,\sin(\mu\pi/2)\,\Gamma(1+\mu).
\label{lmu_asymp}
\end{equation}
In particular, for $\mu=1$, the scaling function ${\cal L}_1(y)$ is the Cauchy density ${\cal L}_1(y)= \frac{1}{\pi}\,\frac{1}{1+y^2}\,$.

\subsection{Sparre Andersen theorem and persistence of a symmetric random walk}

For such a random walk (\ref{eq:markov}), the persistence -- or equivalently in this case the survival probability --  is defined as the probability $Q(x_0,n)$ that the particle, starting at $x_0$, stays positive (i.e. survives) up to step $n$, no matter what the final position is. Thus
\begin{eqnarray}\label{def_Q_rw}
Q(x_0,n) = {\rm Prob.}\,[x_n\geq 0, x_{n-1}\geq 0, \cdots , x_1 \geq 0 | x_0] \; \,
\end{eqnarray}
where we use the notation ${\rm Prob.}\,[A|B]$ to denote the (conditional) probability of the event $A$, given the event $B$. 

It is possible to write a backward equation for $Q(x_0,n)$ in (\ref{def_Q_rw}) by considering the stochastic jump 
$x_0\to x_0'$ at the first step and then subsequently evolves for $(n-1)$ steps starting from this new initial position 
$x_0'$ while staying positive all along (for a recent review see \cite{satya_leuven}). 
Using the Markov property of the evolution (\ref{def_Q_rw}) the backward equation for $Q(x_0,n)$ reads
\begin{eqnarray}\label{int_eq_Q}
Q(x_0,n) = \int_0^\infty Q(x'_0,n-1) \phi(x'_0 - x_0) dx'_0 \;,
\end{eqnarray}
with the initial condition
\begin{eqnarray}\label{Q_IC}
Q(x_0,0) = 1 \;, \forall \; x_0 \geq 0 \;,
\end{eqnarray}
which follows from the fact that the walker does not cross the origin in $0$ step. Even though the integral equation in Eq. (\ref{int_eq_Q}) has a convolution form, the limits of integration over $x'_0$ is over the half-space $[0, + \infty)$, and not the full space $(-\infty, + \infty)$ as in Eq. (\ref{ffp.1}) and hence the Fourier transform is of little use in this case. In fact, such half-space integral equations (\ref{int_eq_Q}) are know as Wiener-Hopf integral equations \cite{MF53}. While for a general kernel $\phi(x)$ this type of equation (\ref{int_eq_Q}) is highly difficult to solve, for the particular case where $\phi(x)$ is continuous [like in the 4 first cases in (\ref{eq:jump_dist})] and has the interpretation of a probability density function (i.e., non-negative and normalized) one can obtain an explicit solution to (\ref{int_eq_Q}). This solution was first found by Pollaczeck in \cite{P52} and later on, in a more combinatorial way, by Spitzer \cite{S56-57a,S56-57b}. The same integral equation also appeared previously in a variety of half-space transport  problems in physics and astrophysics and several other derivations of the solution of this equation (\ref{int_eq_Q}), mostly algebraic, exist in the literature \cite{I94}. We refer the reader to \cite{MCZ06} for a pedagogical description of the solution of (\ref{int_eq_Q}). The solution of Eq. (\ref{def_Q_rw}) with the initial condition in (\ref{Q_IC}) reads, in terms of double Laplace transform of $Q(x_0,n)$
\begin{eqnarray}\label{pollaczeck_spitzer}
\int_0^\infty dx_0 \left[\sum_{n=0}^\infty Q(x_0,n) s^n \right]e^{- p x_0} dx_0 = \frac{1}{p \sqrt{1-s}} \exp{\left[-\frac{p}{\pi}\int_0^\infty \frac{\ln{1- s \hat \phi(k)}}{p^2+k^2} \, dk \right]}\;, \nonumber \\
\end{eqnarray}
where we remind that $\hat \phi(k) = \int_{-\infty}^\infty \phi(\eta) e^{i k\eta} d \eta$ is the Fourier transform of the jump length distribution. Although the survival probability $Q(x_0,n)$ depends explicitly on the jump distribution $\phi(\eta)$ -- as it is obvious on Eq. (\ref{pollaczeck_spitzer}) -- it turns out that $Q(0,n)$ becomes universal, i.e. independent  of $\phi(\eta)$. From Eq. (\ref{pollaczeck_spitzer}) one can indeed show that if $\phi(\eta)$ is not only symmetric but also continuous one has
\begin{eqnarray}\label{eq:SA_laplace}
\sum_{n=0}^\infty Q(0,n) s^n = \frac{1}{\sqrt{1-s}} \;,
\end{eqnarray}
which leads to the famous Sparre Andersen theorem \cite{SparreAndersen}
\begin{eqnarray}\label{SA_th}
q(n) = Q(0,n) = {2n \choose n} 2^{-2n} \;.
\end{eqnarray}
We emphasize that this result (\ref{SA_th}), which holds for any $n$ (and not just for large $n$), states that the survival probability $q(n) = Q(0,n)$ starting from the origin is the same no matter whether the jump length distribution is exponential, Gaussian, uniform or has an algebraic tail $\phi(\eta) \propto \eta^{-1 - \mu}$, including also $\mu < 2 $. Since the original derivation of Sparre Andersen, relying on a combinatorial approach, various derivations of this result have been proposed in the literature \cite{FF94,BGL99}, all of them remaining relatively complicated. In the limit of large $n$, the survival probability $q(n)$ in (\ref{SA_th}) behaves like
\begin{eqnarray}\label{persistence_sym_rw}
q(n) = Q(0,n) \sim \frac{1}{\sqrt{\pi n}} \;,
\end{eqnarray} 
and hence the persistence exponent $\theta$ associated to the symmetric random walk is $\theta = 1/2$. We emphasize again that this result (\ref{persistence_sym_rw}) holds for arbitrary continuous jump distribution, including L\'evy random walks. It turns out that $q(n)$'s in  (\ref{persistence_sym_rw}) also play as basic building blocks for the calculation
of statistics of records of random walks and L\'evy flights which
also turn out to be universal \cite{MZ08}. Note that for arbitrary initial position $x_0$, $Q(x_0,n)$ will depend explicitly on $\phi(\eta)$ (\ref{pollaczeck_spitzer}) but the persistence exponent $\theta = 1/2$ (\ref{persistence_sym_rw}) will remain universal. 

For a lattice random walk [the fifth example in (\ref{eq:jump_dist})] the generalization of (\ref{eq:SA_laplace}) reads
\begin{eqnarray}
\sum_{n=0}^\infty Q(0,n) s^n = \frac{1}{1-s} - \frac{1-\sqrt{1-s^2}}{s(1-s)} \propto \frac{\sqrt{2}}{\sqrt{1-s}} \;, \; s \to 1^- \;,
\end{eqnarray}
which leads, in this case, to 
\begin{eqnarray}
q(n) = Q(0,n) \sim \frac{\sqrt{2}}{\sqrt{\pi n}} \;,
\end{eqnarray}
which differs by a factor of $\sqrt{2}$ from the result (\ref{persistence_sym_rw}) for continuous jump length distribution $\phi(\eta)$. We refer the reader to Ref. \cite{AS12} for more details on the persistence of discrete random walks.

\subsection{Generalized Sparre Andersen theorem and persistence of a random walk with a drift}

The above result for the survival probability starting from the origin (\ref{SA_th}) holds for continuous and symmetric jump length distribution $\phi(\eta)$. There however exists a generalization of this result for non-symmetric (but still continuous) $\phi(\eta)$. For asymmetric jump distribution of a random walk starting from $x_0=0$, the probability that the walker is on the positive side up to $n$ steps is different from the probability that it is on the negative side up to $n$ steps. Hence we need to define two different survival probabilities 
\begin{eqnarray}
&&q_+(n) = {\rm Prob.}\,[x_n \geq 0, \cdots, x_1\geq 0 | x_0 = 0] \;, \label{p_q} \\
&& q_-(n) = {\rm Prob.}\,[x_n \leq 0, \cdots, x_1\leq 0 | x_0 = 0] \;. \label{m_q}
\end{eqnarray}
Of course, for symmetric jump distributions $q_+(n) = q_-(n)$. In the asymmetric case, the generalization of the Sparre Andersen theorem (\ref{SA_th}) reads \cite{SA54}
\begin{eqnarray}
&&\tilde q_+(s) = \sum_{n=0}^\infty q_+(n) s^n = \exp{\left[\sum_{n=1}^\infty \frac{p_n^+}{n} s^n \right]} \;, \; p^+_n = {\rm Prob.}(x_n \geq 0) \;, \label{SA_th_asymm_pos} \\
&&\tilde q_-(s) = \sum_{n=0}^\infty q_-(n) s^n = \exp{\left[\sum_{n=1}^\infty \frac{p_n^-}{n} s^n \right]} \;, \; p^-_n = {\rm Prob.}(x_n \leq 0) \;, \label{SA_th_asymm_neg}
\end{eqnarray}
where $p^+_n$ and $p^-_n$ are just the probabilities that exactly at the $n^{\rm th}$ step the particle position is positive and negative respectively. For a symmetric random walk, $p^+_n = p^-_n = 1/2$ and both formulae (\ref{SA_th_asymm_pos}, \ref{SA_th_asymm_neg}) reduce to (\ref{eq:SA_laplace}). 

These formulae (\ref{SA_th_asymm_pos}, \ref{SA_th_asymm_neg}) can be used to study the persistence properties of a random walk in the presence of a constant drift $c$, thus evolving via
\begin{eqnarray}\label{markov_drift}
x_{n} = x_{n-1} + c + \eta_n \;,
\end{eqnarray} 
starting from $x_0$=0. The authors of Ref. \cite{MSW12} studied the persistence probability $q_-(n)$ (\ref{m_q}) for a random walk with a drift as in (\ref{markov_drift}) where the jump variables $\eta_n$ are drawn from a L\'evy stable distribution, $\hat \phi(k) = e^{-|l_\mu k|^\mu}$, with $0 < \mu \leq 2$.  
\begin{figure}[ht]
\centering
\includegraphics[width=0.8\textwidth]{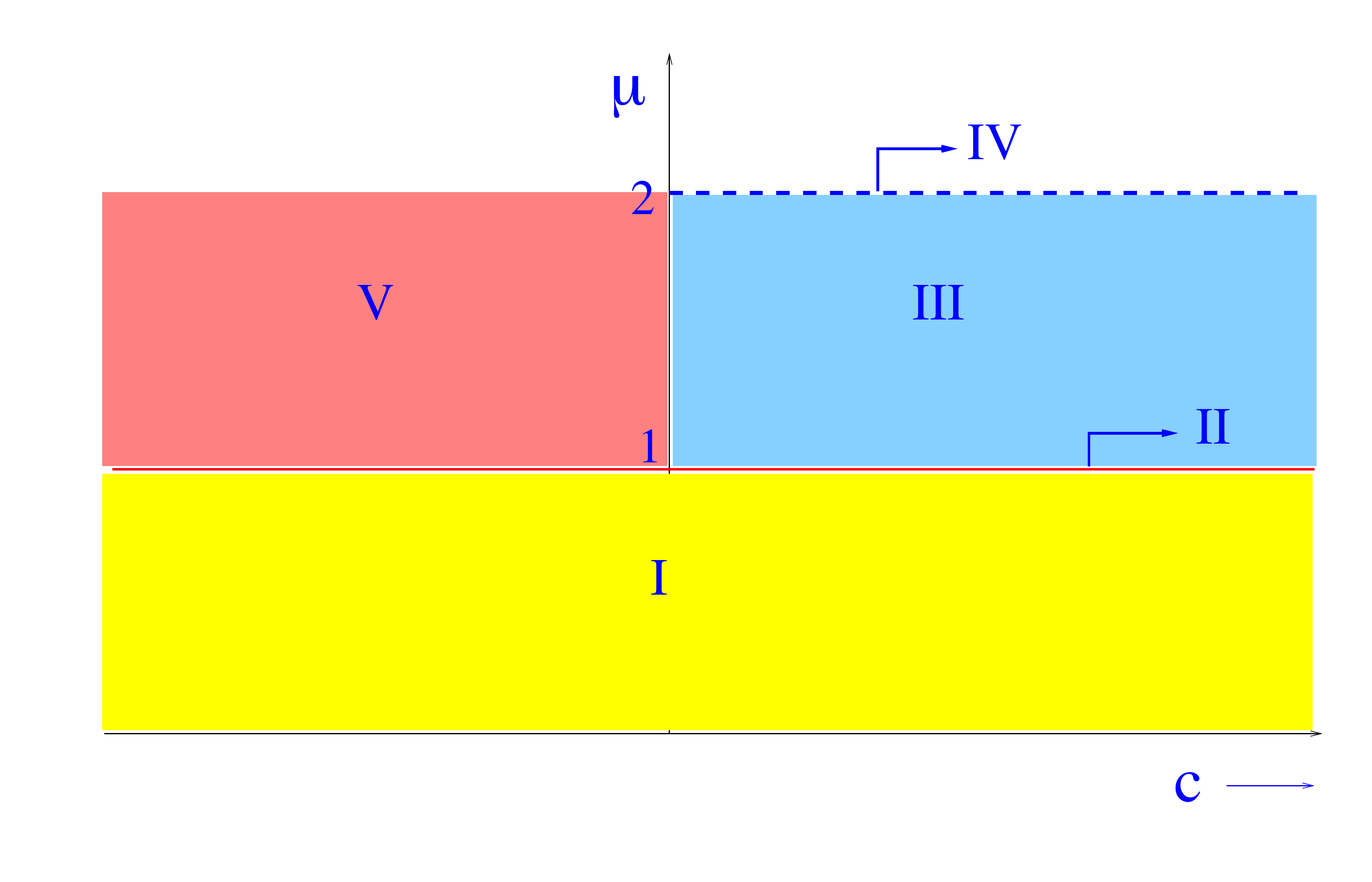}
\caption{ Phase diagram in the $(c,0<\mu\le 2)$ strip
depicting $5$ regimes: (I) $0<\mu<1$ and $c$ arbitrary (II) the line
$\mu=1$ and $c$ arbitrary (III) $1<\mu<2$ and $c>0$ (IV) the semi-infinite line
$\mu=2$ and $c>0$ and (V) $1<\mu\le 2$ and $c<0$. The persistence $q_-(n)$ exhibit different asymptotic behaviors
in these $5$ regimes [see (\ref{qn_asymp}) in the text and Ref. \cite{MSW12}].} 
\label{Fig_phd}
\end{figure}
In this case, the asymptotic behavior of $q_-(n)$ for large $n$ depends on the sign of $c$ and the value of the exponent $\mu$ in the following way (see Fig. \ref{Fig_phd}): 
\begin{eqnarray}
q_-(n) &\sim & B_{\rm I}\, n^{-1/2}\quad {\rm for}\quad 0<\mu<1\,\, {\rm and}\,\, c\,\,{\rm 
arbitrary}\quad ({\rm regime}\,\, {\rm I}) \;,
\nonumber \\
&\sim & B_{\rm II}\, n^{-\theta(c)} \quad {\rm for}\quad \mu=1 \,\, {\rm and}\,\,c \,\,{\rm
arbitrary}  
\quad ({\rm regime}\,\, {\rm II}) \;,
\nonumber \\
&\sim & B_{\rm III}\, n^{-\mu} \quad {\rm for}\quad 1<\mu<2\,\, {\rm and}\,\, c>0 
\quad ({\rm regime}\,\, {\rm III}) \;,
\nonumber \\
&\sim & B_{\rm IV}\, n^{-3/2}\, \exp[-(c^2/{2\sigma^2})\, n] \quad  {\rm for}\quad \mu=2 \,\, 
{\rm and}\,\, c>0 
\quad ({\rm regime}\,\, {\rm IV}) \;,
\nonumber \\
&\sim & \alpha_\mu(c) \quad {\rm for}\quad 1<\mu\le 2 \,\, {\rm and}\,\, c<0 
\quad ({\rm regime}\,\, {\rm V}) \;,
\nonumber \\
\label{qn_asymp}
\end{eqnarray}
where $B_{\rm I}, B_{\rm II}, B_{\rm III}, B_{\rm IV}$ and $\alpha_\mu(c)$ are computable constants and the exponent $\theta(c)$ is given by \cite{LDW09}
\begin{eqnarray}\label{eq:theta_c}
\theta(c) = \frac{1}{2} + \frac{1}{\pi} {\rm arctan}(c) \;,
\end{eqnarray}
which thus depends continuously on $c \in (-\infty, +\infty)$. The behavior of $q^+_n$ (\ref{p_q}) is directly obtained from (\ref{qn_asymp}) by transposing
$c$ to $-c$.

\subsection{Continuous-time random walk (CTRW)}

We end this section on random walks by discussing briefly the so called continuous-time random walk (CTRW), which was initially introduced by Montroll and Weiss in Ref.~\cite{MW65}. Within this model, the walker performs a usual random walk as in (\ref{eq:markov}) but has to wait a certain "trapping time" $\tau$ before each jump. The trapping times between each jump are i.i.d. random variables with a common pdf $\psi(\tau)$ which has a power laws tail $\psi(\tau)\propto \tau^{-1 - \alpha}$, $\alpha > 0$. This type of model was suggested by Scher and Montroll \cite{SM75} to model non-Gaussian transport of electrons in disordered systems and, since, it has been widely used to describe phenomenologically anomalous dynamics in various complex systems \cite{BG90,MK00}. Indeed, for $\alpha < 1$, the mean trapping time between two successive jumps is infinite and CTRW is characterized by a sub-diffusive behavior, with a dynamical exponent $z = 2/\alpha > 2$ and non-Gaussian statistics. The CTRW is thus a renewal process and, as shown originally in Ref.~\cite{SM75} (see also~\cite{FM2012}), if one knows explicitly the expression of the expectation value of any observable in the discrete-time jump process (\ref{eq:markov}), the corresponding expectation value for the CTRW can be derived straightforwardly: this fact goes by the name "subordination property" in the literature \cite{MK00}. In particular, the persistence (or survival) probability $Q(x_0,t)$ in a 
given time interval $[0,t]$ for CTRW can be obtained from the corresponding expression for RW in (\ref{pollaczeck_spitzer}). In particular for large $t$, one has $Q(x_0,t) \sim x_0/t^{\theta}$, with $\theta = \alpha/2$ \cite{MK00b}. The value $\theta=\alpha/2$ can be obtained by a simple scaling argument, by combining the result for the persistence of a RW after $n$ steps $q(n) \sim n^{-1/2}$ (\ref{persistence_sym_rw}) and the fact that a CTRW performs typically $n \sim t^{\alpha}$ such steps in the time interval $[0,t]$.

\section{Persistence in single particle systems: Markov and non-Markov processes in continuous time}\label{section:bfp}
In  this section we investigate the first-passage properties of a single particle which evolves via Markovian as well as non-Markovian
dynamics in continuous time. In particular, we 
%
%
introduce a simple yet powerful  method -- the  ``Backward Fokker-Planck'' method -- which  can both
simplify  the calculation  of the  persistence exponent  $\theta$  for already
well-studied models, and  lead to new results for  some apparently nontrivial
models. In particular we  demonstrate the advantages of backward Fokker-Planck
methods over  the more familiar forward Fokker-Planck  equation. We illustrate
the method  via an increasingly complex  set of processes,  beginning with the
simplest Markovian process: the one-dimensional Brownian walk.

\subsection{The Brownian walker: the simplest Markov process}\label{section:RW}

A  Brownian  walk (an unbiased random walk in continuous time and continuous 
space) in one  dimension is described  by  the Langevin  equation, which is the continuous version of Eq. (\ref{eq:markov}), 
\begin{eqnarray}\label{eq:BM}
\dot{x} =  \eta(t),
\end{eqnarray} 
where $\eta(t)$ is a Gaussian white noise  with mean zero and  correlator   
$\langle  \eta(t)\,\eta(t')\rangle  =   2D\delta(t-t')$. It is clear that this Langevin equation (\ref{eq:BM}) defines a Markov
process as the position of the walker at time $t+\Delta t$, $x(t+\Delta t) \approx x(t) + (\Delta t) \eta(t)$ depends only on the position at just the previous 
time $x(t)$ and on the noise $\eta(t)$. We consider the  
case where there is an absorbing boundary at $x=0$, and the walker  
starts at some  position $x>0$ (see Fig. \ref{Fig_rw}).  
The  persistence probability  $Q(x,t)$  is the
probability that the walker {\em survives} (i.e.\ has not  yet reached the 
absorbing boundary)  at  time  $t$, {\em  starting}  from  position  $x$ 
at  time  zero.
Before turning to the BFP equation, we remind the reader how this problem 
is conventionally solved, using the usual (forward) FP equation, 
$\partial_t P = D \partial_{yy}P$, where $P(y,t|x,0)$ is the probability
density to find the walker at position $y$ at time $t$. Ignoring the 
absorbing boundary, the solution for initial condition $P(y,0)=\delta(y-x)$ 
is given by Greens function (or ``heat kernel'')
\begin{equation}
G(y,t|x,0) = \frac{\exp(-[y-x]^2/4Dt)}{(4\pi Dt)^{1/2}}\ .
\end{equation}
The absorbing boundary condition enforces $P(0,t|x,0) = 0$ for all $x$ and 
$t$. This boundary condition can be implemented using the ``image method'' 
\cite{Redner}, 
writing 
\begin{equation}\label{eq:images}
P(y,t|x,0) = G(y,t|x,0) - G(y,t|-x,0)\ ,
\end{equation}
which clearly satisfies the differential equation, the boundary condition 
and the initial condition in the physical region $x\ge 0$. Finally, the 
persistence probability, $Q(x,t)$, that the particle has not reached the 
absorbing boundary at time $t$, given that it started at $x$, is obtained 
by integrating over all $y>0$: $Q(x,t) = \int_0^\infty dy\,P(y,t|x,0)$.
The result is $Q(x,t) = {\rm erf}\,(x/\sqrt{4Dt})$, where ${\rm erf}(z)$ 
is the error function. For large $t$, $Q$ decreases as $t^{-1/2}$, i.e.\ 
the persistence exponent is $\theta=1/2$. 

\subsubsection{Brownian walker in one dimension: the backward Fokker-Planck method}
The BFP equation provides an elegant method for computing the survival 
probability in which no integral over the final coordinate is required.
The method has a large range of applications, many of which we will 
explore here. We begin with the one-dimensional Brownian walk to show 
the great simplifications which this method provides. These 
simplifications pay dividends when we come to more difficult problems. 

Integrating the Langevin equation from $t=0$ to $t=\Delta t$, where 
$\Delta t$ is infinitesimal, we obtain the obvious identity
\begin{equation}
Q(x,t) = \langle Q(x+\Delta x,t-\Delta t) \rangle,\ {\rm for \; all}\ x>0\ ,
\label{BFPE1}
\end{equation}
where the average is over the displacement, $\Delta x$, that occurs in
the first  time interval $\Delta  t$. We emphasize that, to derive this equation (\ref{BFPE1}), we have explicitly used
the Markov property of the process $x(t)$ 
which allows in particular to treat its fluctuations on the time intervals $[0, \Delta t]$ and $[\Delta t, t]$ as statistically independent. Expanding Eq.\  (\ref{BFPE1}) to
first order  in $\Delta t$ and  second order in $\Delta  x$, and using
$\langle  (\Delta x)^2  \rangle =  2D\,\Delta t$  gives the BFP equation
\begin{equation} 
\partial_t Q = D \partial_{xx} Q\ ,
\label{BFPE1a}
\end{equation}
with  initial condition  $Q(x,0) =  1$,  for all  $x>0$, and  boundary
condition  $Q(0,t) =  0$  for  all $t$.  On  dimensional grounds,  the
solution  has the  scaling form  $Q(x,t) =  f(x/\sqrt{Dt})$. Inserting
this form  into Eq.\ (\ref{BFPE1a}), yields  the ordinary differential
equation  $f_{uu}  +  (u/2)f_u   =  0$, where $u=x/\sqrt{Dt}$,  with  
boundary  conditions $f(0)=0$,  $f(\infty)=1$. The  solution  is 
$f(u)  = {\rm  erf}(u/2)$, that is 
\begin{equation}
Q(x,t)  = {\rm erf}(x/\sqrt{4Dt})\ . 
\label{1DBrownian}
\end{equation}
The behaviour at large $t$ is $Q(x,t)  \sim x/\sqrt{\pi  D t} \propto  
t^{-\theta}$, with  $\theta = 1/2$.

This result  is well  known, of course,  but let  us recap how  it was
obtained.  We  started with a  partial differential equation  (PDE) in
two  variables, and  used  dimensional  analysis to  reduce  it to  an
ordinary   differential   equation   (ODE)   which   could   then   be
solved. However, if  we are interested only in the  value of the decay
exponent  $\theta$  a further  simplification  is  possible, using  an
approach which,  to our knowledge,  was first introduced  by Burkhardt
\cite{BurkhardtPSa} in his study  of the random acceleration process
(of which more later).

We  first  illustrate  this  approach  for  the  Brownian  walker.  On
dimensional grounds  we know that  the persistence probability  has the
form  $Q(x,t)  =  f(x/\sqrt{Dt})$.  For  large  $t$  we  anticipate  a
power-law decay  with time, $Q(x,t) \sim (x^2/Dt)^\theta$.  Now we can
insert this form into Eq.\  (\ref{BFPE1a}). Doing this, we notice that
the time  derivative term on the left  is smaller by one  power of $t$
than  the term  on the  right, and  is therefore  negligible  at large
time. This leads to  the remarkably simple equation $2\theta(2\theta -
1)=0$,  leading to $\theta  = 1/2$  as before  (but without  having to
solve any differential equation at all!). Note that the second solution, 
$\theta = 0$, is unphysical, since  the persistence probability must  
decay to zero at long times.

To illustrate  the power  of this  method we apply  it to a few more 
simple problems related to the Brownian walker before embarking on some nontrivial applications. 

\subsubsection{Brownian walker in higher dimensions}
The first extension is to analyse the Brownian walk in general dimension 
$d$, with an absorbing boundary at the origin. The BFP equation reads 
$\partial_t Q = \nabla^2 Q$, a natural extension of equation (\ref{BFPE1a}). 
By rotational symmetry, the survival probability, $Q(r,t)$, only depends on 
the radial coordinate, $r$, of the  initial position, so we can write the 
equation in the form 
\begin{equation}
\partial_t Q = D\left(\partial_{rr} Q + \frac{d-1}{r}\,\partial_r Q\right)\ .
\label{BFPE1b}
\end{equation}
By dimensional analysis, $Q(r,t) = f(r/\sqrt{Dt})$, and for 
$Dt \gg r^2$ we expect the power-law form, $Q(r,t) \sim (r^2/Dt)^\theta$. 
Inserting this in the BFP equation, we see again that the left-hand side 
becomes asymptotically negligible, giving $2\theta(2\theta + d -2)=0$, 
which implies $\theta = (2-d)/2$. Since, however, $\theta$ cannot be negative 
(the survival probability cannot {\em grow} with time), this result is 
restricted to dimensions $1 \le d<2$. The reader may reasonably protest 
that non-integer dimensions are unphysical. There is, however, a physical 
interpretation of this result. Suppose the particle moves in one dimension, 
but is subject to a radial force equal to $A/r$. Then the BFP equation 
reads $\partial_t Q = D\partial_{rr} Q + (A/r) \partial_r Q$, which has 
same form as equation(\ref{BFPE1b}), with $d-1=A/D$. So motion in one 
dimension, with a repulsive force $A/r$ is equivalent to free motion in 
dimension $1+A/D$.   
 
\subsubsection{Brownian walker in a wedge}
\label{sec:wedge}
\begin{figure}
\centering
\includegraphics[width = 0.55\linewidth]{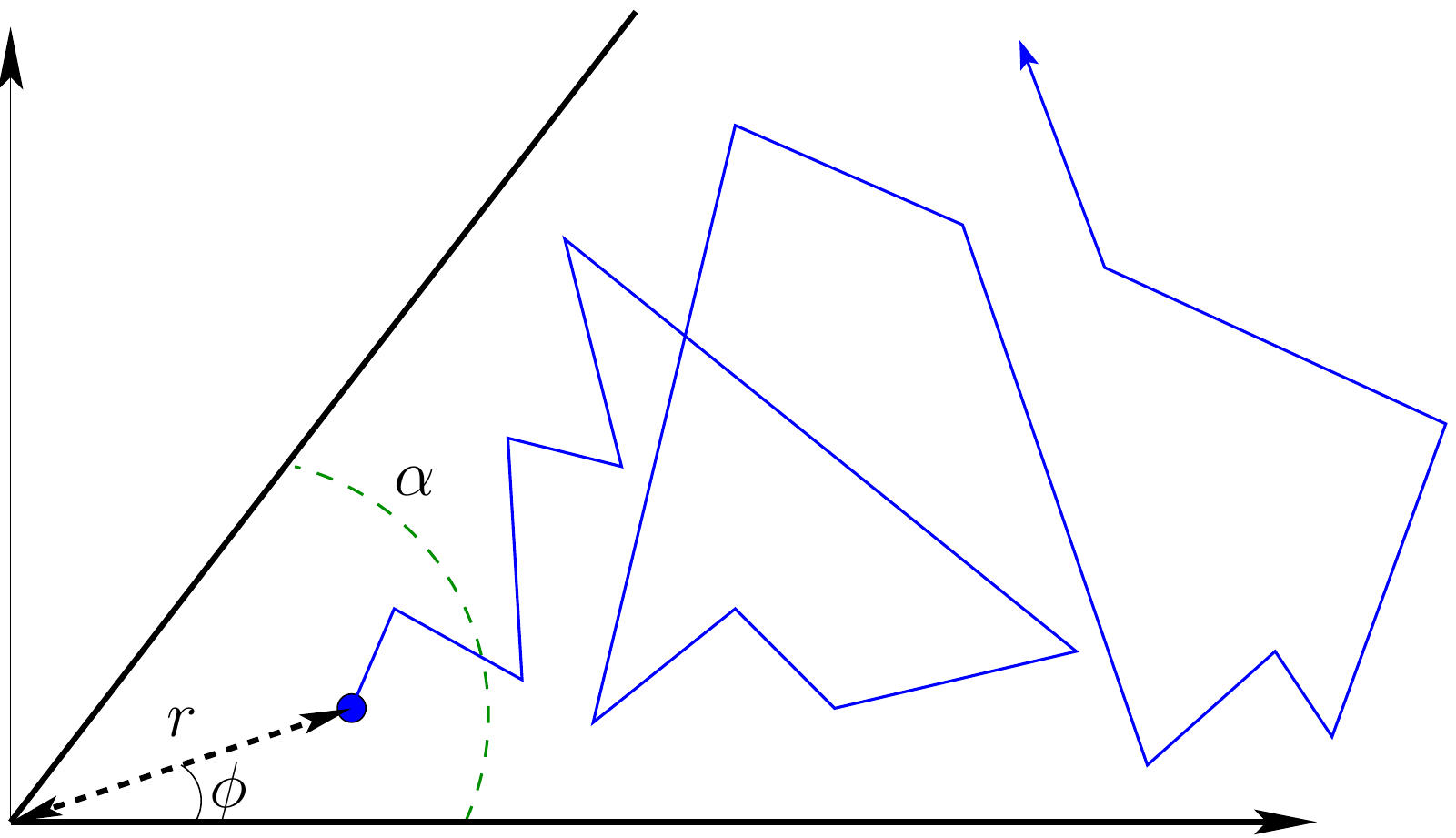}
\caption{Trajectory (in blue) of planar random walk in a wedge of angle $\alpha$, such that the   lines  $\phi=0$  and
$\phi=\alpha$ are absorbing boundaries. The depicted trajectory contributes to $Q(r,\phi)$.}\label{Fig_wedge}
\end{figure}

We consider a random walker moving in a two-dimensional wedge of angle
$\alpha$ (Fig.~\ref{Fig_wedge}).  The BFP  equation reads,  by an  obvious extension  of Eq.\
(\ref{BFPE1a}),
\begin{equation}
\partial_t Q  = D\nabla^2 Q  = D\left(\partial_{rr}Q+\frac{1}{r}\partial_rQ +
\frac{1}{r^2}\partial_{\phi\phi}Q\right) \:,
\label{BFPE2a}
\end{equation}
in plane polar coordinates (where we use $\phi$ for the polar angle to
avoid confusion with the  persistence exponent $\theta$). The wedge is
defined  by  $0   \le  \phi  \le  \alpha$,  the   lines  $\phi=0$  and
$\phi=\alpha$ being  absorbing boundaries.  

On  dimensional grounds we
have $Q(r,\phi,t) = f(r/\sqrt{Dt}, \phi)$. The boundary conditions are
$f=0$ when  $\phi=0$ and  $\phi=\alpha$. For large  $t$ we  expect the
power-law decay $Q \sim  (r^2/Dt)^\theta g(\phi)$. Inserting this form
into Eq.\  (\ref{BFPE2a}) we  see that, once  again, the left  side is
negligible compared  to the  right for large  $t$, leaving us  with the
simple  equation  $g''(\phi)  +  4\theta^2 g(\phi)=0$,  with  boundary
conditions  $g(0)=0=g(\alpha)$. The solution satisfying $g(0)=0$ is 
$g(\phi) = c\sin(2\theta\phi)$, where $c$ is a constant. The  boundary 
condition $g(\alpha)=0$  quantises the allowed  values of  $\theta$: 
$\theta_n  = n\pi/2\alpha$,  with  $n$ a positive integer. The general  
solution for $Q(r,t)$ therefore has the asymptotic form
\begin{equation}
Q(r,t) \sim \sum_{n=1}^\infty c_n (r^2/Dt)^{\theta_n} \sin(2\theta_n\phi)\ .
\end{equation}
This is dominated by the $n=1$  term for large times, so the persistence
exponent is 
\begin{equation}
\theta = \theta_1 = \pi/2\alpha\ . 
\label{theta-wedge}
\end{equation}
The usual derivation of this result \cite{Redner,FisherGelfand} is somewhat longer.

\subsubsection{Brownian walker in a cone}
We  now pursue with a simple application of the  BFP method to  a
three-dimensional example -- the Brownian  walker confined to a cone of
semi-angle  $\alpha$  whose  surface  is  an  absorbing  boundary \cite{Redner,BenKrap2010}.  By
symmetry, the  survival probability $Q$  depends only on  the distance
$r$  of the  initial  position from  the  apex of  the  cone, and  the
principal polar angle $\phi$ of  the initial position (once more using
$\phi$  instead of $\theta$  to avoid  confusion with  the persistence
exponent). Thus $Q=Q(r,\phi,t)$. The  BFP equation reads $\partial_t Q
= D\nabla^2   Q$,  as  before,   with  the  boundary   condition  that
$Q(r,\alpha,t)=0$ for all $r$ and $t$.

In spherical polar coordinates we have
\begin{equation}
\partial_t   Q   =   D\left(\partial_{rr}Q   +   \frac{2}{r}\partial_rQ   +
\frac{1}{r^2\sin\phi}\partial_\phi(\sin\phi\ \partial_\phi Q)\right)\ .
\label{BFPE3a}
\end{equation}
For large $t$ we expect,  once again, the power-law decay $Q(r,\phi,t)
\sim  (r^2/Dt)^\theta  g(\phi)$. Inserting  this  form  into the  Eq.\
(\ref{BFPE3a})  and noting,  once again,  that the  left-hand  side is
negligible for large $t$, we obtain the ODE
\begin{equation}
\frac{1}{\sin\phi}\,\frac{d}{d\phi}\left(\sin\phi\,\frac{dg}{d\phi}\right)
+ 2\theta(2\theta + 1)g=0\ .
\end{equation}
The change  of variable $\cos\theta  = x$ converts this  equation into
Legendre's differential equation:
\begin{equation}
(1-x^2)g''(x) - 2xg'(x) + 2\theta(2\theta+1) g(x) = 0.
\end{equation}
The boundary conditions are  $g(\cos\alpha)=0$, and the condition that
$g(x)$ is  regular at $x=\pm  1$. The latter boundary  condition fixes
the solution as the Legendre function, $P_l(x)$, with $l=2\theta$, and
combined with the former gives the final condition
\begin{equation} 
P_{2\theta}(\cos\alpha) = 0\ .
\end{equation}
For general $\alpha$ this equation has to be solved numerically (with,
as ever, the smallest value of $\theta$  being the physical one). For
certain values of $\alpha$ however, for which $2\theta$ is an integer,
the  Legendre function becomes  a Legendre  polynomial. The  first few
cases are $\theta=1/2$ for $\alpha=\pi/2$,  $\theta = 1$ for $\alpha =
\cos^{-1}(1/\sqrt{3})$    and   $\theta    =3/2$    for   $\alpha    =
\cos^{-1}(\sqrt{3/5})$. We refer the reader to \cite{BenKrap2010} for more details
on this problem. 

A closely related problem is the survival probability $Q(t)$ of a particle inside a paraboloidal domain, of equation $y = a(x_1^2+ \cdots + x_{d-1}^2)^{p/2}$, in $d$ dimensions and with $p>1$, studied in Ref. \cite{KrapRedner2010}. 
When the particle is inside the domain, it was shown that $Q(t)$ generically decays as a stretched exponential $\ln Q(t) \sim -t^{(p-1)/(p+1)}$, independently of $d$ \cite{KrapRedner2010}. See also \cite{LS2002} for a rigorous treatment of this problem.  

Our next group of applications reveals the full force of the BFP method. 

\subsubsection{Brownian walker in an expanding cage}\label{subsection:cage}

\begin{figure}[h]
\centering
\includegraphics[width=0.55\linewidth]{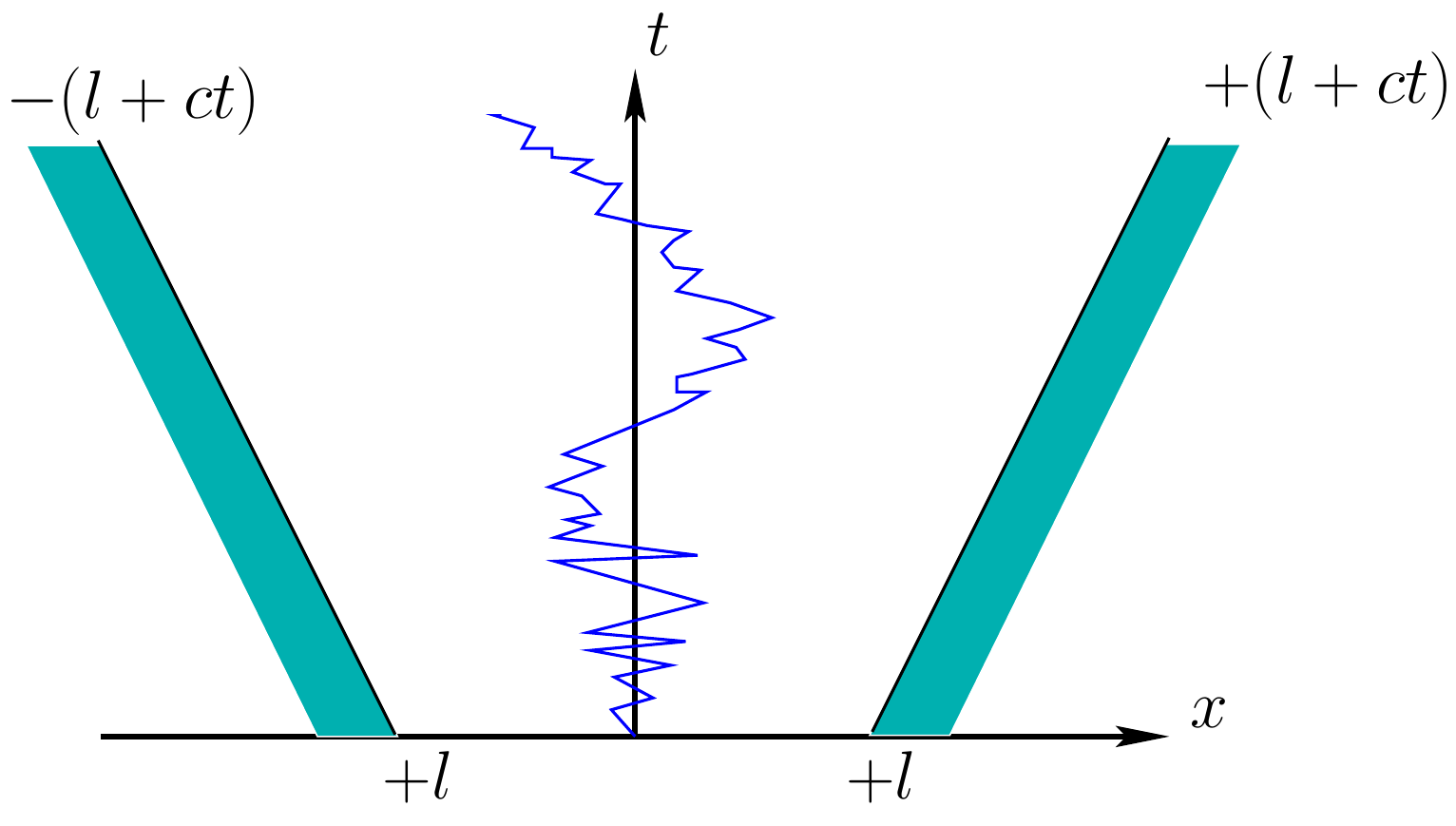}
\caption{Brownian walker, starting from the origin, in a linearly expanding cage}\label{fig:expanding-cage}
\end{figure}

Consider now a particle which diffuses, with a diffusion constant $D$, within a one-dimensional "cage" $[-L(t),L(t)]$ and is absorbed when it touches
the wall (Fig. \ref{fig:expanding-cage}). We are interested in determining the probability $Q(t)$ that such a particle survives up to
time $t$ \cite{Uchiyama1980,Turban1992,Igloi1992,Krapivsky-Redner-Cage}. In a cage of fixed length $2L$, $Q(t)$ decays exponentially, $Q(t) \sim \exp{(-\pi^2 D t/4L^2)}$ at large time $t$. The behavior becomes more interesting when the particle is helped to survive by allowing the cage walls to recede by choosing $L(t) = (A t)^\alpha$. In such a situation, there are three distinct regimes for the behavior of $Q(t)$ which are determined by the competition between the rate at which the cage grows, $\sim t^\alpha$, and the rate at which diffusion brings the particle to the cage walls, $\sim t^{1/2}$:
\begin{itemize}
\item[$({\rm i})$] For $\alpha < 1/2$, the cage grows more slowly than the typical displacement of a freely diffusing particle. This leads to a stretched exponential decay of the survival probability $Q(t) \sim \exp{[-a_\alpha\, t^{\gamma}]}$, with $\gamma = 1 - 2 \alpha$, and $a_\alpha$ is some constant.
\item[$({\rm ii})$] For $\alpha > 1/2$, the cage grows more rapidly than the particle is diffusing and in this case $Q(t)$ goes to a non zero limiting value $Q(t) \sim Q_\infty$ as $t \to \infty$. Below, this limiting value is computed exactly in the case $\alpha = 1$ using the Backward Fokker-Planck method. 
\item[$({\rm iii})$]{For the marginal situation $\alpha=1/2$ a richer behavior arises in which the competition between the cage length $L(t) = (A t)^{1/2}$ and the diffusion length $(D t)^{1/2}$ plays a crucial role. In this case $Q(t)$ decays algebraically at large time, $Q(t) \sim t^{-\theta}$ where $\theta$ depends on the ratio $A/D$ and is determined by the smallest solution on the positive real axis of the following equation 
\begin{eqnarray}
{\cal D}_{2 \theta}(\sqrt{A/2D}) + {\cal D}_{2 \theta}(-\sqrt{A/2D}) = 0 \;,
\end{eqnarray}
where ${\cal D}_\nu(x)$ is a parabolic cylinder function of order $\nu$. This problem was revisited in detail in Ref. \cite{Chicheportiche-Bouchaud} in connection with a generalization of the Kolmogorov-Smirnov ("goodness-of-fit") test. In particular, the prefactor of the power law decay of $Q(t)$ was computed. 
} 
\end{itemize}  

In the marginal situation, $L(t) = (A t)^{1/2}$, one can show, as expected, that $\theta \to 0$ as $A/D \to \infty$ \cite{Krapivsky-Redner-Cage}. On the other hand, when $L(t) \propto t^{\alpha}$ with $\alpha > 1/2$, $Q(t)$ remains finite when $t \to \infty$. One may then naturally wonder: what is the transition between certain death ($Q_\infty = 0$) and finite survival ($Q_\infty > 0$) ? One can actually show (see for instance \cite{Redner}) that $Q_\infty$ is finite provided that 
$L(t)$ grows faster than $L^*(t)$ given by
\begin{eqnarray}\label{log_iterated}
L^*(t) = \sqrt{4D t \left(\log \log t + \frac{3}{2} \log \log \log (t) + \cdots... \right)} \;,
\end{eqnarray}
where higher corrections involve higher iterations of the logarithms. The first term in (\ref{log_iterated}) is known under the name of Khintchine's law of iterated logarithm \cite{Khintchine,Feller}, which is usually stated as follows ($x(t)$ denoting Brownian motion with diffusion constant $D$)
\begin{eqnarray}\label{iterated_sup}
\limsup_{t \to \infty} \frac{x(t)}{\sqrt{4 D t \log \log t}} = 1 \;,
\end{eqnarray} 
where $\limsup_{t \to \infty} f(t) \equiv \lim_{t \to \infty} \sup_{u \geq t} f(u)$ for some function $f$.

Here we show that, for the case of linearly receding walls (corresponding to $\alpha = 1$), the asymptotic 
survival probability can be easily calculated using BFP methods 
\cite{BraySmith2007a}. Let the 
walls be located at positions $\pm (l+ct)$, i.e.\ the walls start at 
positions $\pm l$ and recede at speed $c$ (Fig. \ref{fig:expanding-cage}). 
The Brownian walker is initially 
located at position $x$ in the interval $(-l,l)$. The probability 
$Q(x,l,t)$ that the particle survives till time $t$ satisfies the identity
$Q(x,l,t) = \langle Q(x+\Delta x, l+c\Delta t, t - \Delta t)\rangle$, 
where the average is over the displacement $\Delta x$ that occurs in the 
infinitesimal time interval $\Delta t$. Expanding to first order in $\Delta t$,
using $\langle \Delta x \rangle=0$ and $\langle (\Delta x)^2\rangle 
= 2D\Delta t$, gives 
\begin{equation}  
\partial_t Q = D\,\partial_{xx}Q + c\,\partial_lQ\ .
\end{equation}
Here we focus on infinite-time survival probability, 
$Q(x,l,\infty)$. Introducing dimensionless variables $y = cx/D$, 
$\lambda=cl/D$, we obtain that $Q = \hat Q(y=cx/D,\lambda = cl/D)$ where $\hat Q$ satisfies
\begin{equation}
\partial_{yy}\hat Q + \partial_\lambda \hat Q=0\ ,
\end{equation}
with $-\lambda \le y \le \lambda$. The boundary conditions are $\hat Q(\pm\lambda,\lambda) = 0$ for all $\lambda$, 
and $\hat Q(y,\infty)=1$ for all $y$. A solution satisfying the differential 
equation and the boundary conditions can be written down by inspection:
\begin{equation}
\hat Q(y,\lambda) = \sum_{n=-\infty}^\infty (-1)^n \cosh(ny)e^{-n^2\lambda}\ .
\end{equation}
For a complete solution of this problem in any dimension (where the 
expanding cage becomes an expanding circle, sphere of hypersphere), it seems that
one has to apply the conventional (i.e.\ forward) Fokker-Planck equation
(see \cite{BraySmith2007b}). 

\subsubsection{Maximum excursion of a Brownian walker}
\begin{figure}[hh]
\centering
\includegraphics[width = 0.55\linewidth]{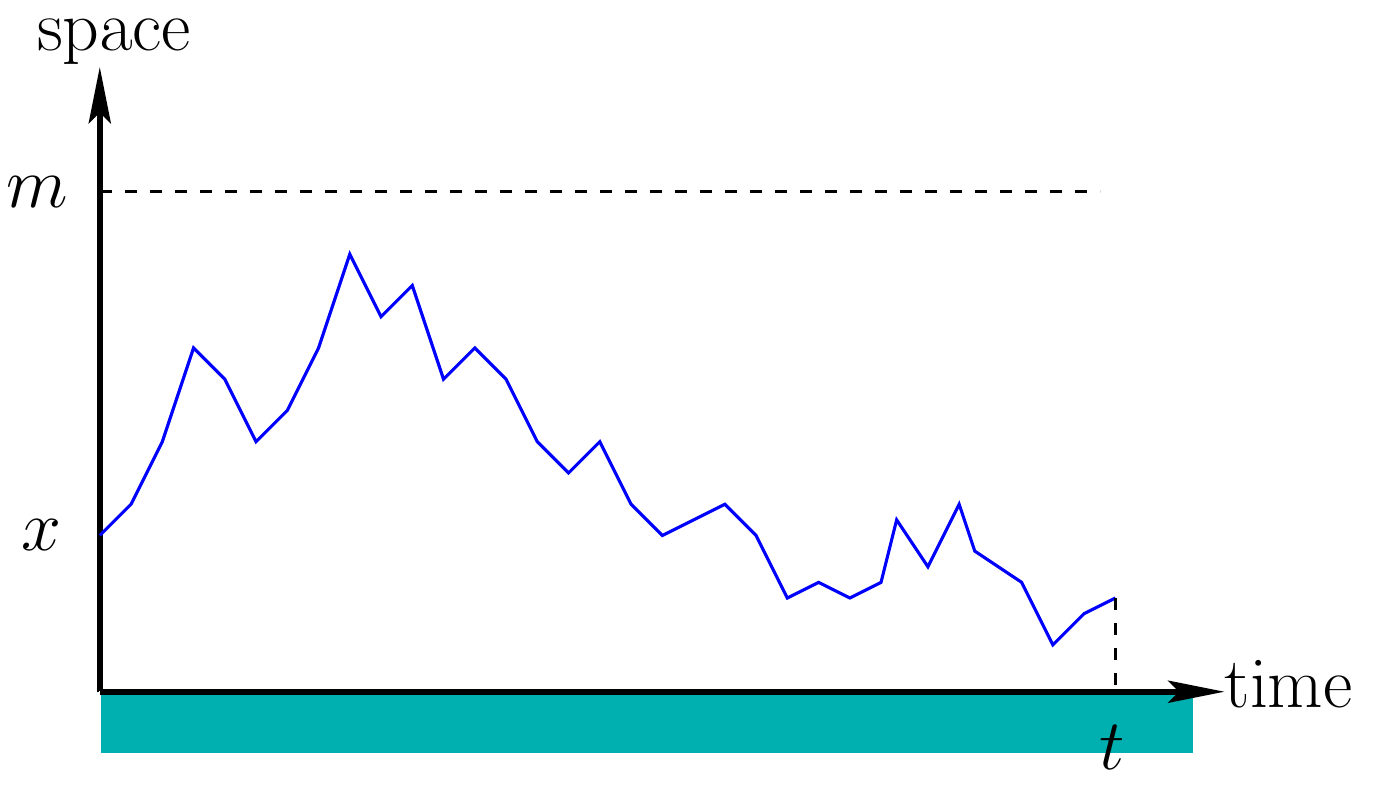}
\caption{Excursion of a random walk with an absorbing boundary at the origin, starting from $x$, and surviving up to time $t$. It has its maximum below the value $m$ and hence
contributes to $P_1(m,t|x)$ (see the text).}\label{fig:max_excursion}
\end{figure}

As a further example of the power of the BFP approach, we consider a Brownian 
walker moving on a semi-infinite line with an absorbing boundary at the 
origin. We compute the probability $P(m|x)$, that the maximum position   
reached by the walker, given that it started at $x$, is smaller than $m$ (see Fig. \ref{fig:max_excursion}). 
To obtain an equation for this quantity we first compute the probability, 
$P_1(m,t|x)$, that the maximum position reached by the walker 
up to time $t$ is smaller than $m$. 
 
By considering the changes that occur in the first time interval, 
$\Delta t$, we obtain $P_1(m,t|x) = 
\langle P_1(m,t-\Delta t|x+\Delta x )\rangle$, where the average is over 
the displacement $\Delta x$, that occurs in time $\Delta t$. Using, as usual, 
$\langle \Delta x \rangle=0$ and $\langle (\Delta x)^2\rangle = 2D\,\Delta t$, 
and expanding to first order in $\Delta t$, one obtains 
$\partial_t P_1 = D\partial_{xx}P_1$. The desired 
quantity $P(m|x)$ is just the infinite-time limit, 
$P(m|x) = P_1(m,\infty|x)$. It obeys Laplace's equation, $\partial_{xx}P=0$,
with the obvious boundary conditions $P(m|0)=1$, $P(m|m)=0$. The solution 
is $P(m|x) = 1-x/m$. The probability density, $R(m|x)$, of the largest 
excursion, $m$, made by the walker is given by 
\begin{equation}
R(m|x) = -\partial_mP(m|x) = (x/m^2)\,\theta(m-x)\,
\end{equation}
where $\theta(x)$ is the step function.
 
This approach is easily generalised to the calculation of the maximum 
position reached by any of $N$ independent random walkers, starting at 
positions $x_1,x_2,\ldots,x_n$, with an absorbing boundary at the origin. 
The probability $P(m|x_1,\ldots,x_N)$ that the maximum position of any 
of the walkers is less than $m$ is simply the product 
\begin{equation}
 P(m|x_1,\ldots,x_N) = \prod_{i=1}^N (1 - x_i/m) \ ,
\end{equation}
and the corresponding probability density of the maximum position is 
\begin{equation}
R(m|x_1,\ldots,x_N) = \frac{1}{m}\,\left(\sum_{i=1}^N\frac{x_i}{m-x_i}\right)
\,\left(\prod_{j=1}^N\left(1-\frac{x_j}{m}\right)\right)\,
\theta(m-{\rm max}[x_1,\ldots,x_N])\ .
\end{equation}

\subsubsection{Mean first-passage times of a Brownian walker}
Our final application of the BFP method to Brownian walker is to the computation of mean first 
passage times (or mean exit times from an interval or domain). This is a large 
field in its own right and we will just give a couple of simple examples 
to illustrate the method. A more complete treatment can be found, for example, 
in \cite{Redner}. 

Consider a Brownian walker, with diffusion constant $D$, moving in $d$ 
space dimensions. Let $T({\bf x})$ be the mean exit time from a domain 
${\cal D}$, given that the walker's initial position is ${\bf x}$. 
Let $Q({\bf x},t)$ be the probability that the walker has remained within 
${\cal D}$ up to time $t$. Then the probability distribution, 
$P_1({\bf x},t)$, of the time of first exit from ${\cal D}$ is given by 
$P_1(x,t) = -dQ({\bf x},t)/dt$. It follows that the mean exit time, 
$T({\bf x})$ is given by 
\begin{eqnarray}
T({\bf x}) &=& -\int_0^\infty t\,dt\,dQ({\bf x},t)/dt \nonumber \\
           &=& \int_0^\infty dt\,Q({\bf x},t)\ ,
\label{exit-time}
\end{eqnarray}
where the second line follows after integration by parts, 
using $Q({\bf x},0) = 1$ and $tQ({\bf x},t) \to 0$ for $t \to \infty$, 
the latter condition being necessary for a finite mean exit time. 
To derive a BFP equation for $T({\bf x})$, we act on both sides of 
Eq.\ (\ref{exit-time}) with the Laplacian operator, and use 
$D \nabla^2 Q = \partial Q/\partial t$, to obtain 
\begin{equation}
D \nabla^2 T = -1\ .
\end{equation}
with boundary condition $T({\bf x}) = 0$ for all ${\bf x}$ on the 
boundary of ${\cal D}$. 

As a simple example we can consider a 1-d Brownian walker moving in 
the interval $(0,L)$. The mean exit time from the interval, starting at $x$,
obeys the equation $D d^2T/dx^2 = -1$, with $T(0)=0=T(L)$. The solution is
$T(x) = x(L-x)/2D$. As a 2-d example, we may consider a Brownian walker 
moving in an infinite wedge, of opening angle $\alpha$,  whose vertex is 
at the origin of a plane polar coordinate system and whose edges are given 
by the lines $\theta = 0$ and $\theta = \alpha$ (see Fig. \ref{Fig_wedge}). The persistence properties
of this Brownian walker were discussed in section \ref{sec:wedge}. 
The mean exit time, $T(r,\theta)$, satisfies the equation 
\begin{equation}
D \left(\frac{\partial^2 T}{\partial r^2} + 
\frac{1}{r} \frac{\partial T}{\partial r}
+ \frac{1}{r^2} \frac{\partial^2 T}{\partial \theta^2}\right) = -1\ ,
\end{equation}
with boundary conditions $T(r,0) = 0 = T(r,\alpha)$. On dimensional 
grounds, the solution must have the form $T(r,\theta) = (r^2/D) f(\theta)$.
Then $f(\theta)$ satisfies the equation $f'' + 4f = -1$, with 
$f(0)=0=f(\alpha)$. The solution is 
\begin{equation}
T(r,\theta) = \frac{r^2}{2D}\,
\frac{\sin\theta\,\sin(\theta-\alpha)}{\cos\alpha}\ .
\end{equation}
Notice that $T$ diverges when $\alpha \to \pi/2$. This is to be expected, 
since the persistence exponent in Eq.\ (\ref{theta-wedge}) approaches unity in 
this limit. The mean exit time can also be calculated for a finite wedge 
(or ``pie wedge'') where the wedge domain is terminated by a boundary 
at $r=L$, and for an infinite cone. We refer the reader to \cite{Redner} 
for more details.

\subsection{The Random acceleration process: the simplest non-Markovian process}\label{subsection:RAP}
The random acceleration process is defined by the equation $\ddot{x} =
\eta(t)$,  where $\eta(t)$  is  Gaussian white  noise  as usual. Using a discretization of the second time derivative, 
this equation of motion reads $x(t) \approx 2 x(t-\Delta t) - x(t-2 \Delta t) + (\Delta t)^2 \eta(t)$. Therefore
it is clear that the position at time $t$, $x(t)$ depends not only on the position at the time right before $x(t-\Delta t)$ but also on $x(t-2 \Delta t)$. As a consequence,
the random acceleration process is a non-Markovian process. However, if one considers the dynamics of the particle in the two-dimensional phase space
$(x,v=\dot x)$, the equation of motion reads:  $\dot{v}=\eta(t)$,
$\dot{x}=v$. This implies that this two-dimensional process is Markovian, and the BFP can thus be 
applied, as in Eq. (\ref{BFPE1}), to compute the survival probability $Q(x,v,t)$ for the random acceleration process with 
an  absorbing  boundary at the origin. Here,  $x$  and  $v$  are  the {\em  initial}
position and  velocity of the particle. One thus  
obtains $Q(x,v,t) = \langle
Q(x+\Delta  x, v+  \Delta  v, t-\Delta  t)  \rangle$, for  infinitesimal
$\Delta x$, $\Delta  v$ and $\Delta t$, where the  average is over the
displacement  $\Delta x$  and  velocity increment  $\Delta  v$ for  an
initial time interval $\Delta t$. Expanding to second order in $\Delta
v$ and  first order in $\Delta  x$, using $\langle \Delta  v \rangle =
0$,  $\langle (\Delta  v)^2 \rangle  = 2D\,\Delta  t$, and  $\Delta  x =
v\Delta t$ gives the BFP equation
\begin{equation}
\partial_t Q = D \partial_{vv}Q + v\partial_x Q\ .
\label{RA}
\end{equation} 
The  absorbing  boundary at  $x=0$  leads  to  the boundary  condition
$Q(0,v,t) =  0$ for $v<0$.  The initial condition is  $Q(x,v,0)=1$ for
$x>0$.

It   is  convenient   to   work  with   the  dimensionless   variables
$(x^{2/3}/D^{1/3}t)$ and $(v^3/Dx)$.  For asymptotically large $t$, we
expect a power-law time dependence of the form
\begin{equation}
Q(x,v,t)      \sim      \left(\frac{x^{2/3}}{D^{1/3}t}\right)^\theta\,
F\left(\frac{v^3}{Dx}\right)\ ,
\label{scalingform}
\end{equation}
where $\theta$  is the persistence  exponent as usual.  Inserting this
form into Eq.\  (\ref{RA}) we find that, as  usual, the left-hand side
becomes negligible at large $t$, and the function $F(z)$ satisfies the
ODE
\begin{equation}
zF''(z) + (2/3 - z/9)F'(z) + (2\theta/27)F(z) = 0\ .
\label{Kummerrandomacc}
\end{equation}
Changing  variables   to  $u=z/9$,  this   equation  becomes  Kummer's
equation,  with independent  solutions $M(-2\theta/3,  2/3,  z/9)$ and
$U(-2\theta/3,   2/3,   z/9)$.   The function $M(a,b,x)$   diverges
exponentially  for $x \to  \infty$, and  is therefore  unphysical. The
desired solution  is, therefore,  $F(z) = A\,U(-2\theta/3,  2/3, z/9)$
where $A$ is an arbitrary constant.  The boundary condition, $Q(0,v,t)
=  0$  for  $v<0$,  implies  that  $F(z)$ should  vanish  for  $z  \to
-\infty$. In this limit,
\begin{equation}
F(z)  \propto \sin[\pi(1-4\theta)/6]\,(-z)^{2\theta/3}\  , \  \  z \to
-\infty\ .
\end{equation}
The condition that the prefactor of $(-z)^{2\theta/3}$ vanishes gives
\begin{equation}
\theta = 1/4 \;,
\end{equation}
(recalling  that  the  smallest  positive  value of  $\theta$  is  the
physical value). The result $\theta=1/4$ for the random acceleration 
process is in accord with previous rigorous 
results \cite{McKean,Goldman,Sinai1992a,Sinai1992b,Burkhardt1993}.

To complete this section, we discuss two problems in which the same
BFP machinery can  be exploited to obtain exact  results. The first of
these is the random  acceleration process with {\em partial survival}.
Partial survival  means that, on reaching the  absorbing boundary, the
particle  is  absorbed  with   probability  $1-p$  and  survives  with
probability $p$.   We then expect  \cite{MajumdarBrayPartial} that the
persistence  exponent  for the  survival  probability  will become  a
function of $p$,  i.e.\ $\theta = \theta(p)$. The  concept of partial
survival will  be discussed  in more general  terms in section \ref{subsection:partial}, but  the  random  acceleration  process  provides  a  simple
introduction to it.

\subsubsection{Random acceleration process with partial survival}\label{subsection:RAP_partial}
For  the process with  survival probability  $p$ associated  with each
crossing of the boundary $x=0$,  the BFP equation takes the same form,
Eq.\  (\ref{RA}),   as  before,   with  the  same   initial  condition
$Q(x,v,0)=1$.  However the boundary condition now becomes
\begin{equation}
Q(0,-v,t) = pQ(0,v,t), \ \ \ v>0\ .
\label{boundarycond}
\end{equation}
 As   before   the   physical   solution   has  the   form   of   Eq.\
(\ref{scalingform}),  with  $F(z)$   given  by  the  Kummer  function,
$F(z)=B U(-2\theta/3,2/3,z/9)$, where $B$ is an arbitrary constant.  To exploit the boundary condition, Eq.\
(\ref{boundarycond}), we need the behaviour of $F(z)$ in the limits $z
\to \pm \infty$. The  corresponding asymptotic behaviour of the Kummer
function~is
\begin{eqnarray}
U\left(-\frac{2\theta}{3},  \frac{2}{3},  \frac{z}{9}\right)  
\sim
\begin{cases}
& \left(\dfrac{z}{9}\right)^{\frac{2\theta}{3}}, \  \ z \to  \infty, \\
& \\
& \dfrac{\sin\left[\frac{\pi}{6}(1-4\theta)\right]}{\sin[\frac{\pi}{6}]}
\left(-\frac{z}{9}\right)^{\frac{2\theta}{3}}, z \to -\infty \;.
\end{cases}
\end{eqnarray} 
These  results, combined  with Eq.\  (\ref{scalingform}), lead  to the
following large-$t$ results for $x=0$ and $v>0$,
\begin{eqnarray}
Q(0,v,t) & = & C\,\left(\frac{v^2}{Dt}\right)^\theta, \\ Q(0,-v,t) & =
&         C\,         \frac{\sin\left[\frac{\pi}{6}(1-4\theta)\right]}
{\sin[\frac{\pi}{6}]}\ \left(\frac{v^2}{Dt}\right)^\theta\ ,
\end{eqnarray}
where  $C$ is  a constant.  Inserting  these forms  into the  boundary
condition (\ref{boundarycond}) gives
\begin{equation}
\theta(p)       =       \frac{1}{4}\left[1       -       \frac{6}{\pi}
\sin^{-1}\left(\frac{p}{2}\right)\right]\ ,
\label{randaccpartialsurvival}
\end{equation}
a result first obtained independently by Burkhardt~\cite{BurkhardtPSa,BurkhardtPSb}
and de Smedt {\it et. al.}~\cite{deSmedtPS}.

\subsubsection{The `windy cliff'}\label{subsection:windy}
\begin{figure}[h]
\centering
\includegraphics[width=0.55\linewidth]{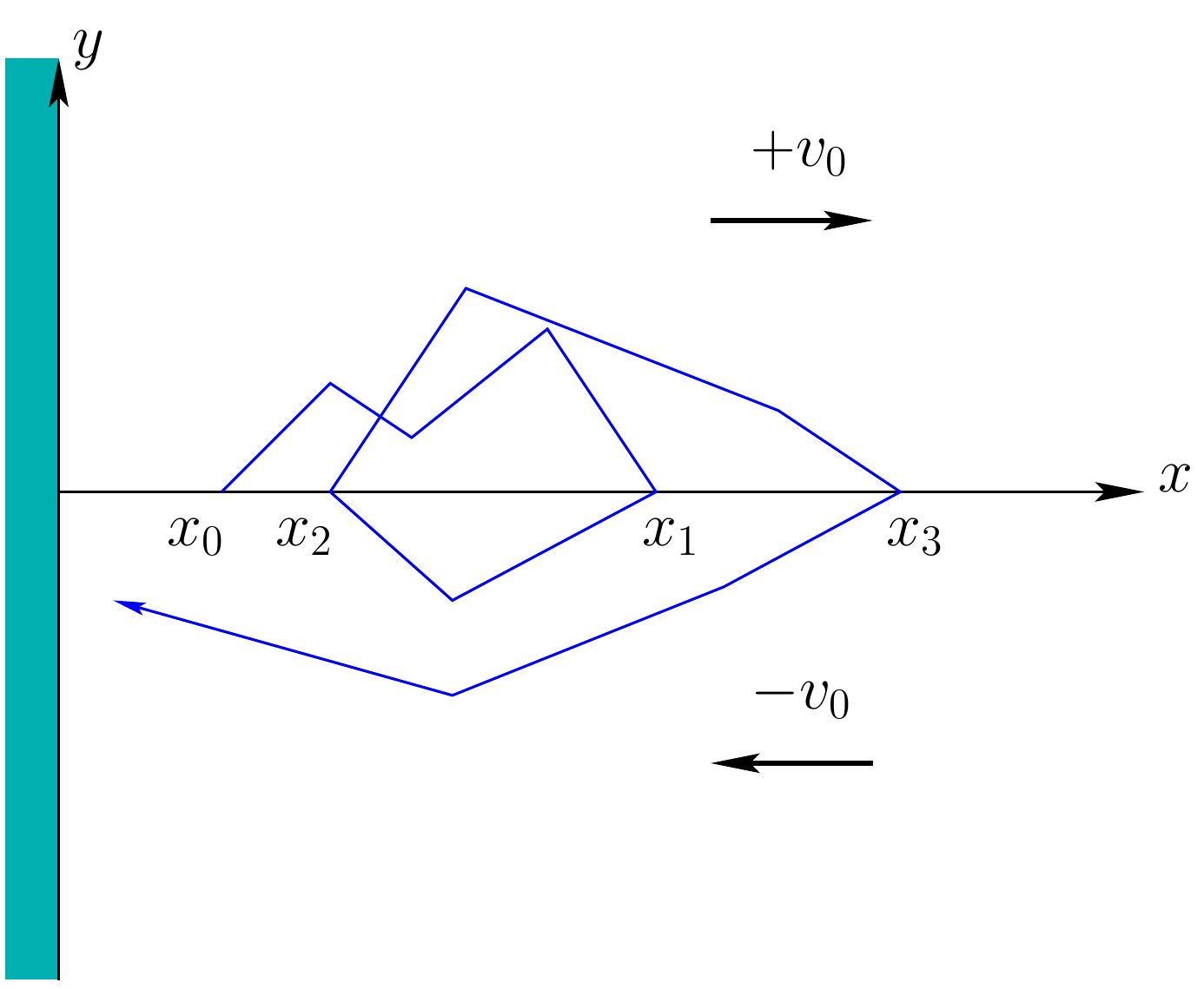}
\caption{Wind shear in two dimensions with a cliff -- an absorbing boundary condition (possibly partially absorbing) -- in $x=0$.}\label{fig:windy-cliff}
\end{figure}

This section deals  with a class of models  that generalise the random
acceleration process. Recall that  this latter process, defined by the
Langevin equation  $\ddot{x}=\eta(t)$, can  be represented by  the two
equations $\dot{v} = \eta(t)$, $\dot{x}  = v$, where $v$ has a natural
interpretation as  a ``velocity''.  We  can, however, regard  the same
equations   as  describing   the   motion  of   a   particle  in   the
two-dimensional space $(x,y)$, with $\dot{y} = \eta(t)$ and $\dot{x} =
y$. In this interpretation, the particle executes a random walk in the
$y$  direction  while   subject  to  a  uniform  shear   flow  in  the
$x$-direction.   The model then  lends itself  to some  rather natural
generalisations,  in  which  the  ``shear-flow''  is  modified.   Most
generally, one  can replace $\dot{x}=y$ by  $\dot{x}=f(y)$. Redner and
Krapivsky   \cite{SidPaulWindyCliff}   introduced   the   model   with
$f(y)=v_0\,{\rm sgn}(y)$, and with  an absorbing boundary at $x=0$, in
the context  of diffusion  near a `windy  cliff' (Fig. \ref{fig:windy-cliff}). 
They found  that the
same exponent, $\theta = 1/4$,  also seemed to describe this model. In
fact  one can show  \cite{BrayGonos}, as we shall see, that  $\theta=1/4$ 
holds  for any such model provided $f(y)$ is  an odd function, with $f(y)$ 
taking the same sign as $y$.

In the following we do not restrict ourselves to the class of odd functions 
$f(y)$, but consider the more general class 
$f(y) = v_{\pm} {\rm sgn}(y)\,|y|^\alpha$, where the upper (lower) sign 
refers to the cases $y>0$ and $y<0$ respectively. These models are thus 
defined by the equations of motion
\begin{eqnarray}
\dot{y} & = & \eta(t)\ , \\
\dot{x} & = & v_{\pm} {\rm sgn}(y)\,|y|^\alpha\ \;,
\end{eqnarray}
where $\langle \eta(t)\eta(t')\rangle = 2D\,\delta(t-t')$ as usual. The 
corresponding BFP reads
\begin{equation}
\partial_t Q = D\partial_{yy}Q \pm v_\pm(\pm y)^\alpha \partial_xQ\ \;.
\label{BFPcliff}
\end{equation}
We include a partially absorbing boundary at $x=0$. When the particle 
crosses this boundary, it survives with probability $p$. The corresponding 
boundary conditions are
\begin{eqnarray}
Q(0+,-y,t) &=& p\,\tilde{Q}(0+,y,t),\ \ \ y>0\ ,\nonumber\\
\tilde{Q}(0+,-y,t) &=& p\,Q(0+,y,t),\ \ \ y>0\ ,
\label{boundarycondcliff}
\end{eqnarray}
where $\tilde{Q}(x,y,t)$ is the survival probability for a model in which 
$v_+$ and $v_-$ are interchanged.  It is clear that $Q$ and $\tilde{Q}$ are 
described by the same value of $\theta$. The initial condition is 
$Q(x,y,0)$ = 1 = $\tilde{Q}(x,y,0)$.

Using dimensional analysis, as before, we expect $Q(x,y,t)$ to have the
large-$t$ form
\begin{equation}
Q(x,y,t) \sim \left(\frac{x^{2\beta}}{t}\right)^\theta
\,F_\pm\left(\pm \frac{v_\pm(\pm y)^{1/\beta}}{Dx}\right)\ ,
\label{cliffscaling}
\end{equation}
where
\begin{equation}
\beta = \frac{1}{2+\alpha}\ ,
\end{equation}
and a similar form for $\tilde{Q}(x,y,t)$ with scaling functions 
$\tilde{F}_{\pm}$. Here the subscripts $\pm$ refer, as before, to the regions
$y>0$ and $y<0$. 

Inserting the form (\ref{cliffscaling}) into Eq.\,(\ref{BFPcliff}) we see 
that the term $\partial_tQ$ is of order $t^{-(1+\theta)}$, which is 
subdominant for large $t$ and can be dropped. The remaining terms give
\begin{equation}
zF_{\pm}''(z) + (1-\beta-\beta^2z)F_\pm'(z) + 2\beta^3\theta\,F_\pm(z)=0\ .
\label{Kummercliff}
\end{equation}
Before continuing we note that the random acceleration process corresponds 
to the special case $\beta = 1/3$ (i.e.\ $\alpha=1$), and one can check 
that Eq.\ (\ref{Kummercliff}) reduces to Eq.\ (\ref{Kummerrandomacc}) in this 
case.

Expressed  in  terms of  the  variable  $u=\beta^2  z$, this  equation
reduces   to    Kummer's   equation,   with    independent   solutions
$M(-2\beta\theta,\beta^2  z)$  and  $U(-2\beta\theta,\beta^2 z)$.  The
first of these functions diverges exponentially for $z \to \infty$, so
it must be  absent from the linear combination  for $y>0$. The general
solution involves the $U$ function, with coefficient $A$, for $y>0$ and
a linear  combination for $U$ and  $M$, with coefficients  $B$ and $C$
respectively, for  $y<0$. Relations between the  coefficient $A$, $B$,
and $C$,  and the  corresponding tilded variables,  can be  obtained by
imposing the boundary conditions, Eqs.\,(\ref{boundarycondcliff}), and 
requiring continuity of $Q$, $\partial_yQ$, $\tilde{Q}$ and 
$\partial_y\tilde{Q}$ at $y=0$. These boundary and continuity conditions 
lead to a consistency condition on $\theta$, leading to the final result
\cite{majumdarandbray}.
\begin{equation}
\theta(p) = \frac{1}{4} - \frac{1}{2\pi\beta}
\sin^{-1}\left(\sqrt{\delta}\sin\left(\frac{\pi\beta}{2}\right)\right)\ ,
\label{theta(p)result}
\end{equation}
where
\begin{equation}
\delta = \frac{2p^2\cos^2\left(\frac{\pi\beta}{2}\right)
+ 2 \sinh^2\left(\frac{1}{2}\ln\gamma\right)}
{\cos(\pi\beta) + \cosh(\ln\gamma)}\ ,
\label{finalresult}
\end{equation}
and
\begin{equation}
\gamma = \left(\frac{v_+}{v_-}\right)^\beta\ , 
\end{equation}
and we recall that $\beta = 1/(2+\alpha)$.

This very general result has a number of interesting special cases: 

(i) $p=1$: In this case $\delta=1$ and $\theta(1)=0$. This is exactly as 
expected, since there is no absorption for $p=1$. \\

(ii) $p=0$: In this case, corresponding to absorption at first crossing, 
$\delta = 
(\sqrt{\gamma}-1/\sqrt{\gamma})^2/[\gamma + 1/\gamma + 2\cos(\pi\beta)]$.
Inserting this into Eq.\ (\ref{finalresult}) one recovers the result 
of Bray and Gonos \cite{BrayGonos}:
\begin{equation}
\theta(0) = \frac{1}{4} - \frac{1}{2\pi\beta}\tan^{-1}
\left[\frac{\gamma - 1}{\gamma + 1}
\tan\left(\frac{\pi\beta}{2}\right)\right]\ .
\label{p=0}
\end{equation}

(iii) $\gamma=1$. This gives $\delta = p^2$, and
\begin{equation} 
\theta(p)       =        \frac{1}{4}       -       \frac{1}{2\pi\beta}
\sin^{-1}\left(p\sin\left(\frac{\pi\beta}{2}\right)\right) \;.
\label{gamma=1}
\end{equation}
For  the special  case  $\beta=1/3$, corresponding  to $\alpha=1$,  we
recover the result (\ref{randaccpartialsurvival}).

The $p=0$  result, Eq.\ (\ref{p=0}), has the  interesting feature that
for  $\gamma=1$ one obtains  $\theta =  1/4$, independently  of $\beta$.
Note  that $\gamma=1$ corresponds  to the  cases where  the amplitudes
$v_+$ and  $v_-$ are equal. In  such cases, the function  $f(y)$ is an
odd function. Bray and Gonos  have argued that this result generalises
to any  odd function $f(y)$ for  which $f(y)$ has  everywhere the same
sign   as  $y$  (a   result  conjectured   by  Krapivsky   and  Redner
\cite{SidPaulWindyCliff}).   We associate  successive crossings  of the
$x$-axis by the particle with steps of a random walk, i.e.\ successive
crossings at  $x=x_n$ and $x=x_{n+1}$  correspond to a step  of length
$x_{n+1}-x_n$ in an effective  random walk along the $x$-axis (Fig.~\ref{fig:windy-cliff}). Clearly
the distribution of step sizes is continuous and symmetric around zero.
According to the Sparre Andersen theorem  \cite{SparreAndersen}, for
any  such  random walk,  with  an  absorbing  boundary at  $x=0$,  the
probability that  the walker  has not crossed  the boundary  after $N$
steps  decreases as  $N^{-1/2}$ for  large  $N$. Since  the number  of
crossings,  $N$,  of the  $x$  axis  in time  $t$  scales  as $N  \sim
t^{1/2}$,  the survival probability  decays as  $Q \sim  N^{-1/2} \sim
t^{-1/4}$.  If $v_+  \ne v_-$  the distribution  of step  sizes  is no
longer symmetric,  the Sparre Andersen theorem no  longer applies, and
$\theta \ne 1/4$. The general $\gamma=1$ result, (\ref{gamma=1}) shows 
that $\theta$ does depend on $\beta$ for all $p$ in (0,1). It is only at 
the endpoints, $p=0$ ($\theta=1/2$) and $p=1$ ($\theta=0$) that $\theta$ 
becomes independent of $\beta$.

A final point worth noting is that these processes are (apart from the
case $\alpha  =1$), {\em  nonlinear}. Linear stochastic  processes are
Gaussian,  which  implies that  all  their  properties are  determined
implicitly  by  the   two-time  correlation  function.  The  processes
discussed above  are non-Gaussian, yet it  is still possible to
compute their persistence exponents exactly.

\subsection{Higher-order processes}\label{subsection:bfp_higher_order}

Until now, we have discussed the persistence properties of the Brownian motion (in section \ref{section:RW}), described by $dx(t)/dt = \eta(t)$ and the random acceleration process described by $d^2x(t)/dt^2 = \eta(t)$ (in section \ref{subsection:RAP}), where $\eta(t)$ is a Gaussian white noise with zero mean and a correlator $\langle \eta(t)\eta(t')\rangle =\delta(t-t')$. Natural extension of these processes are "higher-order" processes, governed by the following equation of motion
\begin{equation}
\frac{d^n x}{dt^n } =\eta(t) \;,
\label{hop.1a}
\end{equation}
with $n>2$. Such processes were introduced first in the mathematics and statistics literature~\cite{Shepp66,Wahba78}. 
In the physics literature, the persistence properties of this process for general $n$ was initiated
in Ref.~\cite{MajumdarSireBrayCornell}.
We will see later in section \ref{section:interfaces} that these processes, with higher values of $n$, appear naturally in the steady state measure of a class
of fluctuating linear interfaces~\cite{MB01}. No exact results are known for the persistence exponents of these higher-order 
processes in (\ref{hop.1a}) with $n>2$, although approximate 
results, using independent interval approximation (to be discussed in section~\ref{subsection:higher_order}), are known with reasonable precision for small values of $n$: $\theta (3) \approx 0.2202$, $\theta(4) \approx 0.2096$ and $\theta(5) 
\approx 0.2042$ \cite{EhrhardtMajumdarBray04}. We also refer the reader to Ref. \cite{SM01} where another approach, based on L\'evy flights, was proposed to approximate the exponent $\theta(n)$ (albeit less accurate than the independent interval approximation).  

We end up this section by mentioning that the study of integrated processes, including integrated L\'evy processes, has recently attracted much attention in the mathematics literature \cite{DemboGao} and we refer the reader to Ref. \cite{AS12} for a review on them.

\section{Persistence of multi-particle systems}\label{section:RD}

Up to now we have considered single-particle systems, both Markovian -- the Brownian walker in section \ref{section:RW} -- and non-Markovian -- the random acceleration process in section \ref{subsection:RAP} and its generalizations to higher-order processes in section \ref{subsection:bfp_higher_order}. Here we generalize the notion
of persistence to multi-particle systems, with and without interactions. The most natural generalization of persistence is the probability that two independent one-dimensional Brownian walkers $x_1$ and $x_2$ do not intersect up to time $t$. 
However, by considering the relative coordinate $y=x_2 -x_1$, one is then back to a single-particle persistence problem for the stochastic
process $y$, which is itself a Brownian motion. Hence the simplest nontrivial generalization of persistence concerns the case
of three Brownian particles, which we discuss below.      


\subsection{Three-walker problems}\label{subsection:three_walkers}

The first interesting generalisation of persistence to multi-particle systems is provided by a reaction-diffusion process. 
The field of reaction-diffusion is a vast one, and here we will 
focus specifically on first-passage aspects that can be addressed using 
the techniques that we have developed earlier in this article.   
Later, we will consider the reactions $A + A \to 0$ and $A + A \to A$ 
in the context of the persistence properties of the one-dimensional Ising 
model and infinite-state Potts model respectively. 
In that context  the reactants $A$ correspond to domain walls (see Fig.~\ref{fig:domain_walls_ising}), and we 
will be interested primarily in the ``site-persistence'' aspects, 
i.e.\ the fraction of sites that had not yet been traversed by a wall. 

In this part of the paper we present some exact results for the persistence 
properties of three Brownian walkers, building on the BFP methods we 
developed earlier. In the context of the $A+A\to A$ problem, we are 
interested in the survival of the {\em central} particle 
of the three, i.e.\ the probability that neither of the two outer particles 
has touched the central particle up to time $t$. The methods employed 
readily generalise, however, to calculating the survival probabilities 
of the outer particles. For generality, we allow arbitrary values of the 
three diffusion constants. The equations of motion are 
$\dot{x_i} = \eta_i(t)$ ($i=1,2,3$) with $\langle \eta_i(t)\eta_j(t') \rangle 
= 2D_i\delta_{ij}\,\delta(t-t')$, and $x_1<x_2<x_3$. We introduce the 
relative coordinates $y_1=x_2-x_1$ and $y_2=x_3-x_2$. The corresponding 
equations of motion are $\dot{y}_1 = \eta_2(t)-\eta_1(t) = \xi_1(t)$, 
and  $\dot{y}_2 = \eta_3(t)-\eta_2(t) = \xi_2(t)$. Then 
\begin{eqnarray}
\langle \xi_1(t) \xi_1(t') \rangle &=& 2(D_1+D_2)\delta(t-t')\ ,\nonumber \\
\langle \xi_2(t) \xi_2(t') \rangle &=& 2(D_2+D_3)\delta(t-t')\ ,\nonumber \\ 
\langle \xi_1(t) \xi_2(t') \rangle &=& -2D_2 \delta(t-t')\ .
\end{eqnarray}
The problem is now equivalent to that of a single particle, with coordinates 
$(y_1,y_2)$, moving in the two-dimensional quadrant $y_1>0$, $y_2>0$, with 
absorbing boundaries at $y_1=0$ and $y_2=0$. The BFP equation reads
\begin{equation}
\partial_t Q = (D_1+D_2)\partial_{y_1y_1}Q + (D_2+D_3)\partial_{y_2y_2}Q
-2D_2 \partial_{y_1}\partial_{y_2}Q,
\end{equation}
with boundary conditions $Q(y_1,0)=0=Q(0,y_2)$. The problem is then solved 
by a coordinate transformation that makes the Laplacian isotropic but 
changes the angle between the absorbing boundaries. 

The first step is to rescale the coordinates according to 
$y_1 = u/\sqrt{D_1+D_2}$, $y_2 = v/\sqrt{D_2 + D_3}$. The absorbing 
boundaries are located at $u=0$ and $v=0$, and the BFP equation reads
\begin{equation} 
\partial_t Q = \partial_{uu}Q + \partial_{vv}Q 
- 2\gamma\partial_u \partial v Q\ ,
\end{equation}
where
\begin{equation}
\gamma = D_2/\sqrt{(D_1+D_2)(D_2+D_3)}\ .
\label{gamma}
\end{equation}
A second change of variable $w=(u+v)/\sqrt{2}$, $z=(u-v)/\sqrt{2}$ 
gives the equation
\begin{equation}
\partial_t Q = (1-\gamma)\partial_{ww}Q + (1+\gamma)\partial_{zz}Q\ ,
\end{equation}
with absorbing boundaries at $w=\pm z$, such that the  problem is now 
defined in the quadrant $w>0$, $|z| \le w$. A final change of variable 
$w \to w\sqrt{1-\gamma}$, $z \to z\sqrt{1+\gamma}$ leaves an isotropic 
model, defined within an absorbing wedge of angle $\alpha$ given by
$\alpha = 2\tan^{-1}(\sqrt{(1-\gamma)/(1+\gamma)})$. Recalling, from section \ref{sec:wedge}, the general result for such a wedge (\ref{theta-wedge}), gives the persistence 
exponent as
\begin{equation}
\theta =  \frac{\pi}{2\alpha} 
= \frac{\pi}{4\tan^{-1}\left(\sqrt{\frac{1-\gamma}{1+\gamma}}\right)}\ ,
\label{threewalkers}
\end{equation}
where we recall that $\gamma$ is given by Eq.\ (\ref{gamma}). 
This result was first obtained by a slightly longer route in 
\cite{FisherGelfand}. For the case $D_1=D_2=D_3$, corresponding to 
$\gamma = 1/2$ and hence $\alpha = \pi/3$, we obtain the result $\theta_w = 3/2$ 
for the survival probability of the central walker. The persistence 
exponent for the left or right walkers can be calculated in a similar 
fashion.  

Some comments are in order here: (i) For more than three walkers on a line, 
the persistence exponent for a specified walker (e.g. ``second from the 
left'') has not been calculated exactly. One can map the problem to the 
motion of a single particle inside a three-dimensional ``wedge'', with 
absorbing boundaries on the faces of a tetrahedral wedge, but the persistence 
problem for the wedge (as opposed to the cone studied earlier) is so far 
unsolved; (ii) the case of $N$ mutually annihilating walkers with the same 
diffusion constants maps onto the ``vicious walker'' model \cite{FisherHuse}. 
The probability that no pair of walkers have met up to time $t$ decays 
as $t^{-\theta}$ with $\theta = N(N-1)/4$ \cite{FisherHuse}. The case $N=3$ 
reproduces the result $\theta=3/2$ obtained above for the $A + A \to A$ 
process, as expected.  We now take a short detour to show how various 
vicious walker problems can be solved rather neatly by exploiting a mapping
introduced earlier in this article.

\subsection{Persistence exponents for vicious walkers}
\begin{figure}[h]
\centering
\includegraphics[width = 0.55\linewidth]{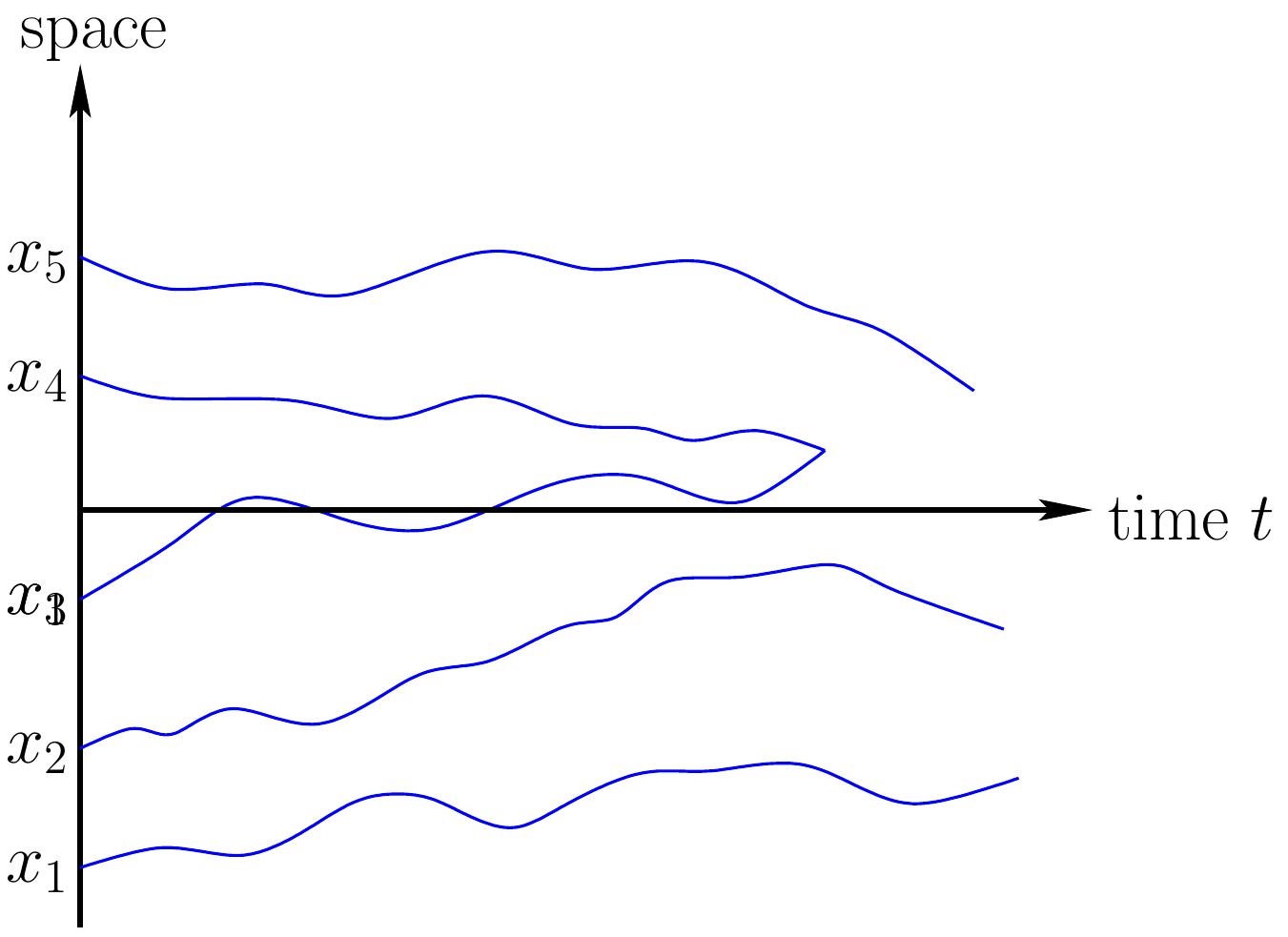}
\caption{A configuration of $N=5$ vicious walkers: when the trajectories of two vicious walkers meet each other, both of them are killed.}\label{fig:vicious}
\end{figure}

Consider $N$ vicious walkers moving on a line, with the same diffusion 
constant $D$ (see Fig. \ref{fig:vicious}). The walker locations are $x_n(t)$, $n=1,\ldots,N$, which 
satisfy the Langevin equations $\dot{x}_n(t) = \eta_n(t)$, with 
$\langle \eta_n(t)\, \eta_m(t') \rangle = 2D\,\delta_{nm}\delta(t-t')$. 

We introduce the following mapping \cite{BrayWinkler}: 
\begin{equation}
X_n = \frac{x_n}{\sqrt{2Dt}},\ \ \ t=t_0\,e^T\ .
\end{equation}
Under this mapping, the Langevin equations transform to 
\begin{equation}
\frac{dX_n}{dT} = -\frac{1}{2}\,X_n(T) + \xi_n(T),\ \ n=1,\ldots,N
\end{equation}
where $\xi_n(T)= (t_0/2D)^{1/2}e^{T/2}\eta_n(t_0e^T)$ is Gaussian white 
noise with mean zero and correlator
\begin{equation}
\langle \xi_n(T)\xi_m(T') \rangle = \delta_{n,m}\delta(T-T')\ .
\end{equation}
The corresponding BFP equation is 
\begin{equation}
\frac{\partial{Q({\bf X},T)}}{\partial T} = \frac{1}{2}\sum_{n=1}^N 
\frac{\partial^2}{\partial X_n^2}Q({\bf X},T) 
-\frac{1}{2}\sum_{n=1}^NX_n \frac{\partial Q({\bf X},T)}{\partial X_n},
\end{equation}
where ${\bf X} = (X_1,\ldots,X_n)$.

In  the new variables,  the problem  is identical  to a  set of  $N$ particles
moving  in a  harmonic  oscillator potential  $V({\bf  X}) =  \frac{1}{2}a{\bf
X}^2$, under the influence of  Gaussian white noise, with $a=1/2$ and $D=1/2$.
The  corresponding relaxation  rates for  a single  particle are  $\lambda_n =
n/2$,  where  $n=0,1,2,\ldots$. Due  to  the  vicious  walker constraint,  the
$N$-particle  probability  distribution, conditioned on all particles 
surviving,  has  to be  antisymmetric  under  the interchange  of  any pair  
of  coordinates.  The  particles therefore  have  a fermionic character, and 
a Pauli exclusion principle operates. The total decay rate for the 
$N$-particle survival probability is therefore 
$\lambda_{\rm TOT}= \sum_{n=0}^{N-1}\frac{n}{2} = N(N-1)/4$, i.e. \ the survival 
probability decays as $Q_N(T) \sim \exp(-\frac{N(N-1)}{4}\,T)$. In the 
original time variable, $t= t_0 e^T$, therefore, the survival probability 
decays as $t^{-N(N-1)/4}$, i.e.\ the persistence exponent is 
\begin{equation}
\theta=\frac{N(N-1)}{4}\ .
\end{equation}
For $N=3$ one obtains $\theta = 3/2$, in agreement with our earlier result.

The method readily generalises to $N$ vicious walkers moving on a 
semi-infinite line with either absorbing or reflecting boundary conditions 
at the origin. For absorbing boundaries, only the odd harmonic oscillator 
states, with relaxation rates $\lambda_n = n/2$ for $n$ odd fit the boundary condition, 
leading to $\lambda_{\rm TOT} = \sum_{k=1}^N (2k-1)/2 = N^2/2$, so the 
persistence  exponent for this case is 
\begin{equation}
\theta_{\rm Abs} = \frac{N^2}{2}\ ,
\label{thetaabs}
\end{equation}
a result first obtained by Krattenthaler et al.\ \cite{Krattenthaler}. 
The special case $N=2$ can be checked against the three-walker result 
(\ref{threewalkers}) for the case $D_1=D_2$, with $D_3=0$ corresponding 
to an absorbing wall. This case corresponds to $\gamma=1/\sqrt{2}$ 
in Eq. (\ref{threewalkers}), which indeed gives $\theta=2$ in agreement 
with the $N=2$ case of (\ref{thetaabs}).

For the case of a reflecting boundary at the origin, only the even 
oscillator states fit the boundary conditions, with relaxation rates 
$\lambda_n = n/2$, with $n$ even. For $N$ particles, 
$\lambda_{\rm TOT} = \sum_{k=0}^{N-1} k = N(N-1)/2$, leading to 
persistence exponent
\begin{equation}
\theta_{\rm Ref} = \frac{N(N-1)}{2}\ ,
\end{equation}
a result derived independently using a different method by Katori and 
Tanemura~\cite{KatoriTanemura}, who also discussed the relation between vicious walkers and random matrix theory.

The computation of the survival probability $Q(t)$ for $N$ vicious walkers has been generalized in several directions. First, the case
of Brownian vicious walkers in higher dimensions $d>1$ has been investigated in \cite{MB93a,MB93b,CK03}. In $d>1$ the mapping between vicious walkers and fermions does not apply and no exact results exist in this case. It is however possible to use renormalization group techniques to obtain the large time behavior of $Q(t)$ depending on the dimension $d$~\cite{MB93a,MB93b,CK03}
\begin{eqnarray}\label{eq:vicious_d}
Q(t) \sim_{t \to \infty}
\begin{cases}
&t^{- \theta} \;, \; 1 \leq d < 2 \;, \\
&(\log{t})^{-\bar{\theta}} \;, \; d = 2 \;, \\
& {\rm const.} \;, \; d > 2 \;,
\end{cases}
\end{eqnarray}
where the persistence exponent $\theta$ has been computed up to two-loop order in $\epsilon = 2 -d$
\begin{eqnarray}\label{eq:exponent_alpha}
\theta = \frac{1}{4} N(N-1) \epsilon + \frac{1}{4} N(N-1)(N-2) \epsilon^2 + {\cal O}(\epsilon^3) \;,
\end{eqnarray}
and where $\bar \theta$ can be computed exactly
\begin{eqnarray}\label{eq:exponent_alphabar}
\bar \theta = \frac{1}{2} N(N-1) \;.
\end{eqnarray}
These results (\ref{eq:exponent_alpha}, \ref{eq:exponent_alphabar}) have been extended to the case where these $N$ vicious walkers
are divided into $p$ different families where, within a particular family, walkers are indifferent to each other (their paths may cross), each family having its
own diffusion constant. These calculations have also been extended in \cite{MuB01,GG10a} to include the effects
of long range interactions between vicious walkers, where the pair potential $V_{\rm LR}(r)$ decays with the distance $r$ between two walkers as $V_{\rm LR}(r) \sim g r^{-\sigma - d}$, $g$ being the coupling constant. The large time decay of $Q(t)$ was found, using also renormalization group techniques, to display different interesting regimes in the $(\sigma,d)$ plane~\cite{MuB01,GG10a} .

This vicious walker problem has been generalized yet in another direction where one considers L\'evy flights, instead of Brownian motions. In Ref. \cite{GG10b}, the authors studied the case where the process terminates upon the first encounter between two walkers (note that one 
could alternatively consider that it terminates upon the first crossing of two walkers, which is a different situation for L\'evy flights \cite{XS11}, see below). 
\begin{figure}
\centering
\includegraphics[width = 0.45\linewidth]{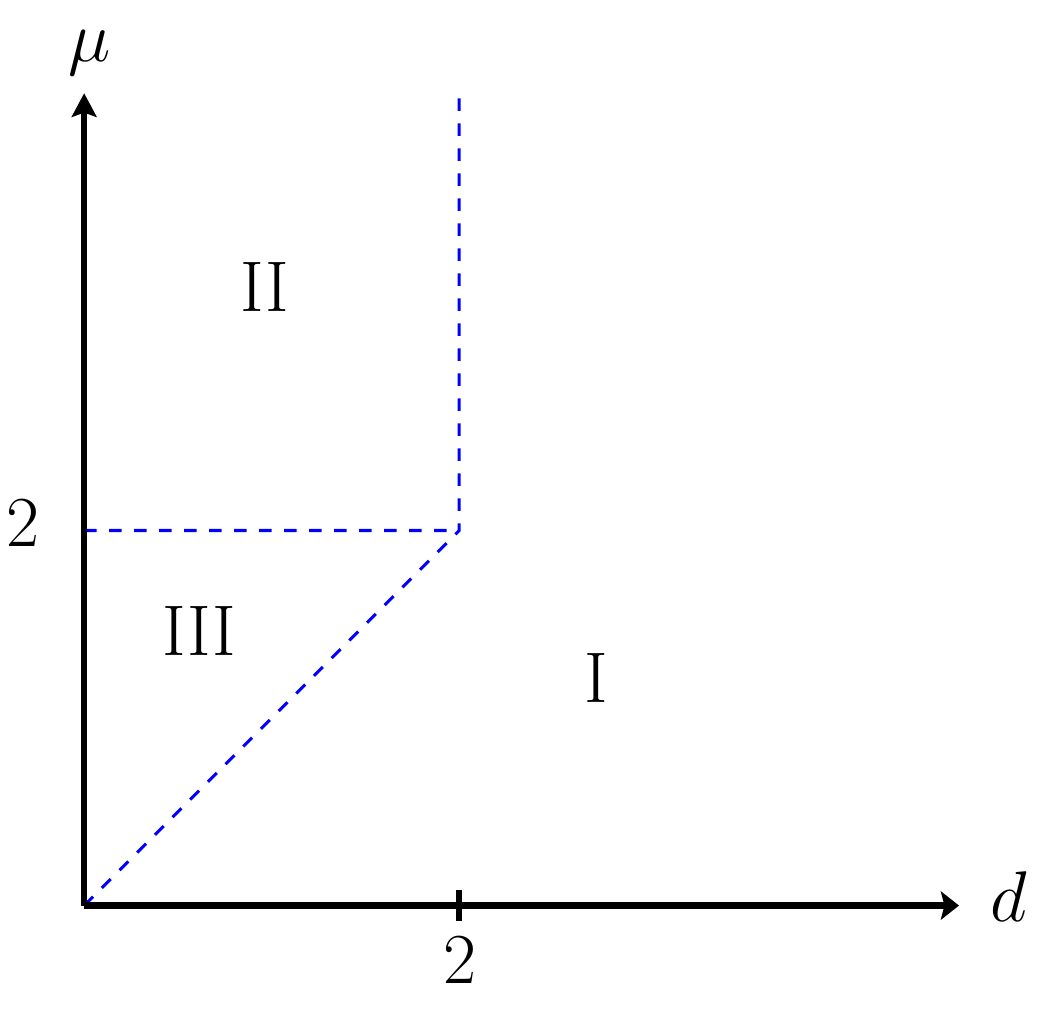}
\caption{Three different regimes for the late time behavior of the survival probability $Q(t)$ for $N$ independent L\'evy walkers in $d$ dimensions
and L\'evy index $\mu$~\cite{GG10b}.}\label{fig:vicious_levy}
\end{figure}
Such L\'evy flights, in $d$ dimensional space, perform Markov random walks where at each time step they jump in a random direction (with an angle which is uniformly distributed between $0$ and $2 \pi$) while the length of the jump $\eta$ is drawn from a pdf $\phi(\eta)$ with an algebraic tail $\phi(\eta) \sim \eta^{-d-\mu}$. In Ref. \cite{GG10b}, the authors found three different regimes for the late time behavior of $Q(t)$ in the $(d,\mu)$ plane (Fig. \ref{fig:vicious_levy}). For fixed $\mu$, the upper critical dimension is $d_c=\mu$ for $\mu < 2$ while $d_c = 2$ for $\mu \geq 2$. For $d>d_c$ (which is region I on Fig. \ref{fig:vicious_levy}), $Q(t) \to {\rm const.}$ as, in this regime, the random walks become non-recurrent and can avoid each other for all time. In the regime II, when $\mu \geq 2$, one recovers the above result for Brownian vicious walkers. In the regime III, when $d < \mu < 2$, the persistence exponent $\theta$ differs from the Brownian case and has been computed perturbatively up to second order in $\epsilon = \mu -d$~\cite{GG10b}.     

Because L\'evy flights are non local, two L\'evy flights may jump over each other without meeting at some exact point. Hence there are two different ways to define vicious L\'evy flights: one may allow jump-overs as discussed before \cite{GG10b} or instead prohibit them as it was studied for $d=1$ in Ref. \cite{XS11}. For $N=2$, the survival probability decays algebraically, $Q(t) \sim t^{-1/2}$, independently of $\mu$ -- as a consequence of the Sparre Andersen theorem \cite{SparreAndersen}, see section \ref{section:rw}. For $N \geq 3$, there is no known exact result but numerical simulations seem to indicate that $\theta$ actually depends on $\mu$. For instance, for $N=4$, 
$\theta = 2.3(1)$ for $\mu = 1$ while $\theta = 2.91(9)$ for $\mu = 2$~\cite{XS11}.  

Finally, this question was extended to the case of $N$ vicious walkers performing random acceleration processes, in dimension $d=1$ \cite{XS11}. For $N=2$, this problem is simply equivalent to a single random walker with an absorbing boundary at the origin and hence $\theta=1/4$ for $N=2$. For $N=3$, it was shown in \cite{XS11} that the survival probability is equivalent to the one of a single particle performing a random acceleration process 
in two dimensions confined in a $60^\circ$ wedge geometry (Fig. \ref{Fig_wedge}), similarly to the case of three Brownian walkers discussed above in section \ref{subsection:three_walkers}. Numerical simulations yield  $\theta = 0.71(1)$ for $N=3$ and 
for generic values of $N$ are consistent with $\theta < N(N-1)/8$~\cite{XS11}. 
     
\subsection{The trapping reaction}
The trapping reaction is defined as follows. Two  species, $A$ and $B$, 
diffuse in space (that is, they execute Brownian walks) and interact 
according to the reaction $A + B \to B$. Thus the $B$ particles act as 
traps for the $A$ particles. The question to be addressed is what is 
the asymptotic behaviour of the concentration of $A$ particles? 
The relation to a first-passage problem is clear when one considers that, 
since the $A$ particles do not interact with each other, it suffices to 
consider a {\em single} $A$ particle moving in a sea of $B$-particles. 
The concentration of $A$ particles at time $t$ is just the initial 
($t=0$) concentration, multiplied by the probability that, in a system 
with only one $A$ particle present initially, this particle survives until 
time $t$. So this system consists of infinitely many particles, but the 
$B$ particles interact separately with the $A$ particles, and not with 
each other. It is this feature that makes the trapping reaction tractable. 

For simplicity we will consider only the 1-d problem in detail, and comment 
later on the differences that appear for higher dimensional systems.
The problem is defined as follows. Initially a single $A$ particle, with 
diffusion constant $D_A$ is located at the origin of an infinite line, 
on which $B$ particles, with diffusion constant $D_B$, are placed at 
random with density $\rho$, i.e.\ any infinitesimal interval of size $dx$ 
contains an $A$ particle with probability $\rho dx$. The calculation of 
the $A$-particle survival probability, $Q(t)$, proceeds in two stages. 
First, we consider a simpler problem, the ``target problem'' (or ``target 
annihilation problem''), in which the $A$ particle does not move 
\cite{BlumenZumofenKlafter,Tachiya,BurlatskyOvchinnikov}.  
Then we show that, as far as the leading asympotics are concerned, 
the target problem gives the correct result for the full problem, in the 
sense that the diffusion constant $D_A$ does not appear in the asymptotic 
large-$t$ result. The calculation of the leading correction to the 
asymptotic behaviour is, however, a challenging open problem.

\subsubsection{The target problem}
Consider a single $A$ particle located at the origin. $B$ particles are 
placed randomly on the line, in the interval $(0,L)$ with density 
$\rho$ as described above. 
\begin{figure}
\centering
\includegraphics[width = 0.45 \linewidth]{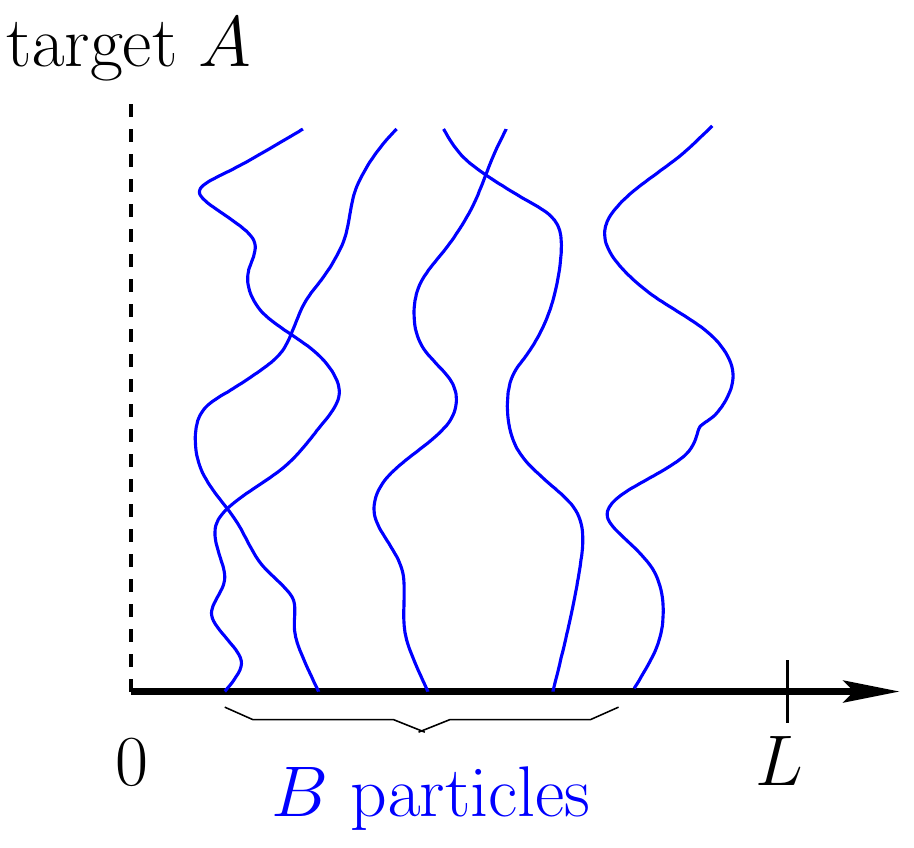}
\caption{Schematic example of a realization where the target, the particle $A$, at the origin is immobile
and there are $N=5$ noninteracting $B$ particles around it undergoing Brownian motion. The traps
are initially placed uniformly in the box $(0,L)$. Eventually we are interested in
the limit when the number of traps $N \to \infty$ and the box size $L \to \infty$ but with the density
of traps $\rho = N/L$ fixed.}\label{Fig_target}
\end{figure}
The probability, $Q_1(t)$, that a given 
$B$ particle, initially located at $x$, has not yet reached the origin 
at time $t$ is given by Eq.~(\ref{1DBrownian}): 
$Q_1(t) = {\rm erf}(x/\sqrt{4D_B t})$. So for $N$ particles at locations 
$\{x_n\}$, the survival probability of the $A$ particle is 
$Q_R(t) = \langle \prod_{n=1}^N {\rm erf}(x_n/\sqrt{4D_Bt})\rangle$, where the 
average is over the initial positions, $\{x_n\}$ of the particles. 
Taking the limit $L \to \infty$, this gives
\begin{eqnarray}
Q_R(t) &=& \lim_{L \to \infty}\left[\int_0^L \frac{dx}{L}\,
{\rm erf}\left(\frac{x}{\sqrt{4D_Bt}}\right)\right]^{\rho L} \nonumber \\
& = & \lim_{L \to \infty}
\left[1-\frac{1}{L}\int_0^L dx\,{\rm erfc}\,\left(\frac{x}{\sqrt{4D_Bt}}\right)\right]^{\rho L} 
\nonumber \\
&=& \exp\left({-\frac{2}{\pi}\rho\sqrt{D_Bt}}\right)\ .
\end{eqnarray}
If we now allow for $B$-particles on both left and right, with the same 
density $\rho$, we obtain for the target survival probability 
\begin{equation}
Q(t) = \exp\left(-\frac{4}{\pi}\rho\sqrt{D_Bt}\right) \ .
\end{equation} 
The target problem can, in fact, be solved exactly in any number of dimensions 
\cite{BlumenZumofenKlafter,Tachiya,BurlatskyOvchinnikov}. Here we will focus 
mainly on the one-dimensional problem, and quote results for higher dimensions. 
In the following section we show that, somewhat surprisingly, the 
leading asymptotics are not affected by allowing a non-zero diffusion 
constant, $D_A$, for the $A$ particle. 

The above result on the average (over initial positions) survival probability of a static target in presence of
a uniformly distributed Brownian walkers has been generalized in a number of ways.
For instance, in Ref.~\cite{EM2011}, while the target $A$ is static at the origin, each
$B$ particle now undergoes slightly different dynamics: In a small time interval $\Delta t$,
each $B$ particle diffuses with probability $(1-r \Delta t)$ and {\em resets} to its
own initial position with probability $r\, \Delta t$. The $B$ particles are
independent and is initially distributed uniformly with density $\rho$. For $r=0$, one goes back to the
standard target problem discussed above. For $r>0$, it was found that the average persistence 
of the $A$ particle decays as a power law at late times~\cite{EM2011}
\begin{equation}
Q(t)\sim t^{-\theta}, \quad \theta= 2\, \rho\, \sqrt{\frac{D}{r}}\, .
\label{reset_pers}
\end{equation}
Thus the persistence exponent $\theta$ depends continuously on the system parameters.
In contrast, the {\em typical} (and not the average) survival probability
of the $A$ particle was found to decay exponentially for all $r>0$~\cite{EM2011}.

Another interesting generalization is to the case when the $A$ particle is still static, but the
$B$ particles undergo independent subdiffusive or superdiffusive motion, often modelled by
a continuous-time random walk (CTRW). The survival probability of the $A$ particle in case
when the $B$ particle is subdiffusive has been studied analytically and numerically
by Yuste and collaborators~\cite{Yuste2005,Yuste2006,Yuste2007,Yuste2008,Abad12}. While a fractional Fokker-Planck approach
could be successfully used to derive asymptotic results for this subdiffusive case, it can not
be easily generalised to  
the superdiffusive case. 
Fairly complete exact asymptotic results for 
the survival probability $Q(t)$ of the $A$ particle were recently derived when the $B$ particle undergoes
CTRW with arbitrary waiting time and jump length distributions~\cite{FM2012}. These results include
the previously known results for the subdiffusive case, but also provide exact asymptotic results
for the superdiffusive case~\cite{FM2012}. This was achieved via an exact and rather general mapping
to an extreme value statistics problem~\cite{FM2012}. 

\subsubsection{The moving target}\label{subsection:moving}

Here we introduce a completely different approach \cite{BrayMajumdarBlythe} 
to the case where both types of particle are mobile by formally treating the 
$A$ and $B$ particles as if they were {\em non-interacting}. 
We exploit the initial condition that each $B$-particle is 
randomly located anywhere in the system 
to show that certain events, where a $B$-particle meets the $A$-particle 
{\em for the first time} (recall that we are treating them as 
non-interacting, so they can meet more than once) have a Poisson distribution.
This means that the probability, $p_n$, that $n$ such events have occurred up 
to time $t$ is given by $p_n = (\mu^n/n!)\exp(-\mu)$, where the mean, $\mu$, 
is a {\em functional}, $\mu[\vec{z}]$ of the trajectory $\vec{z}(\tau)$, 
with $0\le \tau\le t$, of the $A$-particle. Here the notation $\vec{z}$ 
indicates that we are working in general dimension, $d$, so the trajectory
of the $A$-particle is described by a vector $\vec{z}(t)$. The probability 
that the trajectory $\vec{z}(t)$ has survived in the original interacting 
problem is just $p_0(t) = \exp(-\mu[\vec{z}])$. Finally, the $A$-particle 
survival probability $Q(t)$ is obtained by averaging $\exp(-\mu[\vec{z}])$ 
over all $A$-particle trajectories $\vec{z}(t)$ with the appropriate Wiener 
measure, $\exp[-(1/4D_A)\int_0^t d\tau (d\vec{z}/d\tau)^2]$.    

We now derive the Poisson property that plays a central role in our 
subsequent analysis. Consider a finite volume $V$, containing $N=\rho V$ 
$B$-particles, each with diffusion constant $D_B$, randomly distributed 
within $V$, and a single $A$-particle, with diffusion constant $D_A$, 
initially located at the origin, which lies within $V$. Let $P(\vec{x},t)$ 
be the probability that a given $B$-particle, starting at $\vec{x}$, has 
met the $A$-particle before time $t$. The average of this quantity over 
$\vec{x}$ is $(1/V)\int_VdV\,P(\vec{x},t) = R(t)/V$, where $R(t)$ depends  
implicitly on $\vec{z}(t)$. The probability that $N$ distinct $B$-particles  
have met the $A$-particle is 
\begin{equation}
p_n(t) = \frac{N!}{n!(N-n)!}\left(\frac{R}{V}\right)^n\,
\left(1-\frac{R}{V}\right)^{N-n} \;.
\end{equation}
In the limit $N \to \infty$, $V \to \infty$, with $\rho=N/V$ fixed, we 
obtain the Poisson distribution: 
\begin{equation}
p_n = \frac{\mu^n}{n!}e^{-\mu}\ ,
\end{equation}
with $\mu=\rho R$.

We can derive an equation for the functional $\mu[z]$ by calculating, 
in two different ways, the probability density to find a $B$-particle 
at the point $\vec{z}(t)$ at time $t$: (i) since the particles are 
treated as if they are non-interacting, and the $B$-particles start from 
a steady state of uniform density $\rho$, this probability density is just 
$\rho$; (ii) from the Poisson property, the probability that the $A$-particle 
meets a $B$-particle for the first time in the time interval $(t',t'+dt')$ 
is $\dot{\mu}(t')\,dt'$. The probability density for such a particle to 
subsequently arrive at $\vec{z}(t)$ at time $t$ is given by the diffusion 
propagator 
\begin{equation}
G(\vec{z}(t),t|\vec{z}(t'),t') = 
\frac{\exp\{-[\vec{z}(t)-\vec{z}(t')]^2/4D_B(t-t')\}}
{[4\pi D_B(t-t')]^{d/2}} \ . 
\end{equation}
Equating these two results gives the following important equation 
\cite{BrayMajumdarBlythe}:
\begin{equation}
\rho = \int_0^t dt'\,\dot{\mu}(t')\,G(\vec{z}(t),t|\vec{z}(t'),t')\ ,
\label{fundamental}
\end{equation}
which is an implicit equation for the functional $\mu[\vec{z}]$. Note 
that, although this equation formally determines $\dot{\mu}$, the 
condition $\mu(t=0)=0$ (which follows from the fact that no $B$-particle 
can meet the $A$-particle in zero time) means that it also determines 
$\mu$. Finally, the $A$-particle survival probability is given by 
$Q(t) = \langle \exp(-\mu[\vec{z}]) \rangle_z$, where the average is over 
all paths $\vec{z}(t)$ weighted with the Wiener measure, i.e. 
\begin{equation}
Q(t) = N \int {\cal D} \vec{z(t)}\exp\left(-\frac{1}{4D_A}\int_0^t d\tau\,
\left(\frac{d\vec{z}}{d\tau}\right)^2 - \mu[\vec{z}]\right)\ ,
\end{equation}
where the normalisation $N$ is the reciprocal of the same path integral,  
but with $\mu=0$.  

\subsubsection{The ``Pascal Principle'' and an upper bound for $Q(t)$}
As a first application of this approach, we prove that the trajectory 
$\vec{z}=0$, corresponding to a stationary target, is the dominant path, 
i.e.\ that it gives the smallest possible value of $\mu[\vec{z}]$ for all 
$t$. This function, $\mu_0(t)$, is given by Eq.\ (\ref{fundamental}) with 
$\vec{z}=0$, i.e.\ 
\begin{equation}
\rho = \int_0^t dt'\ \dot{\mu}_0(t')[4\pi D(t-t')]^{-d/2}\ .
\label{rho for z=0}
\end{equation}
It is clear that $\mu_0(t)$ must have the form $\mu_0(t)=\lambda_d t^{d/2}$ 
(for $d<2$) in order that the right-hand side of Eq.\ (\ref{rho for z=0}) be 
independent of $t$. Putting this form into (\ref{rho for z=0}) gives
\begin{equation}
\mu_0(t)= \rho\left(\frac{2}{\pi d}\right)\sin\left(\frac{\pi d}{2}\right)\,
(4\pi Dt)^{d/2}\ ,\ \ d<2 ,
\label{mu_0}
\end{equation}
which is an extension to general $d < 2$ of the one-dimensional target 
annihilation result. For the case $d=2$, there is a logarithmic 
correction to this result.

We now show that the target problem, i.e.\ the static trajectory $\vec{z}=0$, 
gives the global minimum of $\mu[\vec{z}]$. To do this we write 
$\mu = \mu_0 + \mu_1$ in Eq.\ (\ref{fundamental}). Using Laplace transform 
methods this equation can be rearranged to give an implicit equation 
for $\mu_1[\vec{z}]$:
\begin{equation}
\mu_1[\vec{z}] = \frac{1}{\pi}\sin\left(\frac{\pi d}{2}\right)
\int_0^t \frac{dt_1}{(t-t_1)^{(2-d)/2}}\,
\int_0^{t_1} \frac{dt_2}{(t_1-t_2)^{d/2}}\,\dot{\mu}(t_2)\,K(t_1,t_2)\ ,
\label{mu_1}
\end{equation}
where
\begin{equation}
K(t_1,t_2) = 1 - \exp\left\{-\frac{[\vec{z}(t_1) - \vec{z}(t_2)]^2}
{4D(t_1-t_2)}\right \}\ .
\end{equation}
Note that Eq.\ (\ref{mu_1}) is an implicit equation  because the full $\mu$ 
appears on the right-hand side. Now observe that $K(t_1,t_2) \ge 0$, and also
$\dot{\mu}(t_2) \ge 0$, because $\mu(t)$ is just the mean number of distinct 
$B$ particles that have met the $A$ particle up to time $t$ -- clearly a 
nondecreasing function. It follows [from Eq.\ (\ref{mu_1})] that 
$\mu_1[\vec{z}] \ge 0$ for all paths $\vec{z}(t)$. It then follows immediately 
that when the average of $\exp(-\mu_0-\mu_1)$ is taken over the Wiener 
measure, one obtains the upper bound 
\begin{equation}
Q_U(t) = \langle\exp(-\mu_0-\mu_1)\rangle_z \le \exp [-\mu_0(t)]. 
\label{upperbound}
\end{equation}
For the one-dimensional case, setting $d=1$ in Eq.\ (\ref{mu_0}) gives
\begin{equation}
Q_U(t) = \exp\left(-\frac{4}{\sqrt{\pi}}\rho\sqrt{D_B t}\right),\ \ d=1.
\end{equation}
This rigorous upper bound, $Q(t) \le \exp[-\mu_0(t)]$, can be combined with 
a rigorous lower bound derived by Bray and Blythe \cite{BrayBlythe2002} 
to obtain the leading asymptotic behaviour of $Q(t)$. 

The observation that a static $A$-particle has the largest survival 
probability has been termed the ``Pascal Principle'' \cite{Moreau2003} (see also Ref. \cite{OBO1989} for a precursor of this idea in a related model of excitations migrating over a disordered array of donor centers), 
after the remark by Blaise Pascal (from his {\em Pens\'ees}) that 
``tout le malheur des hommes vient d'une seule chose, qui est de ne savoir pas 
demeurer en repos dans une chambre'' [All the unhappiness of men comes from 
not knowing how to stay quietly in a room]. 
Moreau et al.\ \cite{Moreau2003,Moreau2004} have generalized to any 
number of space dimensions (for particles moving on a lattice) the result 
that the static $A$-particle has the largest survival probability. 

\subsubsection{A lower bound for $Q(t)$}
\label{LowerBound}
The lower bound is derived as follows. We revisit the target problem, in 
one space dimension, modified by placing absorbing boundaries at 
$x = \pm l/2$, and allowing the $A$ particle, initially at $x=0$, to 
diffuse with diffusion constant $D_A$. Then the $A$ particle will certainly 
survive up to time $t$ if the following sufficient conditions are met: 
(i) there are no $B$ particles initially in the interval $[-l/2,l/2]$; 
(ii) no $B$ particles enter this interval up to time $t$; and (iii) the $A$ 
particle does not leave the interval $[-l/2,l/2]$ up to time $t$. 
The probability for condition (i) to be satisfied is $\exp(-\rho l)$, 
and the probability for condition (ii), given condition (i), is 
$\exp[-4\rho(D_Bt/\pi)^{1/2}]$. Finally, we need to compute the probability, 
$Q_A(x,t)$, that the $A$ particle, starting at $x$, has not yet reached 
$\pm l/2$ at time $t$. It obeys the BFP equation $\partial_t Q_A = 
D_A \partial_{xx}Q_A$, with boundary condition $Q_A(\pm l/2,t)=0$. 
The large $t$ solution is (up to an overall constant), 
$Q_A(x,t) \sim \cos(\pi x/l)\,\exp(-\pi^2\,D_At/l^2)$.
So, setting $x=0$ and combining the three factors above, we find
\begin{equation}
Q(t) \ge {\rm const.}\exp[-4\rho(D_Bt/\pi)^{1/2}-\rho l 
- \pi^2 D_At/l^2]. 
\end{equation} 
The best bound is obtained by maximizing this result with respect 
to $l$, giving the lower bound
\begin{equation}
Q_L(t)={\rm const.}\exp[-4\rho(D_Bt/\pi)^{1/2}-3(\pi^2\rho^2D_At/4)^{1/3}].
\label{lowerbound}
\end{equation}

Comparing the bounds (\ref{upperbound}) and (\ref{lowerbound}), we see 
that the leading asymptotic behaviour is the same for both bounds, in the 
sense that 
\begin{equation}
-\lim_{t \to \infty}\frac{\ln Q(t)}{\rho\sqrt{D_Bt}} 
= \frac{4}{\sqrt{\pi}}
\label{1dasymptotic}
\end{equation}
for both bounds. It is striking that the leading-order asymptotic behaviour 
is completely independent of $D_A$, the diffusion constant of the 
$A$-particle.  

Similar calculations can be performed for continuous dimensions, $d$, in the 
range $1 \le d <2$ \cite{BrayBlythe2003}. Again, both upper and lower bounds 
give the same leading asymptotics, with the result
\begin{equation}
Q(t) \sim \exp[-\mu_0(t)] = \exp\left[-\rho\left(\frac{2}{\pi d}\right)
\sin\left(\frac{\pi d}{2}\right)\,(4\pi D_Bt)^{d/2}\right].
\label{static target d<2}
\end{equation}

The case $d=2$ has to be treated more carefully \cite{BrayBlythe2003} since 
the finite size of the $A$-particle comes into play as a short-distance 
cut-off. We refer the reader to \cite{BrayBlythe2003} for a detailed 
discussion. 
The final conclusion, however, is that upper and lower bounds can still be 
derived, and give the same leading asymptotics,
\begin{equation}
Q(t) \sim \exp\left[-\frac{4\pi\rho D_Bt}{\ln(\rho D_B^2t/D_A)}\right]\ , 
\end{equation} 
in the sense that 
\begin{equation}
\lim_{t \to \infty} -\frac{\ln t\,\ln Q(t)}{\rho D_Bt} = 4\pi\ .
\label{2dasymptotic}
\end{equation}
We note again that this leading asymptotic result is independent of $D_A$.

For dimensions $d>2$, upper and lower bounds can still be derived 
\cite{BrayBlythe2002} but no longer converge. In this case both bounds decay 
exponentially with time, proving that the asymptotic decay is exponential 
in form. The asymptotic forms (\ref{1dasymptotic}) and (\ref{2dasymptotic}) can 
be tested through numerical simulations. A very efficient simulation 
algorithm has been developed by Mehra and Grassberger 
\cite{MehraGrassberger}. However, the approach to the asymptotic limiting 
behaviour is {\it extremely} slow \cite{BrayBlythe2003}. 

The results presented here have been extended in a number of ways. 
Yuste and Lindenberg \cite{Yuste2005} (see also \cite{YusteAcedo2004}) have considered the case 
where the ``traps'' ($B$-particles in our notation) and ``particles'' 
(A-particles) move subdiffusively, with mean-square displacement 
growing as $t^\gamma$ (such that $\gamma=1$ corresponds to standard 
diffusion). Following the method of Bray et al.\ 
\cite{BrayMajumdarBlythe} described earlier in this section, they  
showed that the ``Pascal principle'' holds also for the subdiffusive  
case. In this way they obtained an upper bound for the asymptotic form 
of the particle's survival probability in the form 
$Q_U(t) \sim \exp(-{\rm const.}\ t^{\gamma/2})$ \cite{Yuste2005}. 
Using the method of Bray and Blythe (see section \ref{LowerBound})
they also obtain a lower bound, $Q_L(t)$, for this model and the two bounds 
coincide in the asymptotic large-time limit. Additionally, they consider 
the case where the traps and the particle have different subdiffusive 
behaviour, with exponents $\gamma$ and $\gamma'$ respectively. Again, upper 
and lower bounds can be derived. When $\gamma$ and $\gamma'$ are both less 
than unity, the survival probability is independent of $\gamma'$, and 
determined by the subdiffusive properties of the traps. When the particle 
moves diffusively ($\gamma'$=1), however, the exact asymptotics can only 
be determined for $2/3 < \gamma \le 1$ \cite{Yuste2005}. Recent 
work by Borrego et al.\ \cite{Borrego} extends these results to higher 
dimensions. For related work on various aspects of the trapping reaction see Refs.\ 
\cite{BO1990,Oshanin2002,Anton,AntonBlythe,Ruiz-Lorenzo,YusteLindenberg2007,Yuste2008,OT2008,OKVK2009}.

Before leaving the trapping reaction, it is worth discussing briefly 
the related reaction-diffusion process $A + B \to 0$ (the two-species 
annihilation reaction). This model was introduced by Toussaint and Wilcek 
\cite{ToussaintWilczek}, originally as a model 
of monopole-antimonopole annihilation in the early universe. The $A$ and 
$B$ particles are taken to be randomly distributed in space in the initial 
state (i.e.\ the statistics are Poissonian, as in our discussion of the 
trapping reaction). There are then two cases, namely (i) the initial numbers 
of $A$ and $B$ particles are equal, and (ii) the initial numbers are unequal. 
We consider the case where the $A$ particles are the minority species.
After a long time the $A$ particle density will be very small, while the 
$B$-particle density will be almost constant, with a value that tends to  
the initial difference in densities. This model has been studied in detail 
by Bramson and Lebowitz~\cite{BL88}, who prove the following asymptotic forms for 
the decay of the $A$ particle density, $\rho_A$
\begin{eqnarray}
\rho_A(t) \sim 
\begin{cases}
& \exp(-\lambda_d t^{d/2}) \;, \; d<2 \;,  \\
& \exp(-\lambda_2 t/\ln t) \;, \; d=2  \;,\\
&\exp(-\lambda_dt) \;,\; d>2 \;.
\end{cases}
\end{eqnarray}
The late-time behaviour of this model is equivalent to the trapping reaction,
so our results for that reaction determine the constants $\lambda_d$ as 
follows:
\begin{eqnarray}
\lambda_d & = & \frac{2\rho}{\pi d}\sin\left(\frac{\pi d}{2}\right)
(4\pi D_B)^{d/2},\ \ \ d<2 \, \\
& = & 4\pi\rho D_B,\ \ \ d=2\ , 
\end{eqnarray}
where here $\rho = \rho_B(0)-\rho_A(0) = \rho_B(\infty)$. 
For $d>2$, $\lambda_d$ has not, to our knowledge, been determined exactly. 

\subsubsection{The target problem with a deterministically moving target}
The formalism developed for the trapping reaction can also be applied 
to study the survival of a deterministically moving target. The key result 
is Eq.\ (\ref{fundamental}), which is an implicit equation for the 
functional $\mu[\vec{z}]$, where $\vec{z}(t)$ is the trajectory of 
the target, and the survival probability of the target is given by 
$Q(t) = \exp(-\mu[\vec{z}])$. For the diffusing target considered 
previously, the quantity $\exp(-\mu[\vec{z}])$ has to be averaged 
over the possible trajectories weighted by the Wiener measure, but 
for a deterministically moving target, no such averaging is required. 
An explicit determination of the functional $\mu[\vec{z}]$ is, however, 
necessary. We will consider two examples. The first is a one-dimensional 
system in which the target trajectory is given by $z(t) = \alpha\sqrt{4D_Bt}$. 
Then the function $\mu(t)$  satisfies Eq.\ (\ref{fundamental}) with 
\begin{equation}
G(z(t),t|z(t'),t') = \frac{1}{\sqrt{4\pi D_B(t-t')}}
\exp\left[-\frac{\alpha^2}{t-t'}\left(\sqrt{t}-\sqrt{t'}\right)^2\right]\ .
\end{equation}
Inserting this into Eq.\ (\ref{fundamental}), we see that, on dimensional
grounds, $\mu(t)$ has the form $\mu(t) = c\sqrt{D_Bt}$, where $c$ is a 
constant. Evaluating the integral in (\ref{fundamental}) gives the 
explicit result \cite{BrayMajumdarBlythe}
\begin{equation}
\mu(t) = \frac{4}{\sqrt{\pi}}\rho\sqrt{D_Bt}\,\frac{\exp(-\alpha^2)}
{1 - {\rm erf}^2(\alpha)}\ ,
\end{equation}
and the survival probability of the target is simply $Q(t) = \exp[-\mu(t)]$ 
as usual. This result can be readily generalised to the case where there 
are different trap densities on either side of the target 
\cite{BrayMajumdarBlythe}.

For our second example we consider the survival probability of a ballistically 
moving target. Our starting point is again the fundamental equation 
(\ref{fundamental}). There are three distinct cases: $d<2$, $d=2$ and $d>2$.
For ballistic motion, the trajectory $\vec{z}(t)$ is given by 
$\vec{z}(t) = ct\,\hat{\bf n}$, where $\hat{\bf n}$ is a unit vector, 
and Eq.~(\ref{fundamental}) reads
\begin{equation}
\rho = \int_0^t dt'\,\dot{\mu}(t')\,\frac{\exp[-c^2(t-t')/4D_B]}
{[4\pi D_B(t-t')]^{d/2}}\ .
\label{convolution}
\end{equation}
We first consider the case $d<2$. Since the integral in (\ref{convolution}) 
has the form of a convolution, $\mu(t)$ can be determined by Laplace 
transform methods. We denote $\beta = c^2/4D_B$ and introduce
the Laplace transform $\tilde{\mu}(s) = \int_0^\infty \mu(t)\exp(-st)dt$. 
Taking the Laplace transform of Eq.\ (\ref{convolution}), and using 
$\mu(0)=0$, we obtain 
\begin{equation}
\tilde{\mu}(s) = A[(\beta + s)^{1-d/2}/s^2]\ ,
\label{Laplace}
\end{equation}
where $A = \rho(4\pi D_B)^{d/2}/\Gamma(1-d/2)$ is a constant. Taking the 
inverse Laplace transform of Eq.\ (\ref{Laplace}) gives 
\cite{MajumdarBrayBallistic} 
\begin{equation}
\mu(t) = B[(1+\beta t)\gamma(d/2,\beta t)-\gamma(d/2+1,\beta t)],
\label{ballistic d<2}
\end{equation}
where $B=\rho(4\pi D_B)^{d/2} \sin(\pi d/2))(\pi\beta^{d/2})$, and 
$\gamma(\nu,x) = \int_0^x y^{\nu-1}\exp(-y)dy$ is the incomplete gamma 
function. The result for $\mu(t)$ is valid for all $t$ and $d<2$. The survival 
probability is, as usual, $Q(t) = \exp[-\mu(t)]$. 

We can consider some special cases of Eq.\ (\ref{ballistic d<2}). 
For $c\to 0$ at fixed $t$, it is easily checked that we recover the result 
(\ref{static target d<2}) for a static target. Note that the limit $c \to 0$ 
at fixed $t$ is equivalent to the limit $t \to 0$ at fixed $c$ since $\mu(t)$ 
depends on $c$ and $t$ only through the combination $\beta = c^2/4D$. In the 
opposite limit $t \to \infty$ at fixed $c$, we find from Eq.\ 
(\ref{ballistic d<2}) that $\mu(t) \to \rho(4\pi D_Bt)^{d/2}\sin(\pi d/2)
\Gamma(d/2)/[\pi\beta^{(d-2)/2}]$. This implies an exponential decay for the 
survival probability at late times, $Q(t) \to \exp(-\theta t)$, where the 
decay (or `persistence') exponent $\theta$ is given by
\begin{equation}
\theta = \rho\pi^{d/2-1}(4D_B)^{d-1}\sin(\pi d/2)\Gamma(d/2)c^{2-d}\ .
\end{equation}

The marginal dimension $d=2$ is a special case. We can still use 
Eq.\ (\ref{fundamental}) provided we introduce an ultraviolet cut-off 
reflecting the need to introduce a lattice structure for  $d=2$. 
Alternatively, we can introduce a short-time cut-off, $t_0$, in the diffusion 
propagator:
\begin{equation}
G(\vec{z}(t),t|\vec{z}(t'),t') = 
\frac{\exp\{-[\vec{z}(t)-\vec{z}(t')]^2/4D_B(t-t'+t_0)\}}
{4\pi D_B(t-t'+t_0)}\ . 
\end{equation}
To extract the leading asymptotic behaviour we can put $t_0=0$ in the 
exponential, and retain it only in the denominator of the propagator. 
Putting $\vec{z}(t)=ct{\bf n}$ and taking the Laplace transform as for $d<2$ 
gives
\begin{equation}
\tilde{\mu}(s) = 4\pi\rho D_B/[s^2 \tilde{g}(s)]\ ,
\end{equation}
where
\begin{equation}
\tilde{g}(s) = \int_0^\infty dt\,\exp[-(\beta + s)t]/(t+t_0)\ .
\end{equation}
The large-$t$ form of $\mu(t)$ can easily be extracted from  the small-$s$
behavior of the Laplace transform. The result \cite{MajumdarBrayBallistic} 
is that the survival probability again decays exponentially for large $t$, 
$Q(t) \sim \exp(-\theta t)$, where $\theta$ is now non-universal:
\begin{equation}
\theta = 4\pi\rho D_B/[-\ln(\beta t_0)]\ , 
\end{equation}
this result being valid for $\beta t_0 \ll 1$.

For $d>2$, the calculation is more complex, and we will just give the 
result for the physically relevant dimension, $d=3$, referring the reader 
to \cite{MajumdarBrayBallistic} for the details. We consider a spherical 
particle of radius $a$ moving ballistically 
with constant speed $c$. We again find a power-law decay of the survival 
probability, $Q(t) \sim t^{-\theta}$, with the persistence exponent $\theta$ 
given by an infinite sum:
\begin{equation}
\theta = 2\pi a\rho D_B\left[1-2\pi\sum_{l=0}^\infty(-1)^l
\left(l+\frac{l}{2}\right)
\frac{K'_{l+1/2}(\gamma a)}{K_{l+1/2}(\gamma a)}\,
I^2_{l+1/2}(\gamma a)\right]\ ,
\end{equation}
where $I_\nu(x)$ and $K_\nu(x)$ are modified Bessel functions,
$K'_\nu(x) = d K_\nu(x)/dx$, and $\gamma = c/2D_B$.  
The limiting forms for small and large $c$ are readily recovered: 
for $c \to 0$ one recovers the known result for a static target, 
$\theta = 4\pi\rho a D_B$, while in the limit $c \to \infty$ one obtains
$\theta \rightarrow \pi a^2 c \rho$. The latter result can be understood 
by noting that it has the form $\theta = \rho V/t$, where $V=\pi a^2 ct$
is the volume swept out by the sphere in time $t$. The particle will 
survive until time $t$ if this volume initially contained no traps, 
which occurs with probability $\exp(-\rho V)$.

\subsubsection{The lamb and the $N$ lions}

We end up this section on persistence for multi-particle systems by discussing the problem of the moving target studied above in section \ref{subsection:moving} (see Fig. \ref{Fig_target}) in the case where the number $N$ of $B$-particles is finite -- while in the previous case this number was infinite, with a finite density of $B$-particles. This problem has been 
reformulated as the one of a diffusing prey, say a "lamb" (the $A$-particle), surrounded by $N$ predators, say $lions$ (the $B$-particles) \cite{KR1996, KR1999}. If the lamb meets a lion, the lamb is killed. This problem was first studied in the mathematics literature \cite{BG1991,Kes1992}, where it is sometimes known under the name of "Brownian pursuit" \cite{LS2001}. In the most interesting situation where the lions are all on one
side of the lamb, the survival probability $Q(t)$ of the lamb asymptotically decays as
a power-law in time, $Q(t) \sim t^{-\theta_N}$, with the exponent $\theta_N$ exhibiting a nontrivial
dependence on the number of lions $N$ and also on the diffusivities $D$ of each animal. For
simplicity, the case where the diffusivities of all animals are the same (and set to one)
is normally considered. The initial positions of the lamb and the lions are irrelevant in
this asymptotic behavior. 

For this capture problem, the exponent $\theta_N$ is known exactly for $N=1$ and $N=2$, with the results $\theta_1 = 1/2$ and
$\theta_2 = 3/4$ \cite{BG1991, Kes1992, KR1996, KR1999,LS2001,KMR2010}. For the case $N=3$, a mapping to an equivalent electrostatic
problem leads to the accurate numerical estimate $\theta_3 = 0.91342(8)$ \cite{leader_laggard}, while the inequality $\theta_3<1$ was rigorously established \cite{LS2001}. For $N > 3$, the
value of $\theta_N$ has been estimated with moderate accuracy only for a few values of $N$, for instance, $\theta_4 = 1.032$ and $\theta_{10} = 1.4$ \cite{BG1991}. In Ref. \cite{LS2001}, it was rigorous proved that $\theta_5 > 1$. One may wonder what happens for large $N$. To determine the survival probability of the lamb, it is sufficient to track the position of the closest lion only. For concreteness and
simplicity, suppose initially that all the $N$ lions are at the origin and the lamb is at $x_0 > 0$. For large $N$, the position $x_+(t)$ of the rightmost lion can be determined by the standard argument from extreme value statistics of independent and identical random variables:
\begin{eqnarray}\label{expr_evs}
\int_{x_+(t)}^\infty \frac{1}{\sqrt{4 \pi t}} e^{-\frac{x^2}{4t}} dx = \frac{1}{N} \;, 
\end{eqnarray}
which simply expresses the fact that there should be one lion in the interval $[x_+(t),+\infty)$ out of the group of $N$ lions. For large $N$, the solution of Eq. (\ref{expr_evs}) yields
\begin{equation}\label{eq_x+}
x_+(t) \sim \sqrt{A t} \;, \; A = 	4 \log N \;.
\end{equation}
In addition one can show that for large $N$, the trajectory $x_+(t)$ of the rightmost lion is deterministic (for fixed time $t$, the fluctuations
are of order $1/\sqrt{\log N}$). Hence the survival probability of the lamb can be computed as the persistence probability of a single Brownian motion in the presence of a deterministically moving boundary, evolving like in Eq. (\ref{eq_x+}). Such a problem can be solved using the Backward Fokker-Planck method, as shown in section \ref{subsection:cage}, yielding~\cite{KR1996,KR1999}
\begin{eqnarray}\label{thetaN_largeN}
\theta_N = \frac{\log N}{4} + o(\log N) \;. 
\end{eqnarray}   
This asymptotic result (\ref{thetaN_largeN}) was proved rigorously in Ref. \cite{LS2002}, using completely different methods, namely a comparison inequality which can be viewed as an extension of Slepian's lemma, discussed in section \ref{subsection:GSP}. 

This model was generalized \cite{GMPR2012} to the case where the pursuit takes place on the half-line, with an absorbing boundary condition at the origin, which plays the role of a haven for the lamb. Hence here, the initial position of the lamb is at $x>0$, while the $N$ lions all start at $L>x$. In this case, if the lamb reaches the origin, the haven, before meeting any lion, the lamb survives and the goal is to determine the survival probability $Q_N(x,L)$. 
In the case of one lion, $N=1$, this problem can be mapped onto the diffusion of a Brownian motion in a wedge (Fig. \ref{Fig_wedge}) and it can eventually be solved, using for instance the Backward Fokker-Planck method presented in section \ref{sec:wedge}. For $N>1$, there is no exact result but the large $N$ analysis can be performed along the lines outlined  above (\ref{expr_evs}, \ref{thetaN_largeN}) yielding the rather unusual result, valid for large $N$ \cite{GMPR2012}:
\begin{eqnarray}\label{asympt_lamblion_absorb}
Q_N(x,L) \sim N^{-z^2} \;, z = \frac{x}{L} \;.
\end{eqnarray}  
Note that this behavior (\ref{asympt_lamblion_absorb}) does not
become apparent until N becomes of the order of $10^{500}$, which can be tested using event driven simulations \cite{GMPR2012}. Related questions to this capture problem include the probability that the $k^{\rm th}$ rightmost lion remains in the positive half-line up to time
$t$. These probabilities, for different values of $k$, all decay algebraically, in the long time limit, with a family of nontrivial first passage exponents $\theta_{k,N}$ \cite{BenKrap2010a}. 

We leave this section by mentioning an extension of the standard Brownian pursuit problem to the case where the prey and the $N$ predators are performing fractional Brownian motion (fBm) \cite{Kes1992}, of Hurst index $H$ (see section \ref{section:fBm} for a detailed discussion of fBm). In particular, the fBm with $H=1/2$ corresponds to standard Brownian motion. Here also one expects that the survival probability $Q_N(t)$ decays algebraically at large time $t$, $Q_N(t) \sim t^{-\theta_N}$ \cite{Kes1992}. In this case there is no exact result beyond $N>1$, but Li and Shao in Ref. \cite{LS2002} conjectured, based on rigorous bounds, an analogous result to (\ref{thetaN_largeN}) for large $N$
\begin{equation}\label{largeN_fBm}
\theta_N = \frac{1}{d_H} \log N + o(\log N) \;, \; d_H = 2 \int_0^\infty \left[e^{2 H x} + e^{-2 H x} - (e^x -e^{-x})^{2H} \right] dx\;,
\end{equation}
although there exists no proof (nor physical derivation) of this result (\ref{largeN_fBm}).

\section{Persistence in coarsening phenomena}

The study of persistence in systems with infinitely many degrees of freedom 
began with the coarsening (or `phase-ordering') dynamics of the 1-d Ising 
model \cite{BDG1994a}. For a review of phase-ordering and phase-separation 
we refer the reader to \cite{BrayReview}. Some exact results for the 
persistence properties can be obtained for some one-dimensional models. 



\subsection{Ising and Potts models}\label{subsection:coarsening_ising}
The field of persistence phenomena,  in its modern context, began with
the study of  the 1-d Ising model at  zero temperature \cite{BDG1994a},
with  Hamiltonian $H  = -J  \sum_i S_i  S_{i+1}$. Each  spin  $S_i$ is
initially  assigned either  to the  ``up'' ($S_i=1$)  or  the ``down''
($S_i=-1$) state  at random.  The dynamical  rules employed correspond
to the  zero-temperature limit of  Glauber dynamics, in which  at each
time  step a  randomly chosen  spin is  aligned with  its  two nearest
neighbours when the latter are  in the same state, and randomised when
the neighbours are in different states. Equivalently, one can say that
the spin is aligned with one  of its two nearest neighbours, chosen at
random.  The persistence probability,  $Q(t)$, is the probability that
a given spin has remained in the same state (``up'' or ``down'') up to
time $t$ or,  equivalently, $Q(t)$ is the fraction  of spins that have
not yet flipped at time $t$. Numerical studies  \cite{BDG1994a} show that $Q(t)$ decays  as a power
law, $Q(t) \sim t^{-\theta}$, with $\theta \approx 0.37$. A subsequent
exact calculation \cite{DHP1995a,DHP1995b} showed that $\theta =3/8$.

The simulations are readily extended  to the $q$-state Potts model. In
the  initial state, each  Potts `spin'  is assigned  one of  the $q$
possible states at random. The updating rule is that at each time step
a randomly chosen spin is assigned to the same state as one of its two
neighbours, chosen  at random.  It  was found numerically that  $\theta$ 
increases with  $q$, attaining  a maximum  value of  unity in  the limit  
$q \to \infty$  \cite{BDG1994a}.   In a  {\em  tour  de  force} of  
analysis, Derrida, Hakim and Pasquier \cite{DHP1995a,DHP1995b} obtained the general 
result
\begin{equation}
\theta(q)        =        -\frac{1}{8}        +        \frac{2}{\pi^2}
\left[\cos^{-1}\left(\frac{2-q}{\sqrt{2}q}\right)\right]^2.
\label{theta(q)}
\end{equation}
From   this  general  formula   one  finds   $\theta(2)  =   3/8$  and
$\theta(\infty)=1$.  For  these and other  values of $q$,  the results
obtained  from  numerical  simulations  \cite{BDG1994a}  are  in  good
agreement with Eq.~(\ref{theta(q)}).

It is  instructive to think of the  process in terms of  the motion of
domain walls rather than the  flipping of spins (or Potts states). 
\begin{figure}
\centering
\includegraphics[width = \linewidth]{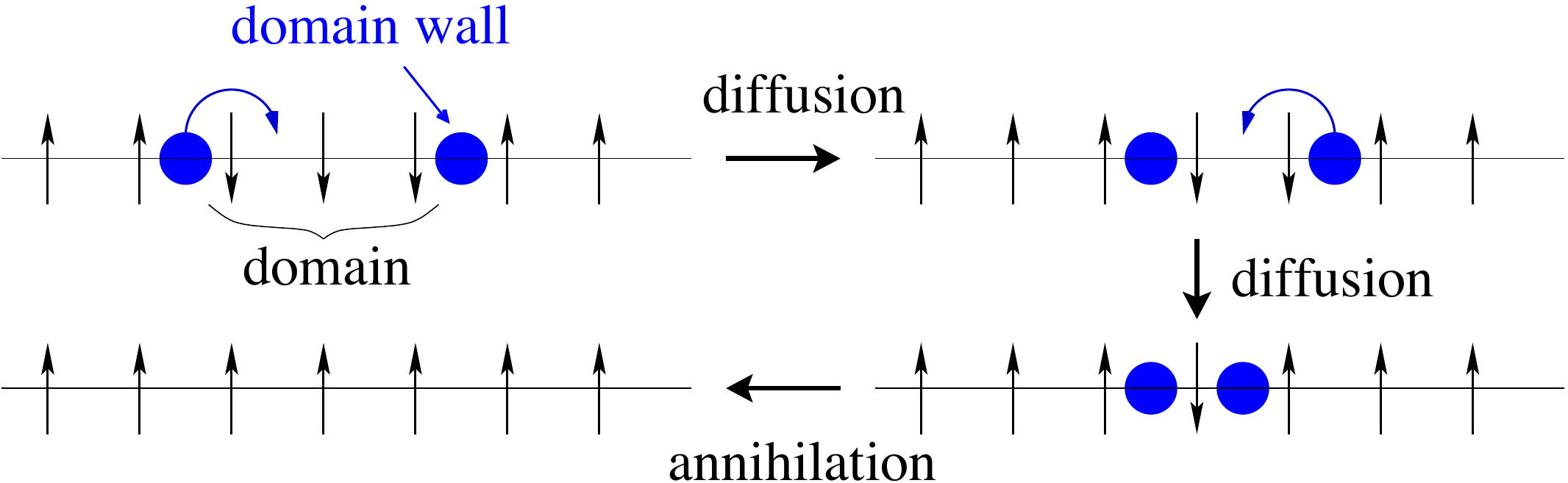}
\caption{Domain walls coarsening in the one-dimensional Ising model ($q=2$) with Glauber dynamics: the blue points (domains walls) perform random walk and eventually annihilate.}\label{fig:domain_walls_ising}
\end{figure}
The
domain walls execute random walks on the lattice. For the Ising model,
domain   walls   annihilate  on   contact   (successive  walls   form,
alternately, kinks  and antikinks in  the spin configuration as in Fig. \ref{fig:domain_walls_ising}),  so the
domain wall density decreases and the mean domain size increases. This
is a  ``coarsening'' process.  For  Potts models with $q>2$  there are
$q(q-1)/2$ different types of wall separating the $q$ different domain
types.   When two  walls  meet,  they can  either  annihilate (if  the
surviving domains  either side  of the merging  walls are of  the same
type, which  occurs with probability $1/(q-1)$), or  coalesce (if the
surviving  domains   are  of   different  types,  which   occurs  with
probability $(q-2)/(q-1)$).  Thus this  process maps onto the chemical
reactions $A+A \to \emptyset$ (annihilation)  and $A + A \to A$ (coalescence),
where the $A$ particles  (the domain walls) perform independent random
walks.

In  the domain  wall  picture, the  probability  that a  spin has  not
flipped is  just the probability  that the site  at which the  spin is
located has not been {\em crossed} by a domain wall. In the limit $q \to
\infty$ domain  walls coalesce with  probability one. It  follows that
only the  nearest walls to  the left and  right of the given  site are
relevant, since other walls that  coalesce with these two do not alter
their trajectories.  The  probability of a given site  not having
been  reached by  a  specified random  walker  in time  $t$ decays, 
as we have seen,  as $t^{-1/2}$, so the probability that  it has not 
been reached by either walker decays as $t^{-1}$, giving 
$\theta(\infty)=1$ in agreement with Eq.~(\ref{theta(q)}).

This class  of problems has  been termed `site persistence',  as the
``survival'' of the spin at a given site is the question  of interest. 
Within the domain  wall  representation, one  can  also  discuss  the 
problem  of `walker persistence',  namely the question of what fraction $S_w(t)$  
of the random walkers (representing the domain walls) has not been touched 
by another walker up  to time $t$.  This problem has  not yet been solved
for general $q$, but there are  some simple special cases. Let us call
the exponent  for walker persistence $\theta_w(q)$, such that $S_w(t) \sim t^{-\theta_w(q)}$ for large $t$. 
For  $q=2$ (the Ising
model) the  number of walkers  at time $t$  that have not  met another
walker is  equal to the total  number of walkers present  at time $t$,
since the walkers  annihilate on contact. The number of walkers is known
to     decrease     as     $t^{-1/2}$  \cite{AmarFamily,Bray89},  so
$\theta_w(2)=1/2$. Another simple limit is $q \to \infty$. As discussed
above, in  this limit one needs only to consider  a given walker  and its
nearest neighbours  to the left  and right. This  three-walker problem
can be solved exactly for any values of the diffusion constants of the
walkers  \cite{Redner,FisherGelfand}. For  the special  case  of equal
diffusion constants  relevant here, the result  is $\theta_w(\infty) =
3/2$, as discussed in subsection \ref{subsection:three_walkers}. For a discussion on the walker persistence problem from a
slightly different angle, see also section \ref{subsection:gen_pers_patterns}.

\subsection{Spin models in higher dimensions}\label{section:higher_dim}

Earlier studies of the persistence properties in Ising spin models of coarsening at $T=0$ \cite{BDG1994a} indicated that the 
large time behavior of the persistence probability $Q(t)$ depends on the dimension $d$ of the system. For $d=2$ it was found
numerically that $Q(t) \sim t^{-\theta}$ with $\theta = 0.22 < 3/8 = 0.375$ \cite{BDG1994a}. This result in $d=2$ was then confirmed
by extensive Monte-Carlo simulations by Stauffer \cite{Sta94} who also estimated that $\theta = 0.166$ in $d=3$. Although there exists no
exact results for $d>1$, approximate analytical methods were developed in Ref. \cite{MajumdarSirePT,MSR00} (explained in detail in section \ref{section:perturbation}) to compute $\theta$ in dimension $d=2, 3$, yielding for instance $\theta \approx 0.19$ in $d=2$. Quite remarkably, the persistence probability for such 
Ising systems in dimension $d=2$ was measured in an experiment on twisted nematic liquid crystals in two dimensions \cite{YPMS97}, as discussed already in the introduction of this review, where the value $\theta = 0.190(31)$ was found in good agreement with the theoretical approaches. 

What happens in larger dimensions ? For $d=4$, numerical simulations of Ising spin systems at $T=0$ \cite{Sta94} indicate
that $Q(t)$ still decays to zero logarithmically but for $d>4$, the persistence probability saturates to a finite value $Q(t) \to Q_\infty$ where $Q_\infty > 0$, which means that there is a finite fraction of spins which never flip. A similar phenomenon of "spin blocking" was also observed in numerical simulations of the zero temperature spinodal decomposition of $q$-states Potts model, on a square lattice, for $q > 4$ \cite{DOS96}. Such freezing phenomena of the persistence probability was shown to be generic for disordered systems \cite{NS99, Jain99}, as discussed later in section~\ref{section:disordered}. Finally, persistence has also been studied for the
vortex dynamics in the $2$-d XY model \cite{BrayLogarithmic}.

\subsection{The 1-d Ginzburg-Landau model}\label{subsection:TDGL}
Another exactly soluble model is the 1-d Ginzburg-Landau model at zero 
temperature. The   Ginzburg-Landau   (or   Ginzburg-Landau-Wilson)   
model is a coarse-grained model suitable for the analysis of phase transitions
using,  in  particular,  renormalisation-group  methods.   As  far  as
critical phenomena  are concerned, the Ginzburg-Landau  model falls in
the   same  university   class  as   spin  models   having   the  same
symmetries. Here  we treat the Ginzburg-Landau  model in one-dimension
and at zero temperature, where (as far as the dynamical properties are
concerned)  the  system  has   {\em  different}  properties  from  the
corresponding spin model (we will discuss the reason for this below).

The Time-Dependent  Ginzburg-Landau (TDGL) model in one  dimension is defined
by the equation of motion
\begin{equation}
\frac{\partial  \phi}{\partial  t}  = \frac{\partial^2  \phi}{\partial
x^2} - \frac{\partial V(\phi)}{\partial \phi},
\label{TDGL}
\end{equation}
where  $V(\phi)$ is  a  symmetric double-well  potential, for  example
$V(\phi) =  (1-\phi^2)^2/4$. The symmetry of the  equation under $\phi
\to -\phi$  reflects the  underlying Ising symmetry  of the  model. At
non-zero  temperature, there  would  be an  additional Langevin  noise
term (this equation is then known in the theory of critical dynamics under the name of model A in the classification of Hohenberg and Halperin \cite{Hohenberg-Halperin}), but here  we focus on the zero-temperature  dynamics. The system
is  evolved from a  random initial  state, such  as a  Gaussian random
field with correlator $\langle \phi(x)\phi(x')\rangle = \delta(x-x')$,
although  the  precise form  of  the  initial  state is  not  relevant
provided that  any spatial correlations are short-ranged  and that the
symmetry  of the  model  under  $\phi \to  -\phi$  is respected.   The
temporal evolution of the system under the dynamics (\ref{TDGL}) leads
to  the  rapid  formation   of  domain  walls  (kinks  and  antikinks)
separating  regions (domains)  where $\phi$  is close  to $\pm  1$. An
isolated  wall satisfies the  equation $\partial_{xx}\phi  = V'(\phi)$,
e.g.\ for  $V(\phi) = (1-\phi^2)^2/4$ one  obtains $\partial_{xx}\phi =
\phi - \phi^3$, with boundary conditions $\phi(\pm\infty) = \pm 1$ for
a kink or $\phi(\pm\infty) = \mp 1$ for an antikink. The kink solution
is $\phi_k(x) = \tanh([x-x_0]/\sqrt{2})$, where $x_0$ is arbitrary and
specifies the  location of the  kink. The kink profile  approaches its
limiting values of $\pm 1$ exponentially. Domain growth, or  coarsening, 
in this model  proceeds in the  late stages  of growth  through the very 
weak interactions through  the exponential tails.  This means  that, at 
late times when the  domain wall density is small, the coarsening 
proceeds through the  annihilation of the  closest kink-antikink pair,  
while the other kinks and antikinks hardly move.

This suggests the following  simplified model. Assign random sizes 
to the domains from some distribution. Flip  the smallest domain
and  combine it  with its  left and  right neighbours,  decreasing the
total  number of  domains  by  two, and  repeat  this process.   Under
iteration  a scaling  regime is  approached in  which  the domain-size
distribution  approaches  a  limiting  form.   How  does  one  discuss
persistence in this model? The site persistence at a given time is the
fraction of the line that has  not yet been traversed by a domain wall
at that time.  The removal of  a domain proceeds by the bounding walls
moving together and annihilating. It follows that the part of the line
formerly   occupied   by    the   removed   domain   is   subsequently
non-persistent. Numerical simulations \cite{BDG1994b} suggest that the
fraction of the line which is persistent, $Q$, decreases as a power of
the  mean domain  size,  $\langle  l \rangle$,  according  to $Q  \sim
\langle l \rangle^{\beta - 1}$. The exponent $\beta$ can be determined
as follows.

For  simplicity of  presentation  we  take the  domain  lengths to  be
integers  and call  the minimum  domain length  $i_0$. Domain  $I$ has
length $l(I)$, of which the persistent part has total length $d(I)$.
At each step, the smallest domain is removed.  So three domains -- the
smallest $I_{\rm min}$ and its neighbours  $I_1$ and $I_2$ -- are replaced
by a single  domain $I$. The total length and  the persistent parts of
$I$ are given by
\begin{eqnarray}
l(I) & =  & l(I_1) + l(I_{\rm min}) +  l(I_2)\ ,\label{l(I)} \\ 
d(I) &  = & d(I_1) +d(I_2)\ . \label{d(I)}
\end{eqnarray}
Since  the domain  lengths  remain uncorrelated  (see  below) one  can
choose  the domains  $I_1$,  $I_2$ randomly  instead  of choosing  the
neighbours of $I_{\rm min}$.  We assume that the total  number of domains,
$N$, is large, that the number of domains of length $i$ is $n_i$, and
that the mean  length of the persistent part of  domains of length $i$
is  $d_i$.  Denoting with  primes the  values of  these quantities
after the $n_{i_0}$ intervals of length $i_0$ have been eliminated, we
obtain
\begin{eqnarray}
N'  &   =  &  N   -  2  n_{i_0}  \;, \\  n_i'  &   =  &  n_i   \left(1  -
\frac{2n_{i_0}}{N}\right)                    +
n_{i_0}\sum_{j=i_0}^{i-2i_0}\frac{n_j}{N}   \frac{n_{i-j-i_0}}{N} \;,   \\
n_i'd_i' & = & n_i d_i \left(1 - \frac{2n_{i_0}}{N}\right)  +
n_{i_0}\sum_{j=i_0}^{i-2i_0}\frac{n_j}{N}
\frac{n_{i-j-i_0}}{N}(d_j+d_{i-j-i_0}) \;.
\end{eqnarray} 
After many iterations, when $i_0$  becomes large, a scaling region is
reached where
\begin{eqnarray}
n_i &  = & \frac{N}{i_0}\,f\left(\frac{i}{i_0}\right),  \\ 
n_i d_i &=& N (i_0)^{\beta-1}\,g\left(\frac{i}{i_0}\right)\ ,
\end{eqnarray}
assuming $n_{i_0}\ll N$, which is valid when $i_0$ becomes large. 
Clearly, $f(x)$ is simply the distribution of interval sizes, where 
the lengths are measured in units of the current minimum length, $i_0$. 

For almost no extra effort, we can also calculate the autocorrelation 
function $A$ for this process, $A$ being, at a given stage in the iteration, 
the overlap of the current state with the initial condition 
\cite{BrayDerrida95}. In addition to the length, $l(I)$, and the 
persistent part, $d(I)$, of interval $I$, 
therefore, we also consider the overlap $a(I)$ of the interval $I$ 
with the initial condition, where initially $a(I)=l(I)$ for all $I$.
The iterative equation for $a(I)$ is
\begin{equation}
a(I) = a(I_1)+a(I_2) - a(I_{\rm min})\,,
\end{equation}  
analogous to (\ref{l(I)}) and (\ref{d(I)}), and we have the additional 
iteration
\begin{eqnarray}
n_i'a_i' & = & n_ia_i\left(1-\frac{2n_{i_0}}{N}\right) + n_{i_0}\sum_{j=i_0}^{i-2i_0}\frac{n_j}{N}\,\frac{n_{i-j-i_0}}{N}
(a_j+a_{i-j-i_0}-a_{i_0})\ .
\end{eqnarray}

We assume that, after many iterations, a scaling regime is reached where
\begin{eqnarray}
n_i & = & \frac{N}{i_0}\ f\left(\frac{i}{i_0}\right)\ , \\
n_id_i & = & \frac{N}{i_0}\ (i_0)^{\beta}\,g\left(\frac{i}{i_0}\right)\ , \\
n_ia_i & = & \frac{N}{i_0}\ (i_0)^{\lambda}\,h\left(\frac{i}{i_0}\right)\ .
\end{eqnarray}
In this limit, $i/i_0$ can be treated as a continuous variable. Using the 
fact  that the functions $f(x)$, $g(x)$ and $h(x)$ should become independent 
of $i_0$ for large $i_0$ leads to the integro-differential equations 
\cite{BDG1994b,BrayDerrida95}
\begin{eqnarray}
f(x)+xf'(x)+\theta(x-3)f(1) \int_1^{x-2}dy\,f(y)f(x-y-1) &=& 0\, , \\
(1-\beta)g(x)+xg'(x)+2\theta(x-3)f(1)  \int_1^{x-2}dy\,g(y)f(x-y-1)& = & 0\, , \\
(1-\lambda)h(x) +xh'(x) + 2\theta(x-3)f(1)\int_1^{x-2}dy\,h(y)f(x-y-1) && \nonumber \\
-h(1)\theta(x-3)\int_1^{x-2}dy\,f(y)f(x-y-1) &=& 0\,.
\end{eqnarray}
To solve these equations one introduces the Laplace transforms 
\begin{eqnarray}
\phi(p)& = &\int_1^\infty e^{-px}f(x)\,dx\ , \label{phidef}\\
\psi(p)& = &\int_1^\infty e^{-px}g(x)\,dx\ , \label{psidef} \\
\chi(p)& = &\int_1^\infty e^{-px}h(x)\,dx\ , \label{chidef}
\end{eqnarray}
which satisfy the equations
\begin{eqnarray}
p\,\phi' &=& -f(1)e^{-p}[1-\phi^2]\ , \label{phi}\\
p\,\psi' &=& -\beta \psi -e^{-p}[g(1)-2f(1)\phi\psi]\ , \label{psi}\\
p\,\chi' &=& -\lambda\psi-e^{-p}[h(1)-2f(1)\phi\chi + h(1)\phi^2]\ .
\label{chi}
\end{eqnarray}
The solutions of these equations can be written as
\begin{eqnarray}
\phi(p) &=& \tanh[r(p)/2]\ , \\
\psi(p) &= & g(1)\int_p^\infty\frac{\cosh^2[r(q)/2]}{\cosh^2[r(p)/2]}
\frac{q^{\beta-1}}{p^\beta} e^{-q}\,dq\ , \label{psi1}\\
\chi(p) &=& 2g(1)\int_p^\infty \frac{\cosh r(q)}{1+\cosh r(p)}
\frac{q^{\lambda-1}}{p^\lambda}\,e^{-q}\,dq\ ,
\end{eqnarray}
where
\begin{equation}
r(p) = 2f(1)\int_p^\infty\frac{e^{-t}}{t}\,dt\ .
\label{r}
\end{equation}
When integrating Eqs.\ (\ref{phi})-(\ref{chi}) the integration constants 
were fixed by the requirement that $\phi$, $\psi$ and $\chi$ are all 
positive functions, as is clear from their definitions.  

So far, the values of the constants $f(1)$, $\beta$ and $\lambda$ are 
arbitrary. They are all fixed, however, by physical considerations. 
Consider the expansion
\begin{equation}
\int_p^\infty \frac{dq}{q}e^{-q} = -\ln p - \gamma - 
\sum_{n=1}^\infty \frac{(-p)^n}{n\,n!}\ ,
\end{equation}
where $\gamma = -\int_0^\infty dt\,e^{-t}\ln t = 0.577215\ldots$ is 
Euler's constant. Then, using Eqs.\ (\ref{phi}) and (\ref{r}) we find
the small-$p$ expansion
\begin{equation} 
\phi(p) = 1- 2e^{2f(1)\gamma}\,p^{2f(1)}\,[1+{\cal O}(p)]\ .
\end{equation}
From the definition of $\phi(p)$, Eq.\ (\ref{phidef}), one has 
$\phi(p) = 1 - \langle x \rangle p + \cdots$, provided that the 
first moment of the interval size distribution, $f(x)$, exists.  
Comparing the two expansions fixes $f(1)=1/2$, and also gives the first 
moment of the scaling function $f(x)$ as 
$\langle x \rangle = 2e^\gamma = 3.56214\ldots$, which is the ratio 
of the mean domain length to the minimum length. Since the first moment 
cannot be zero we have, quite generally, $f(1) \le 1/2$. Cases with 
$f(1)<1/2$ would correspond to models with infinite first moments (see, 
for example, the discussion of a related model by Yekutieli {\it et al.}\ 
\cite{Yekutieli1,Yekutieli2}) and we will consider them no further. 

The exponents $\beta$ and $\lambda$ can be determined in a similar way. 
If we define the function $s(p)=r(p) + \ln p$, a straightforward 
calculation, starting from Eq.\ (\ref{psi1}), gives 
the small-$p$ behaviour of $\psi(p)$ in the form
\begin{equation}
\psi(p) = \frac{g(1)}{1-\beta}\left[1+B(\beta)p^{1-\beta} 
+ {\cal O}(\beta)\right]\ ,
\end{equation}
where 
\begin{equation} 
B(\beta) =e^\gamma \int_0^\infty dq\,q^{\beta-2}\,e^{-q}
\left[(1-q-e^{-q})e^{s(q)} +2(1-\beta)q + (1-\beta)q^2e^{-s(q)}\right]\ .
\end{equation}
If the function $g(x)$ is to have a finite first moment, then $B(\beta)$ 
must vanish. This fixes the value of $\beta$:
\begin{equation}
\beta = 0.82492412\ldots.
\end{equation}
In a similar way, the requirement that $h(x)$ have a finite first moment 
fixes the value of $\lambda$ \cite{BrayDerrida95}:
\begin{equation}
\lambda = 0.39938\ldots.
\end{equation}

We can use, as a proxy for the timescale, the size of the smallest
domain, $L_{\min}$, at a given stage in the coarsening process. 
Then the persistence $Q(L_{\min})$ and the autocorrelation 
function $A(L_{\min})$ decay as $L_{\min}^{\beta-1}$ \cite{BDG1994b} 
and $L_{\min}^{\lambda-1}$ \cite{BrayDerrida95} respectively.

\subsection{Coarsening with a conserved order parameter}

So far we have discussed the case of the coarsening dynamics of a non-conserved order parameter. However, in many physical situations, such
as the spinodal decomposition of binary alloys -- e.g. Ostwald ripening -- or phase separation of fluids or binary liquids, the order parameter is conserved. This yields to an equation of motion for the coarse grained order parameter $\phi(x,t)$ different from the TDGL in Eq. (\ref{TDGL}) and known under the name of Cahn-Hilliard equation \cite{BrayReview}:
\begin{eqnarray}\label{eq:Cahn-Hilliard}
\frac{\partial \phi}{\partial t} = - \nabla^2\left[\frac{\partial V(\phi)}{\partial \phi} - \nabla^2 \phi \right] \;,
\end{eqnarray}
which has thus the form of a continuity equation. Here also $V(\phi)$ has a double-well structure, e. g. $V(\phi) = (1-\phi^2)^2$, with two minima $\phi = \pm 1$ corresponding to the two equilibrium phase. The same equation (\ref{eq:Cahn-Hilliard}) with an additional Langevin noise term on the right-hand
side is known, in theory of critical dynamics, under the name of model B \cite{Hohenberg-Halperin}. In the limit where the volume fraction ${\mathbf \epsilon}$ of the minority phase is small, one can show from Eq. (\ref{eq:Cahn-Hilliard}) \cite{BrayReview} that the dominant growth mechanism is the transport of the order parameter from interfaces of high curvature to regions of low curvature by diffusion through the intervening
bulk phases. In this regime, the theory of Lifshitz, Slyozov \cite{LS61} and Wagner \cite{Wag61} (LSW) demonstrates that the average domain size grows with time $t$ as $t^{1/3}$ (see also Ref. \cite{Hus86} for a more refined calculation of this $1/3$ exponent). In Ref. \cite{LeeRutenberg97}, the persistence probability $Q(t)$ for such a system with a conserved order parameter was studied where $Q(t)$ is the fraction of the system that has not undergone phase change between time $0$ and time $t$. In the limit $\epsilon \to 0$,
using the LSW theory, the authors of \cite{LeeRutenberg97} showed that $Q(t)$ decreases algebraically, $Q(t) \sim t^{-\theta}$ with an exponent $\theta$ which can be computed exactly in the limit $\epsilon \to 0$ as \cite{LeeRutenberg97}
\begin{eqnarray}\label{LeeRuthenberg_theta}
\theta = \gamma_d \epsilon + o(\epsilon) \;,
\end{eqnarray} 	 
where $\gamma_d$ is universal, i.e. does not depend on the surface tension, quench depth, temperature or mobility. This constant $\gamma_d$ can be computed analytically, yielding in particular $\gamma_2 = 0.39008$ and $\gamma_3 = 0.50945$ (the authors also obtained a large $d$ expansion of $\gamma_d$). Remarkably, this exponent $\theta$ (\ref{LeeRuthenberg_theta}) was measured in an experiment on two-dimensional Ostwald ripening \cite{Stavans09} and a very good agreement was found with this theoretical prediction in the limit of small volume fraction. 

The low temperature coarsening dynamics of the one-dimensional Ising ferromagnet with conserved Kawasaki-like dynamics was also studied in Ref. \cite{GonosBray05}. In these models, the domains of size $l$ diffuse with a size-dependent diffusion constant $D(l) \propto l^{\gamma}$, with $\gamma = -1$. In Ref. \cite{GonosBray05} the authors generalized the original model to arbitrary $\gamma$ and, by
using a scaling argument to compute the size distribution of domains, showed that the domain density decreases algebraically as $N(t) \sim t^{-1/(2-\gamma)}$. The persistence probability was shown, numerically, to decay as a power law as $Q(t) \sim t^{-\theta}$, where $\theta$ depends on $\gamma$. We refer the reader to Ref. \cite{GonosBray05} for a more detailed discussion of this model and its relation to so called diffusion-limited cluster-cluster aggregation (DLCA) model. 

We end up this section by mentioning that the local persistence has been studied \cite{SS03} in the case where the order parameter itself is not conserved but it is coupled to an ordering field which is conserved, a situation which corresponds to Model C in the classification of \cite{Hohenberg-Halperin}. In Ref. \cite{SS03} Monte Carlo simulations were performed on antiferromagnetic Ising model, the order parameter being the staggered magnetization, with a conserved global magnetization $M_0 \neq 0$ \cite{SS03} and it was found that the persistence probability $Q(t)$ decays algebraically $Q(t) \sim t^{-\theta}$ at large time $t$, with an exponent $\theta$ which seemingly depends non monotonically on $M_0$.

\section{Persistence of Gaussian sequences and Gaussian processes}\label{section:Gaussian}

In many examples discussed so far in the review (and also to come in
later sections), the underlying stochastic process is Gaussian with
a correlator of the form $a(t_1,t_2)=\langle 
x(t_1)x(t_2)\rangle=t_1\,f(t_1/t_2)$. This is
clearly non-stationary.
An example being the simple Brownian motion $dx/dt=\eta(t)$ where
$\eta(t)$ is the delta correlated Gaussian white noise. 
In this simple Markov case, $f(x)= 2D\, {\rm min} (1,1/x)$. The persistence of such process between two times
$t_1$ and $t_2$ can be mapped, via the change of time variable
$T=\ln (t)$ and rescaling $X(T)=x(t)/\sqrt{\langle x^2(t)\rangle}$ (the so called Lamperti transformation), to the 
problem 
of the persistence of
a stationary Gaussian process $X(T)$ in the new time $T$ with a 
stationary correlator $A(T)= \exp[-|T|/2]$. Another example being
the random acceleration process $d^2x/dt^2=\eta(t)$, perhaps the
simplest non-Markovian process. This process can
again be mapped (via the same change of variables) to
a Gaussian stationary process with the correlator
$A(T)= \frac{3}{2}\exp(-|T|/2)-\frac{1}{2}\exp(-3|T|/2)$.
Other examples include e.g., a higher order process
evolving via $d^nx/dt^n=\eta(t)$ discussed before in section \ref{subsection:bfp_higher_order}, a field evolving 
via
the diffusion equation starting a random initial configuration,
fluctuating interfaces
evolving via linear Langevin equations etc. The last two examples
to be discussed later in the review. Thus the persistence probability
in all these problems can be mapped to that of a Gaussian stationary
process with a prescribed correlator $A(T)$. The precise form of the 
correlator varies from one problem to other.

This then raises the general question: given a Gaussian stationary
process $X(T)$ with a given correlator $A(T)$, what is the
persistence $Q(T)$, i.e., the probability that the process
$X(T)$ stays positive over the interval $[0,T]$ ? 
This general problem has been studied extensively
in the probability literature in the past (we summarize
below some of the salient features of these studies).
One of the main conclusions of these studies is that $Q(T)$, even for large 
$T$, depends crucially on the full functional form of the correlator
$A(T)$ (and not just on its tail properties). Exact results are
known only in very few cases (notably for Markov processes).

Since the literature is a bit sporadic, for the convenience of the readers
we summarize below some basic properties of the Gaussian process
along with some results concerning the zero crossing properties
of Gaussian processes that would be useful for the physics problems
discussed in this review. 

Let us start by recapitulating some basic properties of a Gaussian
stochastic process. A stochastic `process' is defined in continuous time.
But before we define the process in continuous-time, it is conceptually
easier to think in terms of a discrete-time setting. So, we first discuss 
the persistence of a discrete-time Gaussian `sequence' in the next section which will make the ground simpler for the discussion
on the continuous-time Gaussian `process' later.

\subsection{Gaussian sequence}

Consider 
first 
a set of $N$ {\em correlated} random variables 
$\{X_1,X_2,\ldots,X_N\}$, each of zero mean and with a joint distribution 
which is a multivariate Gaussian distribution of the form 
\begin{equation}
P\left(\{X_i\}\right)= B_N\, 
\exp\left[-\frac{1}{2}\sum_{i,j} B_{i,j}\, X_i\, X_j\right]=B_N\, 
\exp\left[-\frac{1}{2} [X]^{t} B [X]\right]
\label{gs.1}
\end{equation}
where $[X]$ represents the column vector with entries $\{X_i\}$ and
$[X]^t$ its transpose.
This joint distribution is thus fully specified by the matrix $[B]$
with symmetric entries $B_{i,j}=B_{j,i}$ and one assumes that all
the eigenvalues of $[B]$ are positive. 
The prefactor is such that the joint distribution is normalized
to unity: $\int_{-\infty}^{\infty} \prod_{i=}^N dX_i P\left(\{X_i\}\right)=1$.
This integral can be performed exactly by making a change
of variable: $[X]= [S][Y]$ where the $(N\times N)$ matrix $S$
diagonalizes the matrix $B$: $[S]^t [B] [S]= [\Lambda]$ where
the entries $\Lambda_i\ge 0$ of the diagonal matrix $[\Lambda]$
are positive. Note that after the change of variables, the limits of integration 
for the $Y_i$ variables are still $-\infty$ and $\infty$. This is
a crucial fact (as we will see later) that allows us to compute
this multidimensional integral exactly  
\begin{eqnarray}
&&\int_{-\infty}^{\infty} \prod_{i=1}^N dX_i\, P\left(\{X_i\}\right)=B_N \int_
{-\infty}^{\infty} \prod_{i=1}^N dY_i \, \exp\left[-\frac{1}{2}\, 
\sum_i \Lambda_i\, Y_i^2\right] \\
&&= B_N \prod_{i=1}^N 
\sqrt{\frac{2\pi}{\Lambda_i}}= B_N 
\frac{(2\pi)^{N/2}}{\sqrt{{\rm det} B}}.
\label{gs.2}
\end{eqnarray}
Setting the right hand side to $1$, one gets the exact prefactor
\begin{equation}
B_N= \frac{\sqrt{ {\rm det} B}}{(2\pi)^{N/2}}.
\label{gs.3}
\end{equation}
   
From the joint distribution in \eqref{gs.1}, one can easily compute all the moments.
For example, it is easy to compute the two point correlation
function (once again using the diagonal basis) and show that
\begin{equation}
A_{i,j}=\langle X_i X_j\rangle = B^{-1}_{i,j} \;,
\label{gs.4}
\end{equation}
where $B^{-1}_{i,j}$ is the $(i,j)$-th entry of the inverse matrix $[B]^{-1}$.
Thus $B_{i,j}= A^{-1}_{i,j}$ and hence once we know the two-point correlation
function $A_{i,j}$ and its inverse, it completely specifies the full joint 
probability
distribution of a Gaussian multivariate distribution
\begin{equation}
P\left(\{X_i\}\right)= \frac{1}{(2\pi)^{N/2}\, \sqrt{{\rm det} A}}\,
\exp\left[-\frac{1}{2}\, \sum_{i,j} A^{-1}_{i,j}\, X_i\, X_j\right].
\label{gs.5}
\end{equation}
Given this joint distribution, one can also compute any marginal.
For example, the one point distribution function $P(X_i)$ can be computed
by fixing $X_i$ and integrating over all the rest of $(N-1)$ variables
\begin{equation}
P(X_i)= \int_{-\infty}^{\infty} \prod_{j\ne i} dX_j\, P\left(\{X_i\}\right).
\label{gs.6}
\end{equation}
Upon carrying out this integration, one recovers the standard Gaussian 
distribution of a single variable
\begin{equation}
P(X_i)= \frac{1}{\sqrt{2\pi A_{i,i}}}\, \exp\left[-X_i^2/{2 A_{i,i}}\right]
\label{gs.7}
\end{equation}
where $A_{i,i}= \langle X_i^2\rangle$ is just the variance of the random
variable $X_i$.

Let us now imagine this multivariate Gaussian set $\{X_1,X_2,\ldots, X_N\}$ 
forms a sequence of length $N$ where $N$ is like a discrete-time.
Its joint distribution is specified in \eqref{gs.5} with a prescribed
correlator $A_{i,j}$. 
We will 
define the persistence $Q_N$  
as the probability that the sequence stays non-negative up to step $N$
\begin{equation}
Q_N = {\rm Prob}\left[X_1\ge 0, X_2\ge 0,\ldots, X_N\ge 0\right].
\label{gs.9}
\end{equation}
Using the joint distribution in \eqref{gs.5} one then has to evaluate the 
following one-sided multiple integral
\begin{equation}
Q_N =  \frac{1}{(2\pi)^{N/2}\, \sqrt{{\rm det} A}}\,\int_0^{\infty}\ldots 
\int_0^{\infty} \prod_{i=1}^N dX_i\, \exp\left[-\frac{1}{2}\, 
\sum_{i,j} A^{-1}_{i,j}\, X_i\, X_j\right].
\label{gs.10}
\end{equation}
Note that unlike in \eqref{gs.2}, where the integral limits were over $(-\infty, 
\infty)$ that fortunately remained unaffected under the change of variables 
$[X]=[S][Y]$
thus enabling us to perform the multiple integral via diagonalisation, here
in \eqref{gs.10} we can no longer use the same trick. This is because the
range of integration now is over $[0,\infty)$ for each variable $X_i$. 
We can still make a change of variable $[X]=[S][Y]$ to diagonalize
the quadratic form inside the exponential, but the limits of integration
over the $Y$ variables now become rather complicated. While the upper
limit of $Y_i$ is still $\infty$, the lower limits $[X]=[0]$, i.e., 
$[S][Y]=0$ become a set of $N$ complicated hyperplanes in the
$N$-dimensional space $\{Y_1,Y_2,\ldots, Y_N\}$~\cite{Slepian}. This is the main
reason why calculating the persistence $Q_N$ of an arbitrary
Gaussian sequence with a given correlator $A_{i,j}$ is a hard 
problem, simply because we do not know in general how to perform
the one-sided multiple integral in \eqref{gs.10} for arbitrary $N$~\cite{Slepian}.
For arbitrary $A_{i,j}$, this integral can be performed in closed form 
only for $N=1$, $N=2$
and $N=3$, but not for $N\ge 4$. For example, for $N=1$, one has trivially
$Q_1=1/2$. For $N=2$ already, the double integral is not so trivial
to perform. However, with a little bit of algebra one can show that~\cite{Slepian}
\begin{equation}
Q_2= \frac{1}{4} + \frac{1}{2\pi}\, 
\sin^{-1}\left[\frac{A_{1,2}}{\sqrt{A_{1,1}\,A_{2,2}}}\right].
\label{gs.11}
\end{equation}
Note that in absence of correlation between the two random variables
$X_1$ and $X_2$ for the $N=2$ case, i.e., when $A_{1,2}=0$ the integrals
get decorrelated and one recovers $Q_2=1/4$.
Similarly for $N=3$ also, one can carry out the triple integral
in closed form. Defining, $r_{i,j}= A_{i,j}/\sqrt{A_{i,i}\,A_{j,j}}$,
it turns out that for $N=3$~\cite{Slepian}
\begin{equation}
Q_3= \frac{1}{8}+ \frac{1}{4\pi}\left[\sin^{-1}(r_{1,2})+ \sin^{-1}(r_{2,3}) 
+\sin^{-1}(r_{3,1})\right].
\label{gs.12}
\end{equation}
Unfortunately our luck runs out for $N\ge 4$ where no closed form
expression is known for $Q_N$! Also, there does not seem to be
any obvious way to derive even the asymptotic behavior for large $N$, in
which we will be primarily interested. There are only few special cases of the correlator $A_{i,j}$ for
which $Q_N$ can be computed exactly for arbitrary $N$ \cite{Slepian,BlakeLindsay73}. In this review, we will discuss another
solvable case in section \ref{section:exact}. 

\subsection{Gaussian process}

A Gaussian `process' is just the continuous-time cousin of the
Gaussian `sequence' discussed above~\cite{Chaturvedi}. We consider a Gaussian process
$\{X(T')\}$ where the continuous time $T'$ runs over a fixed
interval $T'\in [0,T]$. It is useful to think of any realization
of the process $\{X(T')\}$ as a continuous path. Conceptually, it is easier
to discretize the time interval $[0,T]$ into $N$ small intervals
of length $\Delta T$ each: $T=N \Delta T$ and then think of the
path $\{X(T')\}$ as a sequence of multivariate Gaussian variables
$\{X_1,X_2,\ldots, X_N\}$ discussed in the previous section.
Then this Gaussian `sequence' $\{X_1,X_2,\ldots, X_N\}$ converges to the 
Gaussian `process' 
$\{X(T')\}$ in the limit $N\to \infty$, $\Delta T\to 0$ but keeping the
product $T=N \Delta T$ fixed. Following this definition, one can then easily
write down the statistical weight (or probability density) associated
with a path or realization of the Gaussian `process' as a simple
continuous-time analogue of the discrete multivariate joint distribution
in \eqref{gs.5}
\begin{equation}
P\left[\{X(T')\}\right]\propto \exp\left[-\frac{1}{2}\, \int_0^T\int_0^T dT_1\, 
dT_2\, A^{-1}(T_1,T_2)\, X(T_1)\, X(T_2)\right] \;,
\label{gp.1}
\end{equation}
where $A^{-1}(T_1,T_2)$ is again the inverse of the correlation matrix 
$A(T_1,T_2)=\langle X(T_1)\,X(T_2)\rangle$. The statistical weight
in \eqref{gp.1} is to be understood as the probability measure associated
with a path in the standard path-integral or
functional-integral sense. 

The persistence $Q(T)$ of a generic Gaussian process with a prescribed
auto-correlator $A(T_1,T_2)$ can be defined in a similar way as in the case
of a Gaussian sequence in \eqref{gs.9}, namely that it represents
the probability that the process stays non-negative over a fixed
time interval $[0,T]$ 
\begin{equation}
Q(T)= {\rm Prob}\left[ X(T')\ge 0 \;, \; {\rm for\,\, all}\,\, 0\le T'\le T 
\right].
\label{gp.3}
\end{equation}
Thus $Q(T)$ is just the fraction of paths, out of all possible paths in $[0,T]$,
that do not cross the origin over the time interval $[0,T]$. Using the
measure in \eqref{gp.1} one can then formally write $Q(T)$ as a
ratio of two path integrals
\begin{equation}
Q(T)= \frac{\int_{+} {\cal D}X(T') \exp\left[-\frac{1}{2}\, \int_0^T\int_0^T 
dT_1\,
dT_2\, A^{-1}(T_1,T_2)\, X(T_1)\,X(T_2)\right]}{\int {\cal D}X(T') 
\exp\left[-\frac{1}{2}\, \int_0^T\int_0^T    
dT_1\,
dT_2\, A^{-1}(T_1,T_2)\, X(T_1)\,X(T_2)\right]}= \frac{Z_{+}(T)}{Z_0(T)} \;,
\label{gp.4}
\end{equation}
where the subscript $+$ in the numerator indicates that it counts all
paths that stays non-negative over $[0,T]$, while the denominator
counts all possible paths over the interval $[0,T]$ and serves
just as a normalization constant. These two functional integrals
are called the partition functions: $Z_{+}(T)$ for the numerator
and $Z_0(T)$ for the denominator. Clearly, the persistence $Q(T)$
is a function of $T$ and also it depends functionally on the correlator 
$A(T_1,T_2)$. The main problem then is to compute $Q(T)$, given
the correlator $A(T_1,T_2)$. Just as in the case of a Gaussian sequence,
computing $Q(T)$ for a Gaussian process with an arbitrary correlator
$A(T_1,T_2)$ is a very hard problem (for a review see~\cite{BlakeLindsay73}). 

\subsection{Gaussian stationary process}\label{subsection:GSP}

In this review, we will focus on a special subset of Gaussian processes
called Gaussian stationary processes (GSPs) for which the correlator
$A(T_1,T_2)$ actually depends only on the time difference
\begin{equation}
A(T_1,T_2)= A(|T_1-T_2|).
\label{gsp.1}
\end{equation}
The stationarity condition actually makes such processes a bit simpler
to handle. In fact, one can now imagine that the process over the time interval
$(-\infty,\infty)$ with a probability measure as given in 
\eqref{gp.1}, except that the limits of the time integrals now run from
$-\infty$ to $+\infty$. One can now focus on any specific section
$[T_1,T_2]$ of this infinite time axis and define the persistence
$Q(T_1,T_2)$ as the probability that the stationary process stays positive
between time $T_1$ and $T_2$. Formally one can write this as
\begin{equation}
Q(T_1,T_2)= \frac{ \int {\cal D}X(T')\, I\left(T_1,T_2, 
\{X(T')\}\right)\, e^{-\frac{1}{2} 
\int_{-\infty}^{\infty}\int_{-\infty}^{\infty} dT_1'\, dT_2'\, 
A^{-1}(|T_1'-T_2'|)\, X(T_1)\, X(T_2)}}{\int {\cal D}X(T')\,
e^{-\frac{1}{2} \int_{-\infty}^{\infty}\int_{-\infty}^{\infty} dT_1'\, 
dT_2'\,
A^{-1}(|T_1'-T_2'|)\, X(T_1)\, X(T_2)}}
\label{gsp.2}
\end{equation}
where $ I\left(T_1,T_2, 
\{X(T')\}\right)$ is an indicator function that is $1$ if the section
of the path $\{X(T')\}$ between $T_1$ and $T_2$ is positive and 
$0$ if the path crosses zero between $T_1$ and $T_2$. The denominator
is just the normalization factor. Evidently, due to stationarity, the
persistence $Q(T_1,T_2)= Q(|T_2-T_1|=T)$ also depends only on the
time difference $T=|T_2-T_1|$. Thus the problem is well defined: given 
the stationary correlator $A(T)$, 
can one compute
$Q(T)$ in \eqref{gsp.2}? The answer is no for an arbitrary 
correlator 
$A(T)$. However, there are some special cases, such as for Markov 
processes, for which $Q(T)$ is exactly computable (see below). 
However, some general properties of GSP are useful when one 
tries to estimate bounds for $Q(T)$ or even to develop an approximation
method for estimating $Q(T)$ (as discussed at various places of this review).
Below, we summarize a few important results that have been used throughout 
this review.

\vskip 0.2cm

\noindent (i) {\bf Markov property and Doob's theorem:} Consider a GSP
with a given correlator $A(T)$, normalized such that $A(0)=1$. An important 
subclass of these GSP's
are those which satisfy Markovian property. What is the necessary and 
sufficient condition on the correlator $A(T)$ such that the GSP is Markovian?
Doob's theorem (see e.g. ~\cite{Chaturvedi}) answers this question: A GSP is Markovian if and only if
the correlator $A(T)=\exp\left[-\lambda |T|\right]$ for all $T$, i.e.,
the correlator is purely exponential. For any other correlator, the
GSP is non-Markovian. For a Markov GSP, the persistence $Q(T)$ is
exactly known for all $T\ge 0$
\begin{equation}
Q(T) = \frac{2}{\pi}\, \sin^{-1}\left[e^{-\lambda T}\right].
\label{markovp1}
\end{equation}
Note, in particular, that for large $T$, $Q(T)\sim e^{-\theta T}$ where the 
exponent $\theta=\lambda$.

To illustrate this result (\ref{markovp1}), we apply it to one-dimensional Brownian motion (\ref{eq:BM}).  
We recall that the probability that a simple Brownian motion,
starting at $t_1$ at position $x_1$, does not change sign
over the time interval $[t_1,t_2]$ is given by
$Q(x_1,t_2, t_1)= {\rm erf}\left(|x_1|/\sqrt{4D(t_2-t_1)}\right)$ [see Eq. (\ref{eq:images}) and below].
If one averages over the initial position $x_1$ drawn from
the Gaussian distribution $p(x_1,t_1)= e^{-x_1^2/{4Dt_1}}/\sqrt{4Dt_1}$,
one simply obtains the persistence over the time interval $[t_1,t_2]$
(with $t_1\le t_2$ without any loss of generality)
\begin{equation}
Q(t_1,t_2)= \frac{2}{\pi}\sin^{-1}\left[\sqrt{\frac{t_1}{t_2}}\right].
\label{brownpers1}
\end{equation}
Using now the mapping $T=\ln t$ and $X(T)=x(t)/\sqrt{\langle x^2(t)\rangle}$
we map the Markov process $dx/dt=\eta(t)$ to a GSP (Lamperti transformation) with
correlator $A(T)=\exp[-|T|/2]$. Using $t_1=\exp[T_1]$ and $t_2=\exp[T_2]$
in Eq. (\ref{brownpers1}),
we then immediately find that the persistence $Q(T=|T_1-T_2|)$ that the
GSP $X(T)$ does not change sign over the time interval $[T_1,T_2]$
is  given by 
\begin{equation}
Q(T)= \frac{2}{\pi}\, \sin^{-1}\left[e^{-|T|/2}\right] \;,
\label{markovp2}
\end{equation}
which is completely consistent with the exact result for Markov process
in Eq. (\ref{markovp1}) with $\lambda=1/2$.
Note also that for large $T$, $Q(T)\sim \exp[-T/2]$ which, when translated
into the original time variable $t=e^{T}$, reduces
to the standard power-law decay of the persistence of a Brownian motion, 
$Q(t)\sim t^{-1/2}$ for large time $t$.

\vskip 0.2cm

\noindent (ii) {\bf Smooth processes and Rice's formula:} Consider again a 
GSP
with correlator $A(T)$ with the normalization $A(0)=1$. Consider the
short time behaviour of $A(T)$. Quite generically, $A(T)$ behaves
as $T\to 0$~\cite{Slepian,BlakeLindsay73}
\begin{equation}
A(T)= 1 - a |T|^{\alpha} + o\left(|T|^{\alpha}\right) 
\label{stb1}
\end{equation}
where, necessarily, $0\le \alpha\le 2$ and $a>0$ is a constant. If 
$\alpha>2$, one can show that
the Fourier transform ${\tilde A}(\omega)=\int_{-\infty}^{\infty} A(T) 
e^{i\omega T}dT$ is not positive definite, a condition that is necessary
for the normalization of the Gaussian measure in (\ref{gp.1}) with (\ref{gsp.1}). A GSP is called
{\em smooth} if $\alpha=2$. For all $\alpha<2$, the process is called {\em 
rough}. The reason for this is that for $\alpha=2$, the process
$X(T)$ has a finite density of crossings of the origin. To see this,
consider the mean number, $N_0$, of zero crossing of $X(T)$ in a
time interval $T$. This is given by the general formula
\begin{eqnarray}
N_0 & = & \int_0^T dT'\,\langle \delta(X(T'))|\dot{X}(T')|\rangle
\nonumber \\
& = & T \langle \delta(T') \rangle \langle |\dot{X}(T')| \rangle\ ,
\end{eqnarray}
since $X(T)$  and $\dot{X}(T)$ are statistically independent. This follows 
from
$\langle X(T)\,\dot{X}(T) = \frac{1}{2}(d/dT)\langle X^2(T)\rangle = 0$, as
$\langle X^2(T)\rangle  = 1$ is a constant.  Since $X$ has  a normal
distribution  with unit variance, we  have
$\langle \delta(X) \rangle =  1/\sqrt{2\pi}$. Also $\langle
\dot{X}^2 \rangle = (\partial^2/\partial T_1 \partial T_2)_{T_1=T_2}
\langle X(T_1)X(T_2) \rangle = -A''(0)$.   
So the mean crossing density,
$\rho = N_0/T$, is given by the celebrated Rice's formula~\cite{Riceformula}
\begin{equation}
\rho = \frac{1}{\pi}\sqrt{-A''(0)}\ .
\label{Rice}
\end{equation}
From here we see immediately that the crossing density is finite for a smooth
process ($\alpha=2$) and infinite for a rough process ($\alpha<2$).
For example, the Markov GSP with correlator $A(T)=\exp\left[-\lambda 
|T|\right]$ has $\alpha=1$ and hence is a rough process with infinite
density of zero crossings.

\vskip 0.2cm

\noindent (iii) {\bf Slepian's inequality:} In absence of a general 
expression for $Q(T)$ for arbitrary correlator $A(T)$, an inequality
due to Slepian~\cite{Slepian} can be very useful sometimes. Consider two GSP's
with respective correlators $A_1(T)$ and $A_2(T)$, which are normalized, $A_1(0) = A_2(0) = 1$. Let $Q_1(T)$
and $Q_2(T)$ denote their respective persistence probabilities.
Then Slepian's inequality states that if $ A_1(T)\le A_2(T)$ for
all $T$, then $Q_1(T)\le Q_2(T)$ for all $T$. For example,
suppose one comes across a GSP whose correlator satisfies the
property $A(T)\le \exp\left[-\lambda |T|\right]$ for all $T$ and for some
$\lambda>0$.
Then, using the exact result for the Markov GSP in Eq. (\ref{markovp1})
and Slepian's inequality, one can obtain the exact bound  
$Q(T)\le \frac{2}{\pi}\sin^{-1}\left[e^{-\lambda T}\right]$ for all $T$.
Indeed, we have used this inequality to obtain exact bounds on the
persistence exponent for a certain class of fluctuating interfaces (see 
section \ref{section:interfaces}).   

\vskip 0.2cm
\noindent (iv) {\bf Newell-Rosenblatt result:} For a 
GSP with correlator $A(T)$, Newell and Rosenblatt obtained~\cite{Newell}
bounds for the persistence probability $Q(T)$ which
are often very useful. Loosely speaking, their result states that if 
$A(T)\sim T^{-\alpha}$ for large $T$ and for some $\alpha>0$, then $Q(T)$ has 
the
following asymptotic forms for large $T$ depending on the
value of $\alpha$:
\begin{eqnarray}
Q(T) \sim \exp(-K_1 T), \ \ \ \ \ &&\alpha > 1, \\
\exp(-K_2 T^\alpha \ln T) \le Q(T) \le \exp(-K_3 T^\alpha), &&0<\alpha<1\ 
\label{Newell-Rosenblatt}
\end{eqnarray}
where the $K_i$'s are some positive constants.

One of the consequences of the Newell-Rosenblatt result is that if
the correlator decays exponentially for large $T$, $A(T)\sim \exp[-\lambda 
T]$ with some $\lambda$ (as is the
case in many of the physics examples discussed in this review where
a physical process in real time $t$ can be mapped to a GSP in
time $T=\ln t$), then the persistence of this GSP also decays
exponentially for large $T$, $Q(T)\sim \exp[-\theta T]$ with
some decay constant $\theta$ that however depends on the full form of the 
correlator $A(T)$ (and not just on its asymptotic tail
$A(T)\sim \exp[-\lambda T]$). Translated into the real time $t=e^{T}$, this
then proves that for such physical processes, the persistence
in real time $t$ decays as a power law for large $t$, $Q(t)\sim t^{-\theta}$.
Thus the decay constant $\theta$ in $Q(T)$ for the GSP is indeed
the persistence exponent $\theta$ of the underlying process in
real time $t$.

\section{Perturbation theory for non-Markovian Gaussian stationary processes}\label{section:perturbation}
In  this section  we discuss  how  one can  calculate the  persistence
of  a non-Markovian Gaussian stationary process (GSP) that is {\em close}  
to  a  Markovian   process,  using  perturbation  theory 
\cite{MajumdarSirePT,Oerding}. 
We want to calculate the probability that the variable $X(T)$ has not changed 
sign in the time interval $(0,T)$ (or any time interval of length $T$, since 
the process is stationary). This persistence probability can be written 
as the ratio of two path integrals:
\begin{equation}
Q(T) = {\rm Proba.}\, [X(T') \geq 0 \;, \; {\rm for \, all} \, 0 \leq T'\leq T] = \frac{\int_{X>0}\,DX(T)\,e^{-S}}{\int DX(T)\,e^{-S}}\ ,
\label{first line}
\end{equation}
where the path integral in the numerator is restricted to paths $X(T)$ where 
$X(T') \geq 0$ for all $T'$ in $[0,T]$, and where the `action' $S[X]$ has the form
\begin{equation}
S = \frac{1}{2}\int_0^T dT_1 \int_0^T dT_2\, X(T_1)G(T_1,T_2)X(T_2)\ .
\end{equation}
Here $G(T_1,T_2)$ is the matrix inverse of the correlation matrix
$\langle X(T_1)\, X(T_2)\rangle \equiv A(T_1-T_2)$. Notice that $G$ is not 
simply a function of $T_2-T_1$.

Our strategy is to compute the persistence perturbatively, starting from the 
Ornstein-Uhlenbeck process,
\begin{equation}
\frac{dX^0}{dT} = -\mu X^0 + \eta(T)\ ,
\label{OU}
\end{equation}
where $\eta(T)$ is Gaussian white noise with correlator
\begin{equation}
\langle \eta(T)\eta(T') \rangle = 2\mu\,\delta(T-T') \;,
\end{equation}
which defines the (only) Markovian Gaussian stationary process. The 
strength of the noise has been chosen so that in the stationary state 
$\langle (X^0)^2 \rangle = 1$. The autocorrelation for this process is readily 
determined as
\begin{equation}
A^0(T_1-T_2) \equiv \langle X(T_1) X(T_2) \rangle 
= \exp(-\mu|T_1-T_2|)\ .
\end{equation}
Now suppose that the non-Markovian process $X(T)$ is perturbatively close 
to the Markov process $X^0(T)$, such that 
\begin{equation}
G(T_1,T_2) = G^0(T_1,T_2) + \epsilon g(T_1,T_2)\ ,
\end{equation}
where $\epsilon$ is small. Then we can expand the exponentials in the 
path-integrals in Eq.\ (\ref{first line}), and re-exponentiate. To 
${\cal O}(\epsilon)$ the numerator becomes 
\begin{eqnarray}
\int_{\cal C} DX(T)e^{-S} &=& \int_{\cal C}DX(T)\exp\left
(-S^0-\frac{\epsilon}{2}\int_0^TdT_1\int_0^T dT_2 \right. \nonumber \\
&& \left. \times g(T_1,T_2)A_C^0(T_1,T_2) + {\cal O}(\epsilon^2)\right)\ ,
\label{expansion}
\end{eqnarray}
where the subscript $\cal{C}$ indicates that the paths in the integral in 
the numerator of Eq.\ (\ref{first line}) satisfy the constraint $X(T')\geq 0$ 
for $0\leq T' \leq T$, and 
\begin{equation} 
A^0_C(T_1,T_2) = \frac{\int_{\cal C}DX(T)X(T_1)X(T_2)e^{-S_0}}
{\int_{\cal C} DX(T)e^{-S_0}}\ .
\end{equation}

Here  $A^0_C(T_1,T_2)$ is  the correlation  function for  the  Markov process,
$X^0(T)$, averaged  and normalized only  over paths satisfying  the constraint
${\cal  C}$.  The  denominator in  Eq.\  (\ref{first  line})  is given  by  an
identical expression,  except that  $A^0_{\cal C}$ is  replaced by  $A^0$, the
unconstrained correlation function, and the integrals are over all paths.

Due  to  the constraint,  $A^0_{\cal  C}$  will  not be  time  translationally
invariant for  finite $T$. In  the limit $T  \to \infty$, however,  the double
time-integral  in Eq.\  (\ref{expansion})  reduces to  $T$  times an  infinite
integral over the relative time $T_2-T_1$, and $A^0_{\cal C}(T_1,T_2)$ can be 
replaced by its stationary limit, $A^0_{\cal C}(T_2-T_1)$. Similarly, the 
function $g(T_1,T_2)$ will be translationally invariant in this regime, so 
that the required double integral can be written as a single integral in 
Fourier space:
\begin{eqnarray}
\lim_{T \to \infty } \frac{1}{T}\int_0^T dT_1\int_0^T dT_2\, g(T_1,T_2)A^0_{\cal C}(T_1,T_2) = \int_{-\infty}^\infty (d\omega/2\pi)
\tilde{g}(\omega)\tilde{A}^0_{\cal C}(\omega) \;.
\label{Fourier}
\end{eqnarray}
The zeroth order result, 
$\int_{x>0}DX(T)\exp(-S_0)/\int DX(T)\exp(-S_0)$, is just the persistence of 
the GSP $X^0(T)$, which decays as $\exp(-\mu T)$ for large $T$. 
 
From Eqs.\ (\ref{first line}), (\ref{expansion}) and (\ref{Fourier}), the 
persistence exponent can be written as
\begin{eqnarray}
\theta &=& \lim_{T \to \infty} -\frac{1}{T}\ln\left[{\rm Proba.} [X(T') \geq 0 \;, \; 0 \leq T' \leq T] \right] \nonumber \\
&=& \mu + \epsilon \int_0^\infty \frac{d\omega}{2\pi}
[\tilde{A}^0_{\cal C}(\omega) - \tilde{A}^0(\omega)] + {\cal O}(\epsilon^2)\ ,
\label{pert_freq}
\end{eqnarray}
where the term proportional to $\tilde A^0(\omega)$ is the ${\cal O}(\epsilon)$ contribution coming from 
the denominator in Eq.\ (\ref{first line}). 

We now calculate $A^0_{\cal C}(T)$. The conditional probability, 
$P(X,T|X_0,0)$, is the probability that $X(T)$ takes the value $X$ at time $T$ 
given that it took the value $X_0$ at time $0$. It may be directly calculated 
from Eq.\ (\ref{OU}), giving
\begin{eqnarray}
P(X,T|X_0,0) &=& \left[\frac{1}{2\pi(1-e^{-2\mu T})}\right]^{1/2} \exp\left[-\frac{(X-X_0e^{-\mu T})^2}{2(1-e^{-2\mu T})}\right].
\label{conditional}
\end{eqnarray}  
The conditional probability, $P^+(X_2,T_2|X_1,T_1)$ that $X(T)$ has the value 
$X_2>0$ at time $T_2$, given that it had the value $X_1>0$ at time $T_1$ and 
that $X(T)>0$ for all $T$ in $(T_1, T_2)$, is given by the method of images 
\cite{Redner}:
\begin{equation}
P^+(2|1) = P(X_2,T_2|X_1,T_1) - P(X_2,T_2|-X_1,T_1)\ ,
\label{image}
\end{equation}
where we have adopted a natural shorthand notation for the arguments of $P^+$.

To find the joint probability, $P^+(X_1,T_1;X_2,T_2)$, that the process 
has the values $X_1$ at $T_1$ and $X_2$ at $T_2$, averaged only over paths
where $X(T)$ stays positive between an initial time $T_i$ and a final time 
$T_f$, we consider a path starting at $(X_i,T_i)$ and finishing at 
$(X_f,T_f)$, passing through $(X_1,T_1)$ and $(X_2,T_2)$, with $X(T)$ always 
positive. The required stationary limit is given (using Bayes Theorem) as
\begin{equation}
P^+(x_1,T_1;x_2,T_2) = \lim_{T_i \to -\infty,T_f \to \infty}
\frac{P^+(f;2;1|i)}{P^+(f|i)}\ ,
\label{stationary}
\end{equation}
once more, using an obvious shorthand notation. 

The Markov property means that we can write 
$P^+(f;2;1|i) = P^+(f|2)P^+(2|1)P^+(1|i)$. Using Eqs.\ (\ref{conditional}) 
and (\ref{image}) in (\ref{stationary}) we find, after some algebra,
\begin{eqnarray}
P^+(X_1,0;X_2;T)& = &\frac{2}{\pi}(1-e^{-2\mu T})^{-1/2}X_1 \, X_2 \,e^{\mu T}
\nonumber \\
&&\times\exp\left[-\frac{X_1^2+X_2^2}{2(1-e^{-2\mu T})}\right]  \sinh\left(\frac{X_1X_2}{2\sinh\mu T}\right)\ .
\end{eqnarray}

It is now straightforward to calculate the autocorrelation function:
\begin{eqnarray}
A^0_{\cal C}(T)&=& \int_0^\infty dX_1\int_O^\infty dX_2\,X_1X_2P^+(X_1,0;X_2,T)
\nonumber \\
&=& \frac{2}{\pi}\left[3(1-e^{-2\mu T})^{1/2} + (e^{\mu T}+2e^{-\mu T})\,
\sin^{-1}(e^{-\mu T})\right] \;.
\end{eqnarray}
Eq.\ (\ref{pert_freq}) for $\theta$ can be written as a real-time integral 
as follows. We first write $A(T)=A^0(T) + \epsilon \, a(T)$, and we note that, 
since $G(T)$ is the inverse function of $A(T)$, in Fourier space we have 
$[\tilde{A}(\omega)]^{-1} = \tilde{G}(\omega) = \tilde{G}^0(\omega) 
+ \epsilon \tilde{g}(\omega)$, the last equality defining the perturbation 
$\tilde{g}(\omega)$. Using $A^0(T) = \exp(-\mu T)$, we find
\begin{equation} 
\tilde{g}(\omega) = -\tilde{a}(\omega)(\omega^2 + \mu^2)^2/4\mu^2\ .
\end{equation}
Inserting this result in Eq.\ (\ref{pert_freq}) and transforming to real 
time gives
\begin{eqnarray}
\theta&=& \mu - \frac{\epsilon}{4\mu^2}\int_0^\infty dT\,a(T) 
\left(\mu^2-\frac{d^2}{dT^2}\right)^2\,[A^0_{\cal C}(T)-A_0(T)]^2 \nonumber \\
& = & \mu\left(1-\epsilon \frac{2\mu}{\pi}\int_0^\infty a(T)
[1-\exp(-2\mu T)]^{-3/2}\, dT\right) + {\cal O}(\epsilon^2) \;.
\label{pert_time}
\end{eqnarray}
The final result is rather compact. Later in this article, in section \ref{Global}, we will apply 
it to a first-passage problem in critical dynamics. As a trivial example, which also serves as a simple check on the result, consider the 
Markov process with correlator $\exp[-(\mu+\delta\mu)T]$. Clearly, the 
persistence exponent of this process is simply $\theta = \mu + \delta\mu$. 
Using the general result (\ref{pert_time}), we have 
$\epsilon a(T) = -\delta\mu\, T\exp(-\mu T)$. Inserting this into 
(\ref{pert_time}) gives 
\begin{eqnarray}
\theta &=& \mu\left[1+(2\mu\delta\mu/\pi)\int_0^\infty dT\,Te^{-\mu T}\,
[1-e^{-2\mu T}]^{-3/2}\right]\,\nonumber \\
& = & \mu + \frac{2}{\pi}\delta\mu\int_0^\infty dx\,xe^{-x}[1-e^{-2x}]^{-3/2}
\nonumber \\
& = & \mu + \delta\mu
\end{eqnarray}
as required.

The perturbative method (in a different but equivalent form to that 
presented here) was used to obtain approximate results for the persistence 
properties of the coarsening dynamics of the $d$-dimensional Ising model~\cite{MajumdarSirePT}, as discussed earlier in section \ref{section:higher_dim}.


\section{The independent interval approximation}\label{subsection:iia}

Consider a  Gaussian Stationary Process $X(T)$, normalised  such that $\langle
X^2(T) \rangle  = 1$, and  with autocorrelation function $A(T)  \equiv \langle
X(0)\,X(T)  \rangle$. By  definition, $A(0)=1$.  As we have seen before, 
we can  classify  the process
$X(T)$  according to the  small-$T$ behaviour  of $A(T)$.  If, for  small $T$,
$A(T)$ has  the form  $A(T) = 1  - aT^2  +\ldots$, the process  is said  to be
`smooth'.  Such  processes have  a finite  density $\rho$ of  zero
crossings, given by the Rice's formula (\ref{Rice}). If, on  the other  hand, $A(T)  = 1  - bT^\alpha  +  \ldots$, with
$\alpha <2$,  the process $X(T)$ is  said to be  `rough' - it has  an infinite
density of zero crossings (see the discussion in section \ref{subsection:GSP}).

\begin{figure}[h]
\centering
\includegraphics[width = 0.7\linewidth]{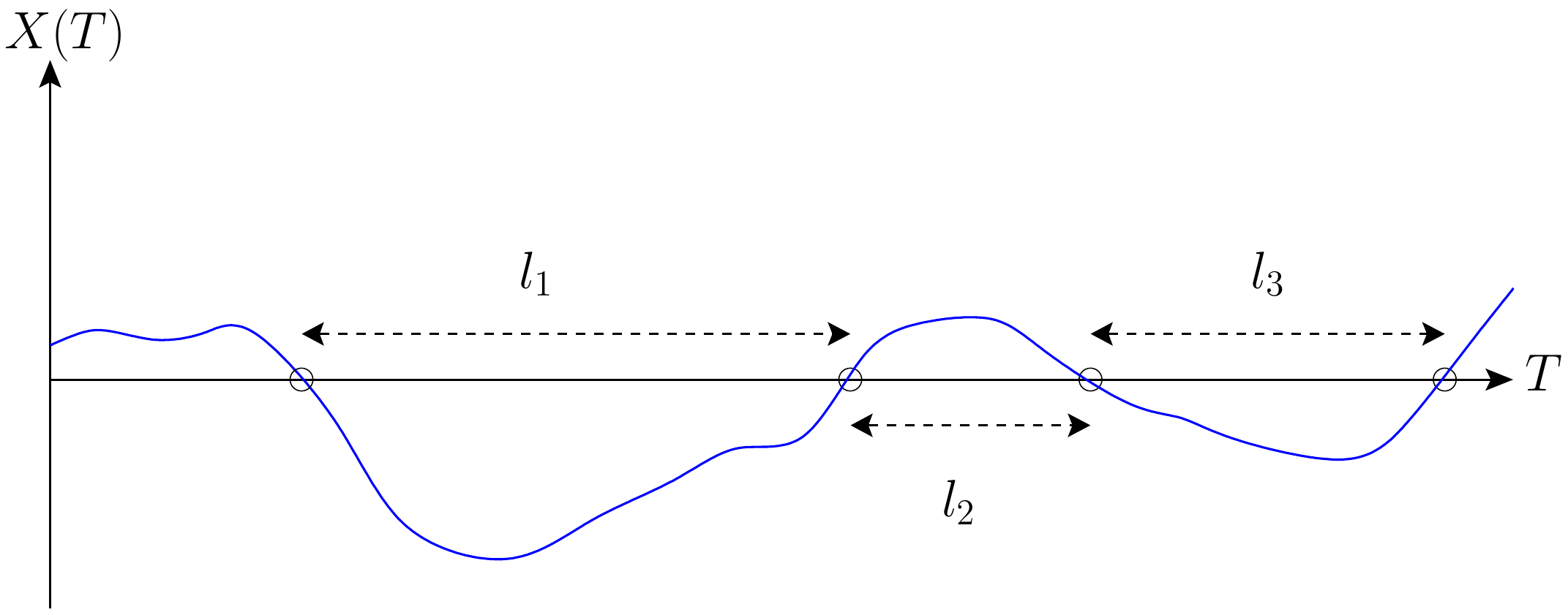}
\caption{The independent interval approximation amounts to assume that the intervals between zero-crossings of $X(T)$, here $l_1, l_2, l_3$, are statistically independent.}\label{fig:iia}
\end{figure}

The basis of the ``independent interval approximation'' (IIA) 
\cite{MajumdarSireBrayCornell,DerridaHakimZeitak,DerridaZeitak}
is to treat the  intervals between successive zero crossings as if they are 
statistically independent, see Fig. \ref{fig:iia}. 
As a preliminary step we introduce 
the ``clipped'' process, described by the variable $\sigma(T) = {\rm sgn}\,X(T)$. 
The correlator $C(T) = \langle \sigma(0)\,\sigma(T)\rangle$ is determined by 
the distribution, $P(T)$, of the intervals between successive zeros of $X(T)$. 
The persistence, $Q(T)\sim \exp(-\theta T)$, is the probability that there 
are no zeros in a randomly chosen interval of length $T$. Clearly it is 
controlled by the distribution of interval sizes, which we expect to have 
the same tail, $P(T) \sim \exp(-\theta T)$. 

The correlator $C(T)$ of the clipped process is  easily computed from $A(T)$
since the process $X(T)$ is Gaussian:
\begin{eqnarray}
C(T) = \langle {\rm sgn}\,X(T)\,{\rm sgn}\,X(0)\rangle = \frac{2}{\pi}
\sin^{-1} A(T)\ .
\label{C}
\end{eqnarray}
But $C(T)$ can also be computed from the interval size distribution $P(T)$.
Clearly 
\begin{equation}
C(T) = \sum_{n=0}^{\infty}(-1)^n\,p_n(T)\ ,
\end{equation}
where $p_n(T)$ is the probability that the interval $T$ contains exactly $n$ zeros 
of $X(T)$. 

We define $Q(T)$ to be the probability that an interval of size $T$ to the 
right or left of a zero contains no further zeros. Then $P(T)=-Q'(T)$. For 
$n \ge 1$, 
\begin{eqnarray}
p_n(T) &=& \langle T \rangle^{-1}\int_0^T dT_1\int_{T_1}^T dT_2 \ldots
\int_{T_{n-1}}^T dT_n Q(T_1) \nonumber \\
&& \times P(T_2-T_1)\ldots P(T_n-T_{n-1})Q(T-T_n),
\end{eqnarray}
where $\langle T \rangle$ is the mean interval size. The key approximation 
here is to write the joint distribution of $n$ successive intervals between 
zero crossings as the product of the distributions of the individual 
intervals, ignoring any correlations (see Fig. \ref{fig:iia}). 

Taking Laplace transforms gives
\begin{equation}
\tilde{p}_n(s) = \frac{1}{\langle T\rangle}\, 
[\tilde{Q}(s)]^2[\tilde{P}(s)]^{n-1}\ .
\label{pntildeofs}
\end{equation} 
But the obvious relation $P(T)=-Q'(T)$, together with $Q(0)=1$, implies 
\begin{equation}
\tilde{P}(s) = 1 - s\tilde{Q}(s)\ .
\end{equation}
Using this to eliminate $\tilde{Q}(s)$ from Eq.\ (\ref{pntildeofs}) yields
\begin{eqnarray}
\tilde{p}_n(s) &=& \frac{1}{\langle T \rangle s^2} [1-\tilde{P}(s)]^2
[\tilde{P}(s)]^{n-1},\ n \ge 1 \label{68} \\
\tilde{p}_0(s) &=&\frac{1}{\langle T \rangle s^2}
[\langle T \rangle s -1 + \tilde{P}(s)]\ ,\label{69}
\end{eqnarray}
where the last result follows from the normalisation condition, 
$\sum_{n=0}^\infty p_n(t)=1$. Finally, the Laplace transform of Eq.\ (\ref{C})
gives $\tilde{C}(s) = \sum_{n=0}^\infty (-1)^n \tilde{p}_n(s)$. Performing 
the sum using (\ref{68}) and (\ref{69}) we can use the result to express 
$\tilde{P}(s)$ in terms of $\tilde{C}(s)$, giving
\begin{equation}
\tilde{P}(s) = [2- F(s)]/F(s),
\label{70}
\end{equation}
where
\begin{equation}
F(s) = 1 + \frac{1}{2}\,s\langle T \rangle[1-s\tilde{C}(s)]\ .
\label{71}
\end{equation}

Let us summarise our main results up to this point. For a GSP $X(T)$, with 
correlator $A(T)$, the correlator $C(T)$ of the clipped process, 
$\sigma(T) = {\rm sgn}\,X(T)$, is given by Eq.\ ({\ref{C}). But $C(T)$ is 
related to the interval-size distribution $P(T)$ through (\ref{70}) and 
(\ref{71}). Finally, $p_0(T)$, the probability that there are no zeros 
in an interval of length $T$, is related to $P(T)$ through (\ref{69}). 
The mean interval length, $\langle T \rangle$, is easily obtained either 
from the correlator $A(T)$, using (\ref{Rice}) and 
$\langle T \rangle = 1/\rho$, or from the correlator $C(T)$ of the clipped 
process, using $C(T) = 1 - 2T/\langle T \rangle + \cdots$ for small $T$.

The asymptotics of $p_0(T)$ are controlled by the singularity of 
$\tilde{p}_0(s)$ with the largest real part, i.e.\ [from Eq.\ (\ref{69})] 
by the corresponding singularity of $\tilde{P}(s)$. In most cases of 
interest, we expect that $p_0(T)$ for a GSP has the form 
$\sim \exp(-\theta T)$, suggesting that the relevant singularity in 
$\tilde{p}_0(s)$ is a simple pole at $s = -\theta$, i.e.\ that $F(s)$ has 
a simple zero at $s = -\theta$. An explicit expression for $F(s)$, in 
terms of the autocorrelation function $A(T)$ is
\begin{equation}
F(s) = 1 + \frac{\pi}{2[-A''(0)]^{1/2}}\,s
\left[1-\frac{2s}{\pi}\int_0^\infty\,e^{-sT}\sin^{-1}A(T) dT\right]\ .
\label{theta}
\end{equation}

Let us suppose, as is generally the case, that $A(T) \sim \exp(-\lambda T)$ 
for large $T$. Then the function $F(s)$ has the following general features:
(i) $F(0)=1$; (ii) $F(s)$ diverges to $-\infty$ for $s \to -\lambda$; (iii)
between these two points $F(s)$ is monotonic, 
implying a single zero at $s = -\theta$, with $\theta \le \lambda$. In the 
following section we give some specific applications of this approach.

But before we do that, let us just mention that in the analysis above we are concerned with
the intervals of zero crossing. The IIA method has been generalised also to the case
of the crossing of a nonzero level~\cite{Sire07,Sire08}. For instance, consider the crossing of a level
at height $M$ by a stationary process $X(T)$.
Let $P_{\pm}(T)$ denote respectively the intervals where $X(T)>M$ and $X(T)<M$, starting at
an initial point $X(0)<M$. For $M>0$, the statistics of $P_{+}(T)$ is different from that
of $P_-(T)$. For the case $M=0$ discussed above, we have seen that $P_+(T)$ has the same statistics as $P_{-}(T)$
and they can be related, within IIA, to the Laplace transform of the autocorrelation
function of the signal as in Eqs. (210) and (211). For $M>0$, to determine $P_+(T)$ and $P_{-}(T)$,
one needs two relations. One of them is the generalisation of the autocorrelation
function to level $M>0$, but the other nontrivial relation can be obtained by relating
the interval size distributions to the average number of crossings of the level $M$~\cite{Sire07,Sire08}.
This IIA result becomes exact in the limit of $M\to \infty$.

\subsection{Scaling phenomenon and Lamperti transformation}\label{section:lamperti}

Many applications of the methods described so far in this article do not 
obviously involve Gaussian stationary processes (GSPs). However, many processes 
of interest can be mapped onto GSPs. These are Gaussian processes which are self-similar, in which the 
two-time correlation function depends fundamentally on {\em ratios} of the 
two time arguments rather than {\em differences}. In many cases, indeed, the two-time $t_1, t_2$ correlation function 
takes the scaling form, when $t_1, t_2 \gg 1$, $a(t_1,t_2) = t_1^\alpha\,f(t_1/t_2)$, 
where $\alpha$ is a scaling exponent. In such cases, especially were the 
processes are Gaussian, it is useful to introduce a new timescale 
$T = \ln t$, and a new variable $X(T) = x(t)/\sqrt{\langle x^2(t) \rangle}$. 
By construction, $\langle X^2(T)\rangle = 1$. This transformation (including normalization of the process and change of time variable) 
is known under the name of the Lamperti transformation \cite{Lamperti}.

\subsection{Application to the Brownian walker and higher order processes}\label{subsection:higher_order}

As a simple example, we begin with Brownian walk encountered earlier, 
$dx/dt = \eta(t)$, with $\langle\eta(t)\eta(t')\rangle = \delta(t-t')$ 
(here, for convenience, we have set the diffusion constant to $D=1/2$). 
If we take the initial condition to be $x(0)=0$, the two-time correlation 
function is readily obtained as $\langle x(t_1) x(t_2) \rangle 
= {\rm min}\,(t_1,t_2)$, a homogeneous function of the two time arguments. 
Introducing the new variable $X(t) = x(t)/\langle x^2(t) \rangle$, such that 
$\langle X^2 \rangle = 1$ for all $t$, the two-time correlator becomes a 
{\em ratio} of the time arguments: $\langle X(t_1)X(t_2) \rangle 
= {\rm min}\,(t_1,t_2)/\sqrt{t_1t_2}$, which we term a scaling form
(such scaling forms, albeit usually more complicated ones, occur naturally 
in many areas of physics). The final step is to introduce the new time 
variable
\begin{equation}
T = \ln t
\end{equation}
With a slight abuse of notation (we use keep the same name, $X$, for the 
$T$-dependent variable) we have $\langle X(T_1) X(T_2) \rangle 
= \exp(-|T_1-T_2|/2)$. The process is now a GSP, with persistence exponent 
$1/2$.


As a second example where IIA can be applied successfully we consider the higher-order processes already introduced briefly
in section \ref{subsection:bfp_higher_order}. More precisely, we consider a generalised Brownian motion in one dimension whose position evolves
with time $t$ via the Langevin equation
\begin{equation}
\frac{d^n x}{dt^n } =\eta(t)
\label{hop.1}
\end{equation}
where $\eta(t)$ is a Gaussian white noise with zero mean and a correlator $\langle \eta(t)\eta(t')\rangle =\delta(t-t')$.
For $n=1$, this process $x(t)$ is the standard Brownian motion and for $n=2$, it represents the random acceleration process
discussed in section \ref{subsection:RAP}. 

Assuming that initially all the derivatives up to order $(n-1)$ are zero, the process $x(t)$
can be represented as an $n$-fold integral
\begin{equation}
x(t)= \int_0^t dt_{n} \int_0^{t_{n}} dt_{n-1}\ldots \int_0^{t_1} dt_1\, \eta(t_1) \,. 
\label{hop.2}
\end{equation}
Another particularly useful representation is~\cite{Shepp66,Lachal97a,Lachal97b}
\begin{equation}
x(t)= \frac{1}{\Gamma(n)}\,\int_0^t \eta(t')\, (t-t')^{n-1}\, dt' \;,
\label{hop.3}
\end{equation}
which can be easily proved by differentiating Eq. (\ref{hop.3}) $n$ times. 
This representation manifestly demonstrates that $x(t)$ is a linear functional of the Gaussian noise $\eta(t)$ and hence
it follows that $x(t)$ is a Gaussian process. 
In fact, the representation in Eq. (\ref{hop.3}) also allows an analytical continuation to a continuous $n>1/2$, not necessarily a positive 
integer~\cite{MB01}.
The persistence $Q(t)$ of this process is defined in the
standard way: the probability that the process does not cross $0$ up to time $t$.
In Ref.~\cite{MajumdarSireBrayCornell},
it was found that $Q(t)\sim t^{-\theta(n)}$ where the persistence exponent $\theta(n)$ 
decreases with increasing value of $n$. For $n=1$ and $2$, one has the analytical values
$\theta(1)=1/2$ and $\theta(2)=1/4$. For higher values of $n$, accurate numerical results are
available~\cite{EhrhardtMajumdarBray04} giving $\theta(3)\approx 0.2202$, $\theta(4)\approx 0.2096$, $\theta(5)\approx 0.2042$
etc. As $n$ becomes large, the exponent seems to saturate to a nonzero value, 
$\theta(n\to \infty)\approx 0.1875$~\cite{MajumdarSireBrayCornell}. 

This Gaussian process $x(t)$ in Eq. (\ref{hop.3}) is non-stationary, as evident from the direct calculation of the
two-time correlation function using Eq. (\ref{hop.3}), which gives
\begin{equation}
\langle x(t_1)x(t_2)\rangle =\frac{1}{\Gamma^2 (n)}\, \int_0^{{\rm min}(t_1,t_2)} (t_1-t')^{n-1}\, (t_2-t')^{n-1}\, dt' \;.
\label{hop_corr.1}  
\end{equation}
However, using the Lamperti transformation mentioned in section \ref{section:lamperti}, one can
map this process to a Gaussian stationary process for any $n>1/2$. To proceed, we make the transformation, $X= x(t)/\sqrt{\langle x^2(t)\rangle}$
and consider it as a function of the logarithmic time $T=\ln t$. It is then easy to show that $X(T)$ becomes
a Gaussian stationary process with auto-correlator $\langle X(T_1)X(T_2)\rangle = A_n(|T_1-T_2|)$ that depends
only on the time difference $T=|T_1-T_2|$ and is given explicitly by~\cite{MajumdarSireBrayCornell,MB01}
\begin{equation}
A_n(T)= \left(2-\frac{1}{n}\right)\, e^{-T/2}\, _2F_1\left(1-n,1;1+n;e^{-T}\right) \;,
\label{hop_corr.2}
\end{equation}
where $_2F_1(a,b;c;,z)$ is the standard hypergeometric function.  
Given this nontrivial form of the autocorrelator, it follows from the discussion in section \ref{subsection:GSP} that the GSP is non-Markovian and the persistence $Q(T)$ of this process will decay as $Q(T)\sim \exp[-\theta(n)\,T]$
for large $T$ where $\theta(n)$ would depend
continuously on $n$. Consequently, in the original time $t$, $Q(t)\sim t^{-\theta(n)}$ for large $t$
where $\theta(n)$ is then the persistence exponent.

The computation of the exponent $\theta(n)$ analytically
for arbitrary $n$ seems difficult (except for $n=1$ and $n=2$). However,
one can determine them fairly accurately using the IIA method discussed in 
section \ref{subsection:iia}. To proceed, let us first expand the correlator in Eq. (\ref{hop_corr.2})
for small $T$ which gives~\cite{MB01}
\begin{eqnarray}
A_n(T)\approx \left\{\begin{array}{ll} 1- a_n T^{2n-1}\quad\quad 1/2<n<3/2 \\
&\\
1+ (T^2/4)\, \ln T  \quad\quad n=3/2 \\
&\\
1- \frac{2n-1}{8(2n-3)}\, T^2 \quad\quad n>3/2
\end{array}
\right.
\label{smallT.1}
\end{eqnarray}
where $a_n= \Gamma(n)\Gamma(2-2n)/\Gamma(1-n)$. Thus for $n>3/2$, the process is ``smooth" (see section \ref{subsection:GSP}) with
a finite density of zero crossings that can be derived from Rice's formula, 
\begin{equation}
\rho=\frac{1}{\pi}\,\sqrt{-A_n''(0)}= \frac{1}{2\pi} \sqrt{\frac{(2n-1)}{2n-3}}\, .
\label{hop_rice}
\end{equation}
For $1/2<n<3/2$, the density is infinite and the zeros are not uniformly distributed, instead they form a
fractal structure with fractal dimension $d_f=n-1/2$. 
Thus, there is a `phase transition' at the critical value $n_c=3/2$ where the process changes from 
being `rough' for $1/2<n<3/2$ to `smooth' for $n>3/2$~\cite{MB01}.
The case $n=3/2$ is marginal, and the result in Eq. (\ref{hop_corr.2})
can be simplified to the form~\cite{MB01}
\begin{equation}
A_{3/2}(T) = {\cosh}(T/2) + \sinh^2(T/2)\, \ln [\tanh(T/4)] \;,
\label{hop_3/2}
\end{equation}
where the density of zeros is still divergent but only logarithmically. A physical example of this marginal case $n=3/2$ is provided
by the steady-state spatial correlation of the $(2+1)$-dimensional linear interface model with dynamical exponent $z=4$ 
and non-conserving noise (see section \ref{subsection:spatial_pers_interfaces}).

For $n>3/2$, where the process is smooth, one can apply the IIA method
discussed in section \ref{subsection:iia}. According to IIA, the exponent $\theta$ is given by the first real negative zero
of the function $F(s)$ defined in Eq. (208). In other words, the exponent $\theta_{\rm IIA}$, within IIA,
is given by the first positive root
of the following equation  
\begin{equation}
1+ \frac{2\theta}{\pi}\, \int_0^{\infty} \sin^{-1}\left[A_n(T)\right]\, e^{\theta\, T}\, dT= \frac{2\rho}{\pi},
\label{hop_iia}
\end{equation}
where $\rho=\sqrt{-A_n''(0)}/\pi$ is the density of zeros computed in Eq. (\ref{hop_rice}) and $A_n(T)$ is given explicitly
in Eq. (\ref{hop_corr.2}). For example,
for $n=2$, Eq. (\ref{hop_iia}) gives $\theta_{\rm IIA}(2)=0.26466\ldots$ which is slightly higher than the exact value
$\theta(2)=1/4$. Similarly, for $n=2$, the IIA estimate from Eq. (\ref{hop_iia}) gives
$\theta_{\rm IIA}(3)= 0.22283\ldots$ to be compared to the numerical value~\cite{EhrhardtMajumdarBray04}
$\theta(3)\approx 0.2202$. Thus one sees that the IIA method provide fairly accurate estimate of the 
exponent $\theta(n)$.

The limit $n\to \infty$ is rather interesting~\cite{MajumdarSireBrayCornell,MB01}. 
Taking $n\to \infty$ limit in Eq. (\ref{hop_corr.2}) gives
\begin{equation}
A_{\infty}(T)= {\rm sech} (T/2)\, .
\label{largen_hop}
\end{equation} 
This also happens exactly to be the correlator of the diffusion process (to be discussed in detail
later in section \ref{subsection:diffusion}) in $d=2$~\cite{MajumdarSireBrayCornell}. 
Thus the $n\to \infty$ limit of the process can be numerically
simulated by simulating the diffusion process in $d=2$, since the two Gaussian processes are
isomorphic. In this limit, one can also obtain
an IIA estimate from Eq. (\ref{hop_iia}) which gives, $\theta_{\rm IIA}(n\to \infty)= 0.1862\ldots$, which is again 
close to the numerically obtained value from simulating diffusion process in $2$-d, $\theta(n\to \infty)\approx   
0.1875$~\cite{MajumdarSireBrayCornell}.

To summarize, the persistence properties of the Gaussian process $d^nx/dt^n=\eta$ is rather rich. It is
a simple example of a non-Markovian Gaussian stochastic process with a nontrivial persistence exponent $\theta(n)$
which can be estimated very accurately for $n>3/2$ by the IIA method. This example then serves as a
nice demonstration of the power and usefulness of the IIA method for smooth Gaussian stationary processes.


\section{Diffusive persistence}\label{subsection:diffusion}

Until now, we have mostly been concerned with persistence in systems 
with a finite number of degrees of freedom, such as the random walk (one 
degree of freedom) and the random acceleration process (two degrees of 
freedom) and its generalisations. Now we consider processes which involve 
infinitely many degrees of freedom. The simplest, perhaps, is the process described by the diffusion equation. We consider a scalar field $\phi({\mathbf x},t)$ in a $d$-dimensional space
which evolves in time under the diffusion equation:
\begin{equation}
\partial_t \phi = \nabla^2 \phi\ ,
\label{diffusion}
\end{equation}
with {\it random} initial conditions $\phi({\mathbf x},t=0) = \psi({\mathbf x})$
where $\psi({\mathbf x})$ is a Gaussian random field of zero mean 
with delta correlations $[\psi({\mathbf x}) \psi({\mathbf
  x'}) ]_{\rm ini} = \delta^{d}({\mathbf x}-{\mathbf x'})$. We use the
notation $[...]_{\rm ini}$ to denote 
an average over the initial condition.  For a system of linear size $L$, the persistence $Q(t,L)$
is the probability that $\phi({\mathbf x},t)$, at some fixed point ${\mathbf x}$ in space, does not change
sign up to time $t$~\cite{MajumdarSireBrayCornell,DerridaHakimZeitak}.  The initial condition being (statistically)
invariant under translation in space, this probability does not depend 
on the position ${\mathbf x}$, provided ${\mathbf x}$ is far enough from the boundary of the system. The diffusion equation is abundant in nature
and the question of persistence is a rather natural question~\cite{Watson_science}.

For a system of linear size $L$, the solution of 
the diffusion equation (\ref{diffusion}) in the bulk of the system is 
\begin{eqnarray}
\phi({\mathbf x},t) = \int_{|{\mathbf y}| \leq L} d {\mathbf y} \; {\cal
  G}({\mathbf x}-{\mathbf y},t) \; \psi({\mathbf y}) \quad, \quad  {\cal
  G}({\mathbf x}) = (4 \pi t)^{-d/2} \exp{(-{\mathbf x}^2/4t)} \, ,
\label{sol_eqdiff} 
\end{eqnarray} 
where $\psi(\mathbf x) = \phi(\mathbf x,0)$ is the initial
uncorrelated Gaussian field. Since Eq.~(\ref{sol_eqdiff}) is linear,
$\phi({\mathbf x},t)$ is a Gaussian variable for all time $t \geq 
0$. Therefore its zero crossing properties are completely determined
by the two time correlator $[ \phi({\mathbf x},t) \phi({\mathbf
  x},t')]_{\rm ini}$.  
To study the persistence probability $Q(t,L)$ we introduce the normalized 
process $X(t) = \phi({\mathbf x},t)/[ \phi({\mathbf x},t)^2]_{\rm ini}^{1/2}$. Its autocorrelation 
function $a(t,t') = [ X(t) X(t') ]_{\rm ini}$ is computed
straightforwardly from the solution in Eq. (\ref{sol_eqdiff}). One obtains
$a(t,t' )\equiv a(\tilde t, \tilde t')$ with $\tilde t= t/L^2$,
$\tilde t' = t'/L^2$ and 
\begin{eqnarray}\label{attprime}
a(\tilde t,\tilde t') = 
\begin{cases}
\left(\frac{4 \tilde t \tilde t'}{(\tilde t+\tilde t')^2}\right)^{d/4} \quad, \quad \tilde t,\tilde t' \ll 1 \\
1 \quad, \quad \tilde t,\tilde t'\gg 1 \, .
\end{cases}
\end{eqnarray}
We first focus on the time regime $\tilde t, \tilde t' \ll 1$.  In terms of  
logarithmic time variable $T = \log \tilde t$, $X(T)$ is a GSP with correlator 
\begin{eqnarray}
a(T,T') = A(T-T') = [{\rm cosh}(|T-T'|/2)]^{-d/2} \; ,
\label{correl_gsp_diff} 
\end{eqnarray}
 which decays exponentially for large $|T-T'|$. Thus the persistence probability $Q(t,L)$,
for $t\ll L^2$, reduces to the computation of the probability ${Q}(T)$
of no zero 
crossing of $X(T)$ in the interval $[0,T]$. From the Newell-Rosenblatt's theorem \cite{Newell} stated above (\ref{Newell-Rosenblatt}), 
one deduces that ${Q}(T)\sim \exp[-\theta(d) T]$
for large $T$ where the decay constant $\theta(d)$ depends
on the full stationary correlator $a(T)$. Reverting back to 
the original time $t=e^T$, one
finds $Q(t,L) \sim 
t^{-\theta(d)}$, for $t \ll L^{2}$. In the opposite limit $t \gg L^2$, one has
$Q(t,L) \to A_L$, a constant which depends on $L$. These two
limiting behaviors 
of $Q(t,L)$ can be combined into a single finite size scaling form \cite{MajumdarSireBrayCornell,DerridaHakimZeitak}
\begin{equation}
Q(t,L) \propto L^{-2\theta(d)} h (L^2/t) \;,
\label{fss1}
\end{equation}
where $h(u) \sim c^{\rm st}$, a constant independent of $L$ and $t$,
for $u \ll  
1$ and $h(u) \propto u^{\theta(d)}$ for $u \gg 1$ where $\theta(d)$ is
a $d$-dependent exponent. This implies that in the $L \to \infty$ limit,
$Q(t) \equiv Q(t,L\to \infty) \sim t^{-\theta(d)}$ for large
$t$. Remarkably, the persistence for $d=1$ was observed in experiments on magnetization of spin polarized Xe gas and the exponent 
$\theta_{\rm exp}(1)=0.12$ was measured \cite{WMWC02}, in good agreement with analytical approximation \cite{MajumdarSireBrayCornell,DerridaHakimZeitak} and numerical simulations \cite{MajumdarSireBrayCornell,DerridaHakimZeitak,NewmanLoinaz}. 

Despite many efforts, there exists no exact result for $\theta(d)$. However various approximation methods have been developed to estimate it
and the most accurate one is certainly the Independent Interval Approximation (IIA) presented in section \ref{subsection:iia}. To compute the
persistence exponent $\theta_{\rm IIA}(d)$ for the diffusion equation within the IIA approximation, one inserts the expression of the correlator (\ref{correl_gsp_diff}) into Eq.\ (\ref{theta})
(using $A''(0)= -d/8)$ and finds the first zero of $F(s)$ on the 
negative $s$-axis, which is located at $s= -\theta_{\rm IIA}(d)$. The results for $\theta_{\rm IIA}(d)$ 
are listed in Table \ref{table_theta_diffusion} for small values of $d$. They are compared to the numerical estimates $\theta_{\rm num}(d)$ obtained in
Ref.\ \cite{NewmanLoinaz}. 
\begin{center}
\begin{table}
\begin{center}
\begin{tabular}{lclclc|c|}
\hline
$d$ & \ \ $\theta_{\rm IIA}(d)$ & \ \ $\theta_{\rm num}(d)$ & \ \ $\theta_{\rm D}(d)$ \\
\hline
1 & \ \ 0.1203 & \ \ 0.12050(5) & \ \ 0.1201(3)\\
2 & \ \ 0.1862 & \ \ 0.1875(1) & \ \ 0.1875(1) \\
3 & \ \ 0.2358 & \ \ 0.2382(1) & \ \ 0.237(1) \\
\hline
\end{tabular}
\caption{Persistence exponents $\theta_{\rm IIA}(d)$ for the diffusion process in 
space dimensions $d=1,2,3$, evaluated within the IIA~\cite{MajumdarSireBrayCornell,DerridaHakimZeitak}. The numerical results, $\theta_{\rm num}$, are taken from Ref.\ \cite{NewmanLoinaz} while $\theta_{\rm D}(d)$ correspond to a Pad\'e resummation (hence the error bars) of a systematic
series expansion introduced in the context of "discrete time persistence"~\cite{EB02}.}\label{table_theta_diffusion}
\end{center}
\end{table}
\end{center}
In the third column of Table \ref{table_theta_diffusion} we have listed the result of a systematic approach in
a series expansion introduced in the context of "discrete time persistence" \cite{EB02}, which is discussed below in section \ref{section:discrete}. We also mention that
there exists yet another systematic approach which consists in performing a small $d$ expansion \cite{HJH00}, relying on the perturbation theory for Non-Markovian Gaussian stationary processes discussed above in section \ref{section:perturbation}, yielding $\theta(d) = d/4 - 0.12065...d^{3/2}+...$, which would
certainly require higher order terms to make it numerically
competitive.

We finally study the persistence exponent $\theta(d)$ in the limit of large dimensions $d$. To do so, it is useful to rewrite the correlator $A(T-T')$ in Eq. (\ref{correl_gsp_diff}) in terms of the rescaled time $T = 2^{3/2} \tilde T/\sqrt{d}$ such that
\begin{eqnarray}\label{correl_larg_d}
A(T-T') = A\left(2^{3/2} \frac{\tilde T-\tilde T'}{\sqrt{d}}\right) \sim \exp{\left[ -\frac{1}{2} (\tilde T - \tilde T')^2\right]} \;, \, d \to \infty \;.
\end{eqnarray}
Therefore in the limit of large dimension $d$, one has 
\begin{eqnarray}\label{theta_large_d}
\theta(d) \sim 2^{3/2} \theta_{\infty} \sqrt{d} \;,
\end{eqnarray}
where $\theta_\infty$ is the decay constant associated with the no zero crossing probability of the GSP with correlator $\exp{[-\frac{1}{2}(T-T')^2]}$. Even in that limit, there is no exact result for $\theta_\infty$. However, it can be approximated using again IIA, yielding $\theta_{\infty, {\rm IIA}} = 0.411497É$ \cite{MajumdarSireBrayCornell,DerridaHakimZeitak} in very good agreement with the numerical simulations, $\theta_{\infty, {\rm sim}} = 0.417(3)$ \cite{NewmanLoinaz}.

\subsection{Application to coarsening dynamics}

The diffusion equation and its  persistence are of more general interest 
than might at first sight be supposed, since it appears in the study of 
the coarsening dynamics of a nonconserved $n$-dimensional field with $O(n)$ 
symmetry, in the limit $n \to \infty$. To see how this comes about, 
consider the Time-Dependent Ginzburg-Landau Equation for the $n$-component 
field $(\phi_i,\ldots, \phi_n)$:
\begin{equation}
\partial_t \phi_i = \nabla^2\,\phi_i + r\phi_i 
- (u/n)\phi_i \sum_{j=1}^n \phi_j^2\ .
\end{equation}
The absence of a thermal noise term indicates that we are working at zero 
temperature. In fact the temperature is irrelevant to the coarsening dynamics 
for temperatures below the critical temperature, $T_c$ \cite{BrayReview}.
The initial condition is given by a random configuration with 
$\langle \phi_i({\bf x},0) \phi_j({\bf x'},0) \rangle 
= \Delta \delta_{ij}\delta({\bf x}-{\bf x'}).$

In the limit $n \to \infty$, one can replace $n^{-1}\sum_j\phi_j^2$ by its 
mean, to give a self-consistent linear equation for any given component of 
the field:
\begin{equation}
\partial_t \phi = \nabla^2\,\phi + r\phi 
- u \langle \phi^2 \rangle \phi\ ,
\end{equation}
where the component index on the field has been dropped. This equation is 
easily solved. Defining $a(t) = r - u\langle \phi^2 \rangle$, one finds, 
in Fourier space
\begin{equation}
\tilde \phi_k(t) = \tilde \phi_k(0) \exp[-k^2t + b(t)]\ ,
\end{equation}
where $b(t) = \int_0^t a(t')dt'$. Averaging over the initial conditions 
gives the two-time correlator in Fourier space,
\begin{equation}
\langle \tilde \phi_k(t_1) \tilde \phi_{-k}(t_2) \rangle = \Delta \exp[-k^2(t_1 + t_2) + b(t_1) 
+ b(t_2)]\ ,
\end{equation}
giving the real-space autocorrelation function
\begin{equation}
C(t_1,t_2) = {\rm const}.\, (t_1+t_2)^{-d/2}\exp[b(t_1)+b(t_2)].
\end{equation}
The constant and the function $b(t)$ can be determined via the 
self-consistency condition $a(t) = db/dt = r-u\langle \phi^2 \rangle$. 
For present purposes, however, we only need the normalised correlator 
\begin{equation}
a(t_1,t_2) = C(t_1,t_2)/[C(t_1,t_1)C(t_2,t_2)]^{1/2} =  
\left(\frac{4t_1t_2}{(t_1 + t_2)^2}\right)^{d/4}\ ,
\end{equation}
which has exactly the same form, (\ref{attprime}), 
as for the diffusion equation. Given the diffusive nature of the 
TDGL equation, this is not so surprising. It follows that the persistence 
properties related to the coarsening dynamics of the TDGL equation in the 
large-$n$ limit are identical to those of the diffusion equation. It is also 
worth noting that the approximate theory of Ohta, Jasnow and Kawasaki 
\cite{OJK} for the coarsening dynamics of a {\em scalar field}, 
corresponding to the case $n=1$, is also described by the diffusion equation.

\subsection{Connections with random polynomials}

A seemingly unrelated topic concerns the study of random algebraic
equations which, since the first work by Bloch and P\'olya \cite{BP32} in the 30's, 
has now a long history \cite{BRS84, F98}. During the last few years
it has attracted a renewed interest  
in the context of probability and number theory \cite{EK95} as well
as in the field of quantum chaos \cite{BBL92a,BBL92b}. 

It was shown \cite{SM07a,SM07b} that there exists a close 
connection between zero crossing properties of the diffusion equation
with random initial  
conditions~(\ref{diffusion}) and the real roots of real
random polynomials ({\it i.e.} polynomials with real random coefficients). In
Ref.~\cite{SM07a,SM07b}, the authors focused on a class of real random polynomials $K_n(x)$ of degree
$n$, the so called generalized Kac polynomials, indexed by an integer~$d$ 
\begin{eqnarray} 
K_n(x) = a_0 + \sum_{i=1}^{n} a_i \; i^{\tfrac{d-2}{4}} x^i \;.
\label{def_kac_poly}  
\end{eqnarray} 
In (\ref{def_kac_poly}) the coefficients $a_i$'s are independent real Gaussian random variables of zero
mean and unit variance, such that $\langle a_i a_j \rangle = \delta_{ij}$ where
we use the notation $\langle ... \rangle$ to denote an average
over the random coefficients $a_i$. In the
case of $d=2$, these polynomials reduce to the standard Kac polynomials
\cite{K43}, which have been extensively studied in the past (see for
instance Ref.~\cite{EK95} for a review). A natural question about these random polynomials concerns the
number of real zeros on a given interval $[a,b]$, denoted as $N_n([a,b])$. 
\begin{figure}
\centering
\includegraphics[angle=0,width=0.6\linewidth]{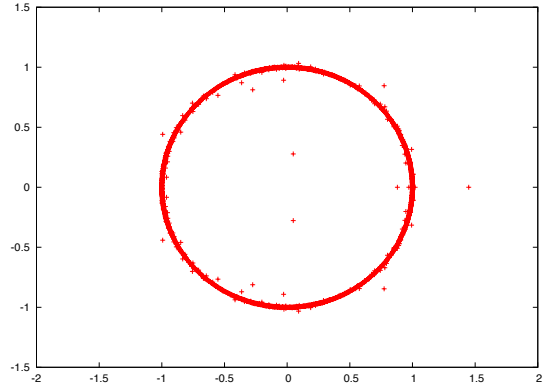}
\caption{Location of the roots of $K_n(x)$ in Eq. (\ref{def_kac_poly}) for $d=2$, $n=100$ and a given realization of the random coefficients $a_i$'s.}\label{Fig_Kac}
\end{figure}
A first well known fact~\cite{K43} is that, in the limit of large 
$n$, the real roots are localized around $x=\pm 1$, in a small region of size $\propto 1/n$ \cite{AF04} [more generally the $n$ complex roots of $K_n(x)$ are localized close to the unit circle \cite{ET50}, see Fig. \ref{Fig_Kac}]. For large $n$, it was shown that~\cite{D72,K43}
\begin{eqnarray}\label{eq:mean_zero}
\langle N_n((-\infty,+\infty)) \rangle = \frac{1}{\pi} \left( 1 + \sqrt{\frac{d}{2}} \right) \log{n} + {\cal O}(1) \;,
\end{eqnarray}
which, for $d=2$, yields back the famous result first obtained by Kac \cite{K43}. In the case $d=2$, it turns out that
the statistics of real roots of $K_n(x)$ is identical in the $4$
sub-intervals $(-\infty, -1], [-1,0], [0,1]$ and
$[1,+\infty)$. Instead, for $d \neq 2$,  the statistical behavior 
of real roots of $K_n(x)$ depend on $d$ in the inner intervals,
while it is identical to the case $d=2$ in the outer
ones. Focusing on the interval $[0,1]$, we consider the probability $Q_0([0,x],n)$, $0 < x < 1$, that $K_n(x)$ has no
real root in the interval $[0,x]$. Such probabilities were often
studied in the context of random matrices, where they are known as
{\it gap probabilities} \cite{mehta} and in Ref. \cite{DPSZ02},
Dembo {\it et al.} showed that, for random polynomials $K_n(x)$ with
$d=2$, $Q_0([0,1],n) \propto n^{-\zeta(2)}$ where the exponent
$\zeta(2) = 0.190(8)$ was computed numerically. In Ref.~\cite{SM07a,SM07b,DPSZ02,DM12}, it was shown that the random
process defined by the random polynomials $K_n(x)$ can be mapped onto a GSP with a correlator
given exactly by (\ref{attprime}), i.e. similar to the one governing the diffusion equation with random initial condition (\ref{diffusion}): it was
indeed shown that $n$, the degree of the random polynomial corresponds to $L^2$, with $L$ the system size in the diffusion equation, while $1-x$ maps onto $1/t$, $t$ being the time variable for the diffusion. Since a Gaussian process is completely characterized by its two point correlator, this result shows that the diffusion
process (\ref{diffusion}) and the random polynomials (\ref{def_kac_poly}) are essentially the
same Gaussian process and hence have the same zero
crossing properties. It follows from this connection that, in the scaling limit where $x \to 1$ (remember that the real roots concentrate close to $x = \pm 1$) and  $n \to \infty$, keeping the product $n(1-x)$ fixed, one has~\cite{SM07a,SM07b}
\begin{eqnarray}\label{persist_poly}
Q_0([0,x],n) \simeq {\cal A}^-_{d,n} n^{-\theta(d)} h^-(n(1-x)) \;,
\end{eqnarray}
where ${\cal A}^-_{d,n}$, which is independent of $x$, is such that $\lim_{n \to \infty} \log{{\cal A}^-_{d,n}}/\log n \to 0$
with $h^-(y) \to 1$ for $y \to 0$ and $h^-(y) \sim y^{\theta(d)}$ for $y \to \infty$, where $\theta(d)$ is the persistence exponent associated to the diffusion equation with random initial conditions, in $d$ spatial 
dimensions. This yields in particular
\begin{eqnarray}\label{persist_poly_full_interval}
Q_0([0,1],n) \propto n^{-\theta(d)} \;.
\end{eqnarray}
By symmetry, the real roots of $K_n(x)$ on the other inner interval $[-1,0]$, have the same statistics as the roots on $[0,1]$ and hence $Q_0([x,0],n)$, for $x \to -1$ and $n$ large, behaves as in Eq. (\ref{persist_poly}). On the other hand, it can be shown \cite{SM07a,SM07b} that the statistics of the real roots of $K_n(x)$ on the outer intervals $(-\infty, -1]$ and $[1,+\infty)$ are governed by the diffusion equation in $d=2$. This implies that $P_0([x,+\infty))$ when $x \to 1$ behaves like
 \begin{eqnarray}\label{persist_poly}
Q_0([x,+\infty),n) \simeq {\cal A}^+_{d,n} n^{-\theta(2)} h^+(n(x-1)) \;,
\end{eqnarray}
where ${\cal A}^+_{d,n}$, which is independent of $x$, is such that $\lim_{n \to \infty} \log{{\cal A}^+_{d,n}}/\log n \to 0$
with $h^+(y) \to 1$ for $y \to 0$ and $h^+(y) \sim y^{\theta(2)}$ for $y \to \infty$. This yields in particular
\begin{eqnarray}\label{persist_poly_full_interval_outer}
Q_0([1,+\infty),n) \propto n^{-\theta(2)} \;,
\end{eqnarray}
and similarly for $Q_0((-\infty-1],n)$. 

It is not surprising that these questions related to the probability $Q_0([a,b],n)$ of no real root in a given interval
$[a,b]$ has generated some interest also in the mathematics literature [see Refs. \cite{AS12,LS05} and references therein]. In particular, the results in Eqs. (\ref{persist_poly_full_interval}, \ref{persist_poly_full_interval_outer}) have been proved rigorously in Refs.~\cite{DPSZ02, DM12}. Rigorous lower and upper bounds for $\theta(2)$ have also been obtained, the best one being $\theta(2) \in (1/(4\sqrt{3}),1/4]$ proved in \cite{Mol12} and \cite{LS02}. We also mention an explicit expression for $Q_0([a,b],n)$ for $d=2$ obtained in Ref. \cite{Zap06} in terms of complicated multiple integrals, whose asymptotic analysis for large $n$ remains however challenging. This connection with random polynomials might open the way, in the future, to exact values of the persistence exponents $\theta(d)$. Besides, in Ref. \cite{SM07b} other types of random polynomials were studied which show a connection with the diffusion equation in the limit of large dimension $d \to \infty$ (\ref{theta_large_d}).

\vspace*{1cm}

\section{Persistence with partial survival}\label{subsection:partial}

We have already met the concept of partial survival briefly in the context
of the random acceleration process and its generalizations in sections \ref{subsection:RAP} and \ref{subsection:RAP_partial}.
Here we show that this concept can be applied more generally to any stochastic process,
in particular including
smooth GSP (where we recall that a ``smooth'' GSP is one for which the
autocorrelator has the small-$T$ form $A(T) = 1-a\,T^2+\cdots$). For smooth GSP's with arbitrary correlator
$A(T)$,
the partial survival probability can in fact be used as a perturbative scheme to obtain
approximate results for its persistence exponent $\theta$.

Let us consider first a general stochastic process $x(t)$.
For simple persistence, we can envisage that the stochastic process stops
(or ``dies'') the first time $x(t)$ crosses zero. The persistence probability
$Q(t)$ is simply the probability that it has survived (not yet died) at time
$t$. A process $x(t)$ with ``partial survival''   is defined as follows
\cite{MajumdarBrayPartial}. Each time the process crosses through $x=0$, it
is defined to survive with probability $p$. If $Q_n(t)$ is the probability
of $n$ zero crossings in time $t$, then the survival probability at time $t$
is simply the generating function
\begin{equation}
Q(p,t) = \sum_{n=0}^\infty p^n\,Q_n(t)\ .
\label{partials_1}
\end{equation}
For $p=0$, $Q(0,t)=Q(t)$, the usual persistence, which decays for large
$t$ as $Q(t)\sim t^{-\theta(0)}$ where $\theta(0)$ is the usual persistence exponent.
In the other limit,
$p=1$, the process always survives, leading to $\theta(1)=0$. One can
also analytically continue to negative $p$. In fact, $Q(-1,p)$ is just the
autocorrelation function, $A(t)$, of the clipped process:
$A(t) = \langle {\rm sgn}\,x(0)\,{\rm sgn}\,x(t)\rangle$, which decays as
$t^{-{\tilde \lambda}}$ where ${\tilde \lambda}$ will be referred to as the autocorrelation exponent.
We will show that for several processes, such as the $1$-d TDGL model
discussed in section \ref{subsection:TDGL} as well as GSP's with smooth correlator,
$Q(p,t) \sim t^{-\theta(p)}$ for large $t$,
where the exponent $\theta(p)$ varies continuously with $p$ for
$-1 \le p \le 1$, interpolating between $\theta(-1)={\tilde \lambda}$ and $\theta(1)=0$.
Moreover, the quantity $A_p(t)=Q(-p,t)/Q(p,t)$ is just the
autocorrelation function computed using only the trajectories that survive
between times $0$ and $t$. So  if $A_p(t) \sim t^{-{\tilde \lambda}_p}$, we have
${\tilde \lambda}_p = \theta(-p)-\theta(p)$. We thus see that the autocorrelation  
and persistence exponents are members of a wider family of exponents.

\vskip 0.3cm 
\noindent{\bf Exactly solvable $\theta(p)$:}
There are few models where the partial survival exponent $\theta(p)$ can be computed exactly. We briefly
mention them here.
As a first example of a solvable model we compute the exponent $\theta(p)$ for the coarsening
dynamics of the the one-dimensional TDGL equation that we first considered in section~\ref{subsection:TDGL}. Note that here the effective process representing a spin at a site
as a function of time is {\em non-Gaussian} and {\em non-Markovian}.
We start with a random distribution of intervals (or domains)
on the line and assign a `fictitious' particle to each point in space. The dynamics   
consists of merging the smallest interval $I_{\rm min}$ with its two neighbours
$I_1$ and $I_2$ to make a single interval $I$. The lengths $l(I)$ and the
persistent part $d(I)$ (i.e.\ the number of live particles in the interval 
$I$) evolve as
\begin{eqnarray}
l(I) &=& l(I_1) + l(I_2) + l(I_{\rm{min}})\ ,\nonumber \\
d(I) &=& d(I_1) + d(I_2) +p\,d(I_{\rm{min}})\ .
\end{eqnarray}
where the final term, absent from our previous treatment in section \ref{subsection:TDGL} [see Eq. (\ref{d(I)})], incorporates
the effect of partial survival.

Just as in section \ref{subsection:TDGL}, we can write down the equations that describe the evolution
of the number of intervals of length $l$, and the average persistent part of
such an interval, and one can solve exactly for the associated scaling
functions by taking Laplace transforms. Demanding, as before, that the first
moments of these scaling functions are finite gives an implicit equation
for the exponent $\theta$ as a function of $p$~\cite{MajumdarBrayPartial}:
\begin{equation}
\int_0^\infty dt\,e^{-t}\,t^{-1-\theta}[(1-p)(1-t-e^{-t})\,e^{r(t)}
+ 2\theta(1+p)t + \theta(1-p)t^2\,e^{-r(t)} = 0\ ,
\end{equation}
where
\begin{equation}
r(t) = -\gamma - \sum_{n=1}^\infty (-t)^n/n\,n!
\end{equation}
and $\gamma$ is Euler's constant. Solving this equation numerically gives   
the function $\theta(p)$ in the whole range $-1\le p\le 1$. 
One can check that in the limit $p\to -1$, we recover the autocorrelation exponent
derived in Ref.~\cite{BrayDerrida95},
$\theta(-1)={\tilde \lambda}=0.6006115\dots$ (note that in section \ref{subsection:TDGL}, the exponent $\lambda=1-{\tilde \lambda})$. 
Similarly, in the limit $p\to 1$, one recovers
$\theta(1)=0$. For $p=0$, one recovers the usual persistence exponent $\theta(0)=0.1750758\dots$, first computed
in Ref.~\cite{BDG1994b}. Note once again that in the notation of section \ref{subsection:TDGL}, $\theta(0)=1-\beta$.
The function $\theta(p)$ decreases monotonically from $\theta(-1)=0.6006115\dots$ to $\theta(1)=0$ as
$p$ increases from $-1$ to $1$ (for a plot of this function, see Fig. 1 of Ref.~\cite{MajumdarBrayPartial}).

There are two other examples of non-Gaussian and non-Markovian processes for which $\theta(p)$ can be
computed exactly. One example will be discussed in detail in section \ref{section:exact}. 
The other interesting example corresponds to the class of nonlinear models
introduced in section \ref{subsection:windy}. In that section we found an exact expression
for $\theta(p)$ for this class. The most general result is given by   
Eq.\ (\ref{theta(p)result}). For the special case $p=-1$, which gives the   
autocorrelation exponent, one finds $\delta = 1$ for all $\beta$ and $\gamma$.
Noting that for negative $p$ one should take the negative square root of
$\delta$ in Eq.\ (\ref{theta(p)result}), i.e.\ $\sqrt{\delta}=-1$, one obtains
$\theta(-1)=1/2$. This means that the normalised variable,
$X(t) = x(t)/\sqrt{x^2(t)}$, has a two-time correlator
$\langle X(t_1) X(t_2)\rangle$ with the asymptotic time-dependence
$t_2^{-1/2}$ for $t_2 \to \infty$ at fixed $t_1$ (i.e. the autocorrelation
exponent is $1/2$), for all models in this class.

\vskip 0.3cm

\noindent{\em Approximate $\theta(p)$ for diffusive persistence:}
We now consider diffusive persistence, introduced in section \ref{subsection:diffusion}. Here the relevant
equation is the deterministic diffusion equation,
$\partial_t\phi = \nabla^2 \phi$,
with a random initial condition. As usual, it is convenient
to make the Lamperti transformation to reduce the process $\phi(x,t)$, at fixed $x$,
to a GSP:
$X(T) = \phi/\langle \phi^2 \rangle^{1/2}$ (where
$T=\ln t$ as usual), with correlator $C(T) = [{\rm sech}(T/2)]^{d/2}$.
If each zero-crossing is survived with probability $p$, the persistence     
$Q(p,T)$, has the asymptotic time-dependence
\begin{equation}
Q(p,T) = \sum_n p^n\, Q_n(T)\sim \exp[-\theta(p)T]\ ,
\label{partials.2}
\end{equation}
where $Q_n(T)$ is the probability of $n$ zero crossings in `time' $T$. 
Eq. (\ref{partials.2}) corresponds to the power law decay,
$Q(p, t)\sim t^{-\theta(p)}$ after the change of variable $T=\ln t$.
The equation for the persistence exponent $\theta(p)$ is easily derived within
the Independent Interval Approximation discussed in section \ref{subsection:iia}. The result is~\cite{MajumdarBrayPartial}
\begin{equation}
\frac{1-p}{1+p} = \theta\pi\sqrt{\frac{2}{d}}\left\{1+\frac{2\theta}{\pi}
\int_0^\infty dT\,\exp(\theta T)\,\sin^{-1}[{\rm sech}^{d/2}(T/2)]\right\}\ .
\end{equation}
The case $p=0$ recovers our previous result (see section \ref{subsection:diffusion}), while $p=1$ gives
$\theta(1)=0$ as expected. For a plot of the function $\theta(p)$ see Fig. 1
of Ref.~\cite{MajumdarBrayPartial}.

\vskip 0.3cm

\noindent{\em Partial survival as a perturbative scheme:} Having computed $\theta(p)$ exactly
in few models and approximately within IIA for the diffusion equation, we now show
how $\theta(p)$ can be computed systematically as a perturbative expansion around
$p=1$ for an arbitrary Gaussian stationary process $X(T)$ with a smooth correlator, $A(T)=\langle X(0)X(T)\rangle$.
The basic idea is simple~\cite{MajumdarBrayPartial}. We start from the definition of 
partial survival probability in $[0,T]$: 
$Q(p,T)= \sum_{n=0}^{\infty} p^n\, Q_n(T)$. We rewrite $p^n= \exp[n\, \ln p]$, expand the
exponential in powers of $\ln p$ and then taking logarithm of $Q(p,T)$, we get the
standard cumulant expansion
\begin{equation}
\ln Q(p,T)= \sum_{r=1}^{\infty} \frac{(\ln p)^r}{r!}\, \langle n^r\rangle_c \;,
\label{partials_cumul}
\end{equation}
where $\langle n^r\rangle_c$ is the $r$-th cumulant of the number $n$ of zeros in $[0,T]$. Substituting $p=1-\epsilon$, we express the right hand side
of Eq. (\ref{partials_cumul}) as a power series in $\epsilon$. Since we expect $Q(p,T)\sim \exp[-\theta(p)\, T]$,
we can then obtain a formal series expansion of $\theta(p)$ by taking the limit
\begin{equation}
\theta(p)= -\lim_{T\to \infty} \frac{1}{T}\, \ln Q(p,T)= \sum_{r=1}^{\infty} a_r\, \epsilon^r \;,
\label{partials_series}
\end{equation}
where the coefficients $a_r$'s involve the cumulants. For example, the first two coefficients are given by
\begin{equation}
a_1= \lim_{T\to \infty} \frac{\langle n\rangle}{T}; \quad a_2= \lim_{T\to \infty}\frac{1}{2\,T}\,\left[\langle n\rangle -\langle 
n^2\rangle_c\right] \;,
\label{partials_coeff}
\end{equation}
where $\langle n^2\rangle_c= \langle n^2\rangle- {\langle n\rangle}^2$. 
Similarly, the higher order coefficient $a_r$ with $r>2$ can also be expressed easily
in terms of cumulants up to order $r$. Hence, if one can compute the coefficients $a_r$'s, in principle
one has an exact series expansion in powers of $\epsilon=(1-p)$. Even though this series
expansion is expected to give accurate answer only near $\epsilon=0$, i.e., near $p=1$, one may,
in the spirit of the $\epsilon$-expansion in critical phenomena, keep terms up to
a certain order and set $\epsilon=1$ to obtain an estimate of the standard persistence exponent $\theta(0)$.
The important point is that unlike the IIA, this method of partial survival provides a systematic
approximation for $\theta(0)$.  

The challenge then is to compute the coefficients $a_r$'s in Eq. (\ref{partials_series}). To compute
them, we need to know
the moments $\langle n^r\rangle$ of the number of zero crossings. Fortunately, this can be done with relative ease, though
it becomes tedious for higher moments. For example, the first moment $\langle n\rangle$, i.e., the mean number of zero
crossings in time $T$ can be easily computed using Rice's formula~\cite{Riceformula}, 
$\langle n\rangle= T\, \sqrt{-A''(0)}/\pi$ [see Eq. (\ref{Rice}) in section \ref{subsection:GSP} where we also gave a simple derivation of this formula].
This gives the exact result
\begin{equation}
a_1= \frac{1}{\pi}\, \sqrt{-A''(0)}\, .
\label{partials_a1}
\end{equation}

Similarly, there is an explicit formula for the second moment $\langle n^2\rangle$ due to Bendat~\cite{Bendat}.
Plugging Bendat's formula in Eq. (\ref{partials_coeff}) and after a few steps of algebra, an exact formula
for $a_2$ can also be derived~\cite{MajumdarBrayPartial}
\begin{equation}
a_2= \frac{1}{\pi^2}\, \int_0^{\infty} \left[S(\infty)-S(T)\right]\, dT \;,
\label{partiala_st}
\end{equation}  
where $S(T)$ is given by a complicated formula
\begin{equation}
S(T)= \frac{\sqrt{M_{22}^2-M_{24}^2}}{\left[1-A^2(T)\right]^{3/2}}\, \left[1+H\, \tan^{-1} H\right] \;,
\label{partials_a2}
\end{equation}
with $H= M_{24}/\sqrt{M_{22}^2-M_{24}^2}$. The $M_{ij}$'s are the cofactors of the $(4\times 4)$ symmetric correlation 
matrix $M$ between four Gaussian variables $\{X(0),X'(0),X(T),X'(T)\}$. The entries $M_{ij}$'s can be explicitly computed 
in terms of the autocorrelation function $A(T)=\langle X(0)X(T)\rangle$. For instance, $M_{11}= \langle X(0)^2\rangle=A(0)=1$, 
$M_{14}= \langle X(0)X'(t)\rangle= A'(T)$, $M_{24}=\langle X'(0)X'(T)\rangle= - A''(T)$ etc.
Even though these expressions look complicated, in many situations the function $S(T)$ and hence the coefficient
$a_2$ can be computed explicitly. Similarly, in principle, one can compute $a_3$ also.

As an example of an explicit evaluation up to ${\cal O}(\epsilon^2)$, one can consider the diffusion equation in $d=2$
where $A(T)= {\rm sech} (T/2)$. In this case, both $a_1$ and $a_2$ can be computed explicitly and one gets~\cite{MajumdarBrayPartial}
\begin{equation}
\theta(p=1-\epsilon)= \frac{1}{2\pi}\, \epsilon +\left(\frac{1}{\pi^2}-\frac{1}{4\,\pi}\right)\, \epsilon^2 + {\cal O}(\epsilon^3)\,.
\label{partials_diff2d}
\end{equation}
Keeping terms up to second order and putting $\epsilon=1$ gives, $\theta(0)= (\pi+4)/(4\pi^2)=0.180899\dots$, just
$3.5\%$ below the numerical simulation value $\theta_{\rm sim}(0)=0.1875\pm 0.0010$~\cite{MajumdarSireBrayCornell,DerridaHakimZeitak}.
Note that although the IIA estimate presented in section \ref{subsection:iia}, $\theta_{\rm IIA}(0)=0.1862\dots$ (see Table \ref{table_theta_diffusion}), is closer to the simulation value, it can not be
improved systematically. In contrast, the series expansion estimate presented above can, in principle, be systematically
improved order by order.

We conclude this section by noting that this series expansion will not work for {\em non-smooth} Gaussian processes where
the moments of zero crossings are infinite. As an example, consider the simple one dimensional Brownian motion, $dx/dt=\eta(t)$
where $\eta(t)$ is a zero mean Gaussian white noise. Under Lamperti transformation, it becomes a 
GSP with correlator, $A(T)=\exp[-T/2]$. In this case, it follows from Rice's formula that the number of zero 
crossings in infinite: if it crosses zero once, it recrosses zero infinitely often immediately afterwards.
As a result, only the $n=0$ term in the expansion $Q(p,T)= \sum_{n=0}^{\infty} p^n\, Q_n(T)$ contributes,
leading to the result, $Q(p,T)\approx Q(0,T)\sim \exp[-\theta(0)\, T]$ with $\theta(0)=1/2$ (see section \ref{subsection:GSP} and Eq. (\ref{markovp1})).
Hence, $\theta(p)=\theta(0)$ for all $0\le p\le 1$, except at $p=1$ where $\theta(1)=0$. Since $\theta(p)$ is discontinuous at $p=1$, 
the expansion around $p=1$ is not possible. Similar conclusion holds, for example, in the case of Glauber dynamics 
of an Ising chain at $T=0$~\cite{BFK96}.

\section{Global persistence}
\label{Global}
Earlier in this article we have been focused in `local' or `site' persistence
properties of coarsening systems, where the persistence of a single, local 
degree of freedom is studied. In
the standard nonconserved dynamics a given spin inside
the sample flips only when a domain wall passes through
it, which happens rather rarely. As a result, persistence,
i.e. the probability that the spin remains unflipped up to
time $t$ decays slowly as a power law, $Q(t) \sim t^{-\theta}$ at late times. 
In contrast, if the spin system is quenched to its
critical temperature $T_c$, $Q(t)$ decays
always exponentially in time due to fast thermal fluctuations. On the other hand, at $T_c$, the equilibrium 
correlation length of the global magnetization is infinite (or say of the order of the system size for a finite system). Hence
this suggests to look at the related 
problem of `global persistence' in which the persistence of a global property, 
such as the magnetization of a ferromagnet, is investigated. This is 
particularly interesting for a quench to the critical point, where the 
nonequilibrium dynamics corresponds to critical coarsening. It turns out 
\cite{MajumdarBrayCornellSire96} that the global persistence $Q(t)$ is described 
by a new critical exponent $\theta_G$, $Q(t) \sim t^{-\theta_G}$, that is unrelated to the standard static and 
dynamical exponents.

One simplifying feature of the global order parameter is that, in the 
thermodynamic limit, it is a Gaussian random variable at all times 
after the quench. This follows from the central limit theorem, when we 
note that the order parameter field $\phi({\bf x},t)$ has a finite correlation 
length, $L(t) \sim t^{1/z}$, where $z$ is the usual dynamic exponent, at 
all finite times $t$ after the quench to the critical point. If the system 
has a volume $V \gg L(t)^d$, the relevant Gaussian field is the $k=0$ 
Fourier component, 
\begin{eqnarray}\label{def_phi_0}
\tilde \phi(t) \equiv \tilde{\phi}(0,t) = \frac{1}{{\sqrt{V}}}\int d^dx\,\phi({\bf x},t) \;.
\end{eqnarray}
For an $n$-component vector field $(\phi_1,\ldots,\phi_n)$, the equation 
of motion reads
\begin{equation}
\partial_t \phi_i = \nabla^2\phi_i - (u/n)\phi_i\sum_{j=1}^n \phi_j^2 + \xi_i\ ,
\label{global1}
\end{equation}
where, as before, $\vec{\xi}({\bf x},t)$ is Gaussian white noise with 
mean zero and correlator $\langle \xi_i({\bf x},t)\xi_j({\bf x'},t')\rangle 
= 2\delta_{ij}\delta^d({\bf x}-{\bf x'})\delta(t-t')$. For $n>1$ we are 
defining the global persistence as the probability that a {\em given component}
of the global order parameter has not changed sign up to time $t$. 

\subsection{Mean-field theory}
Mean-field theory, valid for $d \ge 4$, corresponds to $r=0$ and $u=0$. The 
${\bf k}=0$ Fourier component, $\tilde{\phi}(t)$, (where we have suppressed 
the index $i$), obeys the simple equation 
\begin{equation}
\partial_t\tilde{\phi} = \tilde{\xi}, 
\label{global-meanfield}
\end{equation}
corresponding to a Brownian motion.
It follows that the global persistence exponent is given by $\theta_G=1/2$ 
in mean-field theory. 

\subsection{The large-$n$ limit}

Next we consider the large-$n$ limit. Equation (\ref{global1}) then simplifies 
to a self-consistent linear equation for each component:
\begin{equation}
\partial_t\phi = \nabla^2 \phi - (r+u \langle \phi^2 \rangle)\phi + \xi\ .
\label{global2}
\end{equation}
Defining $a(t) = -r-u\langle \phi^2 \rangle$ and $b(t) = \int_0^t a(t')dt'$, 
Eq.\ (\ref{global2}) has the Fourier-space solution
\begin{equation}
\tilde{\phi}({\bf k},t) = \tilde{\phi}(0,t)\exp[b(t)-k^2t]
+ \int_0^t dt'\tilde{\xi}({\bf k},t')\exp[b(t)-b(t')] - k^2(t-t')]\ .
\end{equation}
One can easily show that the second term, containing the noise, dominates 
at large $t$ \cite{JanssenSchaubSchmittmann}. Retaining only this term, 
computing $\langle \phi^2 \rangle$, and defining $g=\exp(-2b)$ yields the 
equation
\begin{equation}
\partial_t g = 2rg + 4u\int_0^t dt'\,g(t')\sum_{\bf k}\exp[-2k^2(t-t')]\ ,
\end{equation}
which can be solved using Laplace transforms. Putting $r$ equal to its 
critical value, $r_c = -u \langle \phi^2 \rangle = -u\sum_{\bf k}k^{-2}$ 
gives, for the Laplace transform $\bar{g}(s) = \int_0^\infty dt\,
g(t)\exp(-st)$, the result
\begin{equation}
\bar{g}(s) = [s + 4u\{\bar{J}(0) - \bar{J}(s)\}]^{-1}\ ,
\end{equation}
where $\bar{J}(s) = \sum_{\bf k}(s + 2k^2)^{-1}$. From this we deduce that, 
for dimensions in the range $2<d<4$, $\bar{g}(s) \sim s^{(2-d)/2}$ for 
small $s$. Inverting the Laplace transform we obtain 
(with $\epsilon = 4-d$) $g(t) \sim t^{-\epsilon/2}$ for $t \to \infty$, 
whence $b(t) \sim (\epsilon/4) \ln t$ and $a(t) = db/dt \sim \epsilon/4t$. 
The large-$n$ equation of motion thus reduces to the simplified form 
\begin{equation}
\partial_t\tilde{\phi} = (\epsilon/4t)\tilde{\phi} + \tilde{\xi}
\end{equation}
for the ${\bf k}=0$ component of $\phi$. The change of variable
$\tilde{\phi} = t^{\epsilon/4}\psi$ yields the even simpler equation
\begin{equation}
\partial_t\psi = t^{-\epsilon/4}\tilde{\xi}(t)\ .
\end{equation}
Introducing the new time variable $\tau = t^{(1-\epsilon/2)}$, the equation 
reduces to the random walk equation, $\partial_\tau \psi = \eta(\tau)$, 
with the the standard $\tau^{-1/2}$ decay of persistence. In terms of the 
original time variable $t$, one thus obtains (recalling that $\epsilon=4-d$) 
a $t^{-(d-2)/4}$ decay for the global persistence. The global persistence 
exponent for $n=\infty$ is, therefore,
\begin{equation}
\theta_G = (d-2)/4,\ \ \ 2< d \le 4 \;, \; (n=\infty)\ .
\label{theta_Glarge-n}
\end{equation} 
For $d>4$, $\theta_G$ sticks at the mean-field value of $1/2$. 

One can also calculate $\theta_G$ to first order in $\epsilon=4-d$ for 
arbitrary values of $n$, using conventional renormalisation group methods. 
We refer the reader to \cite{MajumdarBrayCornellSire96} and simply quote 
the result:
\begin{equation}
\theta_G = \frac{1}{2} - \frac{1}{4}\left(\frac{n+2}{n+8}\right)\epsilon 
+ {\cal O}(\epsilon^2)\ ,
\end{equation}
which agrees with (\ref{theta_Glarge-n}) for $n=\infty$. 

\subsection{The one-dimensional Ising model}

Another soluble limit is the one-dimensional Ising model with Glauber 
dynamics. For this model, there is no finite-temperature phase transition. 
Instead, the system orders at $T=0$. The dynamics at $T=0$ is governed by 
the motion of the domain walls, equivalent to a set of mutually 
annihilating random walkers. Starting from a completely disordered state 
(each of the $N$ spins independently up or down with probability 1/2), 
the number of surviving walkers at time $t$ is of order $Nt^{-1/2}$
\cite{BrayReview,AmarFamily}. Since the contributions from the different 
walkers add incoherently, the change in $M(t)$ in one time step
is of order $\sqrt{N}t^{-1/4}$. The $k=0$ Fourier component  
$\tilde{\phi}(0,t) = M(0)/\sqrt{N}$ in Eq.~(\ref{def_phi_0}) therefore satisfies the Langevin equation
(up to an overall constant) 
\begin{equation}
\partial_t\tilde{\phi} = t^{-1/4}\xi(t)\ ,
\end{equation} 
where $\xi(t)$ is Gaussian white noise. This equation can be reduced 
to standard random walk dynamics by the change of variable $t=\tau^2$. 
After some straightforward algebra \cite{MajumdarBrayCornellSire96} 
one obtains the final result for the global persistence, 
$Q(t) \sim t^{-1/4}$, i.e.\ $\theta_G = 1/4$ for this model.  

\subsection{$\theta_G$: A new critical exponent}
The results for the global persistence  presented in the preceding sections have one property in common: in each case the underlying dynamics 
is described by a Gaussian Markov process. In such cases one can derive 
\cite{MajumdarBrayCornellSire96} a ``scaling law'' relating $\theta_G$ to the 
other exponents:
\begin{equation}
\theta_G\,z = \lambda_G - d +1 -\eta/2\ ,
\label{putative-scaling-law}
\end{equation}
where $\lambda_G$ describes the asymptotics of the two-time correlation 
function of the global order parameter at the critical point: 
$\langle\tilde{\phi}(t_1)\tilde{\phi}(t_2)\rangle = t_1^{(2-\eta)/z} 
F(t_2/t_1)$, where $F(x) \sim x^{(d-\lambda_G)/z}$ for large $x$.

At this point one might wonder if the exponent $\theta_G$ is related to the 
other (static and dynamic) critical exponents or whether it is an independent 
critical exponent. We remind the reader that there are two independent 
static critical exponents, for example $\nu$ and $\eta$, to which other 
static exponents are related by scaling laws, and the dynamical exponent $z$. 
The exponent $\nu$ describes the divergence of the correlation length $\xi$ 
near the critical temperature, $T_c$: $\xi \sim |T-T_c|^{-\nu}$, while 
$\eta$ characterises the decay of the correlation function, $G(r)$, at 
$T_c$: $G(r) \sim r^{-(d-2+\eta)}$. In addition to the dynamical exponent $z$, 
that relates length scales to time scales via $\tau \sim \xi^z$, there 
is another -- specifically nonequilibrium -- exponent, $\lambda_G$, that 
describes the asymptotics of the two-time autocorrelation function at $T_c$:
\begin{equation}
\langle \tilde{\phi}(t_1)\tilde{\phi}(t_2)\rangle = t_1^{(2-\eta)/z}\,F(t_2/t_1) ,
\label{global-auto}
\end{equation}
with $F(x) \sim x^{(d-\lambda_G)/z}$ for $x \to \infty$. For all the cases 
discussed above (mean-field theory, the large-$n$ limit, the epsilon 
expansion to first order, and the one-dimensional Ising model) one can check 
\cite{MajumdarBrayCornellSire96} that the ``scaling law'' 
(\ref{putative-scaling-law}) is satisfied. 

This raises the question of whether this ``scaling law'' holds generally. 
We believe that it does not. The reason is that the global order parameter 
${\Phi}(t)$ is not a Markov process in general. To see this, we consider 
the autocorrelation function,
$\langle \tilde{\phi}(t_1)\tilde{\phi(t_2)} \rangle$, 
of the $k=0$ mode, $\tilde{\phi}(t)$. It has the scaling form 
displayed in Eq.\ (\ref{global-auto}), with $F(x) \sim x^{(d-\lambda_G)/z}$ 
for large $x$. Now construct the normalised autocorrelation function 
\begin{equation}
a(t_1,t_2) = \frac{\langle\tilde{\phi}(t_1)\tilde{\phi}(t_2)\rangle}
{\langle\tilde{\phi}(t_1)^2\rangle^{1/2}
\langle\tilde{\phi}(t_2)^2\rangle^{1/2}}\ . 
\end{equation}   
This has the scaling form $a(t_1,t_2) = f(t_1/t_2)$ with 
$f(x) \sim x^{(\lambda_G - d +1 - \eta/2)/z}$, for large $x$. Introducing 
the usual log-time variable $T=\ln t$, the normalised autocorrelation 
function has the form $A(T_1,T_2) = g(T_1-T_2)$. This process is thus 
a Gaussian stationary process in the new time variable. Furthermore, 
the function $g(T)$ has the asymptotic form 
$g(T) \sim \exp(-\bar{\lambda}|T|)$, with 
\begin{equation}
\bar{\lambda} = (\lambda_G - d +1 -\eta/2)/z\ .
\end{equation}
If the process were Markovian, $g(T)$ would have this form (pure 
exponential decay) for all $T$ \cite{Chaturvedi}, and the global 
persistence exponent would then be equal to $\bar{\lambda}$ 
\cite{Chaturvedi, Feller, MajumdarSirePT}. We refer the reader to Ref. \cite{CG05} for a review on the analytical and numerical estimates of 
these exponents $\lambda_G$ and $z$. The question of whether the ``scaling law'' (\ref{putative-scaling-law}) 
holds generally thus comes down to whether the function $f(t_2/t_1)$ 
is a simple power-law for {\em all} $t_2>t_1$, not just for $t_2 \gg t_1$ or, 
equivalently, whether the function $g(T)$ is a pure exponential for 
all $T>0$.  

It is easy to check that our result for the large-$n$ limit, the ${\cal O}(\epsilon)$ 
calculation, and the $d=1$ Ising model all satisfy this criterion. 
In a calculation to ${\cal O}(\epsilon^2)$, however, one finds that the function 
$g(T)$ is no longer a simple exponential. Instead one finds \cite{Oerding} 
\begin{equation}
g(T) = \exp(-\bar{\lambda}T)\left[1-\frac{3(n+2)}{4(n+8)^2}]
\epsilon^2 F_A(\exp[T]) + {\cal O}(\epsilon^3)\right] ,
\label{OerdingA}
\end{equation} 
where $F_A(x)$ is rather complicated function \cite{Oerding}. Note 
the subscript `A' denotes that the underlying dynamics we are using, 
described by Eq.\ (\ref{global1}), corresponds to `Model A' of the 
Hohenberg-Halperin classification scheme for dynamic critical phenomena 
\cite{Hohenberg-Halperin}, which corresponds to a relaxational dynamics of a non-conserved order parameter. Using the perturbative result 
given in Eq.\ (\ref{pert_time}) (see section \ref{section:perturbation} for details) we can calulate the global persistence 
exponent to ${\cal O}(\epsilon^2)$. The result is~\cite{Oerding}
\begin{equation}\label{thetaG_twoloop}
\theta_G = \bar{\lambda}\left\{1 + \frac{3(n+2)}{4(n+8)^2} 
\epsilon^2\ \alpha\right\} + {\cal O}(\epsilon^3) \;,
\end{equation}  
where $\alpha$ is given by an explicit but rather lengthy expression 
\cite{Oerding} with numerical value $\alpha = 0.271577604975\ldots$.  

An equivalent result can also be obtained 
\cite{Oerding} for `Model C' dynamics \cite{Hohenberg-Halperin}, which corresponds to the relaxational dynamics of
a non-conserved order parameter coupled to a conserved density. At variance with Model A dynamics, it was shown \cite{Oerding} that
non-Markovian corrections already appear at order ${\cal O}(\epsilon)$. Finally, this perturbative approach, in dimension $d = 4 -\epsilon$, 
was also extended to study the global persistence in the critical dynamics (Model A) of the randomly diluted Ising model \cite{PS05,SP06}. In this case non-Markovian corrections also appear at first order in perturbation theory.

The exponent $\theta_G$ has also been measured in numerical simulations of 
coarsening ferromagnets at $T=T_c$ evolving via Model A dynamics both for the pure system 
\cite{MajumdarBrayCornellSire96,Sta96,SC97,PGS07} in dimension $d=2,3$ and for the diluted Ising model in $d=3$ \cite{PS05} (in $d=2$ the random dilution yields only logarithmic corrections to the pure case). For the pure Ising model in $d=2$, the most precise estimate yields $\theta_G = 0.237(3)$ \cite{SC97} in agreement with other Monte-Carlo estimates \cite{MajumdarBrayCornellSire96,Sta96,PGS07}, which is still slightly larger than the perturbative result (\ref{thetaG_twoloop}) with $\epsilon = 4-d=2$, yielding $\theta_G = 0.218$ (where we have used $\lambda_G = 1.585$ \cite{Gra95}, $\eta = 1/4$ (exact) and $z=2.166$ \cite{NB00}). Note that in this case, it was checked that Metropolis and Heat-Bath algorithms both yield the same value
of $\theta_G$ \cite{SC97}. In dimension $d=3$, the numerical estimate is $\theta_G = 0.41$ \cite{Sta96} (where the author does not provide any estimate of the errorbar), which is also larger than the perturbative result (\ref{thetaG_twoloop}) with $\epsilon = 4-d=1$, yielding $\theta_G = 0.383$ (where we have used $\lambda_G =2.789$, $\eta = 0.032$ and $z=2.032$ \cite{Gra95}). For the randomly diluted Ising model in $d=3$, Monte Carlo simulations yield $\theta_G = 0.35(1)$, which is slightly above the one loop estimate, i.e. the equivalent of (\ref{thetaG_twoloop}), given by $\theta_G = 0.339$. We refer the reader to Ref.~\cite{HenkelPleimlingBook} for a detailed account on the numerical and analytical estimates of $\theta_g$ for various non-equilibrium critical dynamics. 

For Model C dynamics, Monte Carlo simulations were performed on antiferromagnetic Ising model, the order parameter being the staggered magnetization, with a conserved global magnetization $M_0 \neq 0$ \cite{SS03}. According to the analytical predictions \cite{Oerding} the persistence exponent $\theta_G$ was found to be different from the one in model A but with a slight dependence on $m_0$, which is not expected from these analytical results \cite{Oerding}. This discrepancy remains unexplained. Note finally that for 'Model B' dynamics, which corresponds to relaxational dynamics with a conserved order parameter, the global persistence is not defined. 

Global persistence for critical systems has been the subjects of many numerical studies not only for other spin systems, with different kind of interactions or dynamical rules, at a critical point \cite{MO97,DAF03,FF06} but also in a variety of models ranging from genuine non-equilibrium systems at an absorbing phase transition~\cite{HK98,OvW98,AB01,LM02} to polymer systems at the helix-coil transition~\cite{AFH07} or to statistical mechanics models with applications to socio-physics \cite{JY08}.

\subsection{The case of a finite initial magnetization for Model A dynamics}\label{sec:initial_mag}

Up to now, we have considered the case of Model A dynamics for the critical coarsening of a system which is
initially prepared in a completely disordered condition with {\it vanishing} initial
magnetization $M_0=0$. In this case, the average magnetization is zero at all time $t$, $\langle M(t) \rangle = 0$. If, however, one
starts with a non-vanishing initial magnetization $M_0$, it is well known that, after a non-universal transient, the average
magnetization $M(t)$ grows in time as $M(t) \propto M_0 t^{\theta'_{\rm is}}$ for $t
\ll \tau_m \propto M_0^{-1/\kappa}$ whereas, for $t \gg \tau_m$,  
$M(t)$ decays algebraically to zero as $M(t) 
\propto t^{-\beta/(\nu z)}$ (see Fig. \ref{Fig_slip}). 
\begin{figure}[ht]
\centering
\includegraphics[width=0.5\linewidth]{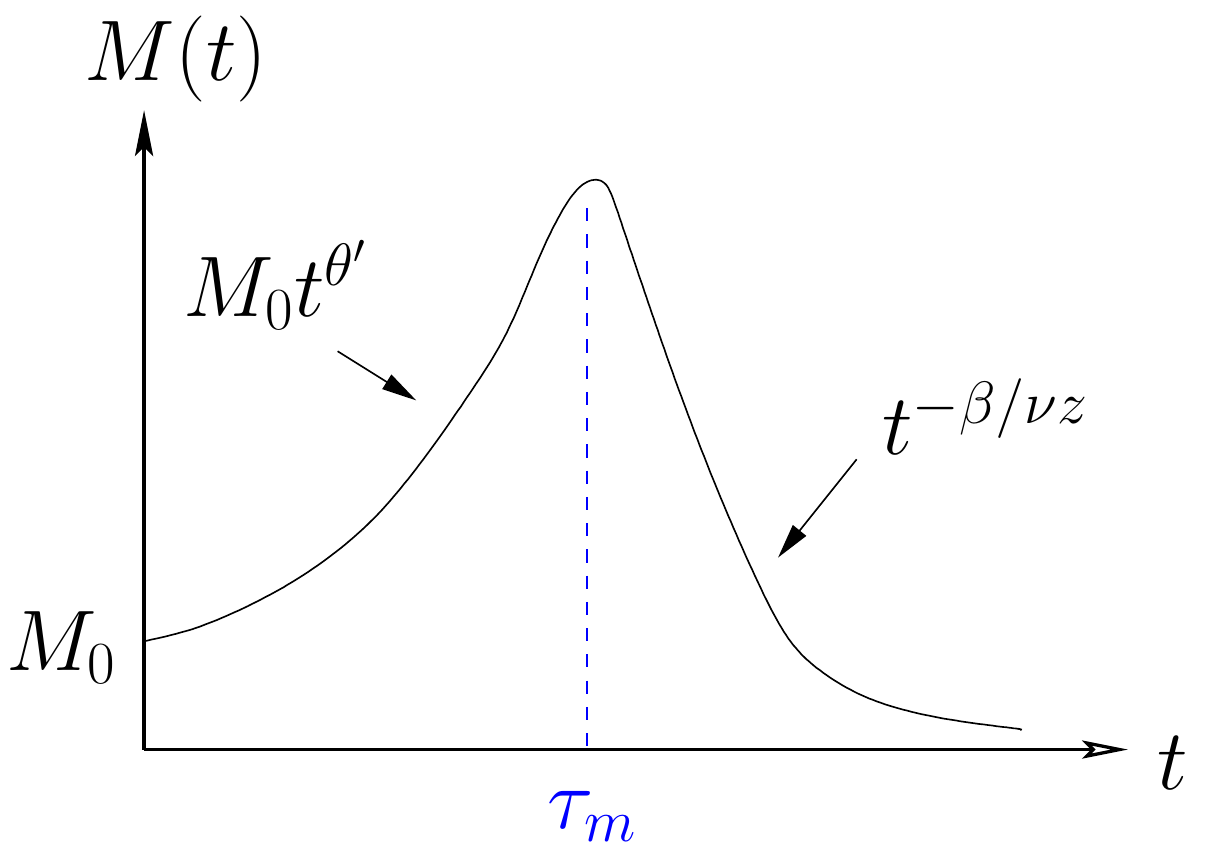}
\caption{Sketch of the time evolution of the global magnetization $M(t)$ after a quench at $T_c$ with a non-vanishing initial magnetization $M_0$.}
\label{Fig_slip}
\end{figure}
These different time dependences are
characterized by the universal exponents  $\theta'_{\rm is}$ (the so-called initial-slip
exponent~\cite{JanssenSchaubSchmittmann}) and $\kappa = \theta'_{\rm is} + \beta/(\nu z)$, where 
$\beta$, $\nu$ and $z$ are the usual static and dynamic (equilibrium) critical
exponents, respectively.

In Ref.~\cite{PGS07}, it was demonstrated that the
persistence probability $Q_G(t)$ of the  thermal fluctuations $\delta M(t) = M(t) - \langle M(t) \rangle$
around $\langle M(t) \rangle$ displays a crossover  between
two different power-law regimes (which is the counterpart of the change of behavior of $\langle M(t)\rangle$ itself): 
\begin{eqnarray}
\label{crossover_dyn}
Q(t) \sim 
\begin{cases}
 t^{-\theta_G}&\quad\mbox{for}\quad t_{\rm micr} \ll t \ll \tau_m \,,\\
t^{-\theta_{G,\infty}}&\quad\mbox{for}\quad t \gg \tau_m\,,
\end{cases}
\end{eqnarray}  
 $t_{\rm micr}$ being a microscopic time scale. In the case $M_0 \to 0$, $\tau_m \to \infty$ and one recovers the behavior discussed above but for finite
 $M_0$, $Q(t)$ eventually decays with an exponent $\theta_{G,\infty} \neq \theta_G$. In the Markovian approximation one finds the equivalent of the above relation (\ref{putative-scaling-law}) which reads here
\begin{eqnarray}
\theta_{G,\infty} z = z + \frac{d}{2} \;,
\end{eqnarray} 
but in general $\theta_{G,\infty}$ is a new independent exponent for non-Markov process. This was shown by a 
perturbative calculation along the same lines yielding (\ref{thetaG_twoloop}) up to one loop in $d=4-\epsilon$ for $O(n)$ models \cite{PGS07}, taking advantage of the Renormalization Group analysis performed in Ref. \cite{CGK06,CG07}. This crossover was also confirmed by numerical simulations of the Ising model in dimension $d=2$, where the value $\theta_{G,\infty} = 1.7(1)$ was found, larger than the one-loop estimate $\theta_{G,\infty} = 1.61$ \cite{PGS07}.

\subsection{Global persistence for $T<T_c$}

A recent study by Henkel and Pleimling \cite{Henkel} extends the study 
of global persistence to the standard ``coarsening'' regime \cite{BrayReview}, 
where the system is quenched to below the critical temperature, starting 
from an  equilibrium state in the high-temperature phase. For cases where 
the process is Markovian (these include the one-dimensional Ising model 
and the large-$n$ limit, as discussed earlier in this section), they argue 
that the global persistence exponent in the coarsening state is given by 
$\theta_{G_0} = (2\lambda_0-d)/2z_0$, where the subscript $0$ here indicates 
coarsening, $\lambda_0$ is the autocorrelation exponent, and $z_0$ is the 
dynamic exponent for the coarsening regime. It is instructive to compare 
this result with Eq.\ (\ref{putative-scaling-law}). In fact the former 
is a special case of the latter, obtained by setting $\eta = 2-d$ in 
(\ref{putative-scaling-law}). The result $\eta=2-d$ for the coarsening 
state follows from the fact that the latter is controlled by a 
zero-temperature fixed point \cite{BrayReview}, another consequence of which 
is that the global persistence exponent should be temperature-independent 
for all $T<T_c$. This prediction is born out by  numerical simulations 
of the two-dimensional Ising model at temperatures $T=1.0$ and $T=1.5$ 
($T_c\approx 2.27$) with results $\theta_{G_0} = 0.062(2)$ and $0.065(2)$,
which are identical within the quoted errors.

\subsection{Block persistence for $T < T_c$}

If one considers the results discussed in the previous sections, at $T = 0$, we see that the global persistence and the local persistence
introduced initially, are characterized by two different decaying exponents $\theta_{G_0} \neq \theta$. For instance, for Ising
systems one has $\theta_{G_0} = \theta_G = 1/4$ \cite{MajumdarBrayCornellSire96} and $\theta = 3/8$ \cite{DHP1995a,DHP1995b} in $d=1$, while $\theta_{G_0} = 0.062(2)$ \cite{Henkel} and $\theta = 0.22$ \cite{BDG1994a,Sta94} in $d=2$. To interpolate
between these two distinct algebraic behaviors, Cueille and Sire have introduced the notion of block persistence \cite{CS97,CS99}. The method is in the spirit of the
real space renormalization group {\it \`a la Kadanoff}. 
\begin{figure}
\centering
\includegraphics[width=\linewidth]{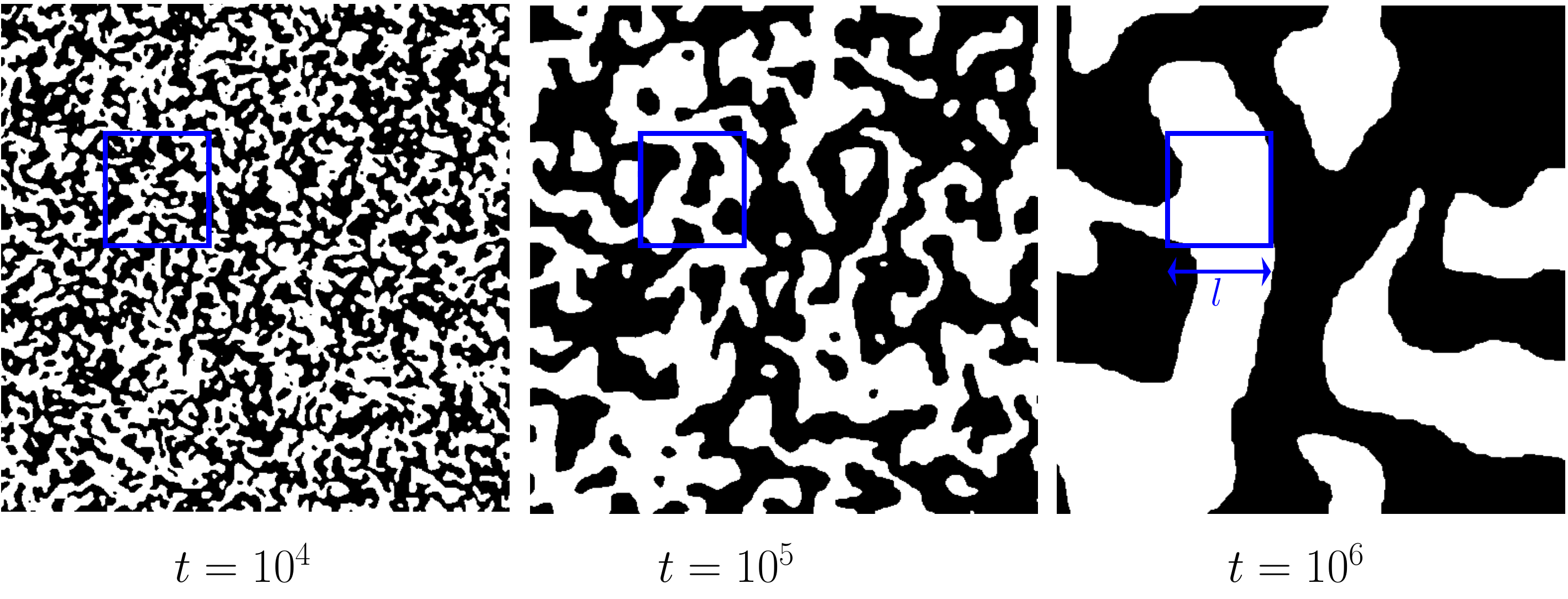}
\caption{Illustration of the block persistence $p_l(t)$ on the 2-d-Ising model at temperature $T=0$: this is the standard persistence probability for the magnetization associate to the square of block in blue, of linear size $l$.}\label{fig:block}
\end{figure}
The block persistence probability $Q_l(t)$ is the standard persistence probability for a
coarse-grained variable obtained by integrating the order parameter on a block of linear size $l$ (Fig. \ref{fig:block}). For $l \to \infty$ (or say $l$ equals the whole system size for a finite system) we recover the global persistence while for $l = 0$ (or $l$ equals the lattice spacing on a lattice), we get the local persistence. Now for finite $l$, the time dependence of $Q_l(t)$ interpolates between the two exponents $\theta_{G_0}$ and $\theta$. Indeed, at early times, when the correlation length $L(t) \sim t^{1/z} \ll l$, the system effectively sees infinite blocks and $Q_l(t) \propto t^{-\theta_{G_0}}$. On the other hand, for $L(t) \sim t^{1/z} \gg l$, the blocks behave effectively as single spins and $Q_l(t) \sim c_l t^{-\theta}$. Therefore $Q_l(t)$ takes the scaling form \cite{CS97,CS99}
\begin{eqnarray}\label{eq:block}
Q_l(t) \sim l^{-z \theta_{G_0}} f(t/l^z) \;,
\end{eqnarray} 
where $f(x)$ is a universal scaling function which behaves like 
$f(x) \propto x^{-\theta_{G_0}}$ when $x \to 0$ and $f(x) \sim x^{-\theta}$ when $x \to \infty$. This scaling behavior in Eq.~(\ref{eq:block}) was
demonstrated analytically for some simplified models of coarsening (namely the diffusion equation and the non-conserved dynamics of 
the $O(n)$ model for $n \to \infty$) and checked numerically for different models, including Ising spin systems and for the Time Dependent Ginzburg-Landau equation in $d=2$ \cite{CS97,CS99}. For Ising systems, for which $z=2$, the measured exponent $\theta_{G_0}$ was found compatible with the value of Ref. \cite{Henkel} discussed above. 

Another important motivation for the introduction of the block persistence $Q_l(t)$ is the extension of the local persistence at finite temperature
$T>0$. Indeed, due to thermal fluctuations, the local persistence decays exponentially at finite $T>0$, while it decays algebraically for $T=0$. This
result seems to be in contradiction with the fact that below $T_c$, the large scale properties (static and dynamical) are governed by a zero temperature fixed point. The block persistence $Q_l(t)$ allows to elucidate this apparent contradiction \cite{CS97,CS99}. Indeed, numerical simulations of Ising models at finite temperature $0<T<T_c$ indicate that in this case, the above scaling form (\ref{eq:block}) becomes
\begin{eqnarray}
Q_l(t) \sim l^{-z \theta_{G_0}} f(t/l^z) \exp{[-t/\tau(l,T)]} \;,
\end{eqnarray}
where the crossover time $\tau(l,T)$ diverges very quickly when $l$ is increased. This can be understood as the effective temperature governing
the thermal fluctuations of a block of size $l$ is $T/l^d$, for a $d$-dimensional system. Consequently, for $l$ of the order of a few lattice spacings, $Q_l(t)$ has the $T=0$ behavior for time scales accessible in numerical simulations \cite{CS97,CS99}. We refer the reader to Ref. \cite{Der97} for an alternative definition of the persistent exponent at finite temperature, its relation to block persistence being discussed in Ref. \cite{CS97,CS99}. Yet another definition of the local persistence exponent at finite temperature, in connection with the notion of "occupation time" \cite{GD98},  is discussed in section \ref{subsection:occupation_time}.

\section{The persistence of manifolds in nonequilibrium critical dynamics}
Up to now we have considered two aspects of persistence: `local' (or `site') 
persistence where the persistence of a localised degree of freedom is 
investigated, and `global persistence', where the global order parameter 
of the system is the quantity studied. Between these two extremes there is 
another class of persistence phenomena involving sets of degrees of freedom 
that are large in number (infinite in the thermodynamic limit) but represent 
a vanishingly small fraction of the total number of degrees of freedom. The 
conceptually simplest such sets consist of manifolds (e.g. lines in two dimensions or planes in three dimensions) 
of degrees of freedom. Here we consider lower-dimensional submanifolds of 
a ferromagnet, with dimension $d'$, undergoing coarsening at its critical 
point, with nonconserved dynamics \cite{BrayReview}.

In a $d$-dimensional sample a single spin is a zero-dimensional manifold, 
$d'=0$, while the global magnetisation is a $d$-dimensional manifold. 
To interpolate between these two limits, we study the persistence of the 
magnetisation of a $d'$-dimensional manifold, with $0 \le d'\le d$. 
So for $d=3$, for example, $d'=1$ corresponds to a line of spins and 
$d'=2$ to a plane. 

For a quench to the critical point, spins fluctuate rapidly due to the 
non-zero temperature, and the persistence of a single spin has an exponential 
tail at long times, while we have seen the the global persistence ($d'=d$) 
has a power-law tail, described by the exponent $\theta_G$. A natural question 
arises if one varies the manifold dimension from $d'=0$ to $d'=d$: How does 
the asymptotic behaviour of the persistence change from an exponential decay 
to a power-law decay as $d'$ increases? Does the behaviour change abruptly 
at some value of $d'$, or is there a range of $d'$ where some other asymptotic 
form is observed? In fact one can show \cite{MajumdarBrayManifolds} that there 
is indeed an intermediate region of $d'$ where the persistence has a 
stretched exponential tail. The results can be summarised in terms of the 
exponent combination 
\begin{equation}
\zeta = (D - 2 + \eta)/z\ ,
\end{equation}
where $D=d-d'$ is the codimension of the manifold, and $\eta$ and $z$ 
are the standard critical exponents. According to the value of $\zeta$, 
the persistence of the magnetisation of a $d'$-dimensional manifold
has the asymptotic forms~\cite{MajumdarBrayManifolds}
\begin{eqnarray}
Q(t) \sim 
\begin{cases}
&t^{-\theta(d',d)} \;, \; \zeta < 0 \;,  \\
&\exp{(-a_1t^\zeta)} \;, \; 0 \leq \zeta < 1 \;,  \\
&\exp{(-b_1t)} \;, \zeta > 1 \;, 
\end{cases}\label{manifolds}
\end{eqnarray}
where $a_1$ and $b_1$ are constants. To be strictly accurate, in the 
intermediate range $0 \le \theta \le 1$, one can show that
$\exp(-a_2 t^\zeta \ln t) \le Q(t) \le \exp(-a_1 t^\zeta)$, where $a_2$ is 
another constant. We obtain the results in Eq.\ (\ref{manifolds}) within the 
mean-field theory, valid for $d>4$, in the $n \to \infty$ limit of the $O(n)$ 
model and, more generally, using scaling arguments.  The results in 
Eq.\ (\ref{manifolds}) hold for all manifolds with dimension $d'>0$. They do 
not hold for $d'=0$, which corresponds to a single degree of freedom. It is 
clear that the persistence of a single spin decays exponentially at $T=T_c$ 
since the flip rate is non-zero. 

Two special cases of the general result (\ref{manifolds}) are the 
persistence of (i) the line magnetisation ($d'=1$) in the $d=2$ Ising model 
at $T_c$, and (ii) the line ($d'=1$) and plane ($d'=2$) magnetisation in the $d=3$ 
Ising model. In case (i) we have $d'=1$, $d=2$, $\eta=1/4$ and $z \approx 
2.172$, to give $\zeta = (d-d'-2+\eta)/z \approx -0.3453 < 0$. Here Eq.\ 
(\ref{manifolds}) predicts a power-law decay for the persistence of the 
line magnetisation. In case (ii) we have $\eta \approx 0.032$ and 
$z \approx 2$, giving $\zeta \approx -0.484<0$ for the plane magnetisation 
but, for the line magnetisation, $\zeta \approx 0.016$, which lies in the 
interval $[0,1]$. Thus Eq.\ (\ref{manifolds}) predicts a power-law decay 
for the persistence of the plane magnetisation, but a stretched-exponential 
decay for the line magnetisation. Numerical simulations 
\cite{MajumdarBrayManifolds} are consistent with these predictions. 
 
We now show how the predictions in Eq.\ (\ref{manifolds}), were obtained. We start 
from the Langevin equation for the vector order parameter 
$\vec{\phi_i} = (\phi_1,\ldots,\phi_n)$. It has the form of a time-dependent 
Ginzburg-Landau equation,  
\begin{equation}
\frac{\partial\phi_i}{\partial t} = \nabla^2\phi_i - r\phi_i 
- \frac{u}{n}\phi_i\sum_{j=1}^n \phi_j^2 + \eta_i(t)\ ,
\label{LangevinManifolds}
\end{equation} 
where $\vec{\eta}({\bf x}, t)$ is Gaussian white noise with mean zero 
and correlator  
\begin{equation}
\langle \eta_i({\bf x},t)\,\eta_j({\bf x'}, t')\rangle 
= 2\delta_{ij} \delta^d({\bf x}-{\bf x'})\,\delta(t-t')\ .
\end{equation}
The magnetisation of a $d'$-dimensional manifold, obtained by integrating 
the order parameter over the $d'$ directions, is given by the vector field
\begin{equation}
\psi_i(x_{d'+1},\ldots,x_d,t) = \int \phi_i(\vec{x},t)
\prod_{j=1}^{d'} \frac{dx_j}{\sqrt{L}}\ ,
\end{equation}
where $L$ is the length of the sample in each direction (i.e.\ the 
sample has volume $L^d$). For this vector order parameter we define the 
persistence, $Q(t)$, of the manifold to be the probability that any given 
component of the manifold magnetisation does not change sign up to 
time $t$. From now on, we drop the subscript $i$ of $\psi_i$ as all spin 
components are equivalent. The manifold magnetisation $\psi$ is a field 
over the remaining $D = d-d'$ dimensional space, specified by coordinates 
$(x_{d'+1},\ldots,x_d)$. For convenience, we relabel these coordinates 
using the vector $\vec{r} = (r_1,\ldots,r_D)$. Just as in our discussion 
of the global persistence in the preceding section, we can infer that the 
manifold magnetisation $\psi(\vec{r},t)$ is a Gaussian field at all times $t$, 
since it is the sum of $L^{d'}$ variables that are correlated over a finite 
correlation length, $\xi(t) \sim t^{1/z}$. The central limit theorem then 
tells us that $\psi(\vec{r},t)$ is a Gaussian field provided $t^{1/z} \ll L$, 
which certainly hold in the thermodynamic limit. It follows that the 
persistence of $\psi(\vec{r},t)$ is determined by the autocorrelation 
function $C(t_1,t_2) = \langle \psi(\vec{r},t_1)\,\psi(\vec{r},t_2) \rangle$. Note, 
however, that our use of the central limit theorem requires $d'>0$, 
which we assume from now on. 

\subsection{Mean-field theory}
We begin with the simplest case, $d \ge 4$, which is described by 
mean-field theory. Here we can set $u=0$ and also $r=0$ (at the critical 
point) in Eq.\ (\ref{LangevinManifolds}). Then we integrate the Langevin 
equation over the $d'$ spatial dimensions, and solve the resulting linear 
equation by transforming to Fourier space. Defining the Fourier transform 
$\tilde{\psi}(\vec{k},t) = 
\int d^Dr\,\psi({\bf r},t)\exp(i\vec{k}\cdot\vec{r})$, the two-time 
correlation function is readily computed as
\begin{eqnarray}
\langle \tilde{\psi}(\vec{k},t_1)\,\tilde{\psi}(\vec{k},t_2) \rangle
& = & \Delta(\vec{k})\exp(-k^2[t_1+t_2]) \nonumber \\
&& + \frac{1}{k^2}\left[\exp(-k^2|t_1-t_2|)-(\exp(-k^2|t_1+t_2|)\right],
\label{manifoldeq}
\end{eqnarray}
where $\Delta(\vec{k})$ is a constant for an uncorrelated initial 
condition. 

At late times, the first term in Eq.\ (\ref{manifoldeq}) is negligible 
compared to the second. The autocorrelation function in this regime 
becomes 
\begin{equation}
C(t_1,t_2) = \langle \psi(\vec{r},t_1)\,\psi(\vec{r},t_2) \rangle = \int d^Dk\exp(-a^2k^2)\langle \tilde{\psi}(\vec{k},t_1) 
\tilde{\psi}(\vec{-k},t_2) \rangle,
\end{equation}
where $a$ is a soft ultraviolet (or short-distance) cut-off.
Evaluating the integral [neglecting the first term on the right in 
Eq.\ (\ref{manifoldeq}] gives
\begin{equation}
C(t_1,t_2) = {\rm const.}\ [(t_1 +t_2 + a^2)^{2\beta} 
- (|t_1-t_2| + a^2)^{2\beta}]\ ,
\label{manifoldautocorrelation}
\end{equation} 
where 
\begin{equation}
\beta = (2-D)/4\ ,
\end{equation}
and the value of the constant prefactor in Eq.\ 
(\ref{manifoldautocorrelation}) is unimportant for our purpose. 
We can consider separately the cases $D<2$ and $D>2$.

\subsubsection{The case $D<2$}
For $D<2$, i.e.\ $\beta>0$, we can set the short-distance cut-off $a$ 
to zero since $\langle \psi^2(\vec{r},t) = C(t,t)$ does not diverge at 
any finite time even for $a=0$. Setting $a=0$ in Eq.\ 
(\ref{manifoldautocorrelation}), we see that $C(t_1,t_2)$ is non-stationary. 
However it can be rendered stationary by using the Lamperti transformation. This amounts to define the 
normalised Gaussian process,
$X = \psi/\sqrt{\langle \psi^2 \rangle}$ which, in terms of the logarithmic
time variable $T = \ln t$, becomes a stationary Gaussian process with 
correlator
\begin{equation}
A(T) = \langle X(0)X(T) \rangle =\cosh(T/2)^{2\beta}-|\sinh(T/2)|^{2\beta}\ .
\label{EWcorrelator} 
\end{equation}
It is interesting that the same correlation function for a Gaussian 
stationary process appears in the context of the persistence of 
rough interfaces in the Edwards-Wilkinson class \cite{Krug}, which will be 
discussed in the following section. 

It is well-known \cite{SatyaReview,Slepian} that for a Gaussian 
stationary process with a correlator decaying exponentially in time for 
large $T$, the persistence also decays exponentially with $T$: $Q(T) 
\sim \exp(-\theta T)$, implying a $t^{-\theta}$ decay in the real time,  
$t=\exp(T)$. For the equivalent interface problem, the 
exponent $\theta$ is known to depend on the parameter $\beta$ \cite{Krug}.
 
\subsubsection{The case $D>2$}

For $D>2$ it is necessary to retain the ultaviolet cut-off $a$ (equivalent 
to a short-time cut-off $a^2$) in order to keep $C(t,t)$ finite.  The correct 
scaling limit here is to take $t_1$ and $t_2$ to infinity keeping 
$|t_1-t_2|$ fixed. In this limit, Eq.\ (\ref{manifoldautocorrelation}) 
reduces to a stationary correlator in the original time variable,
\begin{equation}
A(t_1,t_2) \sim (|t_1-t_2| + a^2)^{-(D-2)/2}\ .
\end{equation}
The calculation of the persistence of a Gaussian stationary process 
with an algebraically decaying correlator is nontrivial. A theorem of 
Newell and RosenBlatt \cite{Newell} (see section \ref{subsection:GSP}) states that if the stationary 
correlator decays as $A(t) \sim t^{-\alpha}$ for large time-difference 
$t = |t_1 -t_2|$, then the persistence $Q(t)$, i.e.\ the probability of 
the process having no zero crossings between $t_1$ and $t_2$, has the 
following asymptotic forms depending on the value of $\alpha$:
\begin{eqnarray}
&Q(t)& \sim \exp(-K_1 t) \;, \; \alpha > 1, \\
\exp(-K_2 t^\alpha \ln t) \le &Q(t)& \le \exp(-K_3 t^\alpha) \;, \; 0<\alpha<1 \;,
\end{eqnarray}
where the $K_i$'s are positive constants. Applying this result to our manifold 
persistence problem, we find
\begin{eqnarray}
Q(t) \sim \exp(-K_1 t), && D>4\ , \\
\exp[-K_2 t^{(D-2)/2} \ln t] \le Q(t) \le \exp[-K_3t^{(D-2)/2}], && 2<D<4,
\end{eqnarray}
with an additional logarithmic correction for $D=4$.

Combining the exact results obtained within the mean-field theory for 
$D<2$ and $D>2$, and using the explicit results $z=2$ and $\eta=0$ valid 
within mean-field theory, we see that the results obtained above are 
special cases of the general result given in Eq.\ (\ref{manifolds}), 
when one uses the mean-field values $\zeta=(D-2)/2$ in Eq.\ (\ref{manifolds}).
  
\subsection{The large-$n$ Limit}
The mean-field theory is valid for $d>4$. For $d<4$, we can still obtain 
exact results if we work in the large-$n$ limit of the $O(n)$ model. 
In this limit, Eq.\ (\ref{LangevinManifolds}) reduces to the simpler equation
\begin{equation} 
\partial_t \phi_i = \nabla^2\phi_i - [r + S(t)]\phi_i + \eta_i\ ,
\label{manifolds-large-n}
\end{equation} 
where $S(t) = u\langle \phi_i^2 \rangle$ has to be determined 
self-consistently. The critical point is determined by $r + S(\infty) = 0$. 
Using standard methods (see, for example, Ref.\ 
\cite{MajumdarBrayCornellSire96}) one can show that 
the long-time behaviour of $S(t)$ has the form 
$S(t) \to S(\infty) - (4-d)/4t$ for $2<d \le 4$. Substituting this 
result into Eq.\ (\ref{manifolds-large-n}), summing over the $d'$ directions 
solving the resulting equation in Fourier space and finally integrating over  
the Fourier space, as in the mean-field theory, we arrive at the following 
autocorrelation function for the manifold magnetisation $\psi(\vec{r},t)$ 
for dimensions $d$ in the range $2<d\le4$:
\begin{equation}
C(t_1,t_2) = A_1(t_1t_2)^{(4-d)/4}\int_0^{t_1}\frac{t'^{(4-d)/2}\,dt'}
{(t_1+t_2-2t'+a^2)^{D/2}},
\label{large-n-manifolds}
\end{equation}
where we have taken $t_1 \le t_2$ without loss of generality.  
In Eq.\ (\ref{large-n-manifolds}), $A_1$ is a constant and $a$ represents 
the soft ultraviolet cut-off as before. 

\subsubsection{The case $D<2$}
For $D<2$ one can set the cut-off $a$ to zero, as in the mean-field theory, 
and the nonstationary correlator becomes a stationary correlator for the 
normalised process $X = \psi/\sqrt{\langle \psi^2 \rangle}$ in logarithmic 
time $T = \ln t$:
\begin{equation}
A(T) = [\cosh(T/2)]^{\mu - D/2}\ \frac{B[\mu,2\beta,2/(1+\exp(T)]}
{B[\mu,2\beta]}\ ,
\end{equation}
where 
\begin{eqnarray}
&&\mu = (d-2)/2\ , \,  \beta = (2-D)/4\ \\
&&B[x;m,n] =  \int_0^x dy\,y^{m-1}(1-y)^{n-1} \ ,
\end{eqnarray}
and $B[x;m,n]$ is the incomplete Beta function. Since the stationary correlator $A(T)$ decays exponentially for large $T$, 
it follows \cite{Slepian} that the persistence also decays 
exponentially for large $T$, $Q(T) \sim \exp(-\theta T)$, and therefore 
as a power law, $Q(t) \sim t^{-\theta}$, in the original time variable 
$t= \exp(T)$. Calculating the exponent $\theta$ explicitly is challenging. One can make 
progress, however, in the limit where the co-dimension $D$ is small. 
For $D=0$, which corresponds to the calculation of the global persistence, 
the autocorrelation function becomes a pure exponential, 
$A(T) = \exp[-(d-2)T/4]$, corresponding to a Markovian process with 
persistence exponent $\theta_0 = (d-2)/4$. This is just equation 
(\ref{theta_Glarge-n}) of the preceding section on global persistence (section \ref{Global}). 
For small $D$ one can use the perturbation theory developed in section \ref{section:perturbation} 
to compute $\theta$ to first order in $D$ \cite{MajumdarBrayManifolds}. 
The result has the form 
\begin{equation}
\theta = \theta_0 + D\theta_0^2I_d/\pi + {\cal O}(D^2)\ ,
\end{equation}
for $2<d\le 4$, where $I_d$ is given by a complicated integral which simplifies 
for special values of $D$, e.g.\ $\theta = 1/2 + (2\sqrt{2}-1)D/4$ for 
$d=4$ and $\theta = 1/4 + 0.183615\ldots D + {\cal O}(D^2)$ for $d=3$.  

\subsubsection{The case $D>2$}
For $D>2$ it is clear that one must retain the cut-off $a$ in order that 
$A(t_1,t_1)$ be finite. The dominant contribution to the integral 
(\ref{large-n-manifolds}), for large $t_1$, comes from the region where 
$t'$ is close to $t_1$. When $t_1$ and $t_2$ are both large with their 
difference fixed, the autocorrelation function, (\ref{large-n-manifolds}), 
becomes a stationary one, $A(t_1,t_2) \approx B_1(|t_1-t_2|+a^2)^{-(D-2)/2}$, 
where $B_1$ is a constant whose value is not important.  
The Newell-Rosenblatt theorem \cite{Newell} then implies that the persistence 
$Q(t)$ decays as $Q(t) \sim \exp(-\kappa_1t)$ for $D>4$, where $\kappa_1$ 
is a constant, while it satisfies the bounds 
\begin{equation}
\exp[-\kappa_2t^{(D-2)/2}\ln t] \le Q(t) \le \exp[-\kappa_3t^{(D-2)/2}]
\end{equation}
for $2<D<4$, where $\kappa_2$ and $\kappa_3$ are constants.  

\subsection{General scaling theory}

Guided by the soluble cases discussed above, it is possible to 
construct \cite{MajumdarBrayManifolds} a general scaling theory 
valid for all $d \ge 2$. At the critical point, the two-point 
order-parameter correlation function has the generic space-time scaling form 
\begin{equation}
\langle \phi(0,t_1)\,\phi({\bf x},t_2)\rangle \sim x^{-(d-2+\eta)}\,,
F(xt_1^{-1/z}, t_2/t_1)
\end{equation} 
for large spatial separation $x = |{\bf x}|$ and large times $t_1$, $t_2$, where 
$\eta$ and $z$ are the standard critical exponents. So in Fourier space 
we have 
\begin{equation}
\langle \tilde{\phi}({\bf K},t_1) \tilde{\phi}({\bf K},t_2)\rangle 
\sim K^{-(2-\eta)}\,G(Kt_1^{1/z}, t_2/t_1), 
\end{equation}
where ${\bf K}$ is a $d$-dimensional vector conjugate to ${\bf x}$, 
and $K = |{\bf K}|$. The manifold magnetisation $\psi$ is obtained by 
summing the order parameter $\phi$ over $d'$ directions. The scaling 
behaviour of the two-point correlator of the manifold magnetisation 
is then obtained as
\begin{equation}
\langle \tilde{\psi}(\vec{k},t_1)\tilde{\psi}(-\vec{k},t_2) \rangle 
\sim k^{-(2-\eta)}g(kt_1^{1/z},t_2/t_1)\ ,
\end{equation}
where $\vec{k}$ is a $D=d-d'$ dimensional vector. One can see, for example, 
that the mean-field theory expression, Eq.\ (\ref{manifoldeq}), is (at late 
times where the term in $\Delta$ can be neglected) precisely of 
this form. The aurocorrelation function, $C(t_1,t_2) = 
\langle\psi({\bf r},t_1)\psi({\bf r},t_2)\rangle$,
is obtained by integrating over ${\vec{k}}$: 
\begin{equation}
C(t_1,t_2) = \int d^D k^{-(2-\eta)}\,g(kt_1^{1/z},t_2/t_1)\,\exp(-k^2a^2),
\label{manifoldauto}
\end{equation}
where $a$ is the usual soft ultraviolet cut-off.

Clearly the form of the correlation function (\ref{manifoldauto}) depends 
on the sign of $D-2+\eta$. When $D-2+\eta<0$, we can set $a=0$ in  
(\ref{manifoldauto}), since the integral converges at the upper limit. 
In the limit $t_1,t_2 \to \infty$, with $t_2/t_1$ arbitrary, this gives  
an autocorrelation function of the form $C(t_1,t_2) = 
t_1^{-(D-2+\eta)/z}\,f(t_2/t_1)$. The function $f(x)$ behaves as 
$f(x) \sim x^{-\lambda_c/z}$ for large $x$, so that 
$C(t_1,t_2) \sim t_2^{-\lambda_c/z}$ for $t_2 \gg t_1$, where $\lambda_c$
is the critical autocorrelation exponent \cite{BrayReview,JanssenSchaubSchmittmann,
Huse}. The non-stationary Gaussian correlator can be transformed, as usual, 
to a stationary one for the normalised variable 
$X=\psi/\sqrt{\langle\psi^2\rangle}$ in the logarithmic time $T= \ln t$. One thus gets
$A(T) = \langle X(T) X(0)\rangle = \exp{\left[ (D-2+\eta T/2z)\right] f(e^T)/f(1)}$ which decays
exponentially for large $T$, $A(T) \propto  \exp{\left[-(\lambda_c- (D-2+\eta)/2) T/z)\right]}$. It thus follows
that $Q(T) \sim \exp{(-\theta T)}$ which yields a power law decay $Q(t) \sim t^{-\theta}$ in the original time variable $t=e^T$.

In the complementary case $D - 2 + \eta > 0$, the integral
in Eq. (\ref{manifoldauto}) is, for $t_1 = t_2$, divergent near the upper
limit without the cut-off (i.e. if we set $a=0$). Hence one needs to keep
a nonzero $a$ and then the appropriate scaling limit is obtained
by taking $t_1, t_2$ both large keeping their difference
$|t_1 - t_2|$ fixed but arbitrary (that is a quasi-equilibrium regime). In this regime, one can replace the scaling
function $g(kt_1^{1/z},t_2/t_1)$ in Eq. (\ref{manifoldauto}) by an other scaling function $g_1(k|t_1-t_2|^{1/z})$. 
Performing the remaining integral over $k$, one then finds $C(t_1,t_2) \sim |t_1-t_2|^{-(D-2+\eta)/z}$, for $|t_1-t_2| \gg a$. 
This correlator is stationary and decays as a power law.
Invoking the Newell-Rosenblatt theorem once more, we
find that $Q(t)$ decays exponentially for $(D-2+\eta)/z > 1$
and as a stretched exponential for $0 < (D-2+\eta)/z < 1$.
Combining this with the result for $D-2+\eta < 0$ outlined in the previous section, one obtains the results in Eq. (\ref{manifolds}).  

One may wonder what would be the effects of a non-vanishing initial magnetization $M_0$ (see Fig. \ref{Fig_slip}), as discussed before in  
section~\ref{sec:initial_mag} on the global persistence, on these results for the persistence of manifold in (\ref{manifolds}). This question has been addressed in Ref. \cite{PGS10} where it was shown that a finite $M_0>0$ only affects the algebraic decay of $Q(t)$ when $D - 2 + \eta < 0$ -- the other regime $D - 2 + \eta > 0$ remaining unaffected by a finite $m_0$. In the regime $D - 2 + \eta < 0$  one can show \cite{PGS10}, using perturbation theory
as well as numerical simulations, that $Q(t)$ exhibits a crossover between two distinct power law regimes, similar to Eq. (\ref{crossover_dyn}) for the global magnetization.

\section{Persistence of fractional Brownian motion and related processes}\label{section:fBm}

Fractional Brownian motion (fBm) is an important generalization of the 
ordinary Brownian motion, first introduced by Mandelbrot and van Ness~\cite{Mandelbrot68}.
The fBm is a Gaussian process $x(t)$ with zero mean and a correlator
that has a special form
\begin{equation}
a(t_1,t_2)= \langle x(t_1)x(t_2)\rangle= K\, 
\left[t_1^{2H}+t_2^{2H}-|t_1-t_2|^{2H}\right] \;,
\label{fbm1}
\end{equation}
where $K$ is a constant and $0\le H\le 1$ is called the Hurst exponent
of the process that parametrizes it. For $H=1/2$, this correlator
reduces to that of the standard Brownian motion, $a(t_1,t_2)= 2K {\rm 
min}(t_1,t_2)$. It turns out that the process is Markovian only for $H=1/2$, 
but all other $H\ne 1/2$, it is non-Markovian. One key feature of this
special form of the correlator is that it has {\it stationary increments}.
By this, one means that the incremental correlation function
\begin{equation}
\sigma^2(t_1,t_2)= \langle \left[x(t_1)-x(t_2)\right]^2\rangle = 2K 
|t_1-t_2|^{2H} \;,
\label{increment1}
\end{equation}
is stationary, i.e., it depends only on the time difference $\tau=|t_1-t_2|$.
In addition, it increases as a power law, $\tau^{2H}$ with the
time difference $\tau$. These two properties, i.e., stationary increments
and the power-law growth, will play a crucial role later.

The fBM appears in many physics problems (some of them are discussed 
in this review) and its persistence and first-passage properties have been 
studied both in the mathematics~\cite{Berman70,Molchan99,Aurzada11} and physics literature~\cite{HEM94,MPB94,DW95,KKMCBS97}. 
For the special Markovian case $H=1/2$, we have seen that the persistence
$Q(t)$, i.e., the probability that the process does not change sign over the
time interval $[0,t]$ decays, for large $t$, as a power law $Q(t)\sim 
t^{-1/2}$. To anticipate what happens for $H\ne 1/2$, it is useful
again to map this process to a GSP using the canonical Lamperti transformation, $T=\ln t$ and $X(T)= x(t)/\sqrt{\langle x^2(t)\rangle}$.
It is easy to see that in variable $T$, $X(T)$ becomes a GSP with a
stationary correlator~\cite{KKMCBS97}
\begin{equation}
A(T)= \cosh(H\,T)-2^{2H-1} |\sinh(T/2)|^{2H}.
\label{fbm2}
\end{equation}
For example, for $H=1/2$, we have the Markov correlator $A(T)=\exp[-|T|/2]$.
For any $H\ne 1/2$, the correlator is different from a pure exponential
and hence, by Doob's theorem discussed in section \ref{subsection:GSP}, the process $X(T)$
is manifestly non-Markovian. However, for arbitrary $0\le H\le 1$, it is 
clear from Eq. (\ref{fbm2}) that $A(T)\sim \exp[-\lambda T]$ for large $T$
with $\lambda= {\min}[H, 1-H]$. Now, from the general result of 
Newell-Rosenblatt in section \ref{subsection:GSP}, we would then expect that the persistence 
$Q(T)$ of this process will decay exponentially for large $T$, $Q(T)\sim 
\exp[-\theta(H) T]$ where the decay constant will depend on $H$. Translating 
back to the real time $t=e^T$, it then follows that the persistence $Q(t)$
should decay at late times as a power law, $Q(t) \sim t^{-\theta(H)}$ with
a persistence exponent $\theta(H)$ that depends on $H$. Clearly, 
$\theta(H=1/2)=1/2$. The question is: can one compute $\theta(H)$ for
general $0\le H\le 1$?

From the general discussion on the persistence of GSP in section \ref{subsection:GSP}, it would 
seem that for a nontrivial correlator as in Eq. (\ref{fbm2}), it is
very hard to compute 
$Q(T)$ and even its asymptotic exponential tail $Q(T)\sim 
\exp[-\theta(H) T]$. 
However, it turns out that for the fBm,  
there
exists a rather general scaling argument~\cite{HEM94,MPB94,DW95,KKMCBS97}
that makes use of two crucial properties, namely the stationary
increments and the power-law growth of the incremental correlation function 
in Eq. (\ref{increment1}), and predicts a very simple
scaling relation
\begin{equation}
\theta(H)= 1-H \;,
\label{fbm3}
\end{equation}
which correctly reproduces $\theta(1/2)=1/2$. Although the scaling
argument giving the result in Eq. (\ref{fbm3}), which is presented below, is not completely rigorous, this relation (\ref{fbm3}) was actually shown rigorously by Molchan  \cite{Molchan99}, using a completely
different approach, who proved that $\log Q(T) = -(1 - H)\log{(T+ o(1))}$, as $T \to \infty$. This estimate on $Q(T)$ for large $T$ was then improved by Aurzada in \cite{Aurzada11}. On the other hand this relation (\ref{fbm3}) has been
numerically verified by several independent groups~\cite{DW95,KKMCBS97}. In addition, even
experimental results reported by the Maryland group~\cite{Dougherty_exp02,Dougherty_exp03} on fluctuating 
crystal steps are consistent with this scaling relation (see also section \ref{section:interfaces}). 

Before we reproduce the scaling argument leading to the relation in Eq. (\ref{fbm3}) below, we note that although
this relation is derived for fBM which is a Gaussian process, the scaling 
argument is more general and seems to be valid even for non-Gaussian
process, the only requirement being the property of stationary increments
and the power-law growth of the incremental correlation function (see e.g. \cite{MPB94,KKMCBS97}).
In addition, the relation $\theta(H)=1-H$ which is valid for symmetric
processes (where the probability distribution of $x$ has the $x\to -x$ 
symmetry) can be generalized to non-symmetric processes~\cite{CDCMD04}.
In fact, below we present the more general result valid for 
non-symmetric processes and recover the result $\theta(H)=1-H$
as a special case when the $x\to -x$ symmetry is restored.

Consider a general stochastic process, {\em not necessarily symmetric} and {\em not 
necessarily Gaussian}, whose incremental correlation function
is stationary and has a power-law form for large time difference 
$\tau=|t_1-t_2|$
\begin{equation}
\sigma^2(\tau)= \langle \left[x(t_1)-x(t_2)\right]^2\rangle \sim \tau^{2H}.
\label{increment2}
\end{equation}
Consider a particular realization of this process with several
crossings at the origin. There are two types of intervals between 
successive zero crossings in time, the `$+$' type (where the process
is above $0$) and `$-$' type (where the process is below $0$).
In absence of the $x\to -x$ symmetry, these two types of intervals
have different statistics and one would expect that the size distribution
of these intervals, denoted by $P_{\pm}(\tau)$ will behave differently.
Since the interval size between zero crossing is simply the derivative
of the corresponding persistence probability (to persist respectively above
or below $0$), we expect a
power-law decay for large $\tau$, $P_{\pm}(\tau)\sim 
\tau^{-1-\theta_{\pm}}$ where the `upper' and `lower' persistence exponents
$\theta_{\pm}$ are in general different. 

Let $P(x,\tau)$ denote the probability density that the process is at 
value $x$ at time $\tau$, given that it starts from its initial value $0$
at time $\tau=0$. The typical width of the process $\sigma(\tau)$
grows with time $\tau$ as a power law, $\sigma(\tau)\sim \tau^{H}$, as 
follows from 
Eq. (\ref{increment2}). 
Then, it is natural to assume that the normalized 
probability density at time $\tau$ has a scaling form for large $\tau$
\begin{equation}
P(x,\tau)\sim \frac{1}{\sigma(\tau)}\, f\left(\frac{x}{\sigma(\tau)}\right) \;,
\label{fbm4}
\end{equation}
where the scaling function $f(z)$ is a constant at $z=0$, $f(0)\sim {\cal O}(1)$.
In general, $f(z)$ need not be a symmetric function of $z$ and should 
decrease to $0$ rapidly as $z\to \pm \infty$. 
Given that a zero occurs initially, the probability density 
$\rho(\tau)=P(0,\tau)$ that the process returns to $0$ after time $\tau$ (not 
necessarily for the first time) scales as [putting $x=0$ in Eq. (\ref{fbm4})]
\begin{equation}
\rho(\tau)\sim \frac{1}{\sigma(\tau)}\sim \tau^{-H} \;,
\label{fbm5}
\end{equation}
as $\tau\to \infty$. The function $\rho(\tau)$ is thus the density of zero 
crossings at time $\tau$ and hence the total number of zeros up to some large 
time $t$ is simply the integral
\begin{equation}
N(t) = \int_0^t \rho(\tau)\,d\tau \sim t^{1-H}.
\label{fbm6}
\end{equation}

Note that the total number of intervals in $[0,t]$ is also $N(t)$, half of 
which 
are `$+$' type and the other half `$-$' type (they alternate), i.e.,
$N_{\pm}(t)= N(t)/2$. Let $n_{\pm}(\tau,t)$ denote the $\pm$ intervals of 
length $\tau$ within the period $[0,t]$. Thus, the fraction of $+$ (or $-$) 
intervals of length $\tau$, $n_+(\tau,t)/N_+(t)$ and 
$n_{-}(\tau,t)/N_{-}(t)$ are precisely the interval
size distributions, $P_{+}(\tau)$ and $P_{-}(\tau)$ defined earlier.
Thus for large $\tau$ and $t$, we have
\begin{equation}
n_{\pm}(\tau,t)= \frac{N(t)}{2}\, Q_{\pm}(\tau)\sim 
N(t)\,\tau^{-1-\theta_{\pm}} \;,
\label{fbm7}
\end{equation}
valid for $1\ll \tau\le t$. On the other hand, we have a sum rule coming from 
the fact that the total length covered by the intervals must be $t$
\begin{equation}
\int_0^t d\tau\, \tau\, \left[n_+(\tau,t)+n_{-}(\tau,t)\right] = t\ .
\label{fbm8}
\end{equation}
Substituting the asymptotic form (\ref{fbm7}) in the sum rule we get
\begin{equation}
N(t)\left[\frac{t^{1-\theta_+}}{1-\theta_+}+ 
\frac{t^{1-\theta_{-}}}{1-\theta_{-}}\right] \propto t\ .
\label{fbm9}
\end{equation}
Next we substitute the result $N(t)\sim t^{1-H}$ for large $t$ from Eq. 
(\ref{fbm6}) giving
\begin{equation}
\left[\frac{t^{1-\theta_+}}{1-\theta_+}+
\frac{t^{1-\theta_{-}}}{1-\theta_{-}}\right] \sim t^{H}\ .
\label{fbm10}
\end{equation}
Finally, taking the limit $t\to \infty$ and matching the power of $t$ on
both sides yields the desired scaling relation~\cite{CDCMD04} 
\begin{equation}
{\rm min} (\theta_+, \theta_{-})= 1-H\ .
\label{fbm11}
\end{equation}
If the process has the $x\to -x$ symmetry, one reproduces
$\theta_+=\theta_{-}=\theta(H)=1-H$ as in Eq. (\ref{fbm3}).
The relation in (\ref{fbm11}) is, however, more general
and has been verified numerically~\cite{CDCMD04} for a class of nonlinear interfaces
in both $(1+1)$ and $(2+1)$ dimensional interfaces which are
in general {\em non Gaussian}, asymmetric but satisfy the two basic properties
(i) stationary increments and (ii) power-law growth of incremental
correlation function. This relation in (\ref{fbm11}) has also been 
used in the analysis of financial data where the time series of stock prices
can be modelled by a fBM~\cite{constantin_finance2005,CCT2009}.
Note that in the above derivation we have implicitly 
assumed
a small time cut-off and focused only on the distribution of large intervals.
Ignoring the short-time behavior of the intervals (in particular
infinite zero crossings of the process) does not seem to affect
this scaling relation. A more rigorous derivation would take into account
these short-time anomalies properly.

Finally, we point out that the first-passage properties of fBm have seen a recent revival in the physics literature~\cite{KK2007,ZRK2007,CKK2008,
ZRM2009,MRZ2010,GRS2010,
AKK2010,SGCSM2010,WMR2011,Sanders2012}, in particular in 
the context of the translocation process of a polymer chain through a nanopore.
The translocation co-ordinate $x(t)$ measuring the number of mononers that are
on one side of the pore at time $t$ seems to be well described by a fBm~\cite{KK2007,ZRM2009}
and the behavior of this anomalous diffusion process, in presence of one or two
absorbing walls (reflecting the finiteness of the size of the polymer chain)
have been studied in detail~\cite{KK2007,ZRK2007,ZRM2009,MRZ2010,WMR2011}. Note also that the survival probability
of a 2-d fBm in a wedge (see Fig. \ref{Fig_wedge}) was studied, mainly numerically, in Ref. \cite{JCM2011}. 

We finally mention a rigorous study \cite{Aurzada_moving} of the persistence for fBm in the presence of a logarithmically moving boundary (see Fig. \ref{fig:expanding-cage} but for fBm), where the authors proved, as expected from scaling argument, that the persistence exponent (\ref{fbm3}) remains unchanged in this case. This result turns out to be relevant for the study of current fluctuations in Sinai type disordered chain, where the disordered potential is itself a fBm trajectory \cite{ORS13}.

\section{Persistence of fluctuating interfaces}\label{section:interfaces}

Stochastic dynamics of fluctuating interfaces have been of extensive interest, both theoretically and experimentally,
over the last four decades~\cite{BS95,HZ95,Krug97}. Such interfaces appear in a variety of growth models, 
such as in tumour growth in biological 
context or in crystals, where they describe the crystal layer boundaries or steps on a vicinal surface of a
crystal, as shown in Fig. \ref{fig:fluctuating} (for a review see~\cite{Krug97}). At a theoretical level, dynamics of such interfaces can be modelled either by discrete atomistic 
growth models
or its coarse grained version by stochastic growth 
equations~\cite{Fam86,WV90,DT91a,DT91b,KK89a,KK89b,KD94a,KDG4b,DGK94,KPK94,MH57a,MH57b,LD91,Vill91}. 
Such noisy growth equations can be linear or nonlinear
depending on the details of the microscopic processes that govern the dynamics. Despite such a wide variation
in dynamics, the long time and large distance properties of such growth models exhibit scale invariance
and universality~\cite{HZ95,Krug97}. Scale invariance is manifest in the power-law behavior as a function of both space and time
of several quantities of physical interest, such as the width of the interface~\cite{HZ95,Krug97,Krug04}. 
If the width of the interface
grows with time, such interfaces are called rough. Otherwise they are smooth. In general, the two-point
height-height correlation function (both in space and time) of rough interfaces exhibit dynamic scaling behavior,
characterized by certain exponents and scaling functions~\cite{FV85,F90}. A lot of studies focused on the classification of
discrete growth models and Langevin equations into different universality classes characterized by these
exponents and the associated scaling functions. A large number of important experimental studies (such as
molecular beam epitaxy growth) have confirmed this dynamic scaling behavior~\cite{Krug97}. 

While the studies of the two-point correlation function and the associated dynamic scaling have provided very important 
insights, the two-point function does not capture the complex history dependence of the temporal evolution of such 
extended objects like interfaces. The simplest and perhaps the most natural probe of the history dependence in such 
spatially extended systems is provided by the studies of the persistence and first-passage properties of such interfaces,
initiated by Krug and coworkers~\cite{KKMCBS97}.  
Roughly speaking, the persistence of the interface height is the probability that the height does not return to its 
initial value up to time $t$. Persistence properties of growing interfaces have been studied extensively, theoretically 
as well as experimentally, over the last few years. Using the scanning tunneling microscopy (STM) technique, one 
can image not just the spatial structures of rough interfaces but also their temporal evolution (Fig. \ref{fig:fluctuating}). As a result, these 
systems constitute a beautiful example where many of the theoretical ideas regarding persistence and first-passage 
properties can be tested experimentally~\cite{Dougherty_exp02,Dougherty_exp03,CCDLW07}. 
For a nice review of the theoretical and experimental results
of persistence and first-passage properties of interfaces, particularly in connection to step edges on crystals, see
Ref.~\cite{CDDDW07}. Apart from step edges on crystals, persistence of fluctuating interfaces have also been measured in
a variety of other
experimental systems, such as in combustion fronts in paper~\cite{MMMTA03}, for interfaces
between phase-separated coloid-polymer mixtures~\cite{van_leeuwen09}, advancing interfaces or fronts in
reactive-wetting systems such as mercury on silver~\cite{Efraim11} and growing droplets of turbulent phase
in nematic liquid crystals~\cite{TS12}.   

Before we define the persistence of such interfaces more precisely, it is useful to briefly review the
different variety of Langevin growth equations for the interface height fluctuations and the associated 
dynamic scaling of the height-height correlation function.
\begin{figure}
\centering
\includegraphics[width=0.4\linewidth]{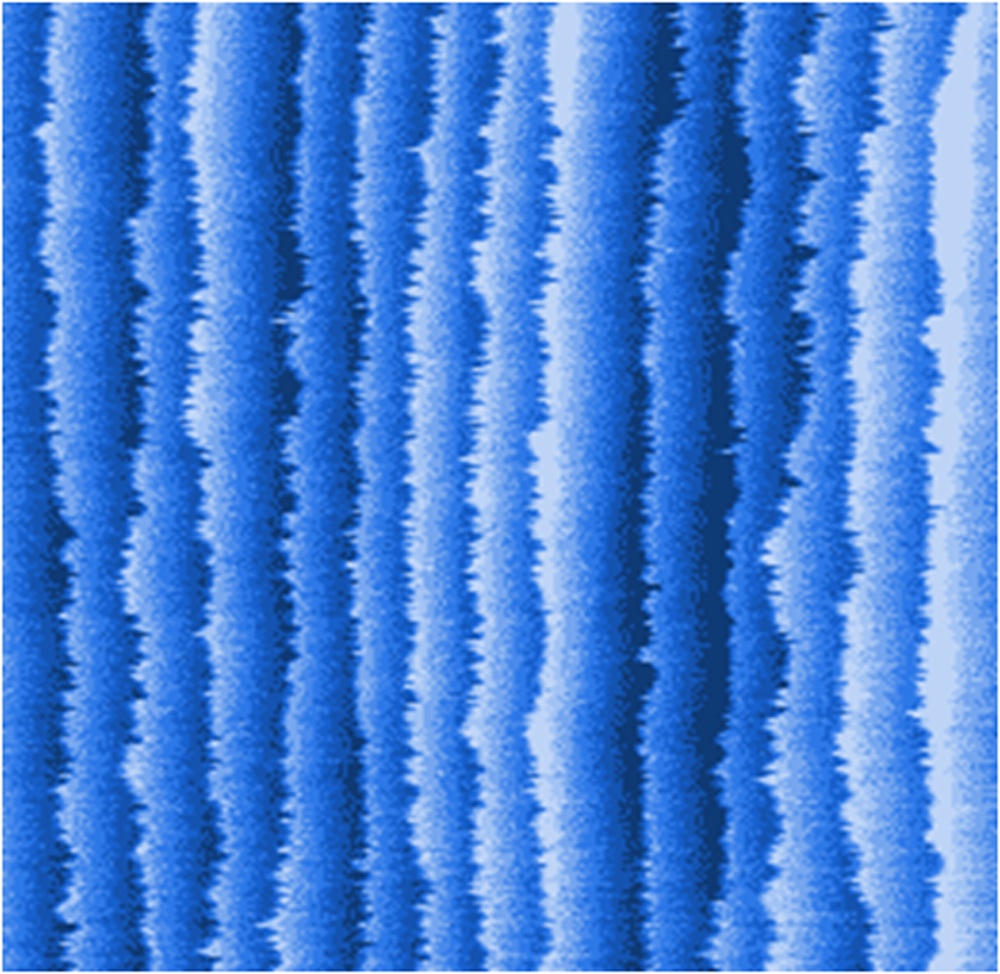}\includegraphics[width=0.5\linewidth]{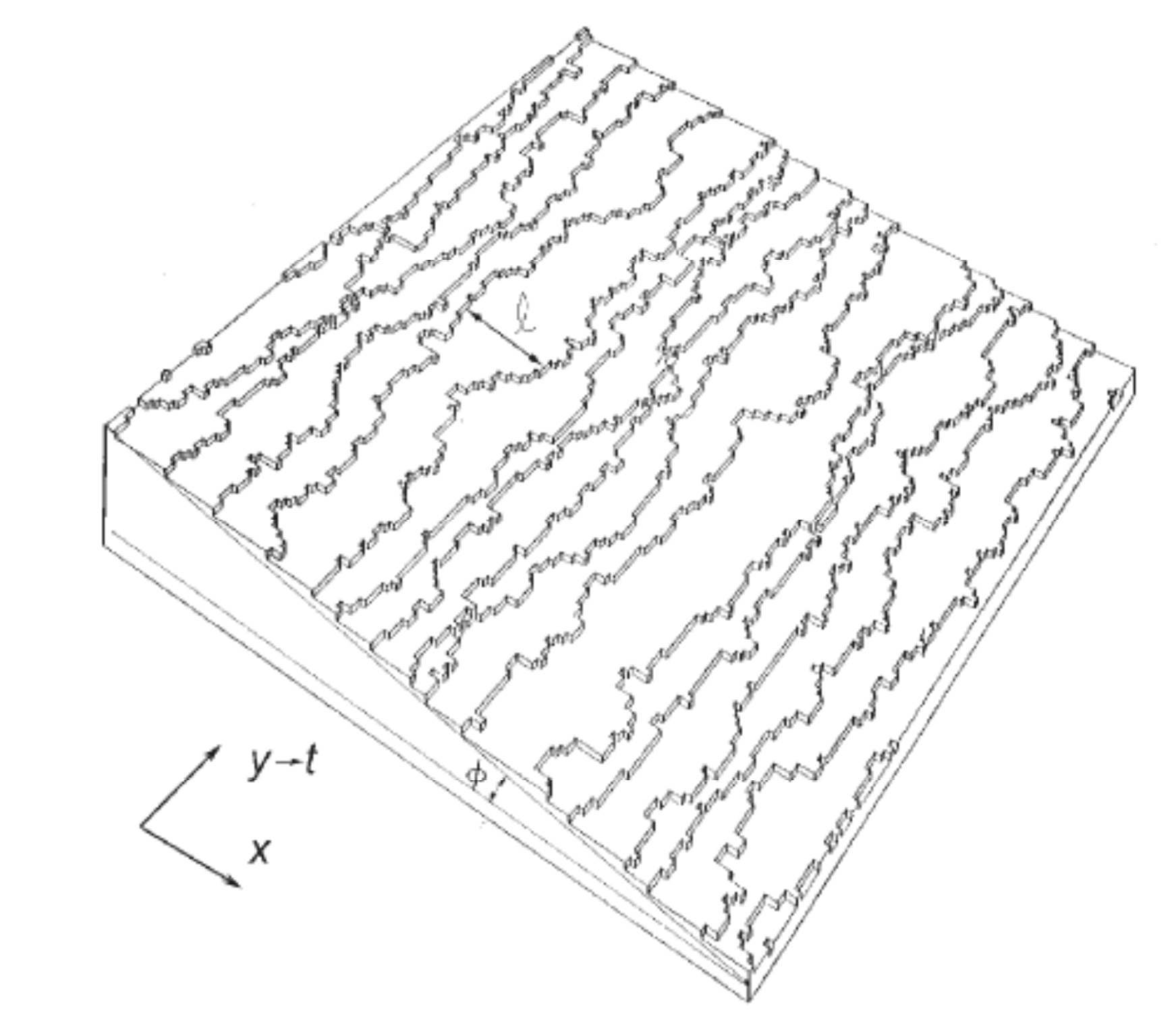}
\caption{{\bf Left:} Experimental STM images of a vicinal surface, here from Cu(111) from Giesen's
group \cite{Giesen2002}. The displayed surface area is $130 \times 130 nm^2$ and the surface height decreases
from left to right. Due to atomic motion at the step edges, the steps do not remain
straight and immobile, but undergo thermal fluctuations around their equilibrium
position. {\bf Right}: Illustration of a vicinal surface obtained in a Monte-Carlo simulation~\cite{Ein1991,Ein2003,Ein2007}. These two figures have been inserted here with the courtesy of M.~Giesen (left panel) and T. L. Einstein (right panel).}\label{fig:fluctuating}
\end{figure}

{\bf Langevin growth equations and dynamic scaling: } A fluctuating interface is characterized by its
height $H(\vec r,t)$ which is a time-dependent single-valued scalar field defined at
each point $\vec r$ of a $d$-dimensional substrate of linear extent $L$. The interfacial width
that characterizes the size of the fluctuations at time $t$ is a function the system size $L$
and time $t$
\begin{equation}
W(L,t)= \left[\langle [H(\vec r, t)- {\bar H}(t)]^2 \rangle\right]^{1/2} \;,
\label{ds1}
\end{equation} 
where ${\bar H}(t)= (1/V) \int H(\vec r,t) d\vec r$ is the spatially averaged height with $V$ being the volume of the
substrate. The width $W(L,t)$ exhibits generically the following scaling behavior
\begin{eqnarray}
W(L,t) \sim 
\begin{cases}
& t^{\beta}\quad\quad {\rm for} \quad 0 \ll t \ll L^z \;, \label{ds_grow.1} \\
& L^{\alpha} \quad\quad {\rm for} \quad t \gg L^z \;,
\end{cases}
\end{eqnarray}
where the three exponents $\alpha$ (roughness exponent), $\beta$ (growth exponent) and $z$ (dynamical exponent)
characterize the universality class of the interface. The regime $t \ll L^{z}$ is called the growing regime where
the width grows (for $\beta>0$), while
$t \gg L^z$ is the steady state regime where the fluctuations become time independent. The two regimes are connected via 
the Family-Vicsek~\cite{FV85} scaling function: $W(L,t) \sim L^\alpha f(t/L^z)$ where $f(x)\sim x^{\beta}$ as $x\to 0$
and $f(x)\to {\rm const.}$ as $x\to \infty$. In order to match the two behaviors in Eq. (\ref{ds_grow.1}) 
requires an additional scaling relation between the three exponents: $\alpha = \beta z$.

The temporal evolution
of the field $H(\vec r,t)$ is usually modelled by a noisy Langevin equation. Depending on the microscopic processes
involved in the growth mechanism, these equations can have different forms. Several such Langevin equations
have been proposed and studied in the literature (see Ref.~\cite{CDCMD04} for an extended review). These equations
can generally be divided into two classes, namely,  linear and nonlinear. 
\vskip 0.3cm

\noindent {\bf Linear interfaces:} Here the height $H(\vec r,t)$ evolves via the linear Langevin equation
\begin{equation}
\frac{\partial H (\vec r,t)}{\partial t}= -(-\nabla^2)^{z/2} H(\vec r,t) + \eta(\vec r,t) \;,
\label{interface.1}
\end{equation}
where the dynamical exponent $z$ (usually $z=2$ or $4$) characterizes the relaxation mechanism of the interface and $\eta(\vec r,t)$
is a Gaussian noise with zero mean. The correlator of this noise depends on whether the noise conserves the total height or not.
It is easier to specify the two-point correlator of the noise in the Fourier space. Defining ${\tilde \eta}(\vec k, t)$ as the
Fourier transform of $\eta(\vec r,t)$, we consider, in general, the correlator of the noise of the form:
$\langle {\tilde \eta}(\vec k, t){\tilde \eta}(\vec k', t')\rangle= D k^{\gamma} \delta(\vec k+\vec k')\delta(t-t')$
where $k= |\vec k|$ and 
the exponent $\gamma\ge 0$ characterizes the conservation property of the noise.
The case $\gamma=0$ corresponds to nonconserving noise. Such linear Langevin equations have been used
widely in the literature to model the stochastic dynamics of interfaces in a wide variety of situations. 
One can consider several special cases of this Langevin dynamics.

\begin{itemize}

\item
The case $z=2$ and $\gamma=0$: this corresponds to the celebrated Edwards-Wilkinson equation~\cite{EW82}
which describes, for instance, the fluctuation of a step edge on a crystal at very high temperature where the dynamics
is governed by random attachment and detachment of atoms at the step edge~\cite{Krug97,CDCMD04}. In this case the noise 
is nonconserving ($\gamma=0$)
as the total number of atoms in a crystal layer that terminates at the step edge is not a constant. 

\item
The case $z=4$ and $\gamma=2$: For step edges on a crystal, at low temperature, the dominant mechanism of 
fluctuations is the step-edge diffusion (SED)~\cite{Giesen01} where
the noise is conserving ($\gamma=2$).

\item
The case $z=4$ and $\gamma=0$: This corresponds to the Mullins-Herring equation for surface growth~\cite{MH57a,MH57b}.

\end{itemize}

\vskip 0.3cm

\noindent {\bf Nonlinear Interfaces:} Among nonlinear interfaces, the two most well known examples are as follows.

\begin{itemize}

\item{ Kardar-Parisi-Zhang (KPZ) interface:} This is a second order nonlinear equation~\cite{KPZ86} \begin{equation} 
\frac{\partial H(\vec r,t)}{\partial t}= \nabla^2 H(\vec r,t) + \lambda |\nabla H(\vec r,t)|^2 + \eta(\vec r,t)  \;,
\label{kpz.1} \end{equation} where $\eta(\vec r,t)$ is a Gaussian non-conserving white noise. In one dimension, the 
exponents are known exactly~\cite{KPZ86,Krug97}: $\alpha=1/2$, $\beta=1/3$ and $z=3/2$.

\item{ Molecular beam epitaxy (MBE) equation:} This is a $4$-th order nonlinear equation
\begin{equation}
\frac{\partial H(\vec r,t)}{\partial t}= -\nabla^4 H + \lambda \nabla^2 \left(|\nabla H|^2\right) +\eta(\vec r, t) \;,
\label{mbe.1}
\end{equation}
where $\eta(\vec r,t)$ is a non-conserving Gaussian white noise.
The exponents in this case are known only numerically in one and two dimensions~\cite{DCT02}. 
For example in $d=1$, $\alpha\approx 1$, 
$\beta\approx 1/3$ and $z\approx 3$~\cite{DP97,PD98,DCT02}.

\end{itemize}

Since the relevant measurable quantity is the deviation of the height from its spatially averaged value, i.e.,
$h(\vec r,t)=H(\vec r,t)-{\bar H}(t)$ (rather than the height $H$ itself), we will henceforth deal
with $h(\vec r, t)$ and with a slight abuse of language refer to the height deviation $h(\vec r,t)$ as 
the height itself.
Note that, by construction, $\int h(\vec r,t) d\vec r=0$ (i.e., the
$\vec k=0$ mode of the Fourier transform of $h(\vec r,t)$ is set to be identically zero). 

For the temporally evolving height field $h(\vec r,t)$ one can define several first-passage quantities of interest. 
Consider the height field $h(\vec r, t)$ at any fixed point in space, say at the origin $\vec r=0$. Due to the 
translational invariance of the system, the temporal properties of the height field do not depend on the choice of this 
point. Let us now monitor the height field $h(0,t)$ at the origin as a function of $t$. Suppose now we measure/observe 
this process after an initial waiting time $t_0$, during the subsequent time interval $t_0< s< t_0+t$. The {\em temporal 
persistence} $Q(t_0,t)$ is defined as the probability that the process $h(0,s)$ does not return to its initial value 
$h(0,t_0)$ during the interval $t_0<s<t_0+t$, i.e., 
\begin{equation} Q(t_0,t)= {\rm Prob.}\,\left[h(0,s)\ne h(0,t_0);\,\, 
{\rm for \,\, all}\,\, t_0<s<t_0+t\right]. 
\label{temporal.0} 
\end{equation} 
The persistence 
$Q(t_0,t)=Q_+(t_0,t)+Q_{-}(t_0,t)$ is the sum of {\em positive} and {\em negative} persistences defined as 
\begin{eqnarray} Q_+(t_0,t) &= & {\rm Prob.}\,\left[h(0,s)> h(0,t_0);\,\, {\rm for \,\, all}\,\, t_0<s<t_0+t\right]  \;,
\label{pers_pos} \\ 
Q_{-}(t_0,t) & =& {\rm Prob.}\,\left[h(0,s)< h(0,t_0);\,\, {\rm for \,\, all}\,\, t_0<s<t_0+t\right] 
\label{pers_neg}.  
\end{eqnarray} 
Clearly, if the probability distribution of the process $h(0,t)$ is invariant under the 
inversion $h\to -h$, the two probabilities would be identical, i.e., $Q_+(t_0,t)=Q_{-}(t_0,t)$. This happens, for 
example, for linear interfaces evolving via Eq. (\ref{interface.1}). However, for nonlinear interfaces, the evolution 
equation does not have the $h\to -h$ symmetry in general and as a result, the positive and negative persistences are 
generically different~\cite{KK99}.

The temporal persistence probabilities $Q_{\pm }(t_0,t)$, in general, depend on both $t_0$ and $t$. 
During the early stage of the growth process
starting from a flat interface when $0\le t_0 \ll L^z$ (called the {\em transient} or growing regime), the interface slowly 
develops roughness and
for times $t_0 \gg L^z$ the roughness becomes fully developed and saturates to a steady state value $\sim L^{\alpha}$ (this is
the {\em steady-state} regime). The large $t$ behavior of the temporal persistence $Q_{\pm}(t_0,t)$ accordingly
depends on whether $t_0$ is chosen from the transient regime or the steady-state regime. From a detailed
analytical and numerical study of temporal persistences for fluctuating interfaces, the following asymptotic
behaviors at large $t$, for a fixed waiting time $t_0$, have emerged~\cite{KKMCBS97}
\begin{eqnarray}
Q_{\pm}(t_0,t) \sim  
\begin{cases}
&t^{-\theta_0^{\pm}} \,\, {\rm for}\,\, 0\le t_0 \ll  L^z \,\,\, {\rm and}\,\, t_0 \ll t \ll L^z \label{trans_pers} \;, \\
& \\
&t^{-\theta_s^{\pm}} \,\, {\rm for}\,\, t_0 \gg L^z  \;,
\end{cases}
\end{eqnarray}
where $\theta_0^{\pm}$ and $\theta_s^{\pm}$ are respectively called the transient and steady-state persistence
exponents. For interfaces respecting the $h\to -h$ symmetry (such as linear interfaces), one gets 
$\theta_0=\theta_0^{+}=\theta_0^{-}$
and similarly, $\theta_s= \theta_s^{+}=\theta_s^{-}$. For nonlinear interfaces on the other hand, the four exponents 
$\theta_0^{+}$, $\theta_0^{-}$,
$\theta_s^{+}$ and $\theta_s^{-}$ are in general different~\cite{KK99}. The only exception is for the KPZ equation 
in Eq. (\ref{kpz.1})
in one dimension. In this case, while the $h\to -h$ symmetry is violated in the transient regime 
indicating $\theta_0^{+}\ne \theta_0^{-}$,
this symmetry is restored in the steady-state regime. This is manifest in the probability 
distribution of the steady-state height profile~\cite{Krug97}
\begin{equation}
{\rm Prob}\left[\{h(x)\}\right]\propto \exp\left[-\frac{1}{4D}\int (\partial_x h)^2\, dx\right] \;,
\label{steady_height1}
\end{equation}
where $\partial_x h= \partial h/{\partial x}$. This steady-state weight of the height profile corresponds 
to that of a Brownian motion in 
space (Wiener measure), i.e., $h(x)$ describes the position of a one dimensional Brownian motion at `time' $x$. 
The $h\to -h$ symmetry is manifest in Eq. (\ref{steady_height1}) indicating 
$\theta_s^{+}=\theta_s^{-}$ for $1$-d KPZ equation. This fact is however purely accidental and 
generically one does not have this
exponent equality.

The challenge then is to determine the persistence exponents, the transient pair $\theta_0^{\pm}$ as well as the steady-state
pair $\theta_s^{\pm}$ for the variety of interfaces described above. Analytical determination of these exponents
are very hard, though some progress can be made for linear interfaces where the height field is a Gaussian 
process~\cite{KKMCBS97}.
Even in this simple linear case, while the two-time correlation function of the height field is easy to compute, the
persistence probability $Q(t_0,t)$ and hence the exponents remain difficult to compute~\cite{KKMCBS97}. In the rest of the section 
we will first describe the linear interfaces
and finally mention some results on nonlinear interfaces.

Another interesting first-passage quantity is called the survival probability $S(t_0,t)$ defined as the probability that 
height field $h(0,t)$ does not return to its {\em average value}, namely, to $0$ in the time interval 
$[t_0,t_0+t]$~\cite{DCDM2004,CDDDW07}. This is different from the temporal persistence $Q(t_0,t)$ where 
one is concerned with the event of not 
returning to the {\em initial value}. While for $t_0=0$, the survival probability is identical to the transient 
persistence $S(0,t)=Q(0,t)$, the two quantities are different in the steady-state regime $S(\infty,t)\ne Q(\infty,t)$. 
The steady-state survival probability $S(\infty,t)$ exhibits rather different asymptotic time dependence~\cite{DCDM2004}, 
namely an 
exponential decay with a system size dependent time scale $S(\infty,t)\sim \exp[-t/\tau_s]$ with $\tau_s\sim L^z$, in 
contrast to the power-law decay of the steady-state persistence $Q(\infty,t)$. We will discuss the survival probability 
in detail for linear interfaces later in the section.
   
\subsection{Linear interfaces: two-time correlation function}

For linear interfaces evolving via Eq. (\ref{interface.1}), the height field $h(0,t)$ is a Gaussian process
in time which is completely characterized by its two-time correlation function. Due to the linear nature of the equation, 
the two-point correlation function can be computed explicitly as demonstrated below.

We start with Eq. (\ref{interface.1}) and consider the deviation $h(\vec r,t)= H(\vec r,t)- {\bar H}(t)$. Since
the Langevin equation is linear in $h$, it is convenient to 
consider the Fourier transform ${\tilde h}(\vec k, t)=\int d\vec r 
h(\vec r,t) e^{i\, {\vec k}{\mathbf .}{\vec r}}$. The Fourier modes get 
decoupled and the $k$-th mode (with $k=|\vec k|>0$) evolves as
\begin{equation}
\frac{\partial {\tilde h}}{\partial t} = -k^z {\tilde h} +{\tilde \eta}(\vec k,t).
\label{interface.2}
\end{equation}
Note that the $k=0$ mode of $h$ is identically zero, ${\tilde h}(k=0,t)=0$.
Assuming that we start at $t=0$ from a flat interface $h(\vec r,0)=0$, one can integrate Eq. (\ref{interface.2})
\begin{equation}
{\tilde h}(\vec k,t) = e^{-k^z t}\, \int_0^t e^{k^z t'} {\tilde \eta}(\vec k, t') dt'.
\label{interface.3}
\end{equation}
Since ${\tilde h}(\vec k,t)$ is a linear combination of the Gaussian fields ${\tilde \eta}(\vec k,t)$, evidently
${\tilde h}(\vec k, t)$ (and hence $h(\vec r, t)$) is also a Gaussian field. In particular, at a fixed position $\vec r$
and as a function of time $t$, $h(\vec r,t)$  is a Gaussian stochastic process. A Gaussian process is fully characterized 
by its mean and the two-time correlation function. Evidently, from Eq. (\ref{interface.3}), the mean of ${\tilde h}(\vec k,t)$
and hence that of $h(\vec r,t)$ is zero. All we need
then is to compute the two-time correlation function that will  fully specify this Gaussian process. 
From Eq. (\ref{interface.3}) it follows that
\begin{equation}
\langle {\tilde h}(\vec k,t_1){\tilde h}(-\vec k,t_2)\rangle = e^{-k^z (t_1+t_2)}\, \int_{0}^{t_1}\int_0^{t_2} e^{k^z (t_1'+t_2')}
\langle {\tilde \eta}(\vec k, t_1'){\tilde \eta}(-\vec k, t_2') dt_1' dt_2'.
\label{interface.4}
\end{equation}
Using the noise correlator 
$\langle {\tilde \eta}(\vec k, t){\tilde \eta}(\vec k', t')\rangle= D k^{\gamma} \delta(\vec k+\vec k')\delta(t-t')$, and 
and performing the time integrals one obtains
\begin{equation}
\langle {\tilde h}(\vec k,t_1){\tilde h}(-\vec k,t_2)\rangle =\frac{D}{2}\, k^{\gamma-z}\,
\left[e^{-k^z |t_1-t_2|}-e^{-k^z (t_1+t_2)}\right].
\label{interface.5}
\end{equation}
Finally the two-time correlator in real space is obtained by integrating over the Fourier modes
\begin{eqnarray}
A(t_1,t_2)\equiv \langle h(\vec r, t_1)h(\vec r, t_2)\rangle 
&= &\int d\vec k \, \langle {\tilde h}(\vec k,t_1){\tilde h}(-\vec k,t_2)\rangle  \\
& = & A_d \int_{k_{\rm min}}^{k_{\rm max}} dk\, k^{d-1+\gamma-z} \left[e^{-k^z |t_1-t_2|}-e^{-k^z (t_1+t_2)}\right] \;, \nonumber
\label{interface.6}
\end{eqnarray}
where $A_d$ is a volume dependent constant and $k_{\rm min}$ and $k_{\rm max}$ are respectively the lower and the upper cut-off
in the $k$-integral, needed to avoid the infrared and ultraviolet singularities when they arise. For example, the lower cut-off can be set to
$2\pi /L$ where $L$ is the linear size of the system and the upper cut-off to $2\pi/a$ where $a$ is the lattice spacing.

\vskip 0.3cm

\noindent {\bf Width of the interface:} At this point it is useful to distinguish between rough and smooth interfaces.
For this, we consider the variance $\langle h^2(\vec r,t) \rangle $ of the height at position $\vec r $ at time $t$, 
or equivalently the width $W= \sqrt{ \langle h^2(\vec r,t)\rangle }$ of the interface. This can be computed easily by putting $t_1=t_2=t$
in Eq. (\ref{interface.6}) and integrating over $k$
\begin{equation}
\langle h^2(\vec r, t)\rangle \propto \int_{k_{\rm min}}^{k_{\rm max}} dk \, k^{\gamma-z+d-1}\, [1- e^{-2 k^z t}].
\label{interface.7}
\end{equation}
Now, depending on the sign of $(\gamma+d-z)$, the following two situations arise.

\begin{itemize}

\item

{\bf Rough interface:} Consider first the case when $\gamma+d-z<0$. For example, the Edwards-Wilkinson interface
(where $\gamma=0$ and $z=2$) is rough when $d<2$. In this case, there is no ultraviolet singularity and one can set
the lattice spacing $a=0$, i.e., the upper cut-off to infinity in Eq. (\ref{interface.7}). By analysing the integral
one finds that
\begin{eqnarray}
W^2(L,t)\equiv \langle h^2(\vec r, t)\rangle \sim 
\begin{cases}
&  t^{2\beta}\quad\quad {\rm for} \quad 0 \ll t \ll L^z  \;, \\
& \\
&L^{2\alpha} \quad\quad {\rm for} \quad t \gg L^z \label{stat.1} \;,
\end{cases}
\end{eqnarray}
where $\beta= (z-d-\gamma)/{2z}>0$ is the growth exponent and $\alpha= (z-d-\gamma)/2>0$ is the roughness exponent. 
Thus the width of the interface initially grows with
time as $t^\beta$ and when $t \gg L^z$, it saturates to a time-independent stationary value $\sim L^\alpha$. The interface is thus rough
and the roughness is characterized by the width. Note also that for linear interfaces the growth exponent $0<\beta=(z-d-\gamma)/{2z}< 1/2$.
However, in principle one can have other interfaces where the growth exponent varies between its minimum value $\beta=0$ and its 
maximum possible value $\beta=1$. In the following discussion, we will assume $\beta$ to be a continuous parameter
in the interval $]0,1[$. 
For one-dimensional fluctuating interfaces, such as in fluctuating step edges on crystals, the three different growth exponents
are given respectively by:
\begin{itemize}

\item[$\circ$] { Edwards-Wilkinson interface:} In this case, setting $d=1$, $z=2$ and $\gamma=0$, one gets $\beta=1/4$.

\item[$\circ$] { Step-Edge diffusion (SED):} Here $d=1$, $z=4$ and $\gamma=2$, indicating $\beta=1/8$.

\item[$\circ$] { Mullins-Herring interface:} This case corresponds to $d=1$, $z=4$, $\gamma=0$, leading to $\beta=3/8$.   

\end{itemize}

\item

{\bf Smooth interface:} In contrast, when $\gamma+d-z>0$, there is no infrared singularity in the $k$-integral in Eq. (\ref{interface.7})
and one can set the lower limit of the integral to $0$, i.e., take the $L\to \infty$ limit. The resulting
integral gives
\begin{equation}
\langle h^2(\vec r, t)\rangle \sim a^{z-d-\gamma} - c\, t^{-(\gamma+d-z/z)}+ \dots
\label{interface.8}
\end{equation}
where $c$ is an unimportant constant.
Thus the width of the interface approaches to a constant (independent of the system size when $L\to \infty$) in the long time limit 
and hence the interface is called smooth.

\end{itemize}

In the rest of the section, we will restrict ourselves only to rough interfaces which exhibit interesting
persistence and first-passage properties. Coming back to the two-time correlation function in Eq. (\ref{interface.6})
and focusing only on rough interfaces (such that $\beta= (z-d-\gamma)/{2z}>0$), it is now easy to carry out 
the integral (setting the upper cut-off to infinity, i.e., $a=0$).
We first rearrange the $k$-integral in Eq.~(\ref{interface.6}) by adding and subtracting $1$ inside the integrand
\begin{eqnarray}
A(t_1,t_2)& =&
A_d \int_{k_{\rm min}=2\pi/L}^{\infty} dk\, k^{d-1+\gamma-z} \left[\left(1-e^{-k^z |t_1+t_2|}\right)-\left(1-e^{-k^z (t_1-t_2)}\right)\right]
\label{auto_inter.1} \nonumber \\
&=& I(t_1+t_2,L)-I(|t_1-t_2|,L) \label{auto_inter.2} \;,
\end{eqnarray}
where the integral $I(t,L)$ is given by
\begin{equation}
I(t,L) = A_d \int_{2\pi/L}^{\infty} dk\, k^{d-1+\gamma-z}\, [1- e^{-k^z t}] \;,
\label{auto_inter.3}
\end{equation}
which is convergent in the upper limit since $d+\gamma-z=-2\beta z<0$. By making a change of variable $k^z t=y$, one can
express $I(t,L)$ in the scaling form (for convenience we rescale $L$ by $L/{2\pi}$)
\begin{equation}
I(t,L) = t^{2\beta} g(t/L^z) \;, \quad {\rm where}\,\, g(x)= C \int_{x}^{\infty} dy\, y^{-1-2\beta} (1-e^{-y}) \;,
\label{auto_inter.4}
\end{equation}
where $C$ is an overall constant. The function $g(x)$ has the following asymptotic behavior
\begin{eqnarray}
g(x) \sim 
\begin{cases}
& {\rm const.}=K \,\,\, {\rm as}\,\, x\to 0 \;, \\
& \\
&\frac{1}{2\beta}\, x^{-2\beta} - x^{-1-2\beta}\, e^{-x} +\dots \, \,\, {\rm as}\,\, x\to \infty \;,
\label{xlarge.1}
\end{cases}
\end{eqnarray}
where $K$ is an unimportant constant.

We now consider first the thermodynamic limit $L\to \infty$, i.e., infinite system. Using the fact that
$g(x)\to K$ as $x\to 0$, we get from Eq. (\ref{auto_inter.2})  
\begin{equation}
A(t_1,t_2)= K\, \left[ (t_1+t_2)^{2\beta} - |t_1-t_2|^{2\beta}\right] \;,
\label{interface.9}
\end{equation}
where $\beta=(z-d-\gamma)/{2z}>0$ is the growth exponent. 
This autocorrelation function of the rough interface is thus parametrized by the single exponent $\beta>0$. Note that
in the limit $z\to \infty$, $\beta\to 1/2$ and the autocorrelation function reduces to that of a
Brownian motion, i.e., as $\beta\to 1/2$
\begin{equation}
A(t_1,t_2) \to 2\,K\, {\rm min}(t_1,t_2) \;,
\label{interface.10}
\end{equation}
the persistence of which has been studied, using various methods, in this review. 

\subsection{Linear interfaces: temporal persistence}

Consider the height field $h(0,t)$
at the origin as a function of $t$. Clearly, $h(0,t)$ is a stochastic Gaussian process in $t$ with zero mean
and the correlator $A(t_1,t_2)= K \left[(t_1+t_2)^{2\beta}-|t_1-t_2|^{2\beta}\right]$ with $\beta>0$. Note
that this process is not stationary, since the auto-correlator depends explicitly on both times $t_1$
and $t_2$ and not just on the time difference $|t_1-t_2|$. 
The {\em temporal persistence} $Q(t_0,t)$, defined before, measures 
the probability that the process $h(0,s)$ does not return to its
initial value $h(0,t_0)$ during the interval $t_0<s<t_0+t$, i.e.,
\begin{equation}
Q(t_0,t)= {\rm Prob.}\, \left[h(0,s)\ne h(0,t_0) \;, \; {\rm for \,\, all}\,\, t_0<s<t_0+t\right].
\label{temporal.1}
\end{equation}
Note that the evolution equation (\ref{interface.1}) respects the $h\to -h$ symmetry, hence the positive and the negative persistences are 
identical for linear interfaces as mentioned before.
The {\em relevant} stochastic process here  is the height difference defined as, 
\begin{equation}
Y(t;t_0)= h(0,t+t_0)-h(0,t_0)
\label{temporal.2}
\end{equation}
for all $t\ge 0$, starting at the initial value $Y(0;t_0)=0$.
The temporal persistence $P(t_0,t)$, defined above, is simply the probability that the relevant Gaussian process
$Y(t;t_0)$, starting at the initial value $0$ at $t=0$, does not return to zero up to time $t$. One can then think of
the waiting time $t_0$ simply as a parameter for this relevant Gaussian process. Clearly, the mean value of $Y(t,t_0)$ is zero for all $t$.
One can easily compute the two-time correlation function of this relevant Gaussian process
\begin{eqnarray}
A_{t_0}(t_1,t_2) &=& \langle Y(t_1;t_0)Y(t_2;t_0)\rangle = \langle [h(0,t_1+t_0)-h(0,t_0)][h(0,t_2+t_0)-h(0,t_0)]\rangle \nonumber \\
&=& [A(t_1+t_0,t_2+t_0)-A(t_1+t_0,t_0)-A(t_2+t_0,t_0)+A(t_0,t_0)] \;,
\label{temporal.3}
\end{eqnarray}
where the autocorelator $A(t_1,t_2)$ of the original height field is given in Eq. (\ref{interface.9}).      
Substituting the result from Eq. (\ref{interface.9}) in Eq. (\ref{temporal.3}) one gets
\begin{eqnarray}
&&A_{t_0}(t_1,t_2)=K\Big[2t_0+t_1+t_2)^{2\beta}-(2t_0+t_1)^{2\beta}-(2t_0+t_2)^{2\beta} \nonumber \\
&&+(2t_0)^{2\beta}+{t_1}^{2\beta}+{t_2}^{2\beta}
-|t_1-t_2|^{2\beta}\Big].
\label{temporal.4}
\end{eqnarray}
The parametric dependence on the waiting time $t_0$ is evident in this formulation. Below, we consider two
limiting situations: (i) $t_0=0$, i.e., we start the measurement right at the very beginning when the interface
is flat--we will call this `transient' regime and (ii) $t_0\to \infty$, i.e., we start measuring the
process only after waiting an infinite time--in other words, we measure the persistence in the `steady-state'
regime. We will see that these two situations give rise to very different behavior of the persistence probability $Q(t_0,t)$.
It turns out that for any finite $t_0$, the asymptotic power-law decay of $Q(t_0,t)$ for large $t$ is 
governed by the `transient' fixed point, i.e., one can effectively set $t_0=0$ as long as $t_0$ is finite~\cite{KKMCBS97}.

\subsubsection{Transient regime: $t_0=0$}

Setting $t_0=0$ in Eq. (\ref{temporal.3}) one gets
\begin{equation}
A_0(t_1,t_2)= A(t_1,t_2)= K \left[(t_1+t_2)^{2\beta}-|t_1-t_2|^{2\beta}\right].
\label{temporal.4a}
\end{equation}
In this case the relevant process $Y\equiv h$, i.e., the original field $h$ itself and we are interested
in the persistence $Q(0,t)$ that the Gaussian process $h(0,t)$, with zero mean and a correlator given
by Eq. (\ref{temporal.4a}), does not return to the origin up to time $t$. This process, though Gaussian, is nonstationary 
since the correlator in Eq. (\ref{temporal.4a}) depends on both times $t_1$ and $t_2$ and not just on their time difference.
However, one can map this process to a Gaussian stationary process (GSP)~\cite{KKMCBS97} by using the same Lamperti 
transformation that has been
repeatedly used in this review. We define a new normalized process $X(t)= h(0,t)/\sqrt{\langle h^2(0,t)\rangle}$
whose mean is zero and whose correlator, using Eq. (\ref{temporal.4a}), is given by
\begin{eqnarray}
\langle X(t_1)X(t_2)\rangle &= & \frac{A(t_1,t_2)}{\sqrt{A(t_1,t_1)A(t_2,t_2)}} \nonumber \\
& =&   \left[\frac{1}{2}\left(\sqrt{\frac{t_1}{t_2}}+\sqrt{\frac{t_2}{t_1}}\right)\right]^{2\beta}-
\left[\frac{1}{2}\left|\sqrt{\frac{t_1}{t_2}}-\sqrt{\frac{t_2}{t_1}}\right|\right]^{2\beta} \;.
\label{temporal.5}
\end{eqnarray}
Next we define the logarithmic time variable $T_1= \ln (t_1)$ and $T_2=\ln t_2$. When measured in this logarithmic time, the process
becomes stationary and its correlator depends only on the time difference $T=T_1-T_2$
\begin{equation}
\langle X(T_1)X(T_2)\rangle = f_0(T)= \left[\cosh(T/2)\right]^{2\beta}-\left|\sinh(T/2)\right|^{2\beta}.
\label{temporal.6}
\end{equation}
For small $T$, $f_0(T)\approx 1- |T/2|^{2\beta}$, indicating that for $\beta<1$ the process is non-smooth (see the general discussion on GSP's
in section \ref{subsection:GSP} with an infinite density of zero crossings).
For large $T$, the correlator decays exponentially $f_0(T)\sim \exp[-(1-\beta)\,T]$. From the general 
discussion on the no-zero crossing properties of GSP's in section \ref{subsection:GSP}, we know that if the correlator
decays exponentially for large $T$, then the corresponding persistence $Q(T)$ (probability of no zero crossing up to time $T$)
also decays exponentially, i.e., $Q(T)\sim \exp[-\theta T]$ with a certain decay constant $\theta$ that
depends on the full functional form of the correlator. This indicates that in our case, with a correlator
$f_0(T)$ in Eq. (\ref{temporal.6}), the persistence will also decay exponentially with $T$, $Q(T)\sim \exp[-\theta_0(\beta)\, T]$
for large $T$. Reverting back to the original time $t=e^T$, this implies that the persistence of the height field
$h(0,t)$ decays as a power law with time $t$, $Q(0,t)\sim t^{-\theta_0(\beta)}$ with a persistence exponent
$\theta_0(\beta)$ that depends on the growth exponent $\beta$~\cite{KKMCBS97}. 

The next challenge is to compute the persistence exponent $\theta_0(\beta)$. As discussed in section \ref{subsection:GSP}, evaluating
the decay constant $\theta$ that characterizes the exponential decay of the persistence $Q(T)\sim \exp[-\theta T]$
for a GSP with an arbitrary correlator $f(T)$ is, in general, an unsolved problem. It can be determined explicitly
only for a Markov process where the correlator is a pure exponential for all $T$, $f(T)=\exp[-\lambda |T|]$ for
which $\theta=\lambda$ (see section \ref{subsection:GSP}). There are few other very special cases of $f(T)$ for which $\theta$ is known. 
Unfortunately, our correlator in Eq. (\ref{temporal.6}) is not one of the exactly solvable ones. Thus one has to 
resort to numerical simulations
or one of the several approximate methods discussed in this review. 
The independent interval approximation (IIA) used in section \ref{subsection:iia} for a smooth GSP can not be applied 
here, since our correlator corresponds to a non-smooth process for $\beta<1$. 
In Ref.~\cite{KKMCBS97}, some rigorous bounds were obtained for $\theta_0(\beta)$. Such bounds can be obtained
by comparing the correlator $f_0(T)$ in Eq. (\ref{temporal.6}) to the correlator of other Markovian GSP's 
and using Slepian's lemma~\cite{Slepian}, discussed in section \ref{subsection:GSP}. For instance, one can show rigorously that (see the Appendix of Ref.~\cite{KKMCBS97})
\begin{eqnarray}
\theta_0(\beta) &\ge & 1-\beta \,\, {\rm for} \,\, \beta<1/2 \label{bound.1} \;, \\
& \le & 1-\beta \,\, {\rm for} \,\, \beta>1/2 \label{bound.2} \;.
\end{eqnarray}
Numerical simulation results for $\theta_0(\beta)$ in Ref.~\cite{KKMCBS97} are consistent with these rigorous bounds.

Note from Eq. (\ref{temporal.6}) that exactly for $\beta=1/2$, $f_0(T)=\exp[-|T|/2]$, i.e., the process
becomes a Markovian GSP for which $\theta_0(\beta=1/2)=1/2$. For $\beta$ close to $1/2$, i.e., when
$\beta=1/2+\epsilon$ where $\epsilon$ is small, one can also determine $\theta_0(\beta=1/2+\epsilon)$
perturbatively for small $\epsilon$, using the perturbation theory developed for GSP's in section \ref{section:perturbation}. 
Indeed, setting $\beta=1/2+\epsilon$ in Eq. (\ref{temporal.6}) one gets
\begin{equation}
f_0(T)= \exp\left(-|T|/2\right) + \epsilon \phi_0(T) + {\cal O}(\epsilon^2) \;,
\label{temporal.7}
\end{equation}
where $\phi_0(T)= 2 \cosh (T/2) \ln \left(\cosh (T/2)\right)- 2 \sinh (|T|/2)\ln \left(\sinh (|T|/2)\right)$. 
According to the perturbation theory discussed in section \ref{section:perturbation} one gets~\cite{KKMCBS97}
\begin{eqnarray}
\theta_0(\beta=1/2+ \epsilon) &= & \frac{1}{2}\left[1-\frac{\epsilon}{\pi}\int_0^{\infty} 
\phi_0(T)\, \left(1-e^{-T}\right)^{-3/2}\, dT \right] + {\cal O}(\epsilon^2) 
\nonumber \\
&=& \frac{1}{2} - (2\sqrt{2}-1)\, \epsilon + {\cal O}(\epsilon^2) \;.
\label{temporal.8}
\end{eqnarray}

\subsubsection{Steady-state regime: $t_0\to \infty$}
 
Setting $t_0\to \infty$ corresponds to an infinite waiting time, i.e., when one measures the process after
it has reached the stationary regime. Taking the limit $t_0\to \infty$ in Eq. (\ref{temporal.3}), one 
finds that the correlator of the relevant Gaussian process $Y(t; \infty)$ is given by
\begin{equation}
A_{\infty}(t_1,t_2)= K \left[ {t_1}^{2\beta} + {t_2}^{2\beta} - |t_1-t_2|^{2\beta}\right].
\label{temporal.9}
\end{equation}
We are then interested in the persistence probability $P(\infty, t)$ that the relevant process $Y(t;\infty)$ does not return 
to $0$ up to time $t$. Note, in particular, that the incremental correlation function of this relevant process behaves as
\begin{equation}
\langle \left[Y(t_1;\infty)-Y(t_2;\infty)\right]^2\rangle =  A_{\infty}(t_1,t_1)+A_{\infty}(t_2,t_2)-2A_{\infty}(t_1,t_2) 
= 2\,K\, |t_1-t_2|^{2\beta} \, .
\label{interface_increment1}
\end{equation}

To make progress, the first observation is that the correlator in Eq. (\ref{temporal.9}) is exactly identical to
the fractional Brownian motion (fBM) discussed in section \ref{section:fBm} with the Hurst exponent $0<H=\beta<1$. 
Equivalently, the incremental correlation function grows as $|t_1-t_2|^{2H}$ with $H=\beta$. In addition,
the probability distribution of the height in the steady state in invariant under the inversion $h\to -h$, indicating
that positive and negative persistence exponents are identical, $\theta_s^{+}=\theta_s^{-}=\theta_s (\beta)$.
From the discussion in
section \ref{section:fBm}, it is then clear that the persistence decays as a power-law in time, $Q(\infty, t)\sim t^{-\theta_s(\beta)}$
as $t\to \infty$ with the exponent given exactly by~\cite{KKMCBS97}
\begin{equation}
\theta_s(\beta)= 1-\beta \, .
\label{temporal.10}
\end{equation}
This exact result is very interesting and has been verified numerically in several cases. It is also consistent with
the following rigorous bounds derived in Ref.~\cite{KKMCBS97}
\begin{eqnarray}
\theta_s(\beta) & \ge & \beta \,\, {\rm for}\,\,  \beta<1/2 \;, \nonumber \\
& \le & \beta \,\, {\rm for}\,\,  \beta<1/2 \;.
\label{bound.3}
\end{eqnarray} 
For one dimensional fluctuating interfaces, it then predicts the following results:

\begin{itemize}

\item {\bf Edwards-Wilkinson interface:} In this case $\beta=1/4$ which then predicts $\theta_s=3/4$. This result was verified 
numerically~\cite{KKMCBS97,CDCMD04}
as well as experimentally by measuring the persistence of steps at high temperatures on the vicinal surface of Si(111) 
surface with Al adsorbed on it~\cite{Dougherty_exp02}. The measured exponent $0.77\pm 0.03$ is in good agreement with the theoretical prediction
$\theta_s=3/4$.
\item{\bf Step edge diffusion (SED):} This case corresponds to $\beta=1/8$ and hence $\theta_s=7/8$. This exponent has also been measured  
experimentally on Pb(111) and Ag(111) surface where the dominant mechanism is indeed step edge diffusion~\cite{Dougherty_exp03}. The measured
values of $\theta_s= 0.88\pm 0.04$ (for Pb (111) surface) and $\theta_s=0.87 \pm 0.02$ (for Ag(111) surface)
are also consistent with the theoretical prediction $\theta_s=7/8$.
\item{\bf Mullins-Herring interface:} Here $\beta=3/8$ and hence the theoretical prediction for $\theta_s=1-\beta=5/8$. As far as we know,
this is yet to be verified experimentally. 

Finally, the relation $\theta_s=1-\beta$ has also been verified recently for nonlinear reacting-wetting
advancing interfaces
in the experimental system of mercury on silver at room temperature~\cite{Efraim11}. In this system, the experimentally measured growth exponent 
$\beta=0.67\pm 0.06$ and the persistence exponent $\theta_s=0.37\pm 0.05$ are consistent with the 
relation $\theta_s=1-\beta$.

\end{itemize}

Note that by the customary transformation, $X(t)= Y(t;\infty)/\sqrt{\langle Y^2(t;\infty)\rangle}$ and $T=\ln (t)$, the
process can again be mapped to a GSP with zero mean and a correlator that is stationary, i.e., only a function
of $T=T_1-T_2$
\begin{equation}
\langle X(T_1)X(T_2)\rangle = f_s(T)= \cosh(\beta T)- \frac{1}{2}\left| 2 \sinh(T/2)\right|^{2\beta} \;,
\label{temporal.11}
\end{equation}
where we have used the expression in Eq. (\ref{temporal.9}). Once again, this GSP $X(T)$ is a non-smooth process, since
$f_s(T) \approx 1- |T|^{2\beta}/2$ as $T\to 0$. Also, for large $T$, $f_s(T)$ decays exponentially, $f_s(T)\sim \exp[-\lambda_s T]$
where $\lambda_s= {\rm min}(\beta,1-\beta)$. Hence, from the general discussion on GSP in section \ref{subsection:GSP}, it follows that
the persistence $Q(T)$ of the GSP will decay exponentially for large $T$, $Q(T)\sim \exp[-\theta_s(\beta) T]$ and hence
as a power law in the original time $t$, $Q(\infty, t)\sim t^{-\theta_s(\beta)}$ for large $t$. Given the nontrivial
form of the stationary correlator $f_s(T)$ in Eq. (\ref{temporal.11}), one would not, in general, be able to determine
$\theta_s(\beta)$ explicitly. Seen from this angle, the fact that one can determine $\theta_s(\beta)=1-\beta$
explicitly is rather surprising. This result then adds to the list of correlators of GSP's for which the
persistence exponent can be determined explicitly.

A further partial confirmation of the result $\theta_s(\beta)=1-\beta$ can be obtained by using perturbation theory around 
$\beta=1/2$ for which the correlator reduces to purely exponential, i.e., the GSP is Markovian with
correlator $f_s(T)=e^{-|T|/2}$ indicating that $\theta_s(1/2)=1/2$. For $\beta=1/2+\epsilon$, one gets
\begin{equation}
f_s(T)= \exp\left(-|T|/2\right) + \epsilon \phi_s(T) + {\cal O}(\epsilon^2) \;,
\label{temporal.12}
\end{equation}
where $\phi_s(T)= \sinh(|T|/2)\, \left[|T|- 2 \ln \left(2\sinh(|T|/2\right)\right]$. Using the perturbation theory
(around a Markov GSP) discussed in section \ref{section:perturbation}, one then gets~\cite{KKMCBS97}
\begin{eqnarray}
\theta_s(\beta=1/2+ \epsilon) &= & \frac{1}{2}\left[1-\frac{\epsilon}{\pi}\int_0^{\infty}
\phi_s(T)\, \left(1-e^{-T}\right)^{-3/2}\, dT \right] + {\cal O}(\epsilon^2)
\nonumber \\
&=& \frac{1}{2} - \epsilon + {\cal O}(\epsilon^2) \;.
\label{temporal.13}
\end{eqnarray}
Thus, to order $\epsilon$ for small $\epsilon$, the perturbation result in Eq. (\ref{temporal.13}) is consistent with the exact 
prediction $\theta_s(\beta)=1-\beta$. 
 
\subsection{Linear interfaces: temporal survival probability}

In the previous section we discussed the temporal persistence $Q(t_0,t)$ defined as the probability that the interface height 
at the origin, measured after a waiting time $t_0$, does not return to its starting value $h(0,t_0)$ up to a subsequent
time $t$ measured since $t_0$. Another natural question is the temporal survival probability $S(t_0,t)$, introduced in 
Ref.~\cite{DCDM2004}, which
denotes the probability that the height does not return to its average value, i.e., to $0$ during the time interval $[t_0,t+t_0]$.
Note that for $t_0=0$, the temporal survival probability is identical to the temporal persistence 
$Q(0,t)$ since the interface starts at time $0$
from a flat initial condition. However, in the opposite stationary regime $t_0\to \infty$, $S(\infty, t)$ is certainly different from
the temporal persistence $Q(\infty, t)$. This is because, the actual height is of $\sim {\cal O}(L^{\alpha})$ in the stationary regime, so
the probability $S(\infty, t)$ that starting from such a large value the height does not return to $0$ must be very small
for large systems and is certainly going to depend on the system size~\cite{DCDM2004}. In contrast, the temporal persistence concerns
the probability of no return to zero of the height difference $h(0,t+t_0)-h(0,t_0)$ (with $t_0\to \infty$). Thus, the relevant
stochastic process is different in the two cases. For the temporal survival probability, the relevant process is the 
original height itself $h(0,t)$. This is a Gaussian process with zero mean. We can compute the correlator
$A(t_0+t_1, t_0+t_2)=\langle h(0,t_0+t_1)h(0,t_0+t_2)\rangle$ from Eq. (\ref{auto_inter.2})
\begin{equation}
A(t_0+t_1, t_0+t_2) = I(2t_0+t_1+t_2,L)-I(|t_1-t_2|,L) \;,
\label{survival.1}
\end{equation}
where $I(t,L)$ is given in Eq. (\ref{auto_inter.4}).
Taking $t_0\to \infty$ while keeping $t_1$ and $t_2$ fixed, and using the asymptotic properties of the integral
$I(t,L)$ in Eq. (\ref{xlarge.1}) we get
\begin{equation}
\lim_{t_0\to \infty} A (t_0+t_1, t_0+t_2)= \frac{C}{2\beta} L^{2\beta z}- |t_1-t_2|^{2\beta} g\left(|t_1-t_2|/L^z\right). 
\label{survival.2}
\end{equation}
The scaling function $g(x)$ in Eq. (\ref{auto_inter.4}) can be expressed as, 
$g(x)= C \int_x^{\infty} dy\, y^{-2\beta-1}\,(1-e^{-y})= C x^{-2\beta}/{2\beta}- g_1(x)$ where $g_1(x)$ is the incomplete Gamma function
\begin{equation}
g_1(x)= C \int_x^{\infty} dy\, y^{-2\beta -1}\, e^{-y}\, .
\label{survival.3}
\end{equation}
Substituting this result for $g(x)$ in Eq. (\ref{survival.2}) gives
\begin{equation}
C(t_1,t_2)\equiv \lim_{t_0\to \infty} A (t_0+t_1, t_0+t_2)= |t_1-t_2|^{2\beta} g_1\left(|t_1-t_2|/L^z\right) \;,
\label{survival.4}
\end{equation}
where $g_1(x)$ is given in Eq. (\ref{survival.3}) and has the following asymptotics
\begin{eqnarray}
g_1(x)\sim 
\begin{cases}
& \frac{C}{2\beta} x^{-2\beta} \,\, {\rm as}\,\, x\to 0 \;, \label{xsmall.2} \\
& \\
& C\, x^{-2\beta-1}\, e^{-x} \,\, {\rm as}\,\, x\to \infty \;.  
\end{cases}
\end{eqnarray}
When $t_1=t_2$, using Eq. (\ref{xsmall.2}), the onsite height variance $\langle h^2(0,t_0+t)\rangle$ approaches to $ \sim L^{2\beta z}$
as $t_0\to \infty$, in
agreement with Eq. (\ref{stat.1}) since $\alpha=\beta z$. It is convenient to consider the normalized process
$X(t)= \lim_{t_0\to \infty} h(0,t_0+t)/\sqrt{\langle h^2(0,t_0+t)\rangle}$. This process is also Gaussian with zero mean
and moreover its correlation function is stationary in the rescaled time $T= t/L^z$
\begin{equation}
\langle X(T_1) X(T_2)\rangle = f(T_1-T_2)\,\, {\rm where}\,\, f(T)= |T|^{2\beta}\, g_1(|T|) \;,
\label{survival.5}
\end{equation}
where $g_1(x)$ is given in Eq. (\ref{survival.3}). Note that, unlike in the case of temporal persistence, here
the relevant process is stationary already in the original time $t$ (just rescaled by $L^{z}$) and the logarithmic transformation
$T=\ln t$ is not necessary to make the process stationary. 
Thus the survival probability $S(\infty,t)$ of the 
height field is precisely the probability that the normalized GSP $X(T)$ (with $T=t/L^z$) with zero mean and a
correlator $f(T)$ does not cross zero up to time $T$. Once again, the problem reduces to the persistence $Q(T)$ of a GSP
with corellator $f(T)$. From our general discussion in section \ref{subsection:GSP}, since the correlator $f(T)\sim e^{-|T|}/|T|$ decays faster 
than exponentially for large $T$ as follows from Eq. (\ref{xsmall.2}), we expect that the persistence also decays
exponentially for large $T$, $Q(T)\sim \exp[-\theta(\beta) |T|]$ with some nontrivial decay constant $\theta(\beta)$ that
depends on the full correlator $f(T)$. Reverting back to the original time, $t= L^z T$, one then predicts
that the equilibrium ($t_0\to \infty$) survival probability behaves for large $t$ as~\cite{DCDM2004}
\begin{equation}
S(\infty, t) \sim \exp[- \theta(\beta) t/L^z] \sim \exp[-t/\tau_s]\,\, {\rm where}\,\, \tau_s= L^z/\theta(\beta) .
\label{survival.6}
\end{equation}
While it is again difficult to estimate the constant $\theta(\beta)$, 
it follows that the typical time scale associated with the exponential decay of the survival probability 
behaves as $\tau_s\sim L^z$ for large $L$.
This is in contrast to the temporal persistence in the stationary regime that decays as a power law for large $t$, 
$Q(\infty, t)\sim t^{-\theta_0(\beta)}$ where $\theta_0(\beta)=1-\beta$ and hence there is no system size
dependent time scale. The theoretical prediction for the asymptotic
behavior in Eq. (\ref{survival.6}) of the equilibrium survival probability, in particular the system size dependence of the time scale 
$\tau_s$ has been verified by extensive
numerical simulations on a class of one dimensional linear interfaces (see Ref.~\cite{CDCMD04} for details).

\subsection{Nonlinear interfaces: temporal persistence}

For nonlinear interfaces evolving via Eq. (\ref{kpz.1}) or (\ref{mbe.1}), one can define, as in the case of linear interfaces, 
the transient and steady-state persistence probabilities. However, unlike the linear interfaces, for nonlinear interfaces one generally
lacks the $h\to -h$ symmetry. As a result, one needs to define positive and negative temporal persistence
probabilities $Q_{\pm}(t_0,t)$ as discussed in the beginning of the section, e.g., see Eqs. (\ref{pers_pos}) and 
(\ref{pers_neg}). 
The corresponding transient (setting $t_0=0$) and steady-state (setting $t_0\to \infty$) persistence probabilities 
decay as power laws for large $t$ as in Eq. (\ref{trans_pers}). In general, one then
needs four exponents to describe the asymptotic power law decay of these probabilities: $\theta_0^{+}$, $\theta_0^{-}$, $\theta_s^{+}$
and $\theta_s^{-}$. 

\vskip 0.3cm

\noindent {\bf Transient persistence:} Unlike in the linear case, the height field in the nonlinear case (such as in Eqs. (\ref{kpz.1})
and (\ref{mbe.1})) is non-Gaussian. Hence one can not use the results of the Gaussian process that were so useful in the linear case.
The transient persistence exponents $\theta_0^{\pm}$ can then be determined only numerically. For example, 
Kallabis and Krug~\cite{KK99} computed the exponents $\theta_0^{\pm}$ numerically for a class of discrete 
nonlinear one dimensional growth models
(which belong to the KPZ universality class as far as the growth and roughness exponents are concerned) and found that     
$\theta_0^{+}= 1.18\pm 0.08$ and $\theta_0^{-}= 1.64\pm 0.08$ and within numerical accuracy, they are universal.
In a recent experiment on liquid crystals (belonging to the KPZ university class)~\cite{TS12}, these exponents were measured
and found to be $\theta_0^+\approx 1.35$ and $\theta_0^{-}\approx 1.85$, not very far from the numerical
results obtained directly by simulating the KPZ equation.

\vskip 0.3cm

\noindent {\bf Steady-state persistence:} As opposed to the transient case where analytical results seem  very difficult to
obtain, it turns out that for the steady-state persistence exponents $\theta_s^{\pm}$, one can obtain at least some partial analytical
information. A very general scaling relation for an arbitrary nonlinear interface was derived in Ref.~\cite{CDCMD04} that relates the
smaller of the pair $\theta_s^{+}$ and $\theta_s^{-}$ to the growth exponent $\beta$ of the interface
\begin{equation}
{\rm min} \left(\theta_s^{+}, \theta_s^{-}\right)=1-\beta .
\label{interface_gen.scaling.1}
\end{equation} 
For the special case where $h\to -h$ symmetry holds (such as for linear interfaces or nonlinear KPZ equation in one dimension), this
generalized scaling relation reduces to $\theta_s= \theta_s^+=\theta_s^{-}= 1-\beta$ that was already derived
for linear interfaces in Eq. (\ref{temporal.10}). Numerical results for several one and two dimensional nonlinear interfaces
are in agreement with this generalized scaling relation~\cite{CDCMD04}. For example, for the one dimensional KPZ equation where $\beta=1/3$, 
one would expect $\theta_s^+=\theta_s^{-}= 2/3$ and the numerical results by Kallabis and Krug~\cite{KK99} $\theta_s^+=\theta_s^{-}=0.66\pm 
0.03$ are consistent with this analytical prediction. In contrast, for the one dimensional MBE equation (\ref{mbe.1}) where
one does not have the $h\to -h$ symmetry even in the steady state, it was found numerically in Ref.~\cite{CDCMD04} that 
$\theta_s^{+}=0.66\pm 0.02$ and $\theta_s^{-}=0.78 \pm 0.02$. For this model, the growth exponent $\beta\approx 1/3$ is known only 
numerically. The numerical value of the smaller of the two exponents $\theta_s^{+} =0.66\pm 0.02$ is consistent
with the scaling relation (\ref{interface_gen.scaling.1}). The relation 
(\ref{interface_gen.scaling.1}) was also verified for a class of other nonlinear discrete growth models. For an extensive review
of the numerical techniques and subtleties associated with such models the reader may consult Ref.~\cite{CDCMD04}.  

The result (\ref{interface_gen.scaling.1}) follows from the following observation. We consider the relevant
process $Y(t;t_0)= h(0,t+t_0)-h(0,t_0)$. Since $h$ is a non-Gaussian process, so is $Y$. Let us now consider
the incremental correlation function
\begin{equation}
\sigma^2(t_1,t_2)= \lim_{t_0\to \infty} \langle [Y(t_1;t_0)-Y(t_2;t_0)]^2\rangle= \lim_{t_0\to \infty}
\langle [h(0,t_1+t_0)-h(0,t_2+t_0)]^2\rangle \, . 
\label{interface_increment2}
\end{equation}
It turns out that for generic self-affine interfaces (which do not have to be necessarily Gaussian), this incremental correlation 
function depends only on the time difference $|t_1-t_2|$ in a power-law fashion for large $|t_1-t_2|$~\cite{Krug97}
\begin{equation}
\sigma^2(t_1,t_2) \sim |t_1-t_2|^{2\beta} \;,
\label{interface_increment3}
\end{equation}
where $\beta$ is the growth exponent. While for linear interfaces one can prove this result explicitly (see Eq. (\ref{interface_increment1}),
for nonlinear interfaces it has been verified numerically~\cite{Krug97}. 
However, this is precisely the defining property of the `generalized' fractional Brownian motion discussed in section \ref{section:fBm}. 
Unlike in the linear case, for nonlinear interfaces, due to the 
generic lack of $h\to -h$ symmetry in the steady state, one would expect the positive and negative persistence exponents
to be generically different $\theta_s^{+}\ne \theta_s^{-}$. The relation (\ref{interface_gen.scaling.1}) then
immediately follows from the scaling argument presented in section \ref{section:fBm} for generic self-affine processes (not necessarily Gaussian)
with incremental correlation function of the type in Eq. (\ref{interface_increment3}).

\subsection{Spatial persistence and spatial survival probability: linear and nonlinear interfaces}\label{subsection:spatial_pers_interfaces} 

So far we have discussed temporal persistence and temporal survival probability of fluctuating interfaces where the
primary issue is to compute the probability of no return to the initial condition or to the average value of the
interface height $h(x,t)$ between time intervals $[t_0,t_0+t]$, but at a fixed point $x$ in space. In an infinite system,
this probability does not depend on $x$ due to translational invariance. Alternatively one can pose similar question
of persistence or survival probability of the height $h(x,t)$ as a function of $x$, but at fixed time $t$. This is
the {\em spatial} counterpart of the temporal first-passage probabilities, first posed and studied in Ref.~\cite{MB01} for linear
interfaces. In exact analogy with the temporal case, the spatial persistence $Q(x_0,x_0+x)$ is the probability
that the height $h(x,t)$, at a fixed time $t$, does not return to its value $h(x_0,t)$ over the {\em spatial} interval
$[x_0,x+x_0]$ along a given direction. A natural choice is to study this quantity in the steady state $t\to \infty$,
so that $Q(x_0,x+x_0)$ is independent of time. In Ref.~\cite{MB01}, $Q(x_0,x+x_0)$ was studied theoretically
for Gaussian linear interfaces described by the Langevin equation (\ref{interface.1}) and it was found that
$Q(x_0,x+x_0)$ decays as a power law for large $x$, $Q(x_0,x+x_0)\sim |x|^{-\theta}$ where the `spatial' persistence exponent $\theta$
depends on the choice of $x_0$. When $x_0$ is sampled uniformly from all points, then the average of $Q(x_0,x_0+x)$ over $x_0$
gives the {\em steady-state} spatial persistence probability $Q_{SS}(x)$ that decays as $Q_{SS}(x)\sim |x|^{-\theta_{\rm SS}}$ for large $x$.
The exponent $\theta_{\rm SS}$ is called the steady-state spatial persistence exponent. In contrast, if $x_0$ is sampled from
a subset of points where the steady state height profile and its derivatives are {\em finite}, then the corresponding
finite-initial-condition (FIC) persistence probability $Q_{FIC}(x)\sim |x|^{-\theta_{\rm FIC}}$ for large $x$ where
$\theta_{\rm FIC}$ is different from $\theta_{\rm SS}$. Indeed, it turns out~\cite{MB01} that $\theta_{\rm FIC}$ and $\theta_{\rm SS}$ are 
respectively the {\em spatial} analogues of the temporal persistence exponents $\theta_0$ (transient) and $\theta_s$ (steady-state) defined
in Eq. (\ref{trans_pers}). 

Let us first briefly discuss the spatial exponent $\theta_{\rm FIC}$.
The exponent $\theta_{\rm FIC}$, just as its temporal counterpart $\theta_0$ turns
out to be hard to compute~\cite{MB01} even for linear interfaces in Eq. (\ref{interface.1}). 
However, for linear interfaces, one can show rigorously~\cite{MB01} that $\theta_{\rm FIC}$ is identical
to the temporal persistence exponent $\theta(n)$ of the generalised random walk equation $d^nx/dt^n=\eta(t)$ (discussed in section \ref{subsection:higher_order})
where $\eta$ is Gaussian white noise and $n= \alpha+1/2$ where $\alpha=(z-d-\gamma)/2$ is the roughness exponent.
For instance, when $\alpha=1/2$ (as in the case of Edwards-Wilikinson equation in $1$-d), $n=1$ and one has 
just the ordinary Brownian motion for which the temporal
persistence exponent is $\theta(1)=1/2$, indicating $\theta_{\rm FIC}(\alpha=1/2)=1/2$. In contrast, 
for the Mullins-Herring equation in $1$-d, where $\gamma=0$ and $z=4$, one has $\alpha=3/2$ and hence $n=2$.
Thus $\theta_{\rm FIC}$ in this case is identical to the persistence exponent for the random acceleration process discussed in section \ref{subsection:RAP},
$\theta_{\rm FIC}= \theta(2)=1/4$. Similarly, by choosing $z$ and $\gamma$ of the underlying interface, one
can engineer higher values of $n$ as well. For higher values of $n$, $\theta(n)$ can not be determined
analytically. However, one can use approximation methods such as IIA discussed in section \ref{subsection:higher_order} to obtain rather accurate
estimates of $\theta(n)$ for $n>2$ and hence of $\theta_{\rm FIC}$~\cite{MB01}. In summary, the exact mapping between the
spatial and the temporal process mentioned above provides a physical realization of the generalized
random walk process $d^nx/dt^n=\eta(t)$ with arbitrary $n$~\cite{MB01}.

We now turn to the steady-state spatial persistence exponent $\theta_{\rm SS}$. It turns out that this can be determined
analytically in terms of the roughness exponent $\alpha$ of the interface. 
Consider a fluctuating interface in its steady state. For the spatial persistence $Q_{SS}(x)$, 
we need to compute the no-zero crossing
probability of the relevant process $Z(x,t)\equiv h(x_0+x,t)-h(x_0,t)$ as a function of $x$.
From generic 
scaling arguments it follows~\cite{Krug97} that for fluctuating interfaces in the steady state ($t\to \infty$), the incremental correlation 
function of the relevant process $Z(x,t)$ between two spatial points $x_1$ and $x_2$ behaves, for large $|x_1-x_2|$, as a power law
\begin{equation}
\langle [Z(x_1,t)-Z(x_2,t)]^2 \rangle \sim |x_1-x_2|^{2\alpha} \;,
\label{increment_rough.1}
\end{equation}
where $\alpha>0$ is the roughness exponent. For linear interfaces in Eq. (\ref{interface.1}), this scaling behavior in 
Eq. (\ref{increment_rough.1}) can be established analytically with roughness exponent $\alpha=(z-d-\gamma)/2$. For nonlinear interfaces 
such as the KPZ or the MBE equation, while one can not show this rigorously one expects this scaling relation to hold on general
grounds (indeed this is just the defining equation for the roughness exponent $\alpha$). Thus, from our discussion
in section \ref{section:fBm}, it follows that the relevant process $Z(x,t)$, as a function of $x$, is a generalised fBM (not necessarily Gaussian)
with Hurst exponent $H=\alpha$ for $0<\alpha<1$. Consequently, from the general result, $\theta=1-H$, of the persistence exponent
of a generalised fBM with Hurst exponent $H$, it follows that for $0<\alpha<1$~\cite{CDCMD04} 
\begin{equation}
\theta_{\rm SS}= 1- \alpha \, .
\label{pers_SS}
\end{equation}
For the case of EW or KPZ interfaces in $d=1$, we have $\alpha=1/2$ and hence $\theta_{\rm SS}=1/2$~\cite{MB01}. This result
has been confirmed~\cite{CDD04} in the numerical simulations of the Family model and 
has also been measured experimentally: for
fluctuating step edges in Al/Si(111) system~\cite{CCDLW07} and fluctuating combustion fronts in paper~\cite{MMMTA03}. 

As in the case of spatial persistence, one can equivalently define~\cite{CDD04} the spatial analogue of the temporal survival probability, 
namely the probability that the interface height $h(x,t)$ stays above its average value $0$ over the spatial
interval $[x_0,x+x_0]$ along a given direction. It is natural to first consider the steady state limit $t \gg L^z$, where sampling 
$x_0$ uniformly from all points in the steady state and averaging
over $x_0$, one obtains the steady state spatial survival probability $S_{\rm SS}(x,L)$. However, 
unlike in the temporal case, numerical simulations~\cite{CDD04}
suggest that $S_{\rm SS}(x,L)$ does not decay with $x$ as a power law for $1 \ll x \ll L$ and instead, it does depend on the system size $L$ 
even for large $L$.
Simulations for various system sizes suggest~\cite{CDD04} instead that $S_{\rm SS}(x,L)$ 
has the scaling behavior: 
$S_{\rm SS}(x,L)= F_{SS}(x/L)$, for large $x$ and large $L$ but with the ratio $z=x/L$ fixed. 
These numerical findings were confirmed later in an analytical study of $S_{\rm SS}(x,L)$ for EW interfaces~\cite{MD06},
where an exact mapping between the spatial statistics of $1$-d EW interfaces at equilibrium and the temporal statistics of $1$-d Brownian
motion was exploited to develop a path integral formalism to compute the nontrivial scaling function $F_{SS}(z)$ analytically.
The expression of this scaling function $F_{SS}(z)$ turns out to be rather complicated involving integrals over special functions~\cite{MD06}.
However, a simpler and more explicit functional form of $F_{SS}(z)$ was derived in Ref.~\cite{MD06} using an 
approximate `deterministic' approach to
evaluate this path integral and the analytical results were found to be in good agreement with simulation results.

One can similarly study the spatial survival probability in the growing regime $t \ll L^z$. In this case, the spatial survival probability 
$S_{\rm gr}(x,t)$, i.e.,
the probability that the height fluctuation around its average (which is $0$) does not change sign over a distance $x$, depends on
$x$ and $t$, but not on $L$ for $t \ll L^z$. The subscript ``${\rm gr}$" in $S_{\rm gr}(x,t)$ denotes that one is in the growing regime $t \ll L^z$.
Given that in the steady state regime $t \gg L^z$, the same quantity exhibits the scaling behavior $S_{\rm SS}(x,L)\sim F_{SS}(x/L)$, one would expect
that in the opposite growing regime $t \ll L^z$, $S_{\rm gr}(x,t)$ should exhibit a similar scaling behavior, $S_{\rm gr}(x,t)\sim F_{\rm gr}(x/t^{1/z})$.
Essentially the effective length scale is $L$ in the steady state regime $t \gg L^z$, while it is $t^{1/z}$ in the growing regime. When
$t\sim L^z$, $S_{\rm gr}(x,t)$ crosses over to $S_{\rm SS}(x,L)$. 
Note that for nonlinear interfaces in the growing regime, one has to distinguish, as before, the positive and negative
excursions due to the lack of $h\to -h$ symmetry. Thus, generically, in the growing regime, one would expect 
two different scaling functions for the positive (negative) spatial survival probability, $S_{\rm gr}^{\pm}(x,t)\sim F_{\rm gr}^{\pm}(x/t^{1/z})$.
Unfortunately, there have not been much theoretical studies
to confirm this scaling behavior in the growing regime. However, recent experimental studies in liquid crystals by Takeuchi and Sano~\cite{TS12} 
did confirm this scaling behavior for the positive (negative) spatial survival probabilities. The authors of Ref.~\cite{TS12} found
that the scaling functions $F_{\rm gr}^{\pm}(z) \sim \exp[- \kappa_{\pm} z]$ for large $z$ with $\kappa_+=1.9(3)$ and $\kappa_{-}=2.0(3)$.
Note however, the authors in Ref.~\cite{TS12} actually measure the spatial {\em survival probability}, though in their
paper they call it spatial {\em persistence}. In Ref. \cite{FF2012} the authors derived an exact formula for the persistence of the so-called
Airy$_1$ process, which describes the fluctuations of KPZ (flat) interfaces in the growing regime. By evaluating numerically their exact formula, expressed in terms of Fredholm determinants, they could evaluate $\kappa_- = 1.83$, in reasonably good agreement with the experimental measurements \cite{TS12}.   
 
\subsection{Persistence properties in flat versus radial geometry}\label{subsection:flat_vs_radial}

In our discussions so far, we have considered fluctuating interfaces in a cylindrical geometry, i.e., interfaces growing
on a $d$-dimensional flat substrate of fixed size $L$ and with periodic boundary conditions in each of the $d$-directions.
In many real systems such as growing bacterial colony, growing tumour, or growing droplet of the turbulent phase
in the recent liquid crystal experiment~\cite{TS12}, the surface grows radially from an initial
seed at the origin and hence the relevant geometry is the radial one. While some of the local scaling properties of the surface
do not depend on the details of the geometry, it turns out that for some observables, such as the autocorrelation
function of the height fluctuation and consequently its persistence properties are qualitatively different in 
the flat and radial geometry~\cite{Singha05}. For instance, the width of the interface in the radial geometry
keeps growing as a power law in time $W(t)\sim t^{\beta}$ and does not saturate to $W\sim L^{\alpha}$ for large $t$
as in the flat case~\cite{Singha05}. In other words, the Family-Vicsek scaling behavior valid for flat geometry
no longer holds for the circular geometry, because in this latter case there is no `steady state' or stationary regime.
One has only the growing regime. 
In Ref.~\cite{Singha05}, Singha studied analytically the 
autocorrelation function of the height fluctuation $A_0(t_1,t_2)$ for linear interfaces in the radial geometry
and found that for large $t_2$ (with fixed $t_1$),
$A_0(t_1, t_2)$ approaches a constant. This is in contrast to the flat geometry, for instance for $1$-d flat geometry in 
Eq. (\ref{temporal.4a}), where it decays as a power law $t_2^{-(1-2\beta)}$ to zero as $t_2\to \infty$. 

Singha also studied the temporal
persistence probability $Q(0,t)$ of linear interfaces in the radial geometry and found~\cite{Singha05} that it decays as a 
power law for large $t$,
$Q(0,t)\sim t^{-\theta_r}$ but the exponent $\theta_r$ is different from $\theta_0$ of the flat geometry.
For nonlinear interfaces, due to the lack of $h\to -h$ symmetry, one needs to define as before a pair of
persistence probabilities, 
$Q^{\pm}(0,t)\sim t^{-\theta_r^{\pm}}$ with a pair of persistence exponents $\theta_r^{\pm}$. Simulations
on the on-lattice Eden model in 2-d show that $\theta_r^{+}=0.88\pm 0.02$ and $\theta_r^{-}=0.80\pm 0.02$~\cite{Singha05}.
Similar results were also obtained for the off-lattice Eden model simulations in 2-d~\cite{Takeuchi12}.
These radial exponents are thus considerably smaller than the corresponding exponents in the $(1+1)$ dimensions with 
a flat substrate~\cite{KK99}: $\theta_0^+=1.18\pm 0.08$
and $\theta_0^{-}=1.64\pm 0.08$.   
In a recent experiment on a growing circular droplet of turbulent phase in a nematic liquid crystal, these radial
persistence exponents were measured~\cite{TS12,Takeuchi12}: $\theta_r^{+}= 0.81(2)$ and $\theta_r^{-}=0.80(2)$.
Thus, both numerics as well as experiments suggest that within numerical or experimental precisions, the
positive and the negative persistence exponents in the circular geometry are very close to each other, 
i.e., $\theta_r^+\approx \theta_r^{-}$, in stark contrast
to the flat geometry. This is a rather surprising result because a priori one would expect $\theta_r^{+}\ne \theta_r^{-}$ due to the
absence of $h\to -h$ symmetry. Somehow in the circular geometry this lack of inversion symmetry does not
seem to have a significant effect on the persistence exponents~\cite{Takeuchi12}.

\section{Discrete persistence}\label{section:discrete}
Up to now, we have been focusing on processes that are continuous in time. 
In this section we want to discuss discrete-time processes. Such processes 
could either be intrinsically discrete, or they could represent data 
sampled at fixed time intervals from a continuous process. The latter could 
arise, for example, from experimental data that is sampled at a finite rate. 
As a by-product of this study we will find that the discrete sampling 
methodology will suggest a method (the ``correlator expansion'' 
\cite{Correlator}) for obtaining improved accuracy for the persistence 
exponent of continuous time processes, such as the diffusion equation. 

To be precise, we will consider a stationary stochastic process in continuous 
time $T$, sampled at discrete times $T_1, T_2, \ldots T_n=T$, separated by 
a uniform window size, $T_i-T_{i-1} = \Delta T$, such that $T=n\Delta T$ (Fig. \ref{fig:discrete}).
\begin{figure}[h]
\centering
\includegraphics[width=0.7\linewidth]{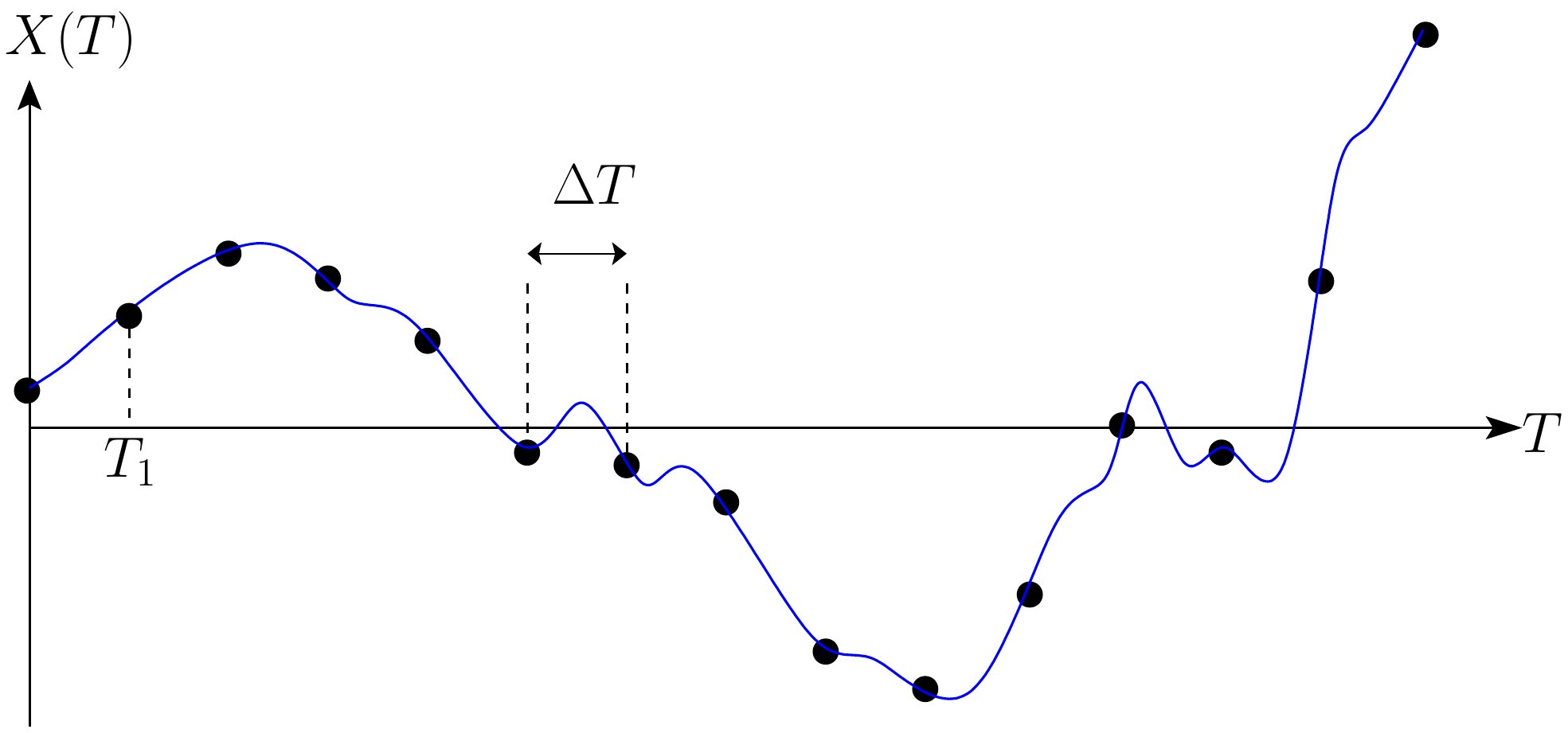}
\caption{Discrete persistence: the stationary stochastic process $X(T)$, which is continuous in time, is sampled at discrete times $T_1, T_2,É$, separated by a uniform time window $\Delta T = T_i - T_{i-1}$.}\label{fig:discrete}
\end{figure}
The persistence $Q(T)$ of the underlying continuous process is approximated 
as $Q(T) \approx Q_n$, where $Q_n$ is the probability that the process is 
positive at all $n$ discrete points. For non-infinitesimal $\Delta T$, 
$Q_n$ is different from $Q(T)$ because the process can change sign more 
than once between any two successive discrete times (Fig. \ref{fig:discrete}). This fact leads to the 
obvious inequality $Q(T) < Q_n$ for all $\Delta T > 0$. However, we note that
the approximation $Q(T) \approx Q_n$ will improve as $\Delta T$ decreases, 
and in the limit $\Delta T \to 0$, $n \to \infty$ keeping $T = n\Delta T$ 
fixed, $Q_n \to Q(T)$. In the opposite limit, $\Delta t \gg \tau$, where 
$\tau$ is the correlation time of the process, the stochastic variables 
at different discrete time points become uncorrelated, and $Q_n \to 2^{-n}$, 
since the value at each point is positive with probability $1/2$. The main  
question we want to address is how the discrete persistence $Q_n$ 
interpolates between its two limiting forms as the time increment $\Delta T$ 
increases from zero to infinity. This problem was first addressed in 
Ref.\ \cite{DiscreteTime}, where it was shown that, for any Gaussian 
Stationary Process, $P_n \sim [\rho(\Delta T)]^n$ for large $n$, where the 
function $\rho(\Delta T)$ is, in general, nontrivial, with the limiting 
behaviour
\begin{eqnarray}
\rho(\Delta T) \sim
\begin{cases}
&1 - \theta \Delta T, \ \ \ \Delta T \to 0\ , \\
&1/2, \ \ \ \ \ \ \ \ \Delta T \to \infty \;,
\end{cases}
\end{eqnarray}
where $\theta$ is the usual (i.e.\ continuous-time) persistence exponent. 
For $\Delta T \to 0$ we recover the continuous persistence,
$Q_n \to (1-\theta\Delta T)^n \to \exp(-\theta T)$, where $T=n\Delta T$. 
The general goal is to compute the function $\rho(\Delta T)$.  

As a simple example we consider a stationary Gaussian Markov process, 
namely the Ornstein-Uhlenbeck process
\begin{equation}
\frac{dX}{dT} = -\mu X + \eta(T) \;,
\label{LangevinMBE}
\end{equation}
where, as usual, $\eta(T)$ is Gaussian white noise with mean zero and 
correlator $\langle \eta(T)\eta(T') \rangle = 2D\delta(T-T')$. This process 
corresponds to the heavily damped motion of a particle moving in the 
potential $V(x) = \mu X^2/2$. The probability, $Q(X,T)$, that the particle 
has not crossed the origin up to time $T$, satisfies the Backward Fokker Planck (BFP) equation
\begin{equation}
\frac{\partial Q}{\partial T} = D \frac{\partial^2 Q}{\partial X^2} 
-\mu X \frac{\partial Q}{\partial X}\ ,
\label{ContinuousBFP}
\end{equation}
with boundary conditions $Q(0,T)=0$ and $Q(\infty, T) = 1$ for all $T$. 
The initial condition is $Q(X,0)=1$ for all $X>0$. The solution is
\begin{equation}
Q(X,T) = {\rm erf}\left[\frac{e^{-\mu T}X}
{\sqrt{2D'(1-e^{-2\mu T})}}\right]\ ,
\end{equation}
where $D'=D/\mu$.

For $\mu>0$ one obtains, for large $T$, the separable form 
$Q(X,T) \sim \exp(-\mu T)X$, an exponential decay with persistence 
exponent $\theta = \mu$.  To compute the discrete persistence, 
we will need the Green's function $G(X_2,T_2|X_1,T_1)$, which gives the 
probability that the particle, starting at $X_1$ at time $T_1$ reaches 
$X_2$ at time $T_2 > T_1$. This is readily computed from the Langevin equation 
(\ref{LangevinMBE}). The result is
\begin{equation}
G(X_2,T_2|X_1,T_1) = \frac{1}{\sqrt{2\pi D'(1-a^2)}}
\exp\left[-\frac{(X_2-aX_1)^2}{2D'(1-a^2)}\right],
\label{OUprop}
\end{equation}
where $a = \exp[-\mu(T_2-T_1)]$.

The discrete persistence $Q_n$ of the continuous process $X(T)$ in Eq.\ 
(\ref{LangevinMBE}) can be computed as follows. Let $Q_n(X)$ be the 
probability that, starting at $X$ at time $T=0$, the process $X(T)$ remains 
positive at all the  discrete times $T_1=\Delta T$, $T_2=2\Delta T$, \ldots, 
$T_n =n\Delta T$, with uniform window size $\Delta T$. The discrete 
persistence is then $Q_n = \int_0^\infty Q_n(X)P_0(X)\,dX$, where $P_0(X)$ 
is the distribution of the initial position of the particle. Using the 
Markov property of the process (\ref{LangevinMBE}) one can readily derive
the following recursion relation for $Q_n(X)$:
\begin{equation}
Q_{n+1}(X) = \int_0^\infty G(Y,\Delta T|X,0)\,Q_n(Y)\,dY\ ,
\label{integral-eigenvalue}
\end{equation}
where $G$ is the propagator given by Eq.\ (\ref{OUprop}), with 
$a=\exp(-\mu\,\Delta T)$ and $Q_0(X)=1$ for all $X>0$. This recurrence 
relation is the discrete analogue of the continuous BFP equation 
(\ref{ContinuousBFP}). 

The presentation can be simplified by introducing the rescaled variable 
$x=X/\sqrt{D'(1-a^2)}$. The recursion relation then reads
\begin{equation}
Q_{n+1}(x) = \frac{1}{\sqrt{2\pi}}\int_0^\infty\exp[-(y-ax)^2/2]\,Q_n(y)\,dy.
\end{equation}
Here we consider only the case $\mu>0$ (for a discussion of the case $\mu<0$ 
see Ref.~\cite{DiscreteTime}). The continuous case suggests that $Q_n(x)$ will
approach the form $Q_n(x) \to \rho^nq(x)$ for $n \to \infty$ at fixed $x$. 
Substituting this form into Eq.\ (\ref{integral-eigenvalue}) yields an 
integral-eigenvalue equation for $q(x)$:
\begin{equation}
\rho q(x) = \frac{1}{\sqrt{2\pi}}\int_0^\infty\exp[-(y-ax)^2/2]\,q(y)\,dy.
\label{eigenvalue-eq}
\end{equation}
We are interested only in the largest eigenvalue, since it determines 
the asymptotic behavior of $Q_n(x)$ for large $n$,  and we shall call 
this eigenvalue $\rho(a)$, since it depends continuously on $a$. 

Eq.\ (\ref{eigenvalue-eq}) has, of course, many eigenvalues, but we are 
interested only in the largest one since it dominates the large-$n$ 
behaviour of $Q_n(x)$. We first consider the limit $a \to 0$, equivalent 
to $\Delta T \to \infty$. For this case Eq.\ (\ref{eigenvalue-eq}) has 
the solution $q(x) = {\rm const.}$, with eigenvalue $\rho=1/2$, implying 
$Q_n(x) \to {\rm const.}\ 2^{-n}$ as expected, with the constant fixed as 
unity by the initial condition $Q_0(x) = 1$ for $x>0$. 

We now show how one can compute $\rho(a)$ perturbatively, as a formal power 
series in the quantity $\epsilon = 2a/(1+a^2)$. First we expand the 
factor $\exp(axy)$ from the exponential in Eq.\ (\ref{eigenvalue-eq}), 
and integrate term by term. This gives 
\begin{equation}
\rho q(x) = \frac{\exp(-a^2x^2/2)}{\sqrt{2\pi}}\,
\sum_{n=0}^{\infty} \frac{b_n}{\sqrt{n!}}(\sqrt{a}x)^n\ ,
\label{9}
\end{equation}
where
\begin{equation}  
b_n = \frac{a^{n/2}}{\sqrt{n!}}\int_0^\infty dy\,y^n\exp(-y^2/2)\,q(y)\ .
\label{10}
\end{equation}
Substituting Eq.\ (\ref{9}) into Eq.\ (\ref{10}) gives the matrix eigenvalue 
equation \cite{DiscreteTime}
\begin{equation}
\rho b_n = \sum_{m=0}^\infty A_{nm}b_m\ ,
\end{equation}
where the matrix elements $A_{nm}$ are given by
\begin{equation}
A_{nm} = \frac{1}{\sqrt{4\pi(1+a^2)}}\,
\epsilon^{(n+m)/2}\,
\Gamma[(n+m+1)/2)]/\sqrt{n!m!}\ ,
\end{equation}
where we recall that $\epsilon = 2a/(1+a^2)$.
This approach enables us to convert an integral eigenvalue equation
to a matrix eigenvalue equation, with matrix elements that decrease 
rapidly as $n$ and $m$ increase. Computing the largest eigenvalue of 
the $N \times N$ submatrix, ($n,m = 1,\ldots,N$), provides a rapidly 
converging sequence of estimates for that largest eigenvalue $\rho$. 
For a given $N$, the result is exact to ${\cal O}(\epsilon^{N-1})$.  
Table \ref{table_rho} gives estimates of the eigenvalue $\rho(a)$, correct to 12 decimal 
places, along with results obtained by numerical integration of 
Eq.\ (\ref{integral-eigenvalue}) correct to seven decimal places 
\cite{DiscreteTime}. 
\begin{table}
\begin{center}
\begin{tabular}{lll}
\hline
\hline
$a$ & \ \ $\rho_{\rm num}$ & \ \ $\rho_{\rm pert}$ \\
\hline
1.0 \ \ & 1 & \ \ 1 \\
0.8 \ \ & 0.852\,454\,7 & \ \ 0.852\,454\,696\,506 \\
0.6 \ \ & 0.740\,595\,9 & \ \ 0.740\,595\,939\,159 \\
0.4 \ \ & 0.647\,766\,6 & \ \ 0.647\,766\,585\,747 \\
0.2 \ \ & 0.5684903 & \ \ 0.568490321623 \\
0.0 \ \ & 1/2 & 1/2 \\
\hline
\hline
\end{tabular}
\end{center}
\caption{Estimates of the eigenvalue $\rho(a)$ 
for $0 \le a \le 1$ (data from Table 1 of Ref.\ \cite{DiscreteTime}).}\label{table_rho}
\end{table}
One can also use a variational approach to estimating $\rho(a)$, but this does 
not seem to be as accurate as the perturbative method \cite{DiscreteTime}. 

It is also possible to investigate discrete persistence for non-Markovian 
processes \cite{EBMnonMarkov}. 
For the simplest non-Markovian process -- the random acceleration 
process -- a  perturbative treatment along the lines used for the 
Ornstein-Uhlenbeck (OU) process is possible \cite{EBMnonMarkov}. The main 
additional complication is that the matrix $A_{nm}$ of the OU process 
becomes an object with four indices and the ``vectors'' $b_n$ of the OU 
process become objects with two indices. For general non-Markovian 
processes, approximation schemes such as the Independent Interval 
Approximation (see section \ref{subsection:iia}) have to be employed. We refer the 
interested reader to Ref.~\cite{EBMnonMarkov}.  

\subsection{The correlator expansion}

In this section, we discuss,
based on the discrete persistence idea discussed above,
yet another powerful approximation scheme to calculate the persistence exponent for a broad class of non-Markovian GSP's. The
same ideas can be used for non-stationary processes using the log-time
transformation (Lamperti transformation) introduced earlier in this work.

We recapitulate the earlier discussion by first considering a non-stationary 
Gaussian variable $x(t)$, and recall that $x(t)$ may be mapped to a Gaussian 
stationary process (GSP) for the variable 
$X(T) = x(t)/\sqrt{\langle x^2(t)\rangle}$ via the log-time transformation 
$T= \ln t$. Then the persistence $Q(T)$ has the asymptotic form 
$Q(T) \sim \exp(-\theta T)$ for large $T$, and the correlator 
$A(T) = \langle X(T)\,X(0)\rangle$ is normalized to unity at $T=0$.
We shall employ the ideas of discrete persistence introduced above, 
in which the process is sampled at discrete times, to derive a perturbative 
scheme for computing persistence exponents \cite{Correlator}. 
If the process $X(T)$ is sampled discretely, then  $X(T)$ may cross and 
recross zero between samplings, leading to an overestimate of the persistence. 
In other words, the persistence exponent, $\theta_{\rm D}$,  of the discrete process 
will be smaller than the continuum exponent $\theta$. If the process is sampled 
uniformly in $T$, we recall that the discrete persistence after $n$ samplings, 
$Q_n$, behaves for large $n$ as $Q_n \sim \rho^n$, where 
$\rho = \exp(-\theta_{\rm D} \Delta T)$. Recall that for $\Delta T \to 0$ we have 
$\rho \to 1$ and $\theta_{\rm D} \to \theta$, while for $\Delta T \to \infty$, 
$\rho \to 1/2$. Our approach \cite{EBMnonMarkov} is to develop a series 
expansion for $\rho = \exp(-\theta_{\rm D}\,T)$ in powers of the correlator, 
$A(\Delta T)$, between neighbouring discrete times. 

The expansion is based on the obvious identity
\begin{equation}
Q_n = \langle \prod_{i=1}^n \Theta[X(i\Delta T)]\rangle\ ,
\end{equation}
where $\Theta(x)$ is the Heaviside step function, and the expectation 
value is taken in the stationary state. Now we can write 
$\Theta(X[i\Delta T]) = (1+\sigma_i)/2$, where 
$\sigma_i = {\rm sgn}[X(i\Delta T)]$, and expand the product to 
give \cite{Correlator} 
\begin{equation}
Q_n = \frac{1}{2^n}\left(1 + \sum^n_{i<j}\langle \sigma_i\sigma_j\rangle
     +  \sum^n_{i<j<k<l}\langle \sigma_i\sigma_j\sigma_k\sigma_l\rangle
     + \cdots \right),
\end{equation} 
where the sums start at $i=1$ and terms with odd numbers of $\sigma$'s 
vanish by symmetry. The terms are evaluated using the representation
\cite{Correlator}
\begin{equation}
\sigma_k = \frac{1}{i\pi}\,\lim_{\epsilon \to 0}\int_{-\infty}^\infty
\frac{dz_k\,z_k\exp(iz_kX_k)}{(z_k-i\epsilon)(z_k+i\epsilon)}\ ,
\end{equation}
where $X_k = X(k\Delta T)$. Evaluating the averages of the Gaussian process 
gives the desired correlation functions:
\begin{equation}
\langle \sigma_{k_1}\ldots\sigma_{k_m}\rangle 
= \int\prod_{j=1}^m\left(\frac{dz_j}{i\pi z_j}\right)
\exp\left(-\frac{1}{2}z_\alpha A_{\alpha\beta} z_\beta\right),
\label{correlator1}
\end{equation}
where $C_{\alpha\beta} = \langle(X(\alpha\,\Delta T)X(\beta\,\Delta T) \rangle
= A(|\alpha - \beta| \Delta T)$. Here there is an implied summation over 
$\alpha$ and $\beta$ from 1 to $m$, and the $A$ is the autocorrelation 
function of the process, as usual. We have already taken the limit 
$\epsilon \to 0$, with the understanding that the integrals are now 
principal part integrals.   

We can now expand the exponential in Eq.\ (\ref{correlator1}) in powers 
of $C_{\alpha\beta}$ (for $\alpha \ne \beta$), leaving the terms with 
$\alpha=\beta$ unexpanded (recalling that $C_{\alpha\alpha}=1$). 
Ref.\ \cite{Correlator} shows how the terms in the expansion can be 
represented diagrammatically. The organisation of the terms in the expansion 
is informed by the fact that for most physical processes (including all those 
considered below) the correlator $A(q\Delta T)$ decreases exponentially 
at large argument. Effectively we are carrying out a large $\Delta T$ 
expansion, so the correlator $A(\Delta T)$ is small and, because of the 
exponential decay at large $\Delta T$, we can treat, in the expansion, 
$A(\Delta T)$ as first order and $A(q\Delta T)$ as $q$th order. 

In Ref.\ \cite{Correlator} all terms up to 14th order in the correlator 
(i.e.\ including all combinations from $C(14 \Delta T)$ to $C(\Delta T)^{14}$),
are included, giving a series expansion for $\rho$. 
The series is summed with the aid of Pad\'e approximants. There are a number 
of subtleties here, and we refer the reader to Ref.\ \cite{Correlator} for 
the technical details. Table \ref{table_correlator} shows the results obtained by this method 
for the random acceleration process (which we regard as a test of the method) 
and the diffusive persistence for $d=1$, 2 and 3.

\begin{table}
\begin{center}
\begin{tabular}{l|l|l}
\hline
\hline
\ & Pad\'e & Numerical \\
\hline 
$\ddot{x} = \eta(t)$ & 0.2506(5) & 1/4 (exact) \\
1d diffusion & 0.1201(3) & 0.12050(5) \\
2d diffusion & 0.1875(1) & 0.1875(1) \\
3d diffusion & 0.237(1)  & 0.2382(1) \\
\hline
\hline
\end{tabular}
\end{center}
\caption{Persistence exponents for the random acceleration process and for the diffusion equation in space dimensions $d=1,2,3$ evaluated
within the correlator expansion \cite{Correlator}.}\label{table_correlator}
\end{table}

\section{Persistence in disordered systems}\label{section:disordered}

So far in this review we have focussed on the persistence in {\em pure} systems (no quenched disorder) and we have seen that quite 
generically the persistence probability $Q(t)$ decays as a power law to zero at long times, $Q(t)\sim t^{-\theta}$ where $\theta$ is the 
persistence exponent. In such pure systems, the relevant local field, such as the spin in $1$-d Ising model undergoing $T=0$ Glauber 
dynamics or the height of a fluctuating interface, changes sign infinitely often, albeit slowly. As we have seen in section \ref{section:higher_dim}, one exception to this generic behavior 
of persistence in pure systems occurs in the $T=0$ Glauber dynamics of Ising model on a $d$-dimensional lattice with 
$d>4$~\cite{Sta94} or for the $q$-state Potts model on a square lattice with $q>4$~\cite{DOS96}. This is due to the existence of the 
so called `blocked' configurations where the system gets trapped at late times and a finite fraction of spins never flip. As a result, 
the persistence $Q(t)$ approaches a finite nonzero constant $Q(\infty)$ as $t\to \infty$. Numerical simulations nevertheless suggest that 
$Q(t)-Q(\infty)$ still decays algebraically with time in such systems~\cite{DOS96}.

What happens to $Q(t)$ when one adds quenched randomness to the system? This is a natural question which has been studied
extensively over the past two decades in a variety of disordered systems. 
In disordered systems, $Q(t)$ will of course vary from one realization of disorder to another. So, one is
interested in the disorder averaged persistence ${\overline {Q(t)}}$ where the overline 
denotes the average over disorder. 
Newman and Stein~\cite{NS99} studied ${\overline {Q(t)}}$ 
in disordered spin systems, e.g., random ferromagnets and spin glasses, undergoing $T=0$ Glauber dynamics
and found several interesting results. 
They showed that generically disorder also leads to `blocking' in metastable configurations
and in many cases, a finite fraction of spins cease to flip, indicating that ${\overline {Q(t)}}$ tends to
a nonzero constant ${\overline {Q(\infty)}}$ as $t\to \infty$. Moreover, in many cases, such as in
the zero field Ising chain where the nearest neighbour couplings $J_{i,i+1}$'s are i.i.d random variables each drawn
from a uniform distribution on $[0,1]$, it was shown that ${\overline {Q(t)}}-{\overline {Q(\infty)}}$
decays to zero exponentially~\cite{NS99}. Numerical simulations in $2$-d strongly diluted random bond Ising  
model on a square lattice confirmed this exponential decay~\cite{Jain99}.

Unfortunately, more explicit analytical results are not available currently for such disordered spin systems.
However, there are few other systems with quenched disorder where the disorder averaged persistence, though highly nontrivial, can
still be computed analytically. In the rest of the section, we will discuss three such examples in some detail.
We will first consider the case of a single particle in one dimension diffusing in a random Brownian potential--the Sinai model
where detailed analytical results can be obtained by a variety of techniques. Next, we will consider a single
particle diffusing in a random layered velocity field--the Matheron- de Marsily model. Finally, we will consider
an extended object, namely a Rouse polymer chain, diffusing in the Matheron-de Marsily velocity fields and we will
see that the persistence of a tagged monomer of this polymer chain can be computed analytically.

\subsection{Persistence in the one dimensional Sinai model}

The Sinai model~\cite{Sinai82} is perhaps one of the simplest one dimensional models with quenched disorder that exhibits a 
rich variety of dynamical behaviors and yet, many nontrivial disorder averaged dynamical observables can
still be computed analytically (for a review see \cite{BCGL90,BG90}). 
Consider a single Brownian particle diffusing on a 
line in presence of an external quenched (time independent) random potential $U(x)$ 
\begin{equation}
\frac{dx}{dt}= - U'(x(t)) + \eta(t) \;,
\label{sinai.1}
\end{equation}
where $\eta(t)$ is a Gaussian white thermal noise with $\langle \eta(t)\rangle=0$ and $\langle \eta(t)\eta(t')\rangle = \delta(t-t')$
and $U'(x)= dU/dx$. In the Sinai model (i.e., the continuous space-time version of the model), 
one chooses the potential $U(x)=\sqrt{\sigma}\, B(x)$ where $B(x)$ represents the trajectory of a
Brownian motion in space, i.e, $B(x)= \int_0^{x} \xi(x')\, dx'$ where $\xi(x)$ is a Gaussian noise with
$\langle \xi(x)\rangle=0$ and $\langle \xi(x)\xi(x')\rangle = \delta (x-x')$ and $\sigma$ just represents the strength
of the disorder. 

In the Sinai model, one has increasingly large barriers and wells in the system since $U(x)\sim \sqrt{|x|}$ for large $|x|$ (see Fig. \ref{fig:sinai}). 
\begin{figure}
\centering
\includegraphics[width=0.6\linewidth]{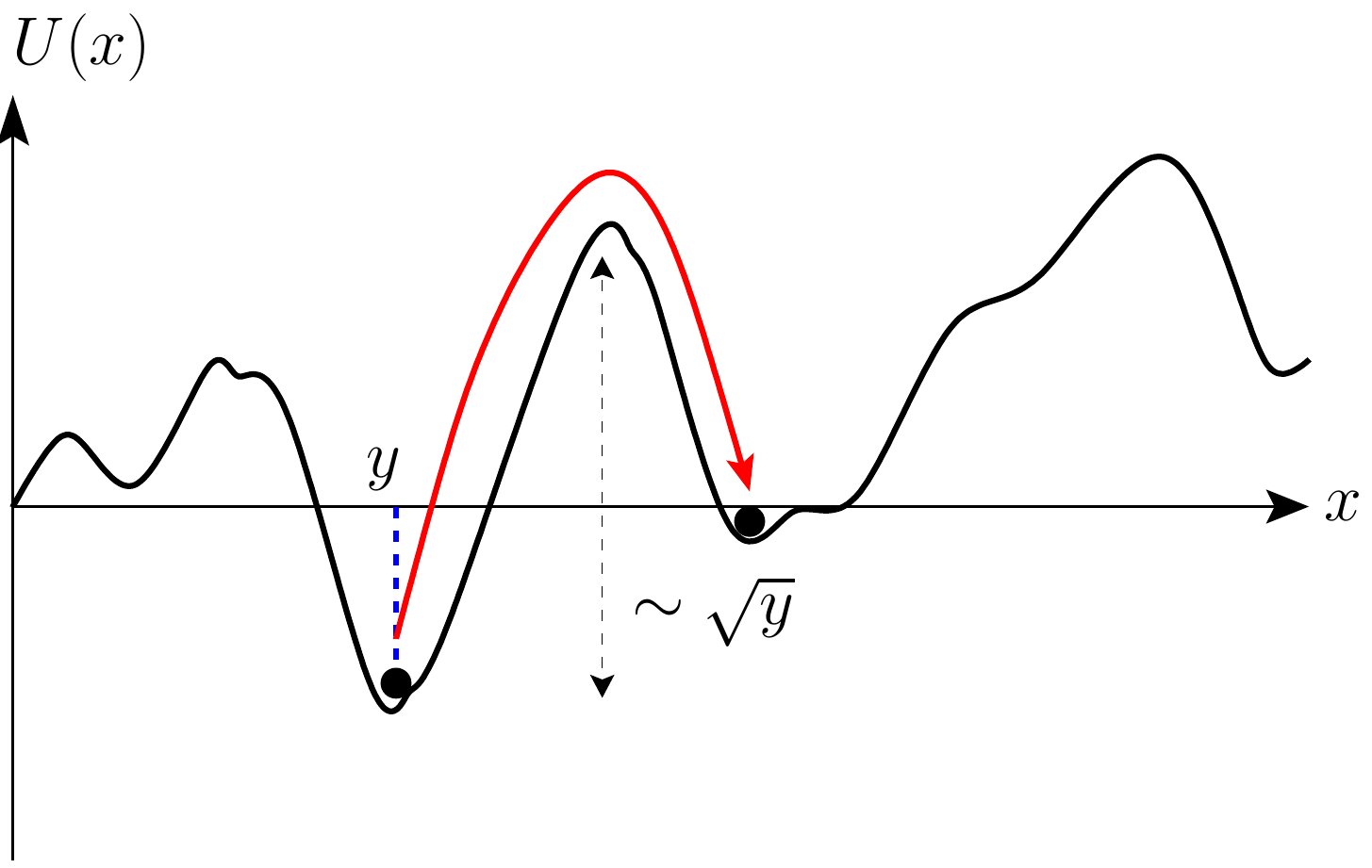}
\caption{Ilustration of the dynamics of a particle in a random potential where the potential $U(x)$ is itself a Brownian motion, the Sinai model (here without drift).}\label{fig:sinai}
\end{figure}
Thus the
particle often gets trapped in local minima and to move, it has to cross increasingly large barriers, leading to an
extremely slow dynamics~\cite{BCGL90,BG90}.
A simple Arrhenius type argument shows that the typical time scale to cross a barrier is, $t\sim \exp[U(x)]\sim \exp[\sqrt{|x|}]$,
indicating that the distance scales ultra slowly with time, $x\sim (\ln t)^2$. 
Consequently many dynamical observables in the Sinai model exhibit anomalous time dependence~\cite{BCGL90,BG90}.
A variety of techniques have been developed to compute different dynamical 
properties in the Sinai model analytically~\cite{BCGL90,BG90,OR2009}. More recently, 
a powerful real space renormalization group technique has been developed~\cite{FLM98,FLM99,IM05} which, besides reproducing already known results,
also gives access to the analytical computation of highly nontrivial quantities.

Let $Q(x_0,t)$ denote, for any fixed realization of the random potential, the persistence probability that the particle starting
at $x_0>0$ does not return to the origin up to time $t$. One is then interested in the disorder averaged persistence
${\overline {Q(x_0,t)}}$ and in particular, its late time properties. 
Comtet and Dean~\cite{ComtetDean98} first computed ${\overline {Q(x_0,t)}}$ for the Sinai model using 
an exact probabilistic approach and found that for large $t$ and fixed $x_0$
\begin{equation}
{\overline {Q(x_0,t)}} \sim   \frac{2\sigma x_0}{\ln t}\, .
\label{sinai.2}
\end{equation} 
Thus, instead of a power law decay with time, it decays anomalously slowly as an inverse logarithm. 
This inverse logarithmic decay was also found in the lattice version of the model
where the hopping rates of a particle are random quenched variables~\cite{IR98,IR99,ITR99}.
In addition, the same exact result was also found from the analysis of the renormalization
group mentioned earlier~\cite{FLM98,FLM99}.
Another interesting quantity is the so called `average persistence'~\cite{IR99} where 
one considers the thermally averaged trajectory of the particle $\langle x(t)\rangle$
(which is deterministic in time for a given disorder realization but the trajectory varies
from one disorder realization to another)
and asks what is the probability (as one varies disorder) that the process $\langle x(t)\rangle$
does not cross zero up to time $t$. In Ref.~\cite{IR99}, this average persistence was numerically found to
decay for large $t$   
as $(\ln t)^{-{\overline {\theta}}}$ with ${\overline {\theta}}=0.191\pm 0.002$. This quantity was also shown to be related to
the magnetization autocorrelation function of a transverse-field Ising chain~\cite{IR99}.

One interesting question is whether one can compute these disorder averaged persistence probabilities for
other types of random potentials apart from the Sinai one. Recently, using an interesting connection to extreme value statistics
followed by robust scaling arguments, the result in 
Eq. (\ref{sinai.2}) 
has been generalised to other self-affine random potentials in one dimension~\cite{MRZ2010}.
As an example, for a potential satisfying $U''(x)=\xi(x)$ (the spatial trajectory of a particle undergoing random acceleration), 
the result of Ref.~\cite{MRZ2010} predicts that
${\overline {Q(x_0,t)}} \sim   (\ln t)^{-1/6}$ for large $t$.

Another interesting generalisation is to the Sinai model as in Eq. (\ref{sinai.1}) but in presence of an additional 
constant drift $\mu$, positive or negative. In other words, the random potential in Eq. (\ref{sinai.1}) is chosen to be, $U(x)= -\mu\, x + \sqrt{\sigma}\, B(x)$ where $B(x)$, as before,
is a Brownian motion in space. The presence of a nonzero drift and the interplay between the drift and disorder qualitatively 
changes the persistence properties
of the particle. One would expect that for positive drift $\mu>0$ away from the origin, the particle, starting at $x_0>0$  will
eventually escape to $+ \infty$ with a finite probability. In contrast, for negative drift $\mu<0$, the particle
starting at $x_0>0$ will definitely cross the origin at some point. It turns out that the theoretical methods developed for the driftless
Sinai model discussed above can be generalised to the drifted case, but only in the limit of {\em vanishing drift}, i.e., when
$\mu\to 0$. For finite $\mu$ these methods can not be easily adapted. However, in Ref.~\cite{MC2002} an alternative backward
Fokker-Planck approach was developed that allowed exact analytical computation of the disorder averaged
persistence ${\overline {Q(x_0,t)}}$ for arbitrary drift $\mu$. In this approach, an exact mapping was found
to a quantum mechanical problem which happened to be integrable~\cite{MC2002}. 
The computation of the disorder averaged persistence required a knowledge of the full spectrum
of the quantum mechanics problem~\cite{MC2002}. 
We skip the details here and just summarize the
main results below. Interested readers may consult Ref.~\cite{MC2002} for details.

\vskip 0.3cm

{\noindent {\bf Positive drift ($\mu>0$):}} In this case, the particle, starting at $x_0>0$, escapes to $+ \infty$
with a finite probability before crossing the origin. In other words, the disorder averaged persistence
${\overline {Q(x_0,t)}}$ approaches a time-independent value as $t\to \infty$. Since this time-independent
value ${\overline {Q(x_0)}}\equiv {\overline {Q(x_0,\infty)}}$ is actually a function of the starting position $x_0>0$, 
we will call this the `persistence profile'. In Ref.~\cite{MC2002} the persistence profile ${\overline {Q(x_0)}}$
was computed exactly for arbitrary positive drift $\mu$. The expression for general $x_0$ is a bit cumbersome
involving special functions. However, the asymptotic behavior for large $x_0$ turns out to be
rather simple and revealing. Defining the ratio $\nu=\mu/\sigma>0$ one obtains~\cite{MC2002}
\begin{eqnarray}
1-{\overline {Q(x_0)}}\sim \left\{\begin{array}{ll} \frac{\nu-2}{\nu-1}\, e^{-2(\nu-1)\,\sigma\, x_0}, \quad\quad \nu>2 \\
&\\
\frac{1}{2\pi \sigma x_0}\, e^{-2\,\sigma\, x_0}, \quad\quad \nu=2 \\
&\\
\frac{A_{\nu}}{(2\sigma x_0)^{3/2}}\, e^{-\nu^2 \sigma\, x_0/2}, \quad\quad 0<\nu<2 \;,
\end{array}
\right.
\label{sinai_drift.1}
\end{eqnarray}
where $A_{\nu}= \pi^{3/2} \Gamma^2(\nu/2)/[\Gamma(\nu)\, (1-\cos \nu \pi)]$.
Evidently, the decay rate associated with the asymptotic exponential shape of the profile in Eq. (\ref{sinai_drift.1}) for large $x_0$ 
changes abruptly as $\nu$ goes through the `critical' value $\nu=\nu_c=2$. The origin of this `phase transition' at $\nu_c=2$
can be traced back to the fact that in the underlying quantum problem, the spectrum has bound states as well as scattering
states for $\nu>2$, while only scattering states for $\nu<2$~\cite{MC2002}. Thus the criticality at $\nu_c=2$ is triggered
by the loss of bound states as $\nu$ decreases from $\nu>2$ to $\nu<2$. Note that this criticality at $\nu_c=2$ could not be derived
by other methods such as the real space renormalization group method. In the limit $\nu\to 0$, the exact result in Ref.~\cite{MC2002}
coincides with the results of the renormalization group calculation~\cite{FLM98,FLM99}.

\vskip 0.3cm  

{\noindent {\bf Negative drift ($\mu<0$):}} In this case, one expects the disorder averaged persistence to decay
with time $t$. The physics is quite different from the $\mu>0$ case. 
In Ref.~\cite{MC2002}, the asymptotic properties of ${\overline {Q(x_0,t)}}$
for large $t$ was computed exactly.  
Once again, the ratio $\nu'= -\mu/\sigma>0$ 
plays the role of a control parameter and one finds very different behaviors for $\nu'>1$ and $\nu'<1$.
Summarizing
\begin{eqnarray}
{\overline {Q(x_0,t)}}\sim \left\{\begin{array}{ll} \theta\left(\frac{x_0}{\sigma(\nu'-1)}-t\right), \quad\quad \nu'>1 \\
&\\
\theta\left(\frac{x_0\, \ln x_0}{\sigma}-t\right), \quad\quad \nu'=1 \\
&\\
\frac{x_0}{t^{\nu'}}, \quad\quad 0<\nu'<1 \;,
\end{array}
\right.
\label{sinai_drift.2}
\end{eqnarray}
where $\theta(z)$ is the Heaviside step function: $\theta(z)=1$ for $z>0$ and $\theta(z)=0$ for $z<0$.
Thus, for $\nu'>1$ (where the drift overwhelms the disorder), the particle essentially moves ballistically towards the origin
and crosses the origin at a finite time $t= x_0/[\sigma(\nu'-1)]$. At the `critical' point $\nu'=1$, there is an additional
logarithmic dependence on $x_0$ of this time scale. But the scenario changes abruptly for $0<\nu'<1$ where
the persistence decays as a power law in time (as in a pure system) with a persistence exponent $\theta=\nu'= -\mu/\sigma$.
This `phase transition' at $\nu'=1$ arises essentially due to the competition between drift and disorder and thus
has a different physical origin than the transition at $\nu=2$ for positive drift discussed above.

Interestingly, ${\overline {Q(x_0,t)}}$ can be computed exactly~\cite{MC2002} for all $x_0>0$ and $t$ for the special value $\nu'=1/2$,
where ${\overline {Q(x_0,t)}}= {\rm erf}(x/\sqrt{2t})$ where ${\rm erf}(z)= (2/\sqrt{\pi})\, \int_0^{z} e^{-u^2}\, du$.
But this is precisely the exact answer for a pure Brownian motion with {\em zero drift and zero disorder}. Thus, it seems,
somewhat strangely, that at this special value $\nu'=1/2$, the effect of disorder and drift somehow exactly cancel each other~\cite{MC2002}.
A similar coincidence at this special value $\nu'=1/2$ was also noted in the context of the computation of 
other observables in the Sinai model, such as the distribution of the occupation time, i.e., the fraction of time
of the interval $[0,t]$ the particle spends on the positive side starting at the origin~\cite{MC_occup2002,SMC_occup2006}.
It would be interesting to further explore the deep reason behind this exact cancellation of drift and disorder
at the special value $\nu'=1/2$, i.e., $\mu=-\sigma/2$.  
 
Finally, in the zero drift limit, the method developed in Ref.~\cite{MC2002} also reproduces the already known results
for the persistence in the driftless Sinai model. 
To summarize, for the Sinai model with arbitrary drift $\mu$, the asymptotics of the disorder averaged persistence undergo
interesting `phase transitions' at the critical values $\mu=2\sigma$ (i.e., $\nu=2$), $\mu=0$ and $\mu=-\sigma$ (i.e., $\nu'=1$)
and somewhat strangely, at the special value $\mu=-\sigma/2$ (i.e., $\nu'=1/2$), the disorder averaged persistence
coincides with the result for the pure case without drift.

\subsection{Persistence of a particle in the Matheron-de Marsily velocity field}

The Matheron-de Marsily (MdM) model, orginally introduced to study the hydrodynamic 
dispersion of a tracer particle in porous rockes~\cite{Matheron1980}, provides
perhaps one of the simplest and rare settings where the persistence probability of
a particle can be computed analytically in a quenched disordered system. 
In the original $(1+1)$-dimensional version of the MdM model, a single
particle diffuses in a layered medium with one transverse ($x$) and one longitudinal ($y$) direction.
While the motion along the transverse $x$ direction is purely Brownian, along the 
longitudinal $y$ direction the particle is advected by a drift velocity $v(x)$ that
is a `quenched' random function of only the transverse coordinate $x$ (see Fig. \ref{mdm.fig1}).  
Even though the velocities in the different $x$ layers are uncorrelated, the motion along 
the $y$ direction gets correlated in time due to the multiple visits to the same 
transverse layer by the particle in a given time $t$. This generates a typical bias in
the $y$ direction giving rise to a super-diffusive longitudinal transport where 
typically
the $y$ coordinate grows with time as $y\sim t^{3/4}$ for large 
$t$~\cite{Bouchaud1990,BG90,Redner1990,Zumofen1990}.

The original $(1+1)$-dimensional MdM model can be easily generalised to $(d+1)$ 
dimensions with $d$ transverse and one longitudinal directions. Let $x_i$ 
$(i=1,2,\ldots,d)$ denote the transverse coordinates of the particle while
$y$ denotes its longitudinal coordinate. The transverse coordinates perform
ordinary Brownian diffusion
\begin{equation}
{\dot x}_i=\eta_i(t) \;,
\label{xeq_mdm.1}
\end{equation}
where $\eta_i$'s are standard zero mean Gaussian white noises with correlators, $\langle 
\eta_i(t)\eta_j(t')\rangle= \delta_{i,j}\delta(t-t')$. The longitudinal coordinate 
$y(t)$, 
in contrast, is driven by a random drift $v[{\bf x}(t)]$ that depends only on the 
transverse coordinates ${\bf x}(t)= \{ x_i(t) \} $,
\begin{equation}
{\dot y}= v[{\bf x}(t)] +\xi(t) \;,
\label{yeq_mdm.1}
\end{equation}
where $\xi(t)$ is again a delta correlated zero mean Gaussian white noise and is 
uncorrelated to the noises $\eta_i(t)$'s. The velocity field $v[{\bf x}]$ is quenched, 
i.e., for a given realization of the function $v$, one first evolves $y(t)$ via
Eq. (\ref{yeq_mdm.1}) and then one needs to `disorder average' (denoted by overline)
over different realizations of the random function $v$. We will choose  
$v[\bf x]$ to be a zero mean Gaussian random field with a short-ranged correlator
\begin{equation}
{\overline {v[{\bf x_1}]v[{\bf x_2}]}}= \frac{1}{(2\pi d a^2)^{d/2}}\, 
\exp\left[-\frac{({\bf {x_1}}-{\bf {x_2}})^2}{2da^2}\right] \;,
\label{vcorr_mdm.1}
\end{equation}
where the short-distance cut-off $a$ represents the correlation length of the velocity 
field in the transverse direction. Physically, this mimics the fact that velocity layers 
have a finite thickness of width $a$. For $d<2$, one can safely take the $a\to 0$ limit
and recover the delta correlated disorder in the original MdM model. However, a nonzero 
cut-off is necessary for $d>2$ in order for the MdM model in continuum space to be well 
defined. For $d>2$, the {\em point} particle will feel the velocity fields in
the transverse layers provided the layers have a nonzero thickness. The choice of a
Gaussian function in Eq. (\ref{vcorr_mdm.1}) just makes the computations simple, but in 
principle one can choose any short range function in Eq. (\ref{vcorr_mdm.1}). The
persistence properties to be discussed below are independent of this choice.
\begin{figure}
\centerline{\includegraphics[width=0.55\linewidth]{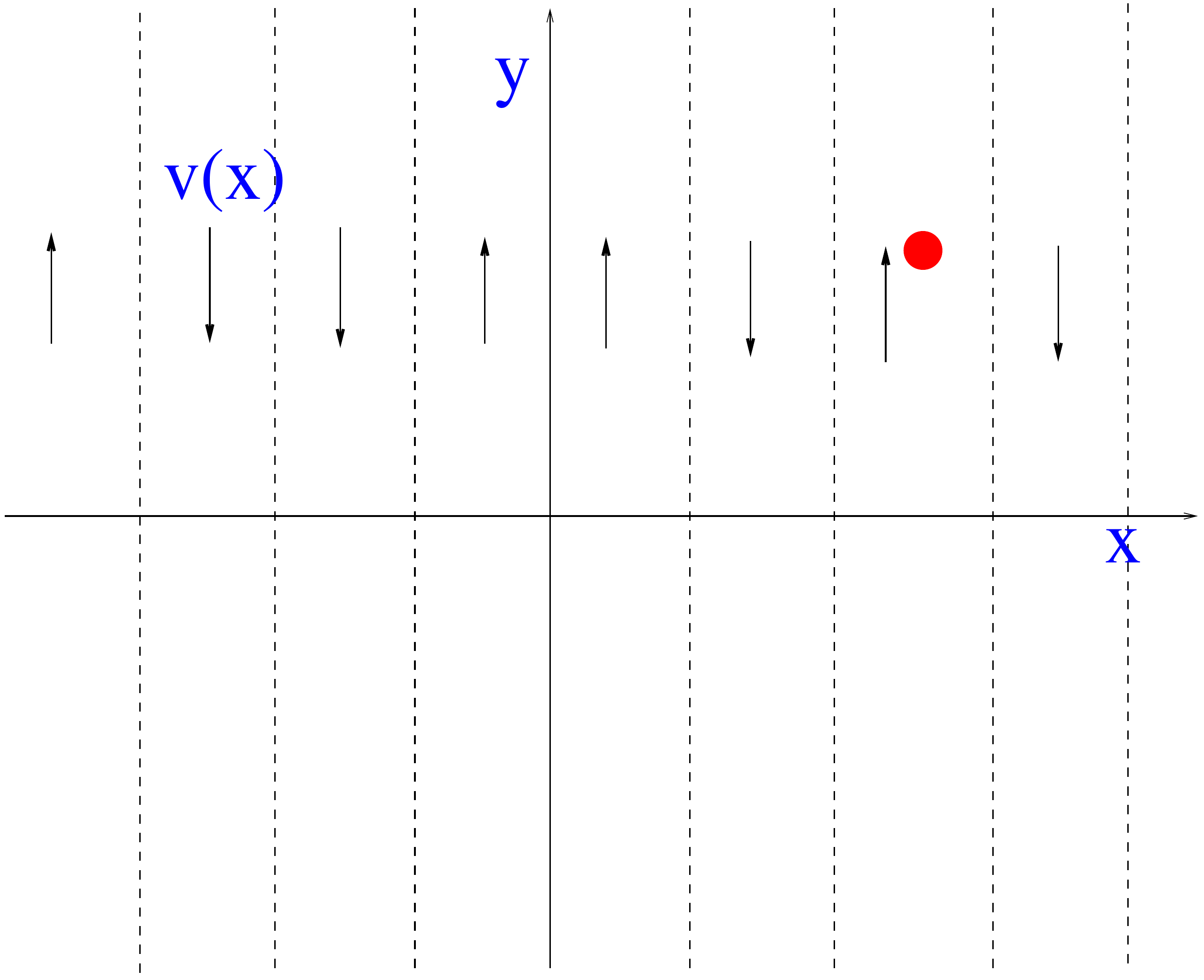}}
\caption{Schematic representation of the MdM model in $(1+1)$-dimensions.
The filled circle (red) represents the particle which
performs ordinary Brownian motion in the $x$-direction, while
in the $y$-direction it gets advected by a random velocity field 
$v(x)$ (represented by arrows)
which depends only on the transverse coordinate $x$ of the particle.}
\label{mdm.fig1}
\end{figure}

The persistence properties of the particle in the transverse directions are trivial
since they represent ordinary Brownian motions. The interesting quantity is
the persistence in the longitudinal direction. More precisely, one defines the
persistence probability $Q(t)$ as the probability that the $y$ coordinate of the
particle does not return to its initial value up to time $t$. Of course, $Q(t)$
will vary from one realization of disorder (the velocity field) to another.
The suitable quantity then is the disorder averaged persistence ${\overline {Q(t)}}$.
This quantity was first studied by Redner~\cite{Redner1997}. Using heuristic physical 
arguments supported by numerical simulations, Redner conjectured the following rather
rich asymptotic behavior~\cite{Redner1997}
\begin{eqnarray}
{\overline {Q(t)}}\sim \left\{\begin{array}{ll} t^{-d/4}, \quad\quad 0<d<2 \\
&\\
({\ln t}/t)^{1/2}, \quad\quad d=2 \\
&\\
t^{-1/2}, \quad\quad d>2 \;,
\end{array}
\right.
\label{redner_mdm.1}
\end{eqnarray}
indicating that $d=2$ is a critical dimension below which the $y$ coordinate of 
the particle survives longer (compared to ordinary diffusion) due to the presence of the
random velocity field. 

The arguments used by Redner in deriving these asymptotic results, though physically 
intuitive, were heuristic. A more rigorous derivation of these results were provided in 
Ref.~\cite{SM2003} where it was shown that the stochastic process $y(t)$, representing 
the longitudinal coordinate of the particle in this $(d+1)$-dimensional MdM model
can be identified as a fractional Brownian motion (fBM). While this process 
is non-Gaussian, it shares one crucial property with the standard Gaussian fBM (defined
in section \ref{section:fBm}), namely its incremental correlation function 
$C(t_1,t_2)=E\left[\left(y(t_1)-y(t_2)\right)^2\right]$ is {\em stationary}, i.e., 
depends only on the time difference $|t_1-t_2|$ and grows as a power law for large
$|t_1-t_2|$, $C(t_1,t_2)\sim |t_1-t_2|^{2H}$ with the Hurst exponent $0<H<1$ that
depends only on $d$~\cite{SM2003}. As argued in section \ref{section:fBm}, only this property is enough
to deduce that the persistence exponent $\theta=1-H$, even if the process is
non-Gaussian. Using this property, one can then derive precisely the
results in Eq. (\ref{redner_mdm.1}), as we briefly outline below.

Integrating Eq. (\ref{yeq_mdm.1}) one gets
\begin{equation}
y(t)= \int_0^{t} \xi(\tau)d\tau + \int_0^t v[{\bf x}(\tau)]\, d\tau \;,
\label{xeq_mdm.2}
\end{equation}
where we assume $y(0)=0$. To relate this process to fBM, one needs to compute
the expectation value $E\left[\left(y(t_1)-y(t_2)\right)^2\right]$ where $E$
denotes an average over all realizations of $y(t)$ arising from thermal noises as well 
as 
the disorder, i.e., $E[\ldots]\equiv {\overline {\langle \ldots\rangle}}$.
Before we do that, it is first instructive to compute the expected correlator
$E\left[y(t_1)y(t_2)\right]$.  
Using Gaussian properties of the velocity field, this 
correlator can be computed in a straightforward manner.
Omitting 
details~\cite{SM2003},
one finds that for all $d\ne 2, 4$ and for all $t_1$ and $t_2$
\begin{equation}
E\left[y(t_1)y(t_2)\right]= A\, {\rm min}(t_1,t_2) + B\, \left[(t_1+a^2)^{\beta}+ 
(t_2+a^2)^{\beta}- (|t_1-t_2|+a^2)^{\beta} - a^{2\beta}\right]
\label{ycorr_mdm.1}
\end{equation}
where $\beta=(4-d)/2$, $A=1- 4 a^{2-d}(2\pi d)^{-d/2}/(2-d)$ and $B=4 (2\pi 
d)^{-d/2}/(2-d)(4-d)$. In the limit $d\to 4$, $B$ diverges but $\beta\to 0$, and the 
second term just becomes a logarithm, but stays finite. Putting $t_1=t_2=t$ in 
Eq. (\ref{ycorr_mdm.1}), one gets the results for the variance for all $t\ge 0$
\begin{equation}
E\left[y^2(t)\right]= A\, t + 2\, B\, \left[(t+a^2)^{\beta}-a^{2\beta}\right].
\label{var_mdm.1}
\end{equation}
This result clearly demonstrates the role of the cut-off $a$ and the critical dimension 
$d=2$. For $d<2$, or equivalently $\beta=(4-d)/2>1$, the second term in 
Eq. (\ref{var_mdm.1}) dominates for large $t$, giving rise to a super-diffusion:
$E[y^2(t)]\approx 2\, B\, t^{(4-d)/2}$. The cut-off $a$ plays no role for $d<2$ and 
one 
can safely take the limit $a\to 0$ in Eq. (\ref{var_mdm.1}). In contrast, for $d>2$, i.e.,
$\beta<1$, the first term in Eq. (\ref{var_mdm.1}) dominates for large $t$ giving rise to 
normal diffusion, $E[y^2(t)]\approx A\, t$ where $A= 1- 4 a^{2-d}(2\pi 
d)^{-d/2}/(2-d)>0$
depends explicitly on the cut-off $a$. 

Using Eq. (\ref{ycorr_mdm.1}), the incremental correlation function can be computed 
explicitly for all $t_1, t_2\ge 0$
\begin{equation}
C(t_1,t_2)= E\left[\left(y(t_1)-y(t_2)\right)^2\right]= A\, |t_1-t_2|+ 2\, B\, 
\left[(|t_1-t_2|+a^2)^{2\beta}- a^{2\beta}\right].
\label{increcorr_mdm.1}
\end{equation}
Thus, in the limit $|t_1-t_2 \gg 1$, the correlator $C(t_1,t_2)$ again has two different 
asymptotic behavior depending on whether $d<2$ or $d>2$. In the former case, one has,
$C(t_1,t_2)\approx 2\, B\, |t_1-t_2|^{\beta}$ where $\beta=(4-d)/2$. Thus, the 
process, 
though non-Gaussian,
shares the same incremental correlator as a fBM with Hurst exponent $H=\beta/2= 1-d/4$.
For $d>2$, in contrast, the first term on the rhs of Eq. (\ref{increcorr_mdm.1}) 
dominates, indicating $C(t_1,t_2)\approx A\, |t_1-t_2|$ which then corresponds to a
fBM with $H=1/2$. This shows that for all $d\ne 2$, the longitudinal coordinate $y(t)$
of the particle is a generalised non-Gaussian fBM with a Hurst exponent
\begin{eqnarray}
H(d)= \left\{\begin{array}{ll} 1-d/4, \quad\quad 0<d<2 \\
&\\
1/2, \quad\quad d>2 \, .
\end{array}
\right.
\label{hurst_mdm.1}
\end{eqnarray}
It then follows immediately from the known first-passage property of the generalised fBM
discussed in section \ref{section:fBm} that the disorder averaged persistence $\overline {Q(t)}$ decays 
as a power law for large $t$, $\overline {Q(t)}\sim t^{-\theta(d)}$ where the persistence 
exponent~\cite{SM2003}
\begin{eqnarray}
\theta(d)= \left\{\begin{array}{ll} d/4, \quad\quad 0<d<2 \\
&\\
1/2, \quad\quad d>2 \, .
\end{array}
\right.
\label{pers_mdm.2} 
\end{eqnarray}

We now turn to the marginal case $d=2$ where the incremental correlator behaves 
as~\cite{SM2003}
\begin{equation}
C(t_1,t_2)= A'\, |t_1-t_2| + 2\, B'\, \left[(|t-1-t_2|+a^2)\ln (|t_1-t_2|+a^2)-a^2\ln 
(a^2)\right] \;,
\label{2d_mdm.1}
\end{equation}
where $A'$ and $B'$ are two computable constants.
Evidently, this correlator is stationary and for $|t_1-t_2| \gg1$, it has a power-law (with 
logarithmic correction) dependence with Hurst exponent $H=1/2$. The analytical scaling 
argument leading to the 
result $\theta=1-H$ discussed in section \ref{section:fBm} can be easily adapted to take into account
this additional logarithmic correction and one finds~\cite{SM2003},
${\overline {Q(t)}}\sim \sqrt{{\ln t}/t}$ for large $t$, thus recovering the
result in Eq. (\ref{redner_mdm.1}) for $d=2$.

In summary, one sees here that the longitudinal position $y(t)$ of a particle in a
$(d+1)$-dimensional MdM model can be exactly represented as a generalised non-Gaussian
fBM with a Hurst exponent $H(d)$ [as in Eq. (\ref{hurst_mdm.1})] that depends on
the dimension $d$. Moreover, this exact connection allows one to use the known 
first-passage results for the fBM and thus derive analytically the asymptotic behavior of 
the disorder averaged persistence in the MdM model that were only known before
via heuristic arguments and simulations. This technique of mapping to a generalised fBM, 
when it is true,
was also used before in section \ref{section:interfaces} in the context of fluctuating interfaces.
Thus this mapping, when valid, seems to be a rather powerful route for computing 
persistence exponents
for complex Gaussian or non-Gaussian processes where there is no other known method 
available for computing the persistence properties. We will see one more example
in the next section where the same method can be used successfully.  

Finally, there have also been parallel developments 
in the mathematics literature
in proving some of these results
in a strictly rigorous sense, see e.g. Refs.~\cite{Molchan99,Aurzada11,Castell12}. 

\subsection{Rouse chain in a Matheron-de Marsily layered medium: persistence of a tagged 
monomer}

In the previous section, we considered a single particle in the Matheron-de Marsily 
layered velocity field. As demonstrated in the previous section, this model is one 
of the rare solvable models with
quenched disorder where persistence exponents can be computed in all dimensions 
exactly by mapping the relevant process (the longitudinal coordinate of the particle)
to a generalised non-Gaussian fBM with a specific Hurst exponent $H$ and then using 
the known persistence property of the fBM, namely the relation $\theta=1-H$ connecting 
the
persistence exponent $\theta$ and the Hurst exponent $H$. 
However, in the standard MdM model, we have a single particle. One interesting 
question is: can one find a solvable model for an extended object with spatial
interaction in a quenched disordered system?

The original MdM model in $(1+1)$ dimensions for a single particle can be 
generalised to the case of an extended object like a
polymer chain with spatial interaction between the monomers of the chain.
The simplest case corresponds to the Rouse polymer chain where the nearest
neighbour monomers have harmonic interactions between them~\cite{Rouse53}. More 
specifically,
we consider a polymer chain embedded in a $(1+1)$ dimensional layered medium as 
before. The chain consists of $N$ beads or monomers connected by harmonic 
springs. In addition, the chain is advected by a random layered velocity field
as shown in Fig. \ref{rouse_chain.fig1}. Let $[x_n(t), y_n(t)]$ denote
the coordinates of the $n$-th bead at time $t$ which evolve
with time according to the following equations of motion
\begin{eqnarray}
\frac{dx_n}{dt} &=& \Gamma\left(x_{n+1}+x_{n-1}-2\,x_n\right) + \eta_1(n,t) \;,
\label{evolx1}\\
\frac{dy_n}{dt} &=& \Gamma\left(y_{n+1}+y_{n-1}-2\,y_n\right) + v\left(x_n(t)\right)+
\eta_2(n,t),
\label{evoly1}
\end{eqnarray}
where $\Gamma$ denotes the strength of the harmonic interaction between nearest
neighbour beads,
$\eta_1(n,t)$ and $\eta_2(n,t)$ represent the zero mean Gaussian thermal noises
along the $x$ and $y$
directions respectively. They are independent of each other and each is delta correlated in time.
The velocity field $v(x)$ is a random quenched function of $x$ taken to be
a Gaussian with the following
moments
\begin{eqnarray}
{\overline {v(x)}}&=& 0 \;, \\
{\overline {v(x)v(x')}}&=& \delta(x-x') \;.
\label{qn1}
\end{eqnarray}
For a finite chain with $N$ beads, the Eqs. (\ref{evolx1}) and (\ref{evoly1}) are
valid only for the $(N-2)$ interior beads. The two boundary beads will have
slightly different equations of motion. However, we will only focus here on an 
infinitely large chain ($N\to \infty$)
so that the system
is translationally invariant along the length of the chain and the boundary conditions 
are
irrelevant.
\begin{figure}
\centerline{\includegraphics[width=0.55\linewidth]{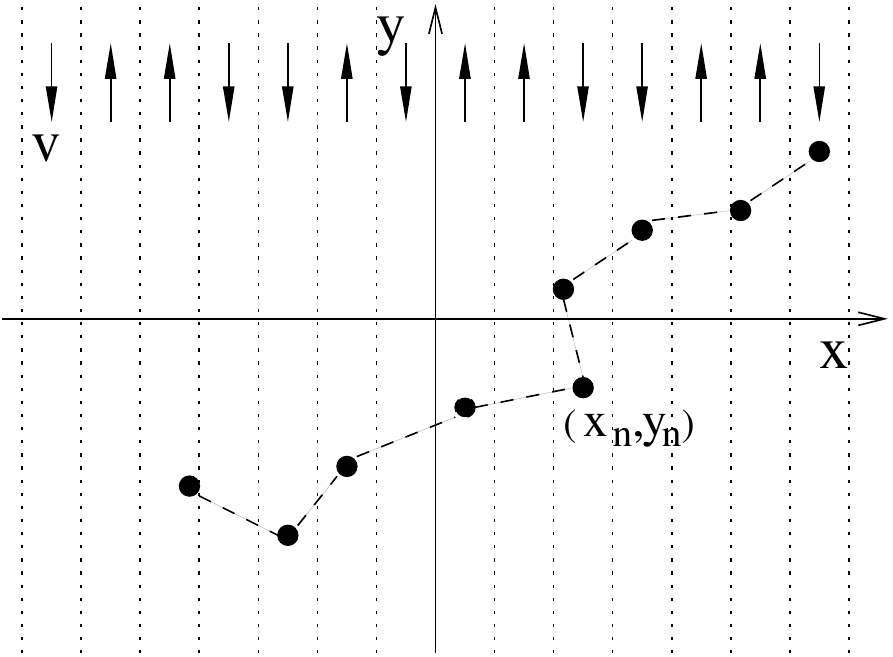}}
\caption{A Rouse chain in a random layered velocity field in $(1+1)$ dimensions.}
\label{rouse_chain.fig1}
\end{figure}

Note that in the absence of the harmonic interaction term, i.e., when $\Gamma=0$, this 
model reduces precisely to a single particle MdM model discussed in the previous section. In presence of the harmonic interaction, various transport properties 
of the chain in 
this
model had been studied previously~\cite{Oshanin94,Oshanin95,Wiese99,Jespersen01}. The
persistence properties of a single tagged monomer in this model was first studied
in Ref.~\cite{SMDD2005}. 
One can define the following persistence probabilities~\cite{SMDD2005},
\begin{eqnarray}
Q_1(t,t_0) &=& {\rm {Prob.}}\, \left[x_n(t')\ne x_n(t_0)
\,\, {\rm for\,\, all\,\,} t': \,\, t_0<t'<t_0+t \right] \;,
\label{perx1}\\
Q_2(t,t_0) &=& {\rm {Prob.}}\, \left[y_n(t')\ne y_n(t_0)
\,\, {\rm for\,\, all\,\,} t': \,\, t_0<t'<t_0+t \right] \:,
\label{pery1} 
\end{eqnarray}
where the former represents the probability that the
$x$ coordinate of a `marked' bead or a `tagged' monomer (say the $n$-th bead)
does not return to its position at time $t_0$ within the
time interval $[t_0, t_0+t]$, while the latter
represents the same probability for the $y$ coordinate of the same bead.
For an infinite chain, the system is
translationally invariant along the length of the chain and
hence these persistence probabilities
do not depend on the bead label $n$. Note that in the 
second case, by $Q_2(t,t_0)$ we mean the already disorder
averaged persistence, i.e., $Q_2(t,t_0)\equiv {\overline {Q_2(t,t_0)}}$.
We use the simple notation $Q_2$ just for convenience.

Since we are interested in the late time properties, one can conveniently
replace the harmonic interaction term in
Eqs. (\ref{evolx1}) and (\ref{evoly1})
by a continuous Laplacian operator
\begin{eqnarray}
\frac{\partial x}{\partial t}&=& \frac{\partial^2 x}{\partial s^2} +
\eta_x(s,t) \;, \label{evolx2}\\
\frac{\partial y}{\partial t}&=& \frac{\partial^2 y}{\partial s^2} +
v\left[x\left(s,t\right)\right]
+\eta_y(s,t),
\label{evoly2}
\end{eqnarray}
where $s$ denotes the distance along the chain and we have rescaled the time to
set the coefficient in front of the Laplacian to be unity. Interpreting $x\equiv h_1$
and $y\equiv h_2$, the equations (\ref{evolx2}) and (\ref{evoly2}) reduce
to the evolution equation of two coupled one dimensional interfaces with heights $h_1$ 
and $h_2$ respectively~\cite{SMDD2005}.

Thus, the persistence $Q_1(t,t_0)$ is precisely the temporal persistence probability
of the $1$- dimensional Edwards-Wilkinson interface discussed in section \ref{section:interfaces}.
In contrast, $Q_2(t,t_0)$ represents the temporal persistence of 
a second interface that is driven by the first one through the coupling term
$v[x(s,t)]$ in Eq. (\ref{evoly2}). As discussed in section \ref{section:interfaces}, the late time behavior 
of $Q_{1,2}(t,t_0)$ depends on whether one is in the {\em transient} regime 
(corresponding 
to the $t_0=0$ fixed point) or in the {\em stationary} regime ($t_0\to \infty$ fixed 
point). Thus, one can define two pairs of persistence exponents
\begin{eqnarray}
Q_1(t,t_0=0) &\sim & t^{-\theta_0^{1}}, \quad {\rm and} \quad Q_2(t,t_0=0)\sim 
t^{-\theta_0^{2}} \;, \label{trans_exp.1} \\
Q_1(t, t_0\to \infty) & \sim & t^{-\theta_s^{1}}, \quad {\rm and}
\quad Q_2(t,t_0\to \infty) \sim t^{-\theta_s^{2}} \;, \label{stat_exp.1}
\end{eqnarray}
where the superscripts $1$ and $2$ refer to the two coordinates $x$ and $y$ 
respectively, while the subscripts $0$ and $s$ refer respectively to transient
and stationary temporal persistence exponents. 

As discussed in section \ref{section:interfaces}, the transient exponents are harder to compute.
Even for the simple EW interface in Eq. (\ref{evolx2}), the exponent 
$\theta_0^1\approx 1.55\pm 0.02$ is known only numerically~\cite{KKMCBS97}.
Similarly, analytical calculation of $\theta_0^2$ seems also very hard.
In contrast, the stationary persistence exponents $\theta_s^1$ and $\theta_s^2$ can be 
computed exactly~\cite{SMDD2005}. In both cases, the strategy again is to map the relevant stochastic 
process to a generalised fBM with the property that its incremental correlation 
function is stationary and grows as a power law, $C(t_1,t_2)\sim |t_1-t_2|^{2H}$ for 
large $|t_1-t_2|$ with a calculable Hurst exponent $H$ and then exploiting the result
$\theta=1-H$ for the fBM. For the $x$ coordinate (pure EW interface), this was already 
discussed in section \ref{section:interfaces} and one finds that the corresponding fBM (Gaussian in this 
case) has $H=1/4$, implying $\theta_s^1=3/4$~\cite{KKMCBS97}. As mentioned in section \ref{section:interfaces}, this analytical 
result was verified experimentally in a system of fluctuating $(1+1)$-dimensional 
steps on Si-Al surfaces~\cite{Dougherty_exp02,Dougherty_exp03}. 
For the $y$ coordinate, one can follow a similar strategy and it was shown by exact 
computation
of the incremental correlation function that the $y$ process corresponds
to a fBM (non-Gaussian) with $H=7/8$, implying the exact result
$\theta_s^2= 1/8$~\cite{SMDD2005}. 
Summarizing, for $(1+1)$-dimensional
Rouse chain in the MdM layers, the exact `stationary' persistence exponents
associated respectively with the $x$ and the $y$ coordinate of a tagged monomer
are given by
\begin{equation}
\theta_s^1= 3/4 \;, \quad {\rm and}\quad \theta_s^2= 1/8 \, .
\label{tagged_mon_exp.1}
\end{equation}  

This effective model of coupled fluctuating interfaces in $(1+1)$-dimensions were 
generalised
in Ref.~\cite{SMDD2005} to $(d+1)$-dimensions and it was shown that for
$d<2$, the stationary persistence exponents are given exactly by~\cite{SMDD2005}
\begin{equation}
\theta_s^1= (2+d)/4 \;, \quad {\rm and}\quad \theta_s^2= (2-d)/8\, .
\label{tagged_mon_exp.2}
\end{equation}
For $d>2$, it was shown~\cite{SMDD2005} that $Q_1(t, t_0\to \infty)$ decays
faster than a power law for large $t$, namely, as A stretched exponential
for $2<d<4$ and exponentially for $d>4$. In contrast, $Q_2(t,t_0\to \infty)\to {\rm 
const.}$ as $t\to \infty$ for $d>2$~\cite{SMDD2005}.

Furthermore, Ref.~\cite{SMDD2005} also considered the general case when the second 
interface evolves via Eq. (\ref{evoly2}), but the first interface may correspond to 
any generic growing interface, evolving not necessarily
by the
Edwards-Wilkinson equation. For example the first interface may evolve by the
Kardar-Parisi-Zhang (KPZ) equation~\cite{KPZ86}. In general, this first interface
will be characterized by a growth exponent $\beta_1$ and a dynamical
exponent $z_1$ defined via the scaling form of the second moment
of the height differences between two points in space,
$\langle \left[h_1({\bf r}_1,{\tau}_1)
- h_1({\bf r}_2,{\tau}_2)\right]^2 \rangle \approx
|\tau_2 - \tau_1|^{2\beta_1} f\left( { {|{\bf r}_1 -
{\bf r}_2|^{z_1}}/{|\tau_2 - \tau_1|}}\right)$.
For example, for the $(1+1)$-dimensional Edwards-Wilkinson equation
one has $\beta_1=1/4$ and $z_1=2$, whereas for the $(1+1)$-dimensional
KPZ equation one has $\beta_1=1/3$ and $z_1=3/2$~\cite{HZ95}. One of the main
results of Ref.~\cite{SMDD2005} was to show that
\begin{equation}
\theta_s^2= \beta_1/2.
\label{thetas2g}
\end{equation}
In particular, Eq. (\ref{thetas2g}) predicts that in $(1+1)$-dimensions, if the
first interface evolves via the KPZ equation, $\theta_s^2=1/6$, a prediction
that was verified numerically~\cite{SMDD2005}.
 
In summary, for an extended Rouse chain in a layered random velocity field, one can make precise
analytical predictions for some of the persistence properties. Let us end this section by
pointing out that persistence properties of Rouse chains
as well as semi-flexible polymer chains have also been studied in presence of a non-random shear velocity 
field and numerical as well as analytical estimates of the persistence exponents are available~\cite{BDM07,Chakraborty}.

\section{Various generalisations of persistence}

In this review, we have so far discussed the time dependence of the persistence probability $Q(t)$, i.e.,
the probability that a fluctuating field $\phi(x,t)$, at a fixed point $x$ and as a function of $t$,
does not change sign up to time $t$. The field $\phi(x,t)$ may represent the spin field in Ising model, or
the height fluctuation of an interface. In many systems of physical interest, we have seen that $Q(t)$
decays as a power law, $Q(t)\sim t^{-\theta}$ for large $t$, with a characteristic persistence exponent $\theta$.
This basic quantity has been generalised in a number of ways and a variety of related observables have been
studied, both theoretically and experimentally, in extended nonequilibrium systems over the past two decades.
In this section, we discuss briefly some of these generalisations.

\subsection{Occupation time and persistent large deviations}\label{subsection:occupation_time}

To start with, consider first a stochastic process $\phi(t)$ that has no $x$ dependence. For example, $\phi(t)$
may represent the position of a single Brownian motion in $1$-d evolving with time $t$ via the stochastic equation,
$d\phi(t)/dt=\eta(t)$ where $\eta(t)$ is a Gaussian, zero mean and delta correlated, white noise. The
{\em occupation time} (or {\em residence time}) fraction of the process $\phi(t)$ is simply the
fraction of time that the process
spends above its mean value (say $0$) when observed over the period $[0,t]$ 
\begin{equation}
T_t= \frac{1}{t}\,\int_0^t \theta\left[\phi(\tau)\right]\, d\tau,
\label{occup.1}
\end{equation}
where $\theta(x)$ is the Heaviside step function, 
and we assume, for simplicity, that the process starts
at the origin, $\phi(0)=0$. Clearly, $0\le T_t\le 1$, with the upper (lower) bound $T_t=1$ ($T_t=0$) achieved when the
process `persists' above (below) $0$ over the full time interval $[0,t]$. Clearly, $T_t$ is a random variable
and its distribution $P(T_t,t)$ provides {\em more} detailed information about the evolution of the process than
the simple persistence $Q(t)$. The occupation time fraction $T_t$ has been studied for a long time in the
probability literature since the seminal work of L\'evy~\cite{Levy39} who computed the distribution
of $T_t$ for a Brownian motion exactly and found that it is independent of $t$ for all $t$ and
is simply given by
\begin{equation}
P(T_t=z,t)= \frac{1}{\pi}\, \frac{1}{\sqrt{z\,(1-z)}}\quad, \; 0\le z\le 1\, . 
\label{levy.1}
\end{equation}
Since then the distribution of $T_t$ has been studied for a variety of stochastic processes in the mathematics~\cite{Kac49,Watanabe95}
as well as physics literature (see e.g. the brief review~\cite{sm_current2005}). 
Exact results are known only in few cases, such as for L\'evy procesess~\cite{Lamperti}
and recently, for a more general class of renewal processes~\cite{BBDG99,GL2001a,BB2011}. Even for simple Markov processes, such as $
d\phi(t)/dt= t^{\alpha-1/2}\, \eta(t)$ with $\alpha>0$, while the persistence exponent is $\theta=\alpha$ (obtained
by the time transformation $t'=t^{2\alpha}$ that reduces it to an ordinary Brownian motion in time $t'$)~\cite{DM99}, 
the distribution $P(T_t,t)$ turns out to be highly nontrivial
and difficult to compute~\cite{DM99,DGL2001}. In many of these symmetric {\em nonstationary} systems, it turns out
that $P(T_t=z,t)$ tends to a time independent limiting form as $t\to \infty$: $P(T_t=z,t\to \infty)=f(z)$
where $f(z)$ has support over $z\in [0,1]$ and is symmetric around $z=1/2$, as in the Brownian case in Eq. (\ref{levy.1}).
While the precise form of $f(z)$ varies from one process to another, one can argue quite generically~\cite{DG98,DM99} that
near the endpoint $z\to 1$ (or equivalently as $z\to 0$), $f(z)$ has the singular behavior, $f(z)\sim (1-z)^{\theta-1}$
(or $z^{\theta-1}$ as $z\to 0$),
where $\theta$ is the persistence exponent of the process. Hence the function $f(z)$ carries the information   
about the persistence exponent $\theta$.  

For {\em stationary} processes, e.g., the Ornstein-Uhlenbeck process, $P(T_t=z,t)$ does not converge
to a limiting distribution. Instead it behaves for large $t$ as, $P(T_t=z,t)\sim \exp\left[-t\, \Phi(z)\right]$
where $\Phi(z)$ is a large deviation function, that has been computed exactly for the 
Ornstein-Uhlenbeck process using a path integral approach~\cite{MB2002}.

The occupation time distribution has also been computed exactly in a class of disordered systems,  
such as for a Brownian particle moving in a random Sinai potential~\cite{MC_occup2002,SMC_occup2006} and also
in models related to spin glasses~\cite{MD2002}. The distribution of $T_t$ has also been studied for one-dimensional Brownian diffusion in an external
field \cite{BarkaiJSP06} where it was shown that there exists a relation between the statistics of $T_t$ and survival probability currents. Finally it has been very useful
in analysing ergodicity properties in blinking quantum dots~\cite{Brokmann03,MB2005,BelBarkai2005}--for a brief
review see~\cite{Stefani2009}. 

\vskip 0.3cm

{\noindent {\bf Occupation time in an extended system and persistent large deviations:}} We discussed above
the occupation time (fraction) $T_t$ of a single stochastic process $\phi(t)$ that has no spatial dependence.
For spatially extended fluctuating field $\phi(x,t)$, one can define an analogous quantity,
$T_t= (1/t) \int_{0}^t \theta\left[\phi(x,\tau)\right]\, d\tau$ and study it, at a fixed point $x$, as a function of $t$.
For spatially extended systems, the study of occupation time (fraction) or equivalently the so called `sign-time' fraction
\begin{equation}
M_t = \frac{1}{t}\, \int_0^{t} {\rm sgn}\left[\phi(x,\tau)\right]\, d\tau
\label{sign1}
\end{equation}
was initiated by Dornic and Godr\`eche~\cite{DG98} and simultaneously by Newman and Toroczkai~\cite{NT98a}.
Clearly, $M_t=2\,T_t-1$ and $-1\le M_t\le 1$ and for translationally invariant (in space) systems, the distribution
of $M_t$ is independent of $x$.  
This distribution $P(M_t=y,t)$ was studied numerically for a variety
of spatially extended systems, e.g., 
for the $T=0$ Glauber dynamics of Ising chain~\cite{DG98}, for a fluctuating
diffusing field~\cite{DG98,NT98a} and for fluctuating interfaces~\cite{NT98b}.
These results indicated that for large $t$, as in the case of a single stochastic process discussed earlier,
the distribution approaches a limiting shape, $P(M_t=y,t)\to f(y)$. The function
$f(y)$ is supported over $-1\le y\le 1$, symmetric around $y=0$ and has a singular behavior at 
the two end-points $y\to \pm 1$, 
 $f(y)\sim (1\mp y)^{\theta-1}$ where $\theta$ again is the associated persistence exponent.
The function $f(y)$ is hard to compute exactly and only approximate estimate is available using IIA
for the diffusion equation in $d$-dimensions~\cite{DG98,NT98a}. Note that for the diffusion equation
$\theta(d)$ increases monotonically with dimension $d$ (see section \ref{subsection:diffusion}). Hence, for $\theta(d)<1$,
the function $f(y)$ diverges at its end-points $y=\pm 1$ and has a $U$ shape over its support. In contrast, for
$\theta(d)>1$, the function $f(y)$ vanishes at the end-points $y=\pm 1$ and $f(y)$ is a bell shaped function.
For the diffusion equation, this transition from the $U$ to bell shape takes place around $d\approx 36$~\cite{NT98a}. The distribution
of $M_t$ (\ref{sign1}) has also been studied for the 2-d Ising model evolving with Glauber dynamics evolving at finite temperature $T>0$ \cite{GD98}. There it was shown, numerically, that, below the critical temperature $T<T_c$, the distribution $P(M_t=y,t)$ approaches a limiting shape $P(M_t=y,t)\to f(y)$ when $t \to \infty$. In this case, the limiting distribution $f(y)$ has support over $[-m_0(T), + m_0(T)]$ where $m_0(T) > 0$ is the equilibrium
magnetization of the 2-d Ising model. Moreover it was shown that $f(y)$ is singular at $\pm m_0(T)$, $f(y) \sim [y\pm m_0(T)]^{\theta-1}$ where $\theta$ was found to be independent of $T$ and given by the persistence exponent of the 2-d Ising model evolving with Glauber dynamics at $T=0$, i.e. $\theta \simeq 0.22$ \cite{BDG1994a,Sta94}. This provides an interesting definition of the persistence exponent for the local magnetization of coarsening ferromagnets at finite temperature, $0<T < T_c$. 
   
Let us come back to the cases mentioned above where the quantity $-1\le M_t\le 1$: $M_t$ can itself be interpreted as `magnetization' and $P(M_t=y, t\to \infty)=f(y)$ is the stationary distribution of the magnetization. However, one can also view $M_t$ as a stochastic process in $t$
that always lies inside the box $-1\le M_t\le 1$. Consider this stochastic process $M_t$ over a time window $[0,t]$.
Dornic and Godr\`eche~\cite{DG98} posed the following interesting question: What is the probability that
the process $M_t$ (starting initially at $M_0=1$) stays above a level $s$ (with $-1\le s\le 1$) up to time $t$, i.e.,
\begin{equation}
R(t,s)= {\rm Prob.}\,[ M_{\tau} \ge s, \quad {\rm for}\,\, {\rm all}\,\, 0\le \tau\le t]\, .
\label{pld.1}
\end{equation}
For instance, if one sets $s=1$, $R(t,1)$ is the probability that $M_{\tau}\ge 1$ for all $0\le \tau\le t$.
It follows from the definition of $M_t$ in Eq. (\ref{sign1}) that for this event to happen, the underlying process
$\phi(x,\tau)$ must stay above $0$ for all $0\le \tau\le t$ which is simply the persistence probability
$Q(t)$ of the underlying process. Hence, $R(t,1)=Q(t)\sim t^{-\theta}$ for large $t$.
In contrast, if one sets $s=-1$, $R(t,-1)$ is the probability that $M_{\tau}\ge -1$ for all $0\le \tau\le t$.
But this is always true since by definition $-1\le M_{\tau}\le 1$ for all $\tau$. Hence $R(t,-1)=1$ for all $t$.
Indeed, Dornic and Godr\`eche found~\cite{DG98} numerically that, $R(t,s)\sim t^{-\theta(s)}$ for large $t$,
where the exponent $\theta(s)$ depends continuously on $s$, interpolating between $\theta(-1)=0$ to
$\theta(1)=\theta$ ($\theta$ being the standard persistence exponent). 
This quantity $R(t,s)$ (named as `persistent large deviation' by the authors of Ref.~\cite{DG98}) 
was also studied for fluctuating interfaces, both numerically for stochastic 
growth equations and also experimentally in Al/Si(111) and Ag(111) surfaces
and the continuous family of exponents $\theta(s)$ was measured~\cite{CDDBDW03}.   

\subsection{Persistence of domains and other patterns}\label{subsection:gen_pers_patterns}

So far we have been discussing the persistence of a field $\phi(x,t)$ at a fixed point $x$, e.g., the persistence
of a spin at a fixed site on a $q$-state Potts chain evolving via the $T=0$ Glauber dynamics starting from
a random initial configuration of spins. A natural
generalisation is to study the persistence or the survival probability of not just the spin at a fixed site,
but a specific extended `pattern' present in the initial random configuration. 
For example, consider the 1-d $q$-state Potts model and focus, say, on a particular
domain (of a given color) present in the initial configuration.
This domain is flanked on either side by a domain wall or a kink. In the $T=0$ dynamics of the Potts chain, the domain walls diffuse with time
and when two walls on either side of the marked domain meet, they either annihilate with probability $1/(q-1)$ (when the
two neighbouring domains of the marked one are of the same color) or aggregate with probability
$(q-2)/(q-1)$ (when the neighbouring domains have different colors)~\cite{Derrida95,MH95,SM95a,MS_largeq95,Monthus96}.
So, a natural question is: what is the probability that the marked domain survives up to time $t$, i.e., what is
the probability that the two domain walls surrounding the marked domain do not collide up to time $t$?
Krapivsky and Ben Naim studied~\cite{KB97} this `domain persistence' $S(t)$ for the $1$-d $q$-state Potts model
and found, both numerically and also analytically using an independent domain approximation, that it decays as a power law for large $t$,
$S(t)\sim t^{-\psi}$, where the exponent $\psi(q)$ is nontrivial and depends on $q$. For example, for $q=2$,
$\psi\approx 0.126$~\cite{KB97}. 

Another interesting question is: what is the probability $S_w(t)$ that two neighboring domains
present in the initial configuration both survive up to time $t$, i.e., the probability that a given domain wall
remains uncollided with other domain walls up to time $t$~\cite{Monthus96,MC97}. 
This is precisely the `walker persistence' problem briefly discussed at the end of section \ref{subsection:coarsening_ising}.
This walker persistence
decays as a power law with a nontrivial $q$-dependent exponent, $S_w(t)\sim t^{-\theta_w(q)}$ for large $t$.
As discussed in section \ref{subsection:coarsening_ising},
for $q=2$ and $q\to \infty$ limit, one can show analytically that $\theta_w(2)=1/2$ and $\theta_w(\infty)=3/2$~\cite{Monthus96,MC97}.
For other intermediate values of $q$, $\theta_w(q)$ is nontrivial. Monthus computed $\theta_w(q)$ within a
perturbation theory for small $(q-1)$ and a rigorous upper bound, $\theta_w(q)\le \ln (q)/[2\,\ln 2]$, was established
in Ref.~\cite{MC97}. The exponent $\theta_w(q)$ in $d<2$ was also computed via a perturbative RG calculation
with $\epsilon=2-d$ being the small parameter~\cite{KRZ2003,RZ2004}.

A related question is the survival probability of a mobile `test' (passive) particle in a fluctuating external field $\phi(x,t)$~\cite{MC97}.
For example, the fluctuating field could be a diffusing field or the height of a fluctuating interface. The field $\phi(x,t)$ evolves via
its own dynamics independent of the test particle.
The test particle moves according to some prescribed deterministic or stochastic rules and survives as long as the
external field $\phi$ that it ``sees' at its own location (moving frame) does not change sign. This is then a
natural generalisation of the `static' persistence $Q(t)$ discussed in this review where the test particle
is `immobile'. In Ref.~\cite{MC97}, two types of test particle motions were studied. In one case the test particle 
adopts a strategy to live longer (by coupling its dynamics to that of the background field $\phi$) and in the other case,
the test particle just undergoes random diffusion irrespective of the background field $\phi$. In both cases, it was found 
numerically and in some special cases analytically that the survival probability of the mobile particle again decays
as a power law at late times with a nontrivial exponent. Moreover, in some special cases, it was shown~\cite{MC97} that given
a pattern present in the initial condition $\phi(x,t=0)$, the persistence of this pattern can be 
computed by the following procedure: launch a mobile test particle into the system with a suitably engineered dynamics that
depends on which pattern one is interested in and then compute the survival probability
of this mobile particle which can then be identified as the persistence probability
of the desired pattern in the initial field configuration~\cite{MC97}.  

\subsection{Spatial structures of persistent sites}

Another interesting generalisation of the standard site persistence is to
investigate the spatial structures of all persistent sites in a given system at a fixed time.
More precisely, consider the $T=0$ Glauber dynamics of Ising or Potts chain, starting from a
random initial configuration. Initially (at $t=0$) we mark all the sites `black'. If the spin
at a site changes at some time, it turns forever `white'. If we now look at the snapshot of the system
at any given time $t$, all sites that are still black are persistent, while the white ones are not.
We have seen before that the fraction of persistent (black) sites decay algebraically with time as
$Q(t)\sim t^{-\theta}$ where $\theta$ is the standard persistence exponent. But it is equally interesting
to study how the persistent (black) sites are distributed in space at a given fixed time $t$. Indeed, this
question was first raised and investigated for the Ising model by Manoj and Ray~\cite{MR2000a,MR2000b,MR2000c}--see also
\cite{JH2000} and for a short review~\cite{Ray2004}. The persistent sites were found to exhibit dynamic
scaling~\cite{MR2000a} in the Ising model. However, in the $q$-state Potts model, the length scale
that governs this dynamic scaling was found to change at some critical value of $q$ where $\theta(q)=1/2$~\cite{BO2000}.  
Recall that for the 1-d Potts model, $\theta(q)= -\frac{1}{8} + \frac{2}{\pi^2}
\left[\cos^{-1}\left(\frac{2-q}{\sqrt{2}q}\right)\right]^2$. For the Ising $q=2$ case, $\theta(2)=3/8$ and
$\theta(q\to \infty)=1$. The value of $q$ corresponding to $\theta(q)=1/2$ is $q_c=2/[1+\sqrt{2}\, \cos(\sqrt{5} \pi/4)]=2.70528\ldots$.
Hence $\theta(q)>1/2$ for all integer $q\ge 3$.

To understand the spatial structure of persistent sites, it is useful to investigate the relevant
length scales. There are two length scales at time $t$~\cite{BO2000}. The first is the typical domain size, or,
equivalently in the domain wall picture, the typical distance between two consecutive domain walls, which grows as 
$L_d(t)\sim t^{1/2}$ for the Potts model for any $q$. In contrast, the fraction of persistent sites decays as
$t^{-\theta(q)}$, hence the mean distance between persistent sites grows as $L_p(t)\sim t^{\theta(q)}$.
This suggests that the spatial structure of the persistent sites for $\theta(q)>1/2$, where $L_p(t)$ is the larger
of the two length scales, will be very different from the case $\theta(q)<1/2$, where $L_w(t)$ is the
larger of the two length scales. 
The characteristic length scale that controls the dynamic scaling of the spatial structure of the
persistent sites turns out to be the larger of the two, $\xi(t)= {\rm max}[L_d(t), L_p(t)]$~\cite{BO2000}.
To test this observation, let us consider the non-persistent
intervals between two consecutive persistent sites and let $n(k,t)$ denote the number of such intervals (per site)
of length $k$. The existence of the two competing length scales manifests itself in an 
unusual dynamic scaling of $n(k,t)$~\cite{BO2000}. 
A detailed investigation, both numerical as well analytical
in the $q\to \infty$ limit, suggested the following dynamic scaling behavior of $n(k,t)$
\begin{equation}
n(k,t) \sim \frac{1}{\xi^2(t)}\, f\left(\frac{k}{\xi(t)}\right) \;,
\label{scaling.1}
\end{equation}
where $\xi(t)= L_p(t)\sim t^{\theta(q)}$ for $\theta(q)>1/2$ and $\xi(t)= L_d(t)\sim t^{1/2}$ for
$\theta(q)<1/2$.   

Another useful insight may be obtained~\cite{MR2000a,MR2000b,JH2000,MR2000c} by studying the function $C(r,t)$ that denotes the 
conditional probability that another site at a distance $r$ is also persistent at time $t$. 
One can conveniently define a binary variable $\rho(x,t)$ such that
$\rho(x,t)=1$ if the site $x$ is persistent at time $t$ and $\rho(x,t)=0$ otherwise. Then
the probability that the site is persistent at time $t$ is simply $Q(t)=\langle \rho(x,t)\rangle$
and $C(r,t)= \langle \rho(x,t)\rho(x+r,t)\rangle/\langle \rho(x,t)\rangle$. Due to the
translational invariance in an infinite system, both $Q(t)$ and  $C(r,t)$ do not depend on $x$. 
Clearly, as $r\to \infty$ the two sites separated by an infinite distance become uncorrelated and
$C(r,t)\to Q(t) \sim t^{-\theta(q)}$. 
This happens for $r \gg \xi(t)$, which scales as $\xi(t)\sim t^{1/2}$,
for any $\theta(q)<1/2$, such as in the Ising case.
Let us first on this $\theta(q)<1/2$ case.
In the opposite limit when $r \ll \xi(t)\sim t^{1/2}$, $C(r,t)$ becomes
time independent and decays algebraically with distance $r$ as, $C(r,t)\sim r^{-a}$ for
$1 \ll  r \ll \xi(t)$, where the exponent $a$ turns out to be nontrivial~\cite{MR2000a}. 
These two limiting behaviors suggest, for $\theta(q)<1/2$, a dynamic scaling of the form~\cite{MR2000a}
\begin{equation}
C(r,t) \sim t^{-\theta(q)}\, G\left( r\, t^{-1/2}\right) \;,
\label{scaling_pers.1}
\end{equation}
such that the scaling function $G(y)\sim y^{-a} $ as $y\to 0$ and $G(y)\sim {\cal O}(1)$ as $y\to \infty$.
The behavior $C(r,t)\sim r^{-a}$ for $r \ll t^{1/2}$ then provides a scaling relation, $a= 2\theta(q) $ needed
to cancel the time dependence. The number of persistent sites within a distance $R$ of a given persistent site
is then estimated as $\int_0^R dr r^{-a}\sim R^{d_f}$ where $d_f=1-a=1-2\theta(q)$.
Consequently, the set of persistent sites form a fractal structure on the line with fractal dimension,
$d_f= 1-2\, \theta(q)$~\cite{BO2000,Ray2004}. Note, however, that this fractal structure occurs only when
$\theta(q)<1/2$, such that $d_f>0$. For $\theta(q)>1/2$, $d_f$ sticks to the value $0$ and the
persistent sites form point like objects, i.e., isolated finite clusters~\cite{BO2000}. 

In higher dimensions, a similar scaling as in Eq. (\ref{scaling_pers.1}) holds, and the persistent sites form 
a fractal with fractal dimension $d_f= d- 2\theta$, which makes sense only when $\theta<d/2$~\cite{BO2000,Ray2004}.  
For $\theta>1/2$, the persistent sites no longer have a fractal structure but become pointlike
objects as in the $1$-d case.
This scaling relation $d_f=d-2\theta$, as well as the conjectured scaling behavior in Eq. (\ref{scaling_pers.1})
were verified for various spins systems in different dimensions via extensive
simulations~\cite{MR2000a,MR2000b,JH2000,MR2000c}. In the 1-d case, the scaling was also studied analytically using an
independent domain approximation~\cite{MR2000b}.

\subsection{Persistence in sequential versus parallel dynamics}

So far we have discussed the persistence of a spin at the $T=0$ Glauber dynamics in the Ising or Potts model, where
the dynamics consists in picking a spin at random and updating its value to one of its neighbours chosen at random.
By definition, this is a sequential or asynchronous dynamics since the spins get updated one at a time. In this case,
the persistence of a spin decays as a power law, $Q(t)\sim t^{-\theta_{\rm seq}}$ for large $t$ where $\theta_{\rm seq}=\theta$
is the standard persistence exponent. For example, as mentioned in the introduction, the persistence exponent for the
$1$-d $q$-state Potts chain is known exactly to be, $\theta_{\rm seq}(q)= -\frac{1}{8} + \frac{2}{\pi^2}
\left[\cos^{-1}\left(\frac{2-q}{\sqrt{2}q}\right)\right]^2$. 
For instance, for the Ising ($q=2$) case, $\theta_{\rm seq}=3/8$.
A natural question is what happens if the spins
are updated not sequentially, but in parallel. At each time step, for each spin one chooses one of its neighbours
at random and registers the spin of the chosen neighbour as its value at the next time step and then all spins are updated simultaneously
at the next step to their new values. This is the parallel dynamics. Does the persistence $Q(t)$ of a spin 
under the parallel dynamics 
decay
differently from that of the sequential dynamics? This question was first investigated for the Ising spin chain
in Ref.~\cite{MRS2000} and it was found that while the persistence under parallel dynamics still decays algebraically with time, it 
does so with a different (larger) exponent, $Q(t)\sim t^{-\theta_{\rm par}}$, where $\theta_{\rm par}\approx 0.75$ numerically, 
approximately twice that of the sequential exponent $\theta_{\rm seq}=3/8$. 
A similar doubling of the exponent value from sequential to parallel dynamics were also found 
numerically for the $q$-state Potts chain~\cite{Menon_Ray2001}. 
A detailed investigation of the 1-d Potts chain showed~\cite{Menon_Ray2001} that 
the parallel dynamics of the spin chain can be effectively divided into the sequential dynamics of 
two independent sub-lattices (respectively at odd and even time steps) and the parallel persistence
of the full chain is just the product of the sequential persistences of the two sub-chains, thus
proving, $\theta_{\rm par}= 2\, \theta_{\rm seq}$.

\section{Persistence in reaction-diffusion models, Voter model, directed percolation}

We have seen that in many nonequilibrium extended systems, in particular in low spatial dimensions, the persistence probability {\em generically}
decays at late times as a power law, $Q(t)\sim t^{-\theta}$. However, other types of {\em non-algebraic} slow decay of $Q(t)$ have
also been observed in a variety of systems some of which we briefly review in this section.

\subsection{Reaction-diffusion models} 

In sections \ref{subsection:coarsening_ising} and \ref{subsection:gen_pers_patterns}, we have seen that the zero temperature
Glauber dynamics of an Ising chain can be conveniently described in a dual picture, where the kinks or domain walls
between $+$ and $-$ phases diffuse and when two domain walls come together, they annihilate~\cite{AmarFamily,Bray89}. 
Denoting the domain walls
by particles of species $A$, this dual picture corresponds to the diffusion-annihilation process $A+A\to \emptyset$ (see Fig. \ref{fig:domain_walls_ising}).
The persistence $Q(t)$ is just the probability that a fixed site on the lattice (say the origin) is not visited 
by any of the particles up to time $t$, starting from a random initial configuration of the particles~\cite{KBR94,BDG1994a}.
The persistence decays as a power law, $Q(t)\sim t^{-\theta}$ at late times, where $\theta=3/8$ exactly~\cite{DHP1995a}.
This dual process of diffusion-annihilation $A+A\to \emptyset$ can be trivially generalised to arbitrary $d$ dimensions.
The average density of particles is known to decay as $\rho(t)\sim t^{-d/2}$ for $d<2$ and $\rho(t)\sim 1/(\lambda\, t)$ for $d>2$,
where $\lambda$ is the reaction rate.
The persistence $Q(t)$ in higher dimensions has been investigated numerically, by a mean-field rate
equation approach for $d>2$, via Smoluchowsky theory in $d\le 2$~\cite{KBR94} and also, by a field theoretic renormalization 
group (RG) approach~\cite{Cardy95}. The main results can be summarized as follows: $Q(t)\sim t^{-\theta(d)}$ for $d<2$ where
the exponent $\theta(d)$ is universal and depends only on $d$, $Q(t)\sim t^{-1/2}$ (with a logarithmic correction)
for $d=2$ and $Q(t) \sim t^{-\theta'}$ for $d>2$ where $\theta'$ is a nonuniversal exponent that depends
on microscopic parameters such as the reaction rate $\lambda$.

This simple two-body reaction-diffusion process $A+A\to \emptyset$ has been generalised to multibody
process where the reaction corresponds to, e.g. trimolecular $A+A+A\to \emptyset$ or in general,
$k$-molecular $k\, A\to \emptyset$. In this general $k$-body reaction process, there is an upper
critical dimension $d_c(k)=2/(k-1)$ such that the average density $\rho(t)\sim t^{-d/2}$ for
$d<d_c(k)$, while $\rho(t)\sim (\lambda t)^{-d_c(k)/2}$ for $d>d_c(k)$~\cite{Cardy95}.
In this multibody case with $k>2$, the persistence $Q(t)$ was studied analytically by Cardy using the RG method
and a variety of late time behaviors was found depending on the dimension $d$~\cite{Cardy95}
\begin{eqnarray}
Q(t)\sim \left\{\begin{array}{ll} \exp \left[-{\rm const.}\, t^{1-d_c(k)/2}\right] \quad\quad d>2 \;, \\
&\\
\exp \left[- {\rm const.}\, t^{1-d_c(k)/2}/\ln t\right] \quad\quad d=2 \;, \\
&\\
\exp\left[- {\rm const.}\, t^{(d-d_c(k))/2}\right]  \quad\quad d_c(k)<d<2 \;, \\
&\\
\exp\left[- {\rm const.}\, (\ln t)^{k/(k-1)}\right]  \quad\quad d=d_c(k) \;, \\
&\\
t^{-\theta}\quad\quad d<d_c(k) \;,
\end{array}
\right.
\label{cardy_result.1}
\end{eqnarray}
where the exponent $\theta$ for $d<d_c(k)$ is nontrivial and is not computable exactly, except for $k=2$ and $d=1$.
One interesting prediction of this general result is, for instance, the fact that for $k=3$ for which $d_c(3)=1$, 
$Q(t)\sim \exp\left[-{\rm const.}\, (\ln t)^{3/2}\right]$ in $d=1$~\cite{Cardy95}.
This single species reaction-diffusion system has been generalised to multi-species case and the persistence 
properties of a fixed site (or equivalently that of an immobile
spectator particle or impurity), as well as that of a single mobile impurity, have been studied~\cite{KBR94}. 
This has been also generalised to the case when there are more than one immobile impurity, for instance
when the immobile impurities form an extended set~\cite{Bennaim96}.

Another low dimensional system which exhibits non-algebraic decay of persistence $Q(t)$ is the axial next nearest neighbour (ANNI) 
chain~\cite{SenDasgupta2004}, with a Hamiltonian $H= - \sum_{i} \left[s_i\,s_{i+1}- {\kappa}\, s_i\,s_{i+2}\right]$
where $s_i=\pm 1$ and $\kappa$ represents the frustration. In this model, the ground state is ferromagnetic for $\kappa<1/2$,
of the antiphase type ($++--$) for $\kappa>1/2$ and is highly degenerate for $\kappa=1/2$. For the zero temperature single spin flip
Glauber dynamics, the persistence $Q(t)$ (the probability that a spin does not flip up to time $t$) was studied numerically
and was found to display different decays depending on the value of $\kappa$~\cite{SenDasgupta2004}. For example,
for $\kappa>1$, $Q(t)\sim t^{-\theta}$ with $\theta\approx 0.69\pm 0.01$. In contrast, for $0<\kappa<1$, the persistence
decays as a stretched exponential, $Q(t)\sim \exp[-{\rm const.}\, t^{0.45}]$, with no special
behavior at the multiphase point $\kappa=1/2$. Exactly at $\kappa=1$, $Q(t)$ again decays as
as a stretched exponential but with a different stretching exponent, $Q(t)\sim \exp[-{\rm const.}\, t^{0.21}]$.

\subsection{Voter model}

Another example where one finds non-algebraic decay of persistence is the $q$-state voter model in $d$ dimensions.
In the voter model, each site on a lattice (representing a voter) can be in one of the $q$ possible `opinions' or `states' and the
stochastic dynamics consists of picking a site at random and changing its opinion to one of its neighbours
chosen at random~\cite{Liggett}. In one dimension, this model is identical to the $q$-state Potts model
undergoing zero temperature Glauber dynamics. However, for $d>1$, the dynamics of the $q$-state voter model differs from
that of the Potts model. It turns out that $d=2$ is a special dimension in the voter model. For $d\le 2$, 
the domains of different opinions coarsen with time and eventually as $t\to \infty$, one of the
opinions wins out and thus the voters reach a consensus. In contrast, for $d>2$, the difference in opinions
persists forever. While many quantities, such as equal and two-time correlation functions can be computed exactly
in the voter model, the persistence properties again turn out to be nontrivial~\cite{BFK96,HG98}.
The persistence $Q(t)$ (probability that a given voter does not change his/her opinion up to time $t$)
has been studied within mean field theory~\cite{BFK96} and also by a field theoretic RG method~\cite{HG98} and
a variety of intriguing behavior for $Q(t)$ has emerged as a function of dimension $d$. For instance, using RG method
Howard and Godr\`eche found the following behaviors for the persistence~\cite{HG98} 
\begin{eqnarray}
Q(t)\sim \left\{\begin{array}{ll} \exp \left[-{\rm const.}\, t\right] \quad\quad d>4 \;, \\
&\\
\exp \left[- {\rm const.}\, t/\ln t\right] \quad\quad d=4 \;, \\
&\\
\exp\left[- {\rm const.}\, t^{(d-2)/2}\right]  \quad\quad 2<d<4 \;, \\
&\\
\exp\left[- {\rm const.}\, (\ln t)^{2}\right]  \quad\quad d=2 \;, \\
&\\
t^{-\theta(q)}\quad\quad d<2 \;,
\end{array}
\right.
\label{hg_result.1}
\end{eqnarray}
where the ${\rm const.}$'s are $q$ dependent. Ben Naim et. al.~\cite{BFK96} also studied $Q_m(t)$, the probability that
a voter changes his/her opinion exactly $m$ times up to time $t$ and found that $Q_m(t)$ displays
very different scaling behavior for $d<2$ and $d>2$.  

\subsection{Directed percolation}

Another important class of reaction-diffusion systems are those with an absorbing phase transition from an active to an
inactive state~\cite{Hinrichsen_review,Odor_review}. Several such systems such as the contact process, Domany-Kinzel cellular automata
models and directed bond and site percolation models on a lattice are characterized by universal critical behavior
belonging to the {\em directed percolation} (DP) universality class~\cite{Hinrichsen_review}. A simple example is
the contact process~\cite{Harris74}, where each site on a $d$-dimensional lattice can be either vacant
or occupied by a single particle. A particle self-annihilates with rate $1$ and with rate $\lambda$ it can create
an additional offspring at a neighbouring site (provided that site is vacant). The system approaches a steady state in the long time 
limit where the density is nonzero for $\lambda>\lambda_c$ (active) and $0$ (inactive) for $\lambda<\lambda_c$.
This and other models belonging to the DP universality class has an upper critical dimension $d_c=4$ above which the
mean field theory holds for standard critical exponents~\cite{Hinrichsen_review,Odor_review}.
The persistence $Q(t)$ is the probability that a site remains inactive up to time $t$, first studied 
by Hinrichsen and Koduvely~\cite{HK98} numerically in $d=1$.
They found that as $t\to \infty$, while $Q(t)\sim {\rm const.}$ for
$\lambda<\lambda_c$, it decays exponentially $Q(t)\sim \exp[-{\rm const.},t]$ for $\lambda>\lambda_c$.
Exactly at the critical point $\lambda=\lambda_c$, $Q(t)\sim t^{-\theta}$ where $\theta\approx 1.5$
seems to be universal for a wide class of one dimensional models that belong to the DP class~\cite{HK98}.
This universal model independent value of the persistence exponent $\theta$ in $d=1$ was later confirmed
numerically in a number of other models, such as in the Ziff-Gulari-Barshad (ZGB) model~\cite{ZGB86} by
Albano and Munoz~\cite{AM2001}
and in models of one dimensional coupled map lattices by Menon et. al.~\cite{MSR2003}. 

Numerical results in higher dimensions also suggest an algebraic decay at the critical point, $Q(t)\sim t^{-\theta}$, in all dimensions.
An early simulation result~\cite{AM2001} on the ZGB model found
$\theta\approx 1.5$ in $d=2$, $\theta\approx 1.33$ ($d=3$) and $\theta\approx 1.15$ ($d=4$).
Based on the observation that $\theta$ in $d=1$ and $d=2$ are both close to $1.5$, it was suggested~\cite{AM2001} 
that $\theta$ might be {\em superuniversal} at least
in low dimensions, i.e., independent of models
belonging to the DP class as well of the spatial dimensionality $d$. This claim was refuted later~\cite{FSGH2008},
where the authors found that $\theta$ depends on $d$ and in particular, for $d>d_c$, it is moreover
nonuniversal, i.e., model dependent. The later results were further confirmed recently by a rather 
sophisticated simulation up to $7$ spatial dimensions by Grassberger~\cite{Grassberger09} (see also Ref.~\cite{Hinrichsen2009}).
Finally, the persistence exponent has been measured experimentally in turbulent liquid crystals which belong to the
DP universality class~\cite{TKCS2009}.  

\subsection{Turbulent fluid in $2$ dimensions}

Persistence properties of a two dimensional fluid (on a thin film) has been investigated recently by numerically
solving the $2$-d Navier-Stokes equation driven by Kolmogorov forcing~\cite{Perlekar11}.
Persistence was studied in both the Eulerian and Lagrangian framework. In the Eulerian framework,
the authors monitored the time series $\Lambda(t)$, at a fixed point $(x,y)$ in space,
where the Okubo-Weiss parameter $\Lambda(t)$ takes the value $+1$ (if the flow at $(x,y)$ is
{\em vortical}) and $-1$ (if the flow at $(x,y)$ is {\em extensional})~\cite{Perlekar11}.
In this case, the $+$ and $-$ persistences $Q_{\pm}^{\rm E}(t)$ (probability that $\Lambda(t)$ does not change sign in
$[0,t]$) both were found to decay exponentially with time. In contrast, in the Lagrangian framework,
one moves with a Lagrangian particle and monitors the probability $Q_{\pm}^{\rm L}(t)$
that the particle stays in a vortical $(+)$ or extensional $(-)$ regime up to time $t$.
In this case, it was found numerically that while $Q_{-}^{\rm L}(t)$ decays exponentially,
$Q_{+}^{\rm L}(t)\sim t^{-\theta}$ with $\theta\approx 2.9\pm 0.2$~\cite{Perlekar11}.
Note that the two persistence probabilities $Q^{\rm E}(t)$ and $Q^{\rm L}(t)$ are
analogues of the `site' persistence and the `walker' persistence in the reaction-diffusion systems
discussed in section \ref{subsection:coarsening_ising} and section \ref{subsection:gen_pers_patterns}.

Another interesting turbulent system in 2-d is the nematic liquid crystal undergoing
electroconvection~\cite{TKCS2007,TS12}. This system exhibits two distinct turbulent states called the
dynamic scattering modes and a detailed investigation of how a cluster of one phase grows in
the other has been studied experimentally~\cite{TS12}. Amongst other quantities, both the
temporal and spatial persistence has been studied (see the discussion in section \ref{subsection:flat_vs_radial}).
Finally, persistence has also been studied in the advection of a passive scalar~\cite{Chakraborty2009}.

To summarize, the {\em local} site persistence $Q(t)$ has been studied extensively, by mean field theory, RG method and also
numerically, in a number of systems undergoing reaction-diffusion processes. The picture that has emerged is that 
in many of these systems, especially in higher dimensions, $Q(t)$ displays a non-algebraic decay with time.
In some cases, it may decay algebraically even in higher dimensions but the persistence exponent in higher dimensions
tend to be nonuniversal and model dependent. In addition to the local persistence, the {\em global} persistence (discussed in 
section \ref{Global}) has also been studied in many of these systems and  
similar nonuniversal power laws and in some cases, non-algebraic decay of the global persistence have been
reported.

\section{Persistence of a stationary non-Markovian non-Gaussian sequence: An exactly solvable case}\label{section:exact}

We have seen that analytical computation of the persistence probability of a stochastic sequence or a process
is, in general, very difficult. The difficulty can be typically traced back to the non-Markovian or the non-Gaussian nature of the 
process and there are very few exact results \cite{DH2000,Deloub2000,Farago2000}. In this section, we discuss a special stationary sequence that
is non-Markovian and, in general, non-Gaussian and yet exactly solvable~\cite{MD2001}. For this sequence,
one can compute exactly not only the simple persistence probability, but even other related quantities
discussed in the review, namely, the persistence with partial survival~\cite{majumdar2002} as well
as the distribution of the occupation time~\cite{MD2002}.

We construct a stationary sequence as follows. Let $\{\eta_0,\eta_1,\eta_2,\ldots\}$ denote an infinite
set of i.i.d random variables (noise), each from a symmetric and continuous distribution $\rho(\eta)$,
normalized to unity, $\int_{-\infty}^{\infty} \rho(\eta)\, d\eta=1$. From this infinite set, we now construct 
a new sequence $\{\phi_i\}$ by summing up the consecutive pairs of noises~\cite{MD2001}
\begin{equation}
\phi_i = \eta_i+ \eta_{i-1}\quad i=1,2,\ldots
\label{solv_seq.1}
\end{equation}  
This toy sequence was originally derived in Ref.~\cite{MD2001} as a limiting case of the diffusion process
(discussed in section \ref{subsection:diffusion}) on a hierarchical lattice. Note that even though $\eta_i$'s are uncorrelated,
the variables $\phi_i$'s are {\em correlated}, as evident from Eq.~(\ref{solv_seq.1}). 
The two point correlation function, $C_{ij}=\langle \phi_i \phi_j\rangle$,
can be easily computed from Eq. (\ref{solv_seq.1}) to give
\begin{equation}
C_{ij}= \sigma^2\, [2\delta_{i,j} + \delta_{i-1,j}+ \delta_{i,j-1}] \;,
\label{solv_corr.1}
\end{equation}
where $\delta_{i,j}$ is the Kronecker delta function and $\sigma^2=\int_{-\infty}^{\infty} \eta^2\, \rho(\eta)\, d\eta$
when it exists. Thus the sequence $\{\phi_i\}$ is stationary with only nearest neighbour correlations.
In addition, the sequence $\{\phi_i\}$ is non-Markovian in the following sense. Let us try to express
a member $\phi_i$ of the sequence in terms of the previous members of the sequence. One gets for all $i\ge 2$~\cite{MD2001}
\begin{equation}
\phi_i= \eta_i + \sum_{k=1}^{i-1} (-1)^{k-1}\, \phi_{i-k} + (-1)^{i-1}\,\eta_0 \;,
\label{solv_nm.1}
\end{equation}
which clearly demonstrates the history dependence of the sequence: $\phi_i$ depends not just on the
local noise $\eta_i$ and its immediate predecessor $\phi_{i-1}$ of the sequence (as would have been the case
for a Markov process such as a random walk where $\phi_i=\eta_i +\phi_{i-1}$), 
but on the full history of the sequence preceding $\phi_i$.

The persistence $Q(n)$ of the sequence $\{\phi_1,\phi_2,\ldots,\}$ up to $n$ steps is defined as
\begin{equation}
Q(n)= {\rm Prob.}\left[\phi_1\ge 0, \phi_2\ge 0,\ldots, \phi_n\ge 0\right] \;,
\label{solv_persdef.1}
\end{equation}
which, using Eq. (\ref{solv_seq.1}), can be expressed as an $(n+1)$-fold integral
\begin{eqnarray}
Q(n)&=& \int_{-\infty}^{\infty} d\eta_0\, \rho(\eta_0)\, {\rm Prob.}\left[\eta_1\ge -\eta_0, \eta_2\ge -\eta_1,
\ldots, \eta_n\ge -\eta_{n-1}\right] \nonumber \\
&=& \int_{-\infty}^{\infty} d\eta_0\, \rho(\eta_0)\, \int_{-\eta_0}^{\infty} d\eta_1\, \rho(\eta_1)\,
\int_{-\eta_1}^{\infty} d\eta_2\, \rho(\eta_2) \ldots \int_{-\eta_{n-1}}^{\infty}d\eta_n\, \rho(\eta_n)\,.
\label{solv_int.1}
\end{eqnarray} 
To evaluate this multiple integral recursively, it is convenient to make the lower limit of the integration
over $\eta_0$ in Eq. (\ref{solv_int.1}) as a variable and define the following function as an
$n$-fold integral
\begin{equation}
q_n(x)=\int_{x}^{\infty} d\eta_0\, \rho(\eta_0)\, \int_{-\eta_0}^{\infty} d\eta_1 \ldots\,  
\int_{-\eta_{n-2}}^{\infty}d\eta_{n-1}\, \rho(\eta_{n-1}) \;,
\label{solv_multint.1}
\end{equation}
such that $Q(n)= q_{n+1}(-\infty)$. Now, differentiating Eq. (\ref{solv_multint.1}) with respect to $x$
gives a recursion relation
\begin{equation}
\frac{dq_n(x)}{dx}= - \rho(x)\, q_{n-1}(-x),\quad n\ge 1,
\label{solve_diff.1}
\end{equation}
starting with $q_0(x)=1$ and the boundary condition $q_n(\infty)=0$ for all $n\ge 1$. 
The generating function $F(x,z)= \sum_{n=1}^{\infty} q_n(x)\, z^n$ then satisfies a first order {\em nonlocal} differential 
equation
\begin{equation}
\frac{\partial F(x,z)}{\partial x}= -\rho(x)\, z\, \left[1+ F(-x,z)\right] \;,
\label{solv_diff.2}
\end{equation}
with the boundary condition $F(\infty, z)= 0$ for any $z$. Once we know the solution $F(x,z)$, the
persistence probability is obtained by inverting the generating function via Cauchy's formula
\begin{equation}
Q(n)= q_{n+1}(-\infty)= \frac{1}{2\pi i}\, \int_{C_0} \frac{F(-\infty, z)}{z^{n+2}}\, dz \;,
\label{solv_cauchy}
\end{equation}
where $C_0$ is a contour in the complex $z$ plane encircling the origin.

Fortunately, the differential equation, though nonlocal, can be solved exactly~\cite{MD2001}. To proceed,
let us first make the change of variable, $u(x)= \int_{0}^x \rho(\eta)\, d\eta$
and ${\tilde F}(u,z)= F(x,z)$.  Note that $u\in [-1/2,1/2]$.
Since $\rho(\eta)$ is symmetric around $\eta=0$, it follows that $u(-x)= -u(x)$
and the function ${\tilde F}(-u,z)= F(-x,z)$. Under this transformation, Eq. (\ref{solv_diff.2})
reduces to
\begin{equation}
\frac{\partial {\tilde F}(u,z)}{\partial u}= - z\, \left[1+{\tilde F}(-u,z)\right],
\label{solv_diff.3}
\end{equation}
where $u\in [-1/2,1/2]$ with the boundary condition ${\tilde F}(1/2,z)=0$ for all $z$.
Note the important fact that the distribution $\rho(\eta)$ has dropped out of 
the equation--this shows that ${\tilde F}(u,z)$ is universal
and independent of the noise distribution $\rho(\eta)$. Eq. (\ref{solv_diff.3}) is still nonlocal,
but can be made local by differentiating once more 
\begin{equation}
\frac{\partial^2 {\tilde F}(u,z)}{\partial u^2}= - z^2\, \left[1+ {\tilde F}(u,z)\right]\,,
\label{solv_local.1}
\end{equation}
whose general solution is given by
\begin{equation}
{\tilde F}(u,z)= -1 + a_0(z)\, [\cos(z\,u)-\sin(z\,u)] \;,
\label{solv_local.2}
\end{equation}
where $a_0(z)$ is fixed by the boundary condition ${\tilde F}(1/2,z)=0$ and in terms of the
original variable $x$ one finally gets~\cite{MD2001}
\begin{equation}
F(x,z)= -1+ \frac{\cos(u(x)\, z)- \sin(u(x)\,z)}{\cos(z/2)-\sin(z/2)}\,.
\label{solv_sol.1}
\end{equation}
Consequently, $F(-\infty, z)= 2/[\cot (z/2)-1]$ which has poles at $z=\pi/2+ 2\, m\,\pi$, where $m$ is an integer.
Substituting $F(-\infty,z)$ in Eq. (\ref{solv_cauchy}) and evaluating the
contour integral, one gets the exact result~\cite{MD2001}
\begin{equation}
Q(n)= 2 \sum_{m=-\infty}^{\infty} \left[\frac{\pi}{2}+ 2 \, m\, \pi\right]^{-n-2} \;,
\label{solv_exact.1}
\end{equation}
valid for all $n\ge 1$. For example, by summing the series, one gets $Q(1)=1/2$, $Q(2)=1/3$,
$Q(3)=5/24$ etc. The {\em remarkable} fact is that the persistence $Q(n)$ is {\em universal} for all
$n$, i.e., independent of the noise distribution $\rho(\eta)$. For large $n$, the leading asymptotic
behavior is governed by the $m=0$ term in Eq. (\ref{solv_exact.1}) and one gets
\begin{equation}
Q(n)\sim \exp[-\theta\, n],\quad \theta= \ln (\pi/2)\, .
\label{solv_exp.1}
\end{equation}
Since the process is stationary with short range correlations, one would have guessed 
that $Q(n)$ decays exponentially. However,
in this case, the `persistence exponent' (the inverse decay constant of the exponential decay) 
is universal and can be exactly computed, $\theta=\ln(\pi/2)$.
This is thus a rare solvable example. Interestingly, $Q(n)$ in this toy sequence is closely related
to the average fraction of metastable configurations at zero temperature of an Ising spin glass on 
a $1$-d lattice with $n$ sites and with Hamiltonian
$H= -\sum_{i} J_{i,i+1} s_i s_{i+1}$ where $s_i=\pm 1$~\cite{MD2001}. 
A configuration is metastable at zero temperature, if the energy change
$\Delta E_i= 2\, s_i\, [J_{i-1,i}\, s_{i-1}+ J_{i,i+1}\, s_{i+1}]\ge 0$ due to the
flip of every spin. In the $1$-d spin glass context, the average number of metastable configurations
$\sim (4/\pi)^n$ for large $n$ was computed in Ref.~\cite{Li81}
and \cite{DG86}, by different methods than the one presented above. 

It turns out that the non-Markovian sequence $\{\phi_i\}$ defined in Eq. (\ref{solv_seq.1}) remains solvable
for other related observables and not just for the persistence. For instance, let $Q_m(n)$
denote the probability that the sequence undergoes $m$ sign changes up to $n$ steps ($0\le m\le n$).
Clearly, the persistence $Q(n)=Q_0(n)$, i.e., the probability of no sign changes up to step $n$. 
The generating function
\begin{equation}
{\tilde Q}(p,n)= \sum_{m=0}^n Q_m(n)\, p^m \;,
\label{solv_partial.1}
\end{equation}
is called the `partial survival' probability introduced in section \ref{subsection:partial}. Physically, one
may interpret ${\tilde Q}(p,n)$ in Eq. (\ref{solv_partial.1}) as follows. Let $\phi_i$
represent the position of a particle at time $i$. Every time the particle crosses
the origin it `survives' with probability $0\le p\le 1$.
Then the survival probability up to step $n$ is given precisely 
by Eq. (\ref{solv_partial.1}).
For a stationary sequence with short range correlations, one expects 
${\tilde Q}(p,n)\sim \exp[-\theta(p)\, n]$ for large $n$, where $\theta(p)$ is called
the partial survival exponent. For the sequence $\{\phi_i\}$ in Eq. (\ref{solv_seq.1}), 
the partial survival probability and hence the statistics of multiple sign changes
can be computed exactly~\cite{majumdar2002} by adapting the method described above for the
computation of $Q_0(n)$. The partial survival exponent $\theta(p)$ depends continuously on $p$~\cite{majumdar2002}
\begin{equation}
\theta(p)= \ln \left[\frac{\sin^{-1}\left(\sqrt{1-p^2}\right)}{\sqrt{1-p^2}}\right],\quad 0\le p\le 1 \;,
\label{solv_ps_exp.1}
\end{equation}
and is universal, i.e., does not depend on the noise distribution $\rho(\eta)$. 
In the limit $p\to 0$, one recovers the usual persistence exponent $\theta(0)= \ln(\pi/2)$.
In the opposite limit $p\to 1$, $\theta(1)=0$ which is consistent
with the fact that ${\tilde Q}(1,n)=1$. In addition, all moments of
the number of sign changes up to step $n$ can also be computed
explicitly from the above exact result~\cite{majumdar2002}.
Thus, once again, this is a rare example of a non-Markovian sequence where
one can compute the partial survival exponent $\theta(p)$ exactly.

Another interesting observable is the distribution of the occupation time discussed in section \ref{subsection:occupation_time}.
For the sequence $\{\phi_i\}$ up to $n$ steps, one can define the occupation time as 
\begin{equation}
R_n = \sum_{i=1}^n \theta(\phi_i) \;,
\label{solv_occup.1}
\end{equation}
where $\theta(z)$ is the Heaviside step function, $\theta(z)=1$ for $z>0$ and $\theta(z)=0$ for $z<0$.
Thus, $R_n$ is a random variable that measures the number of steps (up to $n$) at which the sequence is positive. 
The probability distribution of $R_n$ can be computed exactly~\cite{MD2002} for the sequence 
$\{\phi_i\}$ in Eq. (\ref{solv_seq.1}) and 
it also turns out to be universal, i.e., independent of the noise distribution $\rho(\eta)$
for any $n$. For large $n$ and large $R$, but with the ratio $R/n$ fixed, this distribution
has the form
\begin{equation}
P(R,n)\equiv {\rm Prob.}[R_n=R] \sim \exp\left[-n \, \Psi(R/n)\right] \;,
\label{solv_ld.1}
\end{equation}
where $\Psi(r)$ is a large deviation function that can be computed exactly~\cite{MD2002}
\begin{equation}
\Psi(r)= \max_{0\le y\le 1} \left[ \ln \left( \frac{2\,y^r}{(1-y)}\, \tan^{-1}\left(\frac{1-y}{1+y}\right)\right)\right] \;.
\label{solv_ld.2}
\end{equation}

In summary, the sequence in Eq. (\ref{solv_seq.1}) serves as a rare solvable example of a non-Markovian, non-Gaussian
sequence for which several persistence and related quantities can be computed exactly. Moreover, quite remarkably,
all these properties associated with zero crossings are completely universal.

\section{Summary and conclusion}

In this review we have discussed the persistence properties of a fluctuating field $\phi(x,t)$ in
a variety of many body interacting nonequilibrium systems. This stochastic field $\phi(x,t)$ may represent 
the local spin at site $x$ in a spin model (Ising or Potts) undergoing phase ordering dynamics, the local density fluctuation
in a diffusing system, or the local height of a fluctuating interface. The dynamics in such many body interacting systems
were studied earlier principally by measuring the two-point space-time correlation functions.
However these systems typically have complex history dependence which is not adequately captured by
these two-point correlation functions. Persistence was the answer to the quest of a natural, simple and easily measurable quantity that
would capture the history dependence in such processes. Persistence $Q(t)$ in such systems is simply
the probability that the field $\phi(x,t)$, at a fixed point $x$ in space,  does not change sign 
(or more generally stays in one particular phase) within a time interval $[0,t]$. In many of these systems,
persistence decays algebraically at late times, $Q(t)\sim t^{-\theta}$ where $\theta$ is called the persistence exponent
which happens to be a {\em new} exponent not related to any other known exponents of the dynamics by a simple
scaling relation.
The persistence $Q(t)$ provides nontrivial information about the history dependence in such interacting out of equilibrium systems
and due to the relative ease in measuring this quantity, it has been studied extensively over the past $20$
years both theoretically and experimentally---this review tried to capture some of these developments.

\vskip 0.2cm

While the persistence $Q(t)$ can be measured relatively easily in simulations as well as in real experiments,
its computation is theoretically challenging. The reason for it can be traced back to the fact that due
to the spatial correlations present in the underlying many body system, the effective stochastic process
$\phi(x,t)$, at fixed $x$ but as a function of $t$, is generically a {\em non-Markovian} process. While persistence
and related first-passage probability had been well studied in both physics and mathematics literature before,
very few results were known for non-Markov processes. While $Q(t)$ is easy to compute
for Markov processes thanks to the Fokker-Planck formalism, its computation becomes highly
nontrivial whenever the process deviates from its Markovian nature. This was precisely
the main theoretical challenge behind the computation of $Q(t)$. 
While the latest developments
in the theoretical physics community did not fully succeed in computing $Q(t)$ for arbitrary non-Markov processes,
some new exact solutions for specific cases were found and also
several new approximation techniques were developed, reviewed at length here, that were crucial
in the theoretical understanding of the persistence probability
in several many body systems. The main emphasis of this review had been the discussion
of some of these theoretical developments over the last $20$ years.

\vskip 0.2cm

A major part of this review focused on a special type of stochastic process, namely the Gaussian stationary process (GSP).
There are two reasons for this. First, such processes appear naturally in the description
of many physical systems whose dynamics evolve via a linear equation, e.g., the diffusion equation and the stochastic growth of 
linear interfaces. The second reason is that in some cases the dynamics may not be linear, but
the physical observable of interest is a sum of many random variables
(such as the global magnetization in a spin system) and by virtue of the central limit theorem (as long as it holds),
the observable may be treated as a Gaussian variable. While the persistence probability of a GSP with an
arbitrary correlator still remains an outstanding unsolved problem, there have been major theoretical advances
over the last $20$ years in developing several approximation techniques for the persistence of such
GSP's. In this review we have described these approximation techniques. There have been
$4$ major techniques with different conditions for their validities:

\begin{itemize}

\item Perturbation theory for a non-Markov GSP around a Markov correlator reviewed in section \ref{section:perturbation}.

\item Independent Interval Approximation (IIA) valid for smooth GSP's reviewed in section \ref{subsection:iia}. This technique
provides, somewhat surprisingly, rather accurate estimates for the persistence exponent $\theta$ for
several smooth processes. However, it is not easy to see how to systematically improve the IIA estimate.

\item Persistence with partial survival, also valid for smooth processes, provides a systematic series
expansion for the persistence exponent $\theta$ and is reviewed in section \ref{subsection:partial}.

\item The correlator expansion method, reviewed in section \ref{section:discrete}, provides very good systematic estimates for $\theta$ for several GSP's.

\end{itemize}

\vskip 0.2cm

Another outcome
of these theoretical efforts was to ask new related questions which led to
the generalization of this basic persistence probability. This includes, for instance, the study of global persistence
at the critical point of spin systems (the probability that the total magnetization does not flip sign up to time $t$)
and the fact that the global persistence exponent $\theta_G$ is a new nonequilibrium
critical exponent. The other generalisations include the study of `walker' persistence, the
study of the occupation time distribution etc.---some of these generalizations have been discussed
in some detail in this review. In addition, persistence has also been studied in quenched disordered systems
and several new exact results have been derived. 
Furthermore, inspired by these developments, this basic quantity, i.e., persistence has now 
been studied in various other fields, going far outside the domain of condensed matter systems, such as
in finance, in geology, in ecology etc. Unfortunately, we could not review all these new
applications in the limited space of this review. Our main focus here had been mostly on physical systems.
However, it is important to make a note of the fact that
if a question is simple and natural, it often leads to interesting and
important developments across fields as had been the case for persistence. 
Indeed, we are very happy to note the recent surge of renewed interest in the mathematics community 
on the persistence problem (see for instance the recent review \cite{AS12}).  

\vskip 0.2cm
  
While we focused mostly on physical systems in this review, evidently we have not been able to cover
everything---many important developments in the recent past have been left out due primarily to the
lack of space. Let us briefly mention a few of them below. 

\vskip 0.2cm

An interesting related quantity, not discussed in this review, concerns the mean first-passage time
from a source point to a target point of
a particle undergoing diffusion or subdiffusion in a bounded domain of finite size~\cite{condamin2005,condamin2007}.
This problem is of interest in a number of situations such as in target search problems (e.g. animals searching for food), transport limited
chemical and biochemical reactions etc. The mean first-passage time in a bounded domain (which can be any scale invariant medium
such as a fractal~\cite{condamin2007} or even a complex network~\cite{NohRieger2004}) has been
studied considerably in the recent past and a host of exact and approximate analytical
results are available. It was also shown that first-passage time, even in
bounded domains, exhibits interesting sample-to-sample 
fluctuations~\cite{MOS2011,MMMO2012}. 

\vskip 0.2cm

The mean first-passage time has also played a central role in analysing search strategies.
Search problems are ubiquitous in nature, from a predator searching for a prey~\cite{Bell91} to
proteins searching for a site on a DNA molecule to bind~\cite{AD1968}. Depending on
the specific situation, search strategies can be modelled~\cite{Stone} in a variety of ways (see
also the special issue \cite{special_issue} for a number of articles devoted to this field). One very interesting
strategy is the so called {\em intermittent} strategy which is a combination of
slow moves (allowing detection of the target) and fast moves during which the searcher
relocates to a new area (for a review see \cite{benichou_review}). In all these
problems the mean first-passage time plays a crucial role in characterising the efficiency of the search strategy.
Recently, another interesting
search model, where the searcher diffuses and stochastically resets to its initial position, was introduced 
for which the mean first-passage time is exactly computable~\cite{EM2011,EM2011a,EMM2012,WEM2013}.

\vskip 0.2cm

Another interesting related subject that has seen a lot of renewed interests lately in the physics community
is the extreme value statistics (EVS) of correlated random variables or in a correlated time-series. In EVS,
one is typically interested in the probability distribution of the extremum (maximum or minimum) of
a stochastic process over the time interval $[0,t]$. While the persistence probability
discussed in this review concerns the {\em zero-crossing} properties of a process, the EVS
is closely related to the {\em level-crossing} problem of the process of a level of arbitrary height $H$.
This can be easily seen from the following observation. Consider a stochastic process $X(\tau)$
and let $M(t)= \max_{0\le \tau\le t}\left[ X(\tau) \right]$ denote the maximum of this process in
the time interval $[0,t]$. The cumulative distribution of the maximum,
$Q_H(t)= {\rm Prob.}\left[ M(t)\le H \right]$, is simply the probability that the process
stays below the level $H$ up to time $t$. Thus EVS is a natural generalisation
of the persistence problem to an arbitrary level. There have been considerable theoretical progress lately
in understanding the distribution of the maximum for strongly correlated processes and several
related quantities such as record statistics, order statistics etc. have been studied---but the discussion
of these recent developments in EVS is beyond the scope of this current review and perhaps, by itself,
is a subject of a separate future review.

\section*{Acknowledgements}

The work presented in this review was done over the past several years in collaboration with a large number of
people---we sincerely thank all of them. 
We thank M. Barma, E. Ben Naim,
R. Blythe, 
J.-P. Bouchaud, T. Burkhardt, A. Comtet, M. Constantin, S. J. Cornell, D. Das,
C. Dasgupta, S. Das Sarma, D. S. Dean, B. Derrida, A. Dhar, D. Dhar, I. Dornic, M. R. Evans, D. S. Fisher, J. Franke, A. Gambassi, 
C. Godr\`eche, M. Henkel, H. J. Hilhorst, D. A. Huse, H. Kallabis, P. L. Krapivsky, 
J. Krug, P. Le Doussal,
J.-M. Luck, K. Mallick, C. Monthus, T. J. Newman, G. Oshanin, A. Pargellis, R. Paul, R. Rajesh,  
S. Redner, A. Rosso, S. Sabhapandit, P. Sen,
C. Sire, J. Stavans, G. Wergen, K. J. Wiese, E. Williams, B. Yurke, A. Zoia and R. M. Ziff for collaborations and many fruitful discussions.
S. N. M. thanks, in particular, C. Sire for a long term collaboration on persistence. We would like also to thank M. Giesen and T. L. Einstein for allowing us to use their figures on fluctuating terraces in figure \ref{fig:fluctuating}.    
S. N. M. and G. S. would like to acknowledge support by ANR grant 2011-BS04-013-01 WALKMAT
and the Indo-French Centre for the Promotion of Advanced Research under Project 4604-3.

\end{document}